\documentclass[11pt,english,onecolumn]{IEEEtran}
\usepackage[T1]{fontenc}
\usepackage[latin9]{inputenc}
\usepackage{geometry}
\usepackage{mathrsfs}
\usepackage{enumitem}
\usepackage{amsmath}
\usepackage{amsthm}
\usepackage{amssymb}
\usepackage{graphicx}
\usepackage{setspace}
\PassOptionsToPackage{normalem}{ulem}
\usepackage{ulem}
\makeatletter
\theoremstyle{plain}
\newtheorem{thm}{\protect\theoremname}
\theoremstyle{definition}
\newtheorem{defn}[thm]{\protect\definitionname}
\theoremstyle{plain}
\newtheorem{prop}[thm]{\protect\propositionname}
\theoremstyle{remark}
\newtheorem{rem}[thm]{\protect\remarkname}
\theoremstyle{plain}
\newtheorem{cor}[thm]{\protect\corollaryname}
\theoremstyle{definition}
\newtheorem{example}[thm]{\protect\examplename}
\theoremstyle{plain}
\newtheorem{lem}[thm]{\protect\lemmaname}
\theoremstyle{plain}
\newtheorem{fact}[thm]{\protect\factname}
\theoremstyle{remark}
\newtheorem{claim}[thm]{\protect\claimname}
\newlist{casenv}{enumerate}{4}
\setlist[casenv]{leftmargin=*,align=left,widest={iiii}}
\setlist[casenv,1]{label={{\itshape\ \casename} \arabic*.},ref=\arabic*}
\setlist[casenv,2]{label={{\itshape\ \casename} \roman*.},ref=\roman*}
\setlist[casenv,3]{label={{\itshape\ \casename\ \alph*.}},ref=\alph*}
\setlist[casenv,4]{label={{\itshape\ \casename} \arabic*.},ref=\arabic*}

\usepackage{hyperref}
\usepackage{enumitem}
\usepackage{breqn}
\usepackage{bbm} 
\usepackage{cite}
\setlist[enumerate]{leftmargin=*,wide} 

\DeclareMathOperator*{\argmax}{arg\,max} \DeclareMathOperator*{\argmin}{arg\,min}
\DeclareMathOperator{\supp}{supp}

\allowdisplaybreaks

\global\long\def\s[#1]{\textnormal{\scriptsize #1}}
\global\long\def\st[#1]{\textnormal{\tiny #1}}

\global\long\def\pe{\mathsf{pe}}

\global\long\def\P{\mathbb{P}}
\global\long\def\E{\mathbb{E}}
\global\long\def\V{\mathbb{V}}

\global\long\def\I{\mathbbm{1}}
\global\long\def\m[#1]{\boldsymbol{#1}} 

\global\long\def\r[#1]{#1}

\global\long\def\dfn{:=}

\global\long\def\trre[#1,#2]{\overset{{\scriptstyle (#2)}}{#1}} 

\author{
\IEEEauthorblockN{Nir Weinberger and Neri Merhav}

\IEEEauthorblockA{The Viterbi Faculty of Electrical and Computer Engineering\\
  	    Technion - Israel Institute of Technology\\
Technion City, Haifa 3200004, Israel
} \\
\IEEEauthorblockA{\{nirwein@, merhav@ee\}.technion.ac.il}\\
}

\makeatother

\usepackage{babel}
\providecommand{\casename}{Case}
\providecommand{\claimname}{Claim}
\providecommand{\corollaryname}{Corollary}
\providecommand{\definitionname}{Definition}
\providecommand{\examplename}{Example}
\providecommand{\factname}{Fact}
\providecommand{\lemmaname}{Lemma}
\providecommand{\propositionname}{Proposition}
\providecommand{\remarkname}{Remark}
\providecommand{\theoremname}{Theorem}

\begin{document}
\title{The DNA Storage Channel: Capacity and Error Probability Bounds\thanks{The research of N. Merhav was partly supported by the Israel Science
Foundation (ISF), grant no. 137/18.}}

\maketitle
\renewcommand\[{\begin{equation}}
\renewcommand\]{\end{equation}}
\thispagestyle{empty}
\vspace{-1cm}
\begin{abstract}
The DNA storage channel is considered, in which the $M$ Deoxyribonucleic
acid (DNA) molecules comprising each codeword are stored without order,
sampled $N$ times with replacement, and then sequenced over a discrete
memoryless channel. For a constant coverage depth $M/N$ and molecule
length scaling $\Theta(\log M)$, lower (achievability) and upper
(converse) bounds on the capacity of the channel, as well as a lower
(achievability) bound on the reliability function of the channel are
provided. Both the lower and upper bounds on the capacity generalize
a bound which was previously known to hold only for the binary symmetric
sequencing channel, and only under certain restrictions on the molecule
length scaling and the crossover probability parameters. When specified
to binary symmetric sequencing channel, these restrictions are completely
removed for the lower bound and are significantly relaxed for the
upper bound in the high-noise regime. The lower bound on the reliability
function is achieved under a universal decoder, and reveals that the
dominant error event is that of \emph{outage} -- the event in which
the capacity of the channel induced by the DNA molecule sampling operation
does not support the target rate.
\end{abstract}


\tableofcontents{}

\section{Introduction}

In this paper, we establish new lower and upper bounds on the capacity
and error probability of coded-storage systems based on Deoxyribonucleic
acid (DNA) molecules medium, where in some regime of the system parameters,
those bounds coincide and provide the capacity of the channel. In
principle, since each DNA molecule is comprised of two complementary
strands of four nucleotides (Adenine, Cytosine, Guanine, and Thymine),
it is equivalent to a sequence of a four-letter alphabet sequence
for the purpose of encoding information. However, as explained in
\cite{heckel2019characterization}, current technology is only capable
of synthesizing relatively short strands of DNA -- sequences of one
or two hundred nucleotides. Thus, in practical systems, the stored
information is comprised of a large number of DNA molecules which
are stored in a pool and cannot be spatially ordered. Hence, unlike
ordinary channel coding, in which a codeword is a single sequence
of symbols, the DNA codeword is an unordered multiset of short sequences
of symbols. 

DNA-based storage systems are prone to various of impairments. After
synthesizing molecules based on the encoded data, individual molecules
are either duplicated or completely erased, in a process called Polymerase
Chain Reaction (PCR) amplification. Later on, when the information
is read, molecules are sampled from the amplified pool in a random
manner, without the possibility to choose a specific molecule. Each
chosen molecule is then sequenced to obtain the four-letter alphabet
sequence that it encoded. 

In accordance, DNA storage systems suffer from impairments both on
a molecule level and on a symbol level: In the former, this amounts
to duplications or erasures, resulting from the synthesis and the
amplification or the sampling process. In the latter, this results
in either substitutions, deletions or insertions, resulting from the
synthesis and sequencing operations. In \cite{heckel2019characterization},
a detailed characterization of the impairments of DNA storage systems
was provided, and the parameters were estimated based on data sets
obtained from experimental systems. The conclusion of this survey
reinforced the important role of error-correcting codes in DNA storage
systems, and thus naturally raised the question of fundamental limits
on the rates of such systems, most importantly, their \emph{capacity}. 

In order to characterize the capacity of the DNA storage channel,
Shomorony and Heckel \cite{shomorony2021dna} have proposed a distilled
mathematical model, which both captures the major impairments of the
DNA storage channel, on one hand, and which is reasonably tractable
to analyze, on the other hand. The encoded message is synthesized
to a pool of $M$ molecules, each of length $L$ symbols. Each symbol
is chosen from a given finite alphabet, where an alphabet of size
four is the natural choice. When the message is read, each of the
molecules is sampled a random number of times, and then sequenced,
to obtain the multiset of output molecules. The sequencing operation
is modeled as a discrete memoryless channel (DMC) operating on the
sampled molecules. The decoder decides on the message based on the
sequencing of the sampled molecules. Hence, the non-standard aspects
of this model are the loss of order information of the DNA molecules,
and their possible duplication or erasure by the sampling mechanism.
We mention that this basic model does not include deletions and insertions
during the sequencing operation of a given molecule. The (storage)
capacity $C$ is then defined as the ratio between the number of messages
that can be reliably stored to the total of $ML$ symbols, and it
was studied for a few specific variants of the basic model. 

\subsection{Known Results}

In \cite{shomorony2021dna}, two settings were considered. First,
the model in which the number of samples of each molecule is i.i.d.
according to some given distribution, and the sequencing is perfect
(noiseless). This basic model reveals that the length of the molecule
must scale as $L=\beta\log M$ for some $\beta\geq1$, and then capacity
is $C=(1-\P[\text{erasure}])(1-1/\beta)$ \cite[Theorem 1]{shomorony2021dna},
where $\P[\text{erasure}]$ is the probability that a molecule is
not sampled at all. Compared to a standard erasure channel whose capacity
is $C=1-\P[\text{erasure}]$, the multiplying term $(1-1/\beta)$
can be attributed to the loss of order of the molecules. Second, a
similar model was considered, but with Bernoulli sampling model (each
molecule is sampled with some fixed probability), and with a noisy
memoryless sequencing channel, which is assumed to be a binary symmetric
channel (BSC) with crossover probability $w$ (hence, over input-output
alphabets of size $2$), and for which the rate $(1-\P[\text{erasure}])(1-h_{b}(w)-1/\beta)$
is achievable,\footnote{We mainly use standard notation in the introduction. See Sec. \ref{subsec:Notation-conventions}
for notation conventions. Here $h_{b}(w)$ is the binary entropy function.} and is known to be the capacity only in the regime $w<1/4$ and $\beta>\frac{2}{\log2-h_{b}(2w)}$
(in which a converse holds too). In \cite{lenz2020achievable,lenz2019upper},
the same BSC sequencing channel was assumed, yet with a different
molecule sampling model. In \cite{lenz2020achievable,lenz2019upper},
it is assumed that the $M$ molecules are sampled exactly $N$ times,
uniformly with replacement. In accordance, the samples distribution
of the molecules is multinomial, and the number of times each molecule
is sampled depends on other molecules. When $N=\alpha M$ with a constant
\emph{coverage depth parameter} $\alpha$, and $M$ is asymptotically
large, the empirical count of the number of times each molecule is
sampled is known to tend to a Poisson distribution (an effect called
\emph{Poissonization}). Furthermore, if a molecule is sampled and
sequenced $d$ times, then it effectively undergoes a channel whose
output is a set of $d$ independent observations of its input. The
capacity of this \emph{binomial (multi-draw) }BSC, say $C_{w,d}$,
has a simple closed-form expression \cite{mitzenmacher2006theory}
(more generally, the operation of such channels is termed \emph{information
combining} \cite{sutskover2005extremes,land2005bounds,land2006information}).
In accordance, an upper bound on the capacity (converse) of this DNA
storage channel model that is valid for the regime $w<1/8$ and $\beta>\frac{2}{\log2-h_{b}(4w)}$
was obtained in \cite{lenz2019upper} as 
\begin{equation}
C\geq\sum_{d\in\mathbb{N}}\pi_{\alpha}(d)\cdot C_{w,d}-\frac{1}{\beta}(1-\pi_{\alpha}(0)),\label{eq: capacity DNA Lenz introduction}
\end{equation}
where $\pi_{\alpha}(d)$ is the Poisson probability mass function
(p.m.f.) with parameter $\alpha$. The same expression was found to
be a lower bound on the capacity (achievable), in the same regime
of $(w,\beta)$. 

For general DMCs beyond the BSC, it was described in \cite[Sec. V.A, Thm. 3]{shomorony2021dna},
that the capacity of the noiseless sequencing model can be generalized
to any symmetric DMC (in the sense of \cite[Ch. 7.2]{cover2012elements}),
yet only when $\beta$ is large enough, without specifying how large
$\beta$ should be. There are no claims in \cite{shomorony2021dna}
on general DMCs, and the difficulty of extending the arguments to
that case is explained to stem from the difficulty of bounding entropies
under \emph{general} capacity-achieving input distributions. In \cite{lenz2020achievable,lenz2019upper}
it is mentioned in passing that the results can be generalized to
modulo-additive channels (for which the Hamming distance is an appropriate
measure of similarity).

\subsection{Contributions}

In this paper, we follow the sampling model of \cite{lenz2020achievable,lenz2019upper}
of multinomial sampling with fixed coverage depth $\alpha$, and improve
on the results of \cite{shomorony2021dna,lenz2020achievable,lenz2019upper}
from several aspects:
\begin{itemize}
\item Our results apply to any DMC sequencing channel, including \emph{asymmetric}
ones. 
\item When specified to BSCs, they significantly extend the parameter regime
in which the exact capacity is known, and are the tightest known in
the complementary regimes. 
\item We provide a single-letter lower bound on the reliability function
of the system, which is based on a universal decoding rule. 
\end{itemize}

\paragraph*{Lower bounds (achievability)}

We derive a single-letter capacity lower bound, which naturally generalizes
(\ref{eq: capacity DNA Lenz introduction}) to general DMCs, and prove
that it is a lower bound on the capacity of the DNA storage channel,\emph{
without any restrictions} on $\beta$ or the quality of the sequencing
channel (Theorem \ref{thm: achievable capacity}). En route to the
capacity lower bound, we analyze the error probability of a random
code, and propose a suitable decoder. This decoder is \emph{universal}
-- that is, its decoding rule does not depend on the transition probabilities
of the DMC sequencing channel -- and it can be thought of as a penalized
version of the \emph{maximum mutual information} (MMI) decoder \cite{goppa1975nonprobabilistic,csiszar1977new}.
The analysis of its error probability reveals that the dominating
error event is not an atypical error event in the sequencing procedure,
but rather an atypical sampling event. The sampling empirical distribution,
to wit, the relative fraction of molecules which have been sampled
$d$ times, for $d\in\{0,1,\ldots,N\}$, can be considered as a random
\emph{state} of the DNA storage channel. When sampling is ``good'',
the instantaneous capacity supported by the channel exceeds the target
communication rate, and as we show, the conditional error probability
decays exponentially fast as $e^{-\Theta(M\log M)}$. However, the
probability that the sampling state is ``bad'' and the instantaneous
capacity of the channel does not support the rate, decays slower,
as $e^{-\Theta(M)}$, and thus dominates the error probability. The
latter event is similar to an \emph{outage }event in wireless communication
systems \cite{tse2005fundamentals}, in which various phenomena such
as multipath fading and interference cause the instantaneous capacity
to drop below the required rate, and hence to a high error probability,
while errors due to the additive Gaussian noise are less frequent.
To quantify this effect, we prove a single-letter upper bound on the
exponential decay of the outage probability (Theorem \ref{thm: achievable reliability function}).
Our proof methods are significantly different from the one in \cite{shomorony2021dna},
which relied on explicit molecule indexing, as well as the one in
\cite{lenz2020achieving}, which is based on clustering the $N$ outputs
to (less than $M$) clusters, each pertaining to a different codeword. 

\paragraph*{Upper bounds (converse)}

We derive a single-letter expression similar to that of the lower
bound, and prove it to be an upper bound, but with an additional excess-rate
term (Theorem \ref{thm: converse capacity}). The bound is valid for
any $\beta>1$, and there exists a critical value of $\beta$ such
that the excess-rate term vanishes for all $\beta$ larger than the
critical. In that case, the upper bound on the capacity matches the
lower bound. To prove this result, we follow the principal idea of
\cite{shomorony2021dna} which was later elaborated in \cite{lenz2019upper}.
The core of the argument is that if the codebook is such that the
$M$ molecules comprising each codeword are very similar (or even
identical), then their order is immaterial, and the loss term in rate
associated with the ordering information, namely $\frac{1}{\beta}(1-\pi_{\alpha}(0))$
in (\ref{eq: capacity DNA Lenz introduction}), is eliminated. On
the other hand, for standard channels, which preserve the order of
the symbols, codewords which maximize mutual information should have
independent molecules, not identical. Thus, optimal codebooks should
balance between these two conflicting requirements. The converse argument
of \cite{shomorony2021dna,lenz2019upper} shows that under the specified
conditions on $\beta$ and $w$ (the crossover probability of the
BSC sequencing channel), asymptotically optimal codewords should have
independent molecules. In \cite{shomorony2021dna,lenz2019upper},
the similar-vs.-independent-molecules trade-off is concretely quantified
by the Hamming distance. There, molecules are essentially ``similar''
if the Hamming distance between them is less than $4wL$, and otherwise
``far'' (and thus effectively independent). Thus the distinction
between far and similar molecules is according to a radius which scales
linearly in the molecule length $L$. By contrast, our proof utilizes
a more general distance function, originating from exponential probabilities
of typical sets \cite[Ch. 2]{csiszar2011information}. More importantly,
the radius which distinguishes between similar and far is \emph{sub-linear
}in the molecule length $L$. This enables the aforementioned improvement
of the upper (converse) bound. 

It should also be mentioned that input alphabet size and the possible
asymmetry of the sequencing channel greatly complicates the proof.
In the analysis of BSC sequencing channels, the extermal property
of the i.i.d. uniform $(\frac{1}{2},\frac{1}{2})$ distribution (both
separately for each molecules, and over all molecules) is typically
easily justified. This is not the case for general DMCs. For example,
while the binomial (multi-draw) extension of the BSC is symmetric
for any $d$, this is not true for general symmetric channels (see
Remark \ref{rem:symmetry is not preserved} and Appendix \ref{subsec:The-binomial-extension of symmetric DMC is symmetric}
for details). Moreover, in the standard Fano-based proof of the converse
to capacity of DMCs \cite[Lemma 7.9.2]{cover2012elements}, the output
entropy for blocklength $K>1$ is easily upper bounded by the sum
of marginal entropies, and then the sum of $K$ single-letter mutual
information of each of them is upper bounded by the one obtained by
the capacity achieving input distribution. For the DNA storage channel,
it is not even clear a priori that the $M$ molecules should be \emph{identically}
distributed (even if it is assumed that they are independent, or if
it is somehow proved that no optimality is lost by independence).
In our Fano-based converse argument, this requires to analyze mutual
information for length $L$ vectors, rather than their scalar counterpart
in standard channel coding. See \cite[Sec. V.A]{shomorony2021dna}
for a related discussion. 

\paragraph*{Tightness of capacity bounds for modulo-additive noise channels}

We evaluate our upper and lower bounds on the capacity for modulo-additive
noise channels, which generalize the BSC channel for alphabets larger
than $2$. We provide an explicit sufficient condition on the minimal
$\beta$ required for the lower and upper bound to match. When specified
for the BSC channel, the result displays a significant gain compared
to \cite{lenz2020achievable,lenz2019upper}. For example, for $\mathsf{BSC}(w)$
with $w=0.05$, the minimal $\beta$ required in \cite{lenz2020achievable,lenz2019upper}
is twice as large compared to our condition (see Fig. \ref{fig:Comparison-between-beta for BSC}). 

\paragraph*{Significance}

To begin with, the importance of general DMCs follows from the trivial
fact that the physical DNA storage channel has four-letter input alphabet
$\{\mathsf{A},\mathsf{C},\mathsf{G},\mathsf{T}\}$, and at least $4$
letters in the output alphabet. As we show (Prop. \ref{prop: achievability by a uniform distribution}),
symmetry of the sequencing channel leads to a suitable symmetry in
the DNA storage channel if the output alphabet has size less or equal
to $4$. This is, however, not not true for larger output alphabets
(see a counterexample in Appendix \ref{subsec:The-binomial-extension of symmetric DMC is symmetric}).
Furthermore, the DNA storage channel is also known to be asymmetric
in its nature -- e.g., \cite{gabrys2015asymmetric} states that ``\emph{$\{\mathsf{A},\mathsf{T}\}$
are very likely to be mutually confused during sequencing, while the
bases $\{\mathsf{G},\mathsf{C}\}$ are much less likely to be misinterpreted
for each other.}''. As said, our bounds pertain to any DMC, which
can be asymmetric. 

Next, the importance of improved capacity bounds cannot be overstated,
since the restrictions under which the capacity bounds hold can be
described as \emph{low-error }synthesis/sequencing. As discussed in
\cite{shomorony2021dna}, it is envisioned that next-generation DNA
storage systems will deploy \emph{high-error }synthesis/sequencing
\cite{antkowiak2020low} in order to reduce costs. Thus, it is of
interest to remove, or at least ameliorate, the conditions under which
capacity is known (e.g., $\beta>2/(\log2-h_{b}(4w))$ in the BSC case).
As we next describe, our results remove completely such conditions
for the lower (achievable) bound on capacity, and significantly improve
the qualifying condition for the converse (upper) bound.

Finally, the fact that the sampling mechanism is the dominant error
event, compared to sequencing errors should guide future designs of
coded DNA storage systems. 

\subsection{Other Related Work}

It was recognized long ago \cite{neiman1964some} that DNA molecules
can serve as a medium to data storage, akin to their role in living
organisms as carriers of genetic instructions. As surveyed in \cite{heckel2019characterization},
prototypes of this concept have recently been developed by various
groups of researches, starting with \cite{church2012next,goldman2013towards},
and followed by \cite{grass2015robust} which have deployed error-correcting
codes, \cite{yazdi2015rewritable} which have demonstrated selective
file access, and \cite{organick2018random} which have practically
demonstrated the ability to store over $200$ megabytes of data. Since
DNA storage systems exhibit extreme high density, long durability
\cite{bornholt2016dna}, and low energy consumption \cite{church2012next},
they are competitive candidates for future storage systems. 

Several other papers have studied variations of the DNA storage channel
model \cite{sayir2016codes,erlich2017dna}, and various papers have
proposed and analyzed coding schemes \cite{church2012next,goldman2013towards,grass2015robust,yazdi2015rewritable,kiah2016codes,erlich2017dna,sala2017exact,organick2018random,lenz2019anchor,sima2021coding,tang2021error}.
We refer the reader to \cite[Sec. I.B]{shomorony2021dna} for a short
description. In parallel to the study of fundamental limits of probabilistic
channel models, combinatorial channel models were also studied. In
\cite{kovavcevic2018codes}, a channel model was considered in which,
as for the DNA storage channel, codewords are multisets of unordered
symbols. The model, however, is based on worst-case (adversarial)
insertions, deletions and substitutions errors, and does not explicitly
take into account the probabilistic nature of the sampling and sequencing
mechanism of the DNA channel. In accordance, upper bounds on the cardinality
of optimal codes correcting any given number of errors were derived,
and were asymptotically evaluated in the regime in which the alphabet
size of the molecules is linear with $M$ (this is a slightly different
scaling than what is considered for DNA storage channels).In \cite{sima2021coding}
the redundancy required to be added in order to guarantee full protection
against substitution errors was upper bounded. In \cite{song2020sequence}
a sequence-subset distance has been proposed as a generalization of
the Hamming distance suitable for the analysis of DNA storage channels,
and generalizations of Plotkin and Singleton upper bounds on the maximal
size of the codes were derived. In \cite{lenz2019coding}, Gilbert-Varshamov
lower bounds and sphere packing upper bounds on the achievable cardinality
of DNA storage codes were derived. These bounds complement the Shannon-theoretic
analysis studied in \cite{shomorony2021dna,lenz2019upper,lenz2020achieving}
and in this paper. 

\subsection{Outline}

The rest of the paper is organized as follows. In Sec. \ref{sec:Problem-formulation}
we establish notation conventions and formulate the problem. In Sec.
\ref{sec:Achievability} we present our lower bounds (achievability
results) on the capacity and reliability function, and in Sec. \ref{sec:Converse}
our upper bound on the capacity (converse results). In Sec. \ref{sec:Modulo-additive-sequencing}
we specify our capacity bounds to the case of modulo-additive sequencing
channels (which the BSC is a specific case), and in Sec. \ref{sec:A-numerical-example}
we demonstrate our results via a numerical example. In Sec. \ref{sec:Summary-and-open}
we summarize the paper and discuss open problems. 

\section{Problem Formulation \label{sec:Problem-formulation}}

\subsection{Notation Conventions\label{subsec:Notation-conventions}}

\paragraph*{Random variables and vectors}

Random variables will be denoted by capital letters, specific values
they may take will be denoted by the corresponding lower case letters,
and their alphabets will be denoted by calligraphic letters. Random
vectors and their realizations will be super-scripted by their dimension.
For example, the random vector $A^{K}=(A_{0},\ldots,A_{K-1})\in{\cal A}^{K}$
(where $K\in\mathbb{N}^{+}$), may take a specific vector value $a^{K}=(a_{0},\ldots,a_{K-1})\in{\cal A}^{K}$,
the $K$th order Cartesian power of ${\cal A}$, which is the alphabet
of each component of this vector. The Cartesian product of ${\cal A}$
and ${\cal B}$ (both finite alphabets) will be denoted by ${\cal A}\times{\cal B}$.
The concatenation of two vectors, possibly of different lengths, will
be denote by their juxtaposition, and the superscript of the resulting
vector will be one of two forms -- either by its total dimension,
or by the number of vectors comprising it. An asterisk will be used
in case each of these constituent vectors has a different length.
For example, if $a_{0}^{K_{0}}\in{\cal A}^{K_{0}}$ and $a_{1}^{K_{1}}\in{\cal A}^{K_{1}}$
then their concatenation will be denoted as either $a^{K_{0}+K_{1}}$
or $a^{*2}$. The probability of the event ${\cal {\cal E}}$ will
be denoted by $\P({\cal {\cal E}})$, and its indicator function will
be denoted by $\I({\cal E})$. The expectation operator with respect
to (w.r.t.) a given distribution $P$ will be denoted by $\E_{P}[\cdot]$
where the subscript $P$ will be omitted if the underlying probability
distribution is clear from the context. The empirical count operator
$\mathscr{P}\colon{\cal A}^{K}\to[K+1]^{|{\cal A}|}$ will be defined
as the operator which converts a vector $a^{K}\in{\cal A}^{K}$ to
its empirical count vector $n^{|{\cal A}|}=\mathscr{N}(a^{K})$ so
that for any $a\in{\cal A}$ 
\begin{equation}
n_{a}=\sum_{k\in[K]}\I\{A_{k}=a\},\label{eq: empirical count operator}
\end{equation}
where $[K]\dfn\{0,1,\ldots,K-1\}$. The empirical distribution operator
$\mathscr{P}\colon{\cal A}^{K}\to\frac{1}{K}[K+1]^{|{\cal A}|}$ will
be a normalized version of the empirical count operator, and will
be defined via $\mathscr{N}=\frac{1}{K}\cdot\mathscr{P}$, so that
$p^{|{\cal A}|}=\mathscr{P}(a^{K})$ is such that
\begin{equation}
p_{a}=\frac{1}{K}\sum_{k\in[K]}\I\{A_{k}=a\}=\frac{n_{a}}{K}.\label{eq: empirical distribution operator}
\end{equation}
The composition of two operators $\mathscr{P}_{1}$ and $\mathscr{P}_{2}$
will be denoted by $\mathscr{P}_{1}\circ\mathscr{P}_{2}$, and the
$K$th functional power of $\mathscr{P}$ will be defined via the
recursion $\mathscr{P}^{(k)}\dfn\mathscr{P}\circ\mathscr{P}^{(k-1)}$
with $\mathscr{P}^{(1)}\dfn\mathscr{P}$. 

\paragraph*{Probability distributions, types and typical sets}

We will follow the standard notation conventions for probability distributions,
e.g., $P_{A}(a)$ will denote the probability of the letter $a\in{\cal A}$
under the distribution $P_{A}$ of the random variable $A$. The arguments
will be omitted when we address the entire distribution, e.g., $P_{A}$.
Similarly, joint and conditional distributions of $(A,B)$ will be
denoted by $P_{AB}$ and $P_{B\mid A}$ (respectively). The product
distribution of $P_{A}$ and $P_{B\mid A}$ will be denoted by $P_{A}\times P_{B\mid A}$.
The support of a distribution $P_{A}$ will be denoted by $\supp(P_{A})$.
The set of possible distributions supported on ${\cal A}$ (probability
simplex) will be denoted by ${\cal P}({\cal A})$. The set of conditional
distributions on ${\cal B}$ conditioned on elements of ${\cal A}$
(probability transition matrices) will be denoted by ${\cal P}({\cal B}\mid{\cal A})$.
In what follows, we will extensively utilize the \emph{method of types
}\cite{csiszar2011information,csiszar1998method} and the following
notations. The \emph{type class} of $P_{A}$ at blocklength $K$,
i.e., the set of all $a^{K}\in{\cal A}^{K}$ for which $\mathscr{P}(a^{K})=P_{A}$
will be denoted by ${\cal T}_{K}(P_{A})$. The set of all type classes
of vectors of length $K$ from ${\cal A}^{K}$ will be denoted by
${\cal P}_{K}({\cal A})$ which is a subset of ${\cal P}({\cal A})$.
The $V$-shell (conditional type class) of $a^{K}$ under the DMC
$V\colon{\cal A}\to{\cal B}$, i.e., the set of all $b^{K}\in{\cal B}^{K}$
for which $\{b^{K}\in{\cal B}^{K}\colon\mathscr{P}(a^{K},b^{K})=\mathscr{P}(a^{K})\times V\}$
will be denoted by ${\cal T}_{K}(V\mid a^{K})$. For a given $P_{A}\in{\cal P}_{K}({\cal A})$,
the set of $V$-shells such that ${\cal T}_{K}(V\mid a^{K})$ is not
empty when $a^{K}\in{\cal T}_{K}(P_{A})$ will be denoted by ${\cal P}_{K}({\cal B}\mid P_{A})$.
The notion of typical sets, in the form of \cite{csiszar2011information},
will be used in the proof of the converse. Definitions, notations
and basic results will thus appear before its proof, in Appendix \ref{sec:A-brief-review}. 

\paragraph*{Information measures and probability divergences}

Logarithms and exponents will be understood to be taken to the natural
base. The binary entropy function $h_{b}\colon[0,1]\to[0,1]$ will
be denoted by $h_{b}(a)\dfn-a\log a-(1-a)\log(1-a)$ and the binary
Kullback--Leibler (KL) divergence $d_{b}\colon[0,1]\times(0,1)\to\mathbb{R}^{+}$
by $d_{b}(a||b)\dfn a\log\frac{a}{b}+(1-a)\log\frac{(1-a)}{(1-b)}.$
In general, information-theoretic quantities will be denoted by the
standard notation \cite{cover2012elements}, with subscript indicating
the distribution of the relevant random variables, e.g. $H_{P}(A\mid B),I_{P}(A;B)$
and $I_{P}(A;B\mid C)$, for the random variables $A,B,C$. Alternatively,
the entropy of a distribution $P_{A}\in{\cal P}({\cal A})$ will be
denoted by $H(P_{A})$, and the mutual information for a DMC $V$
with input distribution $P_{A}$ will be denoted by $I(P_{A},V)$.
The KL divergence between $Q_{A}\in{\cal P}({\cal A})$ and $P_{A}\in{\cal P}({\cal A})$
will be denoted by $D(Q_{A}\mid\mid P_{A})$, and the conditional
KL divergence between $Q_{B\mid A}\in{\cal P}({\cal B}\mid{\cal A})$
and $P_{B\mid A}\in{\cal P}({\cal B}\mid{\cal A})$ averaged over
$Q_{A}\in{\cal P}({\cal A})$ will be denoted by $D(Q_{B\mid A}\mid\mid P_{B\mid A}\mid Q_{A})$.
The total variation distance (${\cal L}_{1}$ norm) of $P_{1},P_{2}\in{\cal P}({\cal A})$
will be denoted by $|P_{1}-P_{2}|\dfn\sum_{a\in{\cal A}}|P_{1}(a)-P_{2}(a)|$. 

\paragraph*{General}

The complement of a multiset ${\cal A}$ will be denoted by ${\cal A}^{c}$.
The number of \emph{distinct} elements of a finite multiset ${\cal A}$
will be denoted by $|{\cal A}|$. The equivalence relation will be
denoted by $\equiv$, and will mainly be used to simplify notation
at some parts of the paper (typically, the removal of subscripts/superscripts
in order to avoid cumbersome notation). Asymptotic Bachmann--Landau
notation will be used. Specifically, for a pair of positive sequences
$\{f_{K}\}_{K\in\mathbb{N}},\{g_{K}\}_{K\in\mathbb{N}}$ $f_{K}=O(g_{K})\Leftrightarrow\limsup_{K\to\infty}\frac{f_{K}}{g_{K}}<\infty$,
and $f_{K}=\Theta(g_{K})\Leftrightarrow\{f_{K}=O(g_{K})\text{ and }g_{K}=O(f_{K})\}$,
$f_{K}=o(g_{K})\Leftrightarrow\lim_{K\to\infty}\frac{|f_{K}|}{g_{K}}=0$,
and $f_{K}=\omega(g_{K})\Leftrightarrow\lim_{K\to\infty}\frac{|f_{K}|}{g_{K}}=\infty$.
A variable appearing in the subscript of an asymptotic order term,
e.g., $\alpha$ in $f_{K}=O_{\alpha}(K)$, emphasizes that the constants
involved in the asymptotic relation depend on that variable (possibly,
in addition to other variables). Minimum and maximum will be denoted
as $\min(a,b)\dfn a\wedge b$, $\max(a,b)\dfn a\vee b$, and $a\vee0$
will be denoted by $[a]_{+}$. Throughout, for the sake of brevity,
integer constraints on large numbers which are inconsequential will
be ignored, for example, the number of codewords in a rate $R$ codebook
of dimension $K$ will be simply written as $e^{KR}$ (instead of
$\lceil e^{KR}\rceil$). The Hamming distance between $a^{K},\overline{a}^{K}\in{\cal A}^{K}$
will be denoted by $\rho_{\text{H}}(a^{K},\overline{a}^{K})\dfn\sum_{k\in[K]}\I\{a_{k}\neq\overline{a}_{k}\}$. 

\subsection{Formulation of the DNA Storage Channel \label{subsec:Formulation-of-the}}

In this section, we formulate the DNA storage channel model, or DNA
channel, in short. This channel will be indexed by the number of molecules
$M$ in a codeword, which will be used in what follows to gauge the
dimension of the codewords.

\paragraph*{The encoder}

A DNA molecule is a sequence of $L\equiv L_{M}\in\mathbb{N}^{+}$
nucleotides (symbols) chosen from an alphabet ${\cal X}$, where in
physical DNA storage systems, ${\cal X}=\{\mathsf{A},\mathsf{C},\mathsf{G},\mathsf{T}\}$.
Thus, each molecule is uniquely represented by a sequence $x^{L}\in{\cal X}^{L}$.
A codeword is a sequence of $M$ molecules, $x^{LM}=(x_{0}^{L},\ldots x_{M-1}^{L})$,
where $x_{m}^{L}\in{\cal X}^{L}$ for all $m\in[M]$.\footnote{In principle, the codeword is actually a \emph{multiset }of $M$ molecules,
that is, the order is not specified. However, for analysis, it is
convenient to assume an arbitrary ordering of the molecules. As evident
from the description of the amplification step, this order does not
affect the channel output. In addition, we use the notation $x^{LM}\equiv(x^{L})^{M}$
rather than the equivalent $x^{ML}$ as a mnemonic to the fact that
the codeword is a sequence of $M$ length-$L$ molecules (which the
channel permutes).} Thus a codeword has total $ML$ nucleotides (or symbols) from ${\cal X}$.
A codebook is a set of different codewords, ${\cal C}=\{x^{LM}(j)\}$. 

\paragraph*{The channel model (reading mechanism)}

At the time of reading, the codeword $x^{LM}(j)$ undergoes two stages
which can be considered as a channel operation that produces the output
for the decision on the stored message. The DNA channel is parameterized
by the number of molecule samples $N\equiv N_{M}\in\mathbb{N}^{+}$,
and a sequencing channel $W\colon{\cal X}\to{\cal Y}$. The channel
operates on $x^{LM}(j)$ as follows:
\begin{enumerate}
\item Sampling: $N$ molecules are sampled uniformly from the $M$ molecules
of $x^{LM}(j)$, independently, with replacement. Let $U^{N}\in[M]^{N}$
be such that $U_{n}$ is the sampled molecule at sampling event $n\in[N]$.
We refer to $U^{N}$ as the \emph{molecule index vector}, and it holds
that $U^{N}\sim\text{Uniform}([M]^{N})$. The result of the sampling
stage is thus the vector
\[
(x_{U_{0}}^{L}(j),x_{U_{1}}^{L}(j),\ldots,x_{U_{N-1}}^{L}(j))\in({\cal X}^{L})^{N}.
\]
In what follows, we will use the following additional definitions.
Let $S^{M}\in[N]^{M}$ be such that $S_{m}$ is the number of times
that molecule $m$ was sampled, to wit $S_{m}=\sum_{n\in[N]}\I\{U_{n}=m\}$,
the empirical count of $U^{N}$. It holds that $S^{M}\sim\text{Multinomial}(N;(\frac{1}{M},\frac{1}{M},\ldots\frac{1}{M}))$,
and we refer to $S^{M}$ as the \emph{molecule duplicate vector}.
In a similar fashion, let $Q^{N+1}\in[M+1]^{N+1}$ be such that $Q_{d}$
is the number of molecules that have been sampled $d$ times, to wit
$Q_{d}=\sum_{m\in[M]}\I\{S_{m}=d\}$, the empirical count of $S^{M}$.
We refer to $Q^{N+1}$ as the \emph{amplification vector.} Note that
$\sum_{d\in[N+1]}Q_{d}=M$ and $\sum_{d\in[N]}dQ_{d}=N$ hold with
probability $1$. Also, using the definition of the empirical count
operator in (\ref{eq: empirical count operator}), it holds that $S^{M}=\mathscr{N}(U^{N})$
and $Q^{N+1}=\mathscr{\mathscr{N}}(S^{M})=\mathscr{\mathscr{N}}^{(2)}(U^{N})$.
See Fig. \ref{fig: sampling event} for an illustration. 
\begin{figure}
\begin{centering}
\includegraphics[scale=1.5]{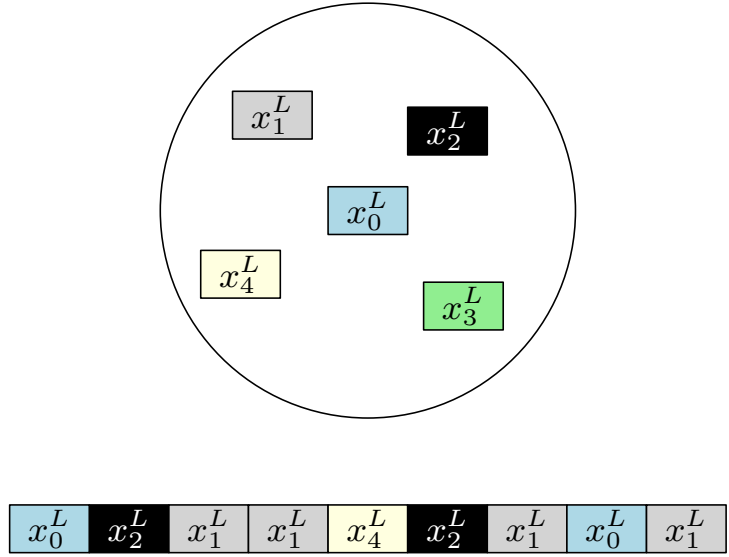}
\par\end{centering}
\caption{Illustration of a sampling event for $M=5$, and $N=9$. Here, $U^{9}=(0,2,1,1,4,2,1,0,1)$,
$S^{5}=(2,4,2,0,1)$ and $Q^{10}=(1,1,2,0,1,0,0,0,0,0)$. \label{fig: sampling event}}

\end{figure}
\item Sequencing: For each $n\in[N]$, $x_{U_{n}}^{L}(j)$ is sequenced
to $Y_{n}^{L}\in{\cal Y}^{L}$, and the sequencing of $x_{U_{n}}^{L}(j)$
is independent for all $n\in[N]$. Denoting the channel output by
$Y^{LN}=(Y_{0}^{L},\ldots,Y_{N-1}^{L})\in({\cal Y}^{L})^{N}$ it thus
holds that 
\[
\P\left[Y^{LN}=y^{LN}\mid x^{LM}(j),\;U^{N}\right]=\prod_{n\in[N]}W^{L}\left(y_{n}^{L}\mid x_{U_{n}}^{L}(j)\right),
\]
where $W^{L}$ is the $L$th product of the DMC $W$, that is, $W^{L}(y^{L}\mid x^{L})=\prod_{i\in[L]}W(y_{i}\mid x_{i})$. 
\end{enumerate}
Due to the random sampling stage, it is clear that the order of $\{Y_{n}^{L}\}_{n\in[N]}$
in $Y^{LN}$ is immaterial. Indeed, the likelihood of $y^{LN}$ conditioned
on an input $x^{LM}$ is given by
\begin{align}
{\cal L}\left[y^{LN}\mid x^{LM}\right] & \dfn\sum_{u^{N}\in[M]^{N}}\P[U^{N}=u^{N}]\prod_{n\in[N]}W^{L}\left[y_{n}^{L}\mid x_{u_{n}}^{L}\right].\label{eq: likelihood DNA channel}\\
 & =\frac{1}{M^{N}}\sum_{u^{N}\in[M]^{N}}\prod_{n\in[N]}W^{L}\left[y_{n}^{L}\mid x_{u_{n}}^{L}\right].
\end{align}

\paragraph*{The decoder}

The decoder $({\cal Y}^{L})^{N}\to[|{\cal C}|]$ maps the channel
output to a codeword index, and with a slight abuse of notation, we
identify the decoder with the set of the decision regions ${\cal D}=\{{\cal D}(j)\}_{j\in[|{\cal C}|]}$,
where ${\cal D}(j)$ is the decision region of the $j$th codeword
${\cal D}(j)\dfn\{y^{LN}\colon\mathsf{{\cal D}}(y^{LN})=j\}$. 

\paragraph*{Global assumptions}

We will assume throughout that:
\begin{enumerate}
\item Molecule length scaling: $L\equiv L_{M}=\beta\log M$ where $\beta>1$
is the molecule length parameter.
\item Coverage depth: $N=\alpha M$ where $\alpha>0$ is the coverage depth
parameter. 
\end{enumerate}
The DNA channel is thus indexed by $M$ and parameterized by $\mathsf{DNA}\dfn(\alpha,\beta,W)$.
The (storage) rate of the codebook ${\cal C}$ is given by 
\begin{equation}
R=\frac{\log|{\cal C}|}{ML}\label{eq: storage rate}
\end{equation}
and by the assumption $|{\cal C}|=e^{RML}=e^{R\beta M\log M}.$ Note
that compared to the scaling $|{\cal C}|=e^{MR}$ in standard channel
coding problems, there is an extra $L=\beta\log M$ factor in the
effective codeword length. The error probability of ${\cal D}$ given
that $x^{LM}(j)\in{\cal C}$ was stored is given by 
\begin{equation}
\pe({\cal C},{\cal D}\mid x^{LM}(j))\dfn\sum_{y^{LN}\in{\cal D}^{c}(j)}{\cal L}\left[y^{LN}\mid x^{LM}(j)\right].\label{eq: error proabability}
\end{equation}

The maximal error probability is the maximum of (\ref{eq: error proabability})
over all codewords $j\in[|{\cal C}|]$, and will be used in the following
definitions. For later derivations, we also denote the average error
probability by $\pe({\cal C},{\cal D})\dfn\frac{1}{|{\cal C}|}\sum_{j\in[|{\cal C}|]}\pe({\cal C},{\cal D}\mid x^{LM}(j))$.

Let a DNA channel $\mathsf{DNA}\dfn(\alpha,\beta,W)$ be given.
\begin{defn}[$(M,\epsilon)$-code]
\label{def: M-epsilon code}${\cal C}$ is an\emph{ $(M,\epsilon)$-code}
for the channel $\mathsf{DNA}$ of dimension $M$ and $\epsilon\in[0,1)$
if there exists a decoder ${\cal D}$ such that $\max_{j\in[|{\cal C}|]}\pe({\cal C},{\cal D}\mid x^{LM}(j))\leq\epsilon$. 
\end{defn}
\begin{defn}[Capacity]
\label{def: capacity}The rate $R\in\mathbb{R}^{+}$ is $\epsilon$-achievable
for the channel $\mathsf{DNA}$ and $\epsilon\in[0,1)$ if for every
$\delta>0$ and every $M$ sufficiently large there exists a codebook
${\cal C}_{M}$ of size $|{\cal C}_{M}|\geq e^{ML(R-\delta)}$ such
that ${\cal C}_{M}$ is an $(M,\epsilon)$-code. The rate $R$ is
achievable if it is an $\epsilon$-achievable rate for all $\epsilon\in(0,1)$.
The supremum of achievable rates is the \emph{capacity} $C\equiv C(\mathsf{DNA})$
of the channel. 
\end{defn}
\begin{defn}[Reliability function]
Let $K_{M}\colon\mathbb{N}^{+}\to\mathbb{N}^{+}$ be monotonic increasing.
An \emph{error exponent $E(R)$ w.r.t. scaling $K_{M}$ }is achievable
for the channel $\mathsf{DNA}$ at rate $R<C$ if there exists a sequence
$\{{\cal C}_{M},{\cal D}_{M}\}_{M\in\mathbb{N}^{+}}$so that\footnote{We define here achievable error exponent with a stringent definition
using limit inferior. The relation (equality/inequality) between the
reliability function defined with limit inferior and the reliability
function defined with limit superior is still unsettled even for standard
DMCs \cite[Problem 10.7]{csiszar2011information}.}
\[
\liminf_{M\to\infty}-\frac{1}{K_{M}}\log\left[\max_{j\in[|{\cal C}|]}\pe({\cal C},{\cal D}\mid x^{LM}(j))\right]\geq E(R).
\]
The supremum of all error exponents $E(R)$ achieved w.r.t. scaling
$K_{M}$ is the reliability function $E^{*}(R)\equiv E^{*}(R,\mathsf{DNA},\{K_{M}\})$
of the channel $\mathsf{DNA}$ w.r.t. to scaling $K_{M}$. 
\end{defn}
In standard channel coding, with the usual codewords of ordered $ML$
symbols, the error exponent decays w.r.t. scaling $K_{M}=ML=M\beta\log M$
\cite[Ch. 5]{gallager1968information} \cite[Ch. 10]{csiszar2011information}.
Here, due to loss of molecules and their order, slower decay rates
of $K_{M}=M$ will also be applicable. 

\section{Lower Bounds \label{sec:Achievability} }

In order to state our results we will need a few notations and definitions.
Let $\tilde{S}\sim\text{Pois}(\alpha)$ and $\pi_{\alpha}(d)\dfn\frac{\alpha^{d}e^{-\alpha}}{d!}$
for $d\in\mathbb{N}$ be the Poisson p.m.f. so that $\P[\tilde{S}=d]=\pi_{\alpha}(d).$
Further, for any given $d\in\mathbb{N}^{+}$, let the \emph{hazard
probability }of $\tilde{S}$ be
\begin{equation}
\pi_{\alpha|\geq d}(d')\dfn\P\left[\tilde{S}=d'\mid\tilde{S}\geq d\right]=\frac{\pi_{\alpha}(d')}{1-\sum_{i\in[d]}\pi_{\alpha}(i)}\label{eq: Poisson hazard probability}
\end{equation}
for all $d\in\mathbb{N}^{+}$ and $d'\geq d$. 
\begin{defn}[The $d$-order binomial extension of a DMC]
\label{def: binomial channel} Let ${\cal A},{\cal B}$ be finite
input and output alphabets (respectively), let $V\colon{\cal A}\to{\cal B}$
be a DMC, and let $d\in\mathbb{N}^{+}$. We call the DMC $V^{\oplus d}\colon{\cal A}\to{\cal B}^{d}$
the \emph{$d$-order binomial extension of $V$ }if $V^{\oplus d}[b^{d}\mid a]=\prod_{i=0}^{d-1}V(b_{i}\mid a)$
for all $a\in{\cal A},b^{d}\in{\cal B}^{d}$. 

Thus the output of the channel $V^{\oplus d}$ is a sequence of $d$
conditionally independent observations of its input $a$. For brevity,
we will refer to $V^{\oplus d}$ as a \emph{$d$-order binomial channel},
whose capacity is $C(V^{\oplus d})=\max_{P_{A}}I(P_{A},V^{\oplus d})$.
By the data processing theorem, it is clear that $I(P_{A},V^{\oplus d})$,
and hence $C(V^{\oplus d})$, are monotonic non-decreasing in $d$. 
\end{defn}

\subsection{A Lower Bound on the Capacity}

We begin with a lower bound on the capacity:
\begin{thm}
\label{thm: achievable capacity}The capacity of the DNA channel,
$\mathsf{DNA}\dfn(\alpha,\beta,W)$, is lower bounded as
\begin{equation}
C(\mathsf{DNA})\geq\max_{P_{X}\in{\cal P}({\cal X})}\sum_{d\in\mathbb{N}^{+}}\pi_{\alpha}(d)\cdot I(P_{X},W^{\oplus d})-\frac{1}{\beta}\left(1-\pi_{\alpha}(0)\right).\label{eq: lower bound on capacity}
\end{equation}
\end{thm}
Theorem \ref{thm: achievable capacity} is a consequence of Theorem
\ref{thm: achievable reliability function} which appears next in
Sec. \ref{subsec:A-Lower-Bound}, and which provides a lower bound
to the reliability function. Its short proof (as a corollary to Theorem
\ref{thm: achievable reliability function}) appears at the beginning
of Appendix \ref{sec: Achievability proofs}. We next highlight a
few features and implications of Theorem \ref{thm: achievable capacity}.

\paragraph*{Interpretation of the capacity bound}

Had the decoder known the molecule index vector $U^{N}$, it would
have matched each of the $M$ input molecules, $X_{m}^{L}$, to the
$S_{m}$ output molecules $Y_{U_{n}}$ for which $U_{n}=m$. The molecule
$X_{m}^{L}$ and those outputs $\{Y_{n}^{L}\}_{n\colon U_{n}=m}$
can then be considered as an input-output pair of a $(S_{m})$-order
binomial extension of $W$. The mutual information for input distribution
$P_{X}$ of this channel is $L\cdot I(P_{X},W^{\oplus d})$. Out of
the $M$ input molecules, there are $Q_{d}$ molecules which were
input to a $d$-order binomial channel. Their total mutual information
is then $LQ_{d}\cdot I(P_{X},W^{\oplus d})$, and summing over all
possible $d\in[N+1]$, the total mutual information for $ML$ channel
uses is $\sum_{d\in[N+1]}LQ_{d}\cdot I(P_{X},W^{\oplus d})$. Now,
since $U^{N}$ is a uniform random vector, its empirical count $S^{M}=\mathscr{\mathscr{N}}(U^{N})$
is distributed as a multinomial, and so $Q^{N+1}=\mathscr{\mathscr{N}}(S^{M})$
is the empirical count of a multinomial distribution. An effect known
as \emph{Poissonization} (see Fact \ref{fact: Poissonization of the multinomial distribution}
in Appendix \ref{subsec:The-proof-of error exponent}) implies that
this distribution tends to a Poisson distribution. Thus $\frac{Q_{d}}{M}\approx\pi_{\alpha}(d)$,
and the total mutual information for $ML$ channel uses is then $ML\sum_{d\in[N+1]}\pi_{\alpha}(d)Q_{d}\cdot I(P_{X},W^{\oplus d})$.
This leads to the first term in the capacity lower bound. The second
term, $\frac{1}{\beta}(1-\pi_{\alpha}(0))$, reduces the bound, and
reflects the fact that the decoder does not know $U^{N}$, and thus
needs to order all non-erased molecules, either implicitly or explicitly.
A reinforcement to this interpretation is offered by simpler schemes
and channel models. Specifically, \cite{shomorony2021dna} analyzed
an indexing-based scheme that uses $\log M$ nats out of the $\beta\log M$
of each of the $M$ molecules to specify its index, to be used over
a noiseless sequencing channel. The capacity was shown to be $1-\pi_{\alpha}(0)-\frac{1}{\beta}(1-\pi_{\alpha}(0))$,
and the rate loss term of $\frac{1}{\beta}(1-\pi_{\alpha}(0))$ in
this model clearly stems from the cost of indexing. By contrast, our
scheme is designed to achieve a non-trivial lower bound on the reliability
function, without relying on sending an explicit indexing information.
Nonetheless, it still suffers the rate loss $\frac{1}{\beta}(1-\pi_{\alpha}(0))$,
which in light of the upper bound (converse) is indeed inevitable
in some regime.

\paragraph*{Comparison to previous results}

In \cite{shomorony2021dna}, the sampling model is such that $S_{m}$
are i.i.d., and that $S_{m}\in\{0,1\}$ where $\P[S_{m}=0]$ is an
erasure probability. That is, any of the $M$ molecules is either
sampled once or none at all. For a $\mathsf{BSC}(w)$ sequencing channel
(cf. Sec. \ref{sec:Modulo-additive-sequencing}) the rate achieved
in \cite[Sec. IV.A]{shomorony2021dna} is 
\begin{equation}
R=\left(1-\P[S_{m}=0]\right)\cdot\left[1-h_{b}(w)-1/\beta\right].\label{eq: Ilan-Reinhard  achievability}
\end{equation}
Per the interpretation of the capacity bound above, $\pi_{\alpha}(d)$
represents the fraction of molecules which have been sampled $d$
times. Thus (\ref{eq: Ilan-Reinhard  achievability}) agrees with
Theorem \ref{thm: achievable capacity} by replacing $\pi_{\alpha}(d)\rightarrow\P[S_{m}=d]$
for $d\in\{0,1\}$ and $\pi_{\alpha}(d)\to0$ for $d>1$, as well
as choosing $P_{X}=(\frac{1}{2},\frac{1}{2})$ (cf. Theorem \ref{thm: error exponent of universal decoder -- ideal amplification}). 

In \cite{lenz2020achievable}, it was shown, for the sampling model
which we adopt here, and for a $\mathsf{BSC}(w)$ sequencing channel,
that the rate 
\[
\sum_{d\in\mathbb{N}^{+}}\pi_{\alpha}(d)\cdot I\left(\left(\tfrac{1}{2},\tfrac{1}{2}\right),\mathsf{BSC}(w)^{\oplus d}\right)-\frac{1}{\beta}\left(1-\pi_{\alpha}(0)\right)
\]
is achievable, but only in the regime $w\leq\frac{1}{8}$ and $\beta>\frac{2}{\log2-h_{b}(4w)}$.
Here the rate implied by (\ref{eq: lower bound on capacity}) is the
same, and holds for any $w\in[0,\frac{1}{2}]$ and $\beta>1$. See
Sec. \ref{sec:Modulo-additive-sequencing} for a more detailed comparison. 

\paragraph*{Choice of input distribution}

The lower bound on capacity in (\ref{eq: lower bound on capacity})
is a weighted sum of mutual information terms $\{I(P_{X},W^{\oplus d})\}_{d\in\mathbb{N}^{+}}$
according to the Poisson distribution. In general, the input distribution
$P_{X,d}$ which maximizes $I(P_{X},W^{\oplus d})$ is different for
each $d$, and in that case the input distribution which maximizes
the objective in (\ref{eq: lower bound on capacity}) would be a compromise
between those input distributions $\{P_{X,d}\}_{d\in\mathbb{N}^{+}}$.
Interestingly, this may occur even for some \emph{symmetric} sequencing
channels (see Remark \ref{rem:symmetry is not preserved} in what
follows). Nonetheless, for some particularly relevant symmetric sequencing
channels $W$, the maximizing input distribution is \emph{provably}
uniform (see Definition \ref{def: symmetric DMC} for a symmetric
channel in Appendix \ref{subsec:The-binomial-extension of symmetric DMC is symmetric}). 
\begin{prop}
\label{prop: achievability by a uniform distribution}If $|{\cal X}|\leq4$,
$|{\cal Y}|\leq|{\cal X}|$, and $W$ is a symmetric channel in Gallager's
sense \cite[p. 94]{gallager1968information} whose transition probability
matrix does not have any identical rows, then the lower bound on the
capacity (\ref{eq: lower bound on capacity}) is achieved by the uniform
input distribution, $P_{X}^{(\text{\emph{unif}})}=(\frac{1}{|{\cal X}|},\ldots\frac{1}{|{\cal X}|})$.
The same holds if $W$ is a modulo-additive channel for any ${\cal X}={\cal Y}$. 
\end{prop}
\begin{IEEEproof}
The claim follows since the mutual information $I(P_{X},W)$ for a
symmetric channel $W$ in Gallager's sense (see Definition \ref{def: symmetric DMC}
in Appendix \ref{subsec:The-binomial-extension of symmetric DMC is symmetric})
is maximized by the uniform input distribution \cite[Thm. 4.5.2]{gallager1968information}.
As we show in Prop. \ref{prop: binomial extension of modulo additive channels}
and Prop. \ref{prop: symmetry of binomial extension} in Appendix
\ref{subsec:The-binomial-extension of symmetric DMC is symmetric},
the binomial extension of any order $d$ of the channels which satisfy
the conditions of the proposition is symmetric in Gallager's sense.
Thus, the uniform input distribution simultaneously maximizes all
terms in the sum (\ref{eq: lower bound on capacity}). 
\end{IEEEproof}
\begin{rem}
\label{rem:symmetry is not preserved}The qualifying conditions in
Prop. \ref{prop: achievability by a uniform distribution} cannot
be refined. There is no general guarantee that if $W$ is symmetric
then $W^{\oplus d}$ is symmetric in Gallager's sense for $d>1$.
There is a symmetric channel with $|{\cal X}|=|{\cal Y}|=5$ that
does not satisfy that. Moreover, if a channel $W$ is only symmetric
in Gallager's sense, there is no such guarantee even if $|{\cal X}|=4$.
There is a symmetric channel in Gallager's sense with $|{\cal X}|=4,|{\cal Y}|=8$
that does not satisfy that. See the end of section \ref{subsec:The-binomial-extension of symmetric DMC is symmetric}
for the specific channel transition probability matrices. Moreover,
the capacity-achieving input distribution of these channels is \emph{not
uniform}, and so the input distribution which maximizes the lower
bound (\ref{eq: lower bound on capacity}) will also \emph{not be
uniform.}
\end{rem}

\subsection{A Lower Bound on the Reliability Function\label{subsec:A-Lower-Bound}}

Theorem \ref{thm: achievable capacity} is a direct implication of
the following stronger result which is a lower bound on the reliability
function of the channel. In short, the next theorem shows that the
error probability decays to zero exponentially fast w.r.t. to the
scaling $K_{M}=M$ for all rates below the lower bound on the capacity
in Theorem \ref{thm: achievable capacity}. 
\begin{thm}
\label{thm: achievable reliability function}The reliability function
of the DNA channel $\mathsf{DNA}\dfn(\alpha,\beta,W)$ w.r.t. scaling
$K_{M}=M$ is lower bounded as
\begin{equation}
E^{*}(R,\mathsf{DNA},\{M\})\geq\max_{P_{X}\in{\cal P}({\cal X})}\inf_{\{\theta_{d}\}_{d\in\mathbb{N}}}\sum_{d\in\mathbb{N}}\left(1-\sum_{i\in[d]}\theta_{i}\right)\cdot d_{b}\left(\frac{\theta_{d}}{1-\sum_{i\in[d]}\theta_{i}}\,\middle\vert\middle\vert\,\pi_{\alpha|\geq d}(d)\right),\label{eq: error exponent for universal decoder}
\end{equation}
where the minimization is subject to:
\begin{equation}
\left\{ \theta_{d}\in(0,1]\quad\forall d\in\mathbb{N},\;\sum_{d\in\mathbb{N}}\theta_{d}=1,\;\sum_{d\in\mathbb{N}}\theta_{d}\cdot I(P_{X},W^{\oplus d})-\frac{1}{\beta}(1-\theta_{0})<R\right\} .\label{eq: minimization set for error exponent}
\end{equation}
\end{thm}
We next highlight a few features and implications of Theorem \ref{thm: achievable reliability function}.

\paragraph*{Choice of decoder}

The error exponent in (\ref{eq: error exponent for universal decoder})
is achieved for a universal decoder which is oblivious to the channel
$W$.\footnote{We nonetheless mention, that, as usual, an optimal choice of the input
distribution, $P_{X}$, depends on the channel $W$.} This universal decoder is a variant of the MMI decoding rule, which
given a channel output, $y^{LM}$, computes a metric for each of the
codewords, $x^{LM}(j)$, as follows. Recall that the molecule index
vector, $u^{N}$, designates a possible sampling event, in which molecule
$u_{n}$ was sampled at the $n$th sampling trial. Considering a candidate
$u^{N}$, the decoder partitions the $ML$ molecule symbols into groups
according to the number of times that molecule has been sampled. A
symbol which belongs to the $d$th group thus has $d$ independent
output symbols, which can be considered a single super-symbol from
${\cal Y}^{d}$. The decoder computes the empirical mutual information
for each group, and then the total empirical mutual information, is
weighted according to the groups size. The decoder repeats this computation
for all possible candidate $u^{N}$ vectors,\footnote{Strictly speaking, for a judiciously chosen subset of all possible
$u^{N}$. } and then chooses a \emph{penalized} maximum of the weighted empirical
mutual information over all $u^{N}$ candidates. The penalty term
is related to the amplification vector $q^{N+1}$ for which $q^{N+1}=\mathscr{\mathscr{N}}^{(2)}(u^{N})$
holds. Specifically, the penalty reduces the metric of those $u^{n}$
for which there is a large number of other candidates $\tilde{u}^{N}$
for which $\mathscr{\mathscr{N}}^{(2)}(\tilde{u}^{N})=\mathscr{\mathscr{N}}^{(2)}(u^{N})=q^{N+1}$
also holds. We term this set the \emph{amplification type class} of
$q^{N+1}$. 

\paragraph*{Proof outline and main ideas}

Appendix \ref{sec: Achievability proofs} is devoted to the proof
of the lower bounds (achievability results). It begins with two preliminary
sections. First, in Appendix  \ref{subsec:Preliminaries:-Sampling-type},
a tight characterization of the asymptotic size of the amplification
type class is provided, which is required for the penalty term of
the universal decoder. Second, in Appendix  \ref{subsec:Preliminaries:-The-DNA},
the likelihood of the DNA  channel is cast as the likelihood of a
repeated mixture -- the mixture over molecule index vectors $u^{N}$,
and the mixture of binomial channels of different orders. 

We then prove Theorems and \ref{thm: achievable capacity} and \ref{thm: achievable reliability function}
by analyzing the error probability of the universal decoder. In Appendix
 \ref{subsec:A-universal-decoder}, we utilize the $u^{N}$ molecule-index/binomial
mixture interpretation of the DNA channel to rigorously define this
universal decoder. In Appendix  \ref{subsec:Random-coding-error},
we condition on specific amplification vector $Q^{N+1}=q^{N}$, and
analyze the average error probability of a codebook randomly chosen
from the standard i.i.d. input ensemble and the universal decoder.
We then prove Theorem \ref{thm: achievable reliability function}
by averaging the error probability w.r.t. $Q^{N+1}$ for the sampling
mechanism of the DNA channel. Here the \emph{Poissonization effect}
(Fact \ref{fact: Poissonization of the multinomial distribution}),
which implies that $\frac{Q_{d}}{M}\approx\pi_{\alpha}(d)$, is used.
However, unlike its usage for capacity analysis in \cite{lenz2020achieving},
here we consider its effect on the error probability, and accordingly,
its effect on tail probabilities. 

\paragraph*{Comparison to previous proof techniques}

As said, there are no claims in \cite{shomorony2021dna,lenz2020achievable}
regarding the decay rate of the error probability. However, it is
still enlightening to compare their coding schemes to ours, as they
are considerably different. 

The coding scheme of \cite{shomorony2021dna} is based on molecule
indexing and on concatenation of inner and outer codes. Recall that
in \cite{shomorony2021dna}, $S_{m}\in\{0,1\}$ with probability $1$.
The outer code is designed to correct erased (non-sampled) and erroneously
decoded molecules, while the inner code is designed to correct the
errors of the sequencing channel ($\mathsf{BSC}(w)$, in case of \cite{shomorony2021dna}),
as well as to identify the index of the molecule, and thus allow the
decoder to order the output molecules. Specifically, the outer code
operates on super-symbols of blocklength $M$, and has rate $1-\P[S_{m}=0]$.
Thus, it can correct up to $M\cdot\P[S_{m}=0]$ erased molecules.
Each super-symbol encodes roughly $L(1-h_{b}(w)-1/\beta)$ bits. Each
of the $M$ molecules of the output alphabet of the outer code, is
appended with $\log_{2}M$ bits which identifies the index $m$ of
the molecule. The resulting $L(1-h_{b}(p))$ bits of each molecule
are encoded to $L$ bits using the inner code, designed to be capacity
achieving for $\mathsf{BSC}(w)$. At reading time, the inner code
is first decoded, \emph{individually} for each molecule. Assuming
a correct inner-code decoding of all sampled molecules, they can be
ordered using the header which identifies their index, and then the
outer erasure-correcting code can correct the erased molecules. By
contrast, our scheme neither uses concatenated codes nor explicit
indexing.

The coding scheme of \cite{lenz2020achieving} is based on the standard
random coding i.i.d. ensemble, and a non-standard decoder, which is
mainly tailored to the BSC case,\footnote{Though it is mentioned in passing that the analysis is also suitable
for symmetric channels.} and is very different from the decoder proposed here. The decoder
employs a clustering algorithm in order to overcome the possible (and
likely) multiple appearances of a single molecule at the output vector
$Y^{LN}$, and lack of any prior ordering information that can match
$Y_{n}^{L}$ with the molecule that generated it. This clustering
algorithm is greedy, and with high probability, it clusters the $N$
outputs to $M(1-\pi_{\alpha}(0))$ clusters, such that an output cluster
is the result of sequencing of one of the $M$ molecules, and where
$M\pi_{\alpha}(0)$ molecules are erased. The clustering algorithm
is based on Hamming distances -- thus it is mainly suitable to a
BSC sequencing channel -- and requires the channel crossover probability
as input. Furthermore, for successful clustering, it is required that
the minimum distance between the molecules of the true stored codeword
(that is, the minimum distance among the $M$ sequences of $L$ bits
representing the codeword) is at least $\approx4wL$. This ``hard''
requirement is  the source of the limited regime of $(w,\beta)$ in
the result -- cf. \cite[Lemma 4]{lenz2020achieving}, which states
the the probability that any two molecules from the same codeword
have sufficiently large Hamming distance tends to $1$. It seems challenging
to extend those arguments of \cite[Lemma 4]{lenz2020achieving} to
soft decoders. Furthermore, the decoded codeword is the unique one
which is weakly jointly typical (in the sense of \cite[Ch. 7]{cover2012elements})
with the output clusters obtained by the clustering algorithm. As
is well known, jointly typical decoding suffices to achieve capacity,
but it is otherwise too weak for obtaining tight error probability
bounds, and so our analysis uses a stronger decoder.

\paragraph*{Outage interpretation}

According to Theorem \ref{thm: achievable reliability function},
there exists a rate $\underline{C}$ (which is a lower bound on the
capacity), such that the error probability decays exponentially in
$M$ for all rates below $\underline{C}$. In the reliability function
bound, the variable $\theta_{d}$ represents $\frac{q_{d}}{M}$ --
the fraction of molecules (out of $M$) that were sampled $d$ times
during the sampling stage. Conditioned on an amplification vector
$q^{N+1}$, or, equivalently, on $\theta$, the term on the left-hand
side of the constraint (\ref{eq: minimization set for error exponent}),
to wit, $\sum_{d\in\mathbb{N}}\theta_{d}\cdot I(P_{X},W^{\oplus d})-\frac{1}{\beta}(1-\theta_{0})$
is thus the conditional ``supported'' rate $R(\{\theta_{d}\})$.
If the random supported rate is above the coding rate $R$, then the
error probability decays at exponential rate w.r.t. scaling $ML$
(as we show in the proof of Theorem \ref{thm: achievable reliability function}).
Otherwise, if the supported rate is below the coding rate, we may
trivially upper bound the error probability by $1$. Thus, the error
probability, averaged over $Q^{N+1}$, is upper bounded as 
\[
\pe({\cal C},{\cal D})\leq\P[R(\{\theta_{d}\})\geq R]\cdot e^{-\Theta(ML)}+\P[R(\{\theta_{d}\})<R]\cdot1.
\]
As shown in the proof of Theorem \ref{thm: achievable reliability function},
the decay rate of $\P[R(\{\theta_{d}\})<R]$ is exponential, yet onlu
in $M$, and thus dominates the error probability. This event can
be thought of as an \emph{outage} event, in which the random state
of the channel (due to the sampling stage) does not allow for coding
at the required rate with low error probability. 

\paragraph*{Sampling versus loss of order}

The DNA channel affects the stored codeword in two non-standard ways:
The lack of molecule order, and the random number of copies of the
molecule present before the sequencing stage -- which can cause either
duplication or erasure of a molecule (beyond the error in sequencing,
which are modeled here as a standard DMC). As Theorem \ref{thm: achievable reliability function}
shows, the resulting error probability is dominated by outage, resulting
from the randomness in the number of copies of the molecule, and is
exponential in $M$. It turns out that when the sampling stage is
ideal, and each molecule is sampled \emph{exactly} $\alpha$ times,
the error probability decays much faster, and it is of exponential
in $ML=\beta M\log M$. Specifically, a slight modification of the
proof of Theorem \ref{thm: achievable reliability function} yields
the following: 
\begin{thm}
\label{thm: error exponent of universal decoder -- ideal amplification}\textbf{
}Consider an idealized sampling DNA channel $\mathsf{DNA}_{\text{ideal}}\dfn(\alpha,\beta,W)$
in which $S_{m}=\alpha$ for all $m\in[M]$ with probability $1$.
Then,
\begin{multline}
E^{*}(R,\mathsf{DNA}_{\text{ideal}},\{ML\})\\
\geq\max_{P_{X}\in{\cal P}({\cal X})}\min_{Q_{XY^{\alpha}}\in{\cal P}({\cal X}\times{\cal Y}^{\alpha})}D(Q_{X}\mid\mid P_{X})+D(Q_{Y^{\alpha}|X}\mid\mid W^{\oplus\alpha}|Q_{X})+\left[D(Q_{A}\mid\mid P_{X})+I_{Q}(X;Y^{\alpha})-\frac{1}{\beta}-R\right]_{+}.\label{eq: error exponent for universal decoder ideal}
\end{multline}
\end{thm}
The proof of Theorem \ref{thm: error exponent of universal decoder -- ideal amplification}
is a simplified and slightly modified version of the proof of Theorem
\ref{thm: achievable reliability function}, and it appears at the
end of Appendix \ref{sec: Achievability proofs}.

\section{An Upper Bound on Capacity \label{sec:Converse}}

We next turn to state our upper bound on the capacity. To this end,
consider a DMC $V\colon{\cal X}\to{\cal Y}$ and $(X,Y,\overline{Y})\in{\cal X}\times{\cal Y}^{2}$
and for which 
\[
\P[X=x,Y=y,\overline{Y}=\overline{y}]=P_{X}(x)\cdot V(y\mid x)\cdot V(\overline{y}\mid x).
\]
That is, $Y$ and $\overline{Y}$ are two conditionally independent
observations given a common input $X$ to the DMC $V$. We then define
the \emph{common-input (mutual information) deficit (CID) $\mathsf{CID}\colon{\cal P}({\cal X})\times{\cal P}({\cal Y}\mid{\cal X})\to\mathbb{R}^{+}$
}as 
\begin{align}
\mathsf{CID}(P_{X},V) & \dfn2\cdot I(X;Y)-I(X;Y,\overline{Y})\\
 & =2\cdot I(P_{X},V)-I(P_{X},V^{\oplus2}).\label{eq: definition common input deficit}
\end{align}
The CID measures the loss in mutual information when the same input
is fed into a pair of independent channels, compared to the case of
independent inputs. It also holds that $\mathsf{CID}(P_{X},V)=I(Y;\overline{Y})$
(see Prop. \ref{prop: common input deficit} in Appendix  \ref{subsec:Preliminaries:-Common-input}).

As in \cite{shomorony2021dna,lenz2019upper}, our upper bound on the
capacity does not match the lower bound for all sequencing channels
and molecule length parameter $\beta$. To present this gap in a concise
manner, we define the \emph{$d$-order excess-rate} term by\textbf{
\begin{align}
\Omega_{d}(\beta,P_{X},W) & \dfn\begin{cases}
\frac{1}{\beta}, & \mathsf{CID}(P_{X},W^{\oplus d})<\frac{1}{\beta}\\
-\mathsf{CID}(P_{X},W^{\oplus d})+\frac{2}{\beta}, & \frac{1}{\beta}\leq\mathsf{CID}(P_{X},W^{\oplus d})<\frac{2}{\beta}\\
0, & \mathsf{CID}(P_{X},W^{\oplus d})\ge\frac{2}{\beta}
\end{cases}\nonumber \\
 & =\left[\frac{1}{\beta}\wedge\left(\frac{2}{\beta}-\mathsf{CID}(P_{X},W^{\oplus d})\right)\right]_{+}.\label{eq: excess term in the capacity upper bound}
\end{align}
}
\begin{thm}
\label{thm: converse capacity}Assume that the DNA channel, $\mathsf{DNA}\dfn(\alpha,\beta,W)$,
satisfies 
\[
\nu_{\text{\emph{min}}}(W)\dfn\max_{x\in{\cal X},\;y\in{\cal Y}}\log\frac{1}{W(y\mid x)}<\infty.
\]
Then, its capacity is upper bounded as
\begin{equation}
C(\mathsf{DNA})\leq\max_{P_{X}\in{\cal P}({\cal X})}\sum_{d\in\mathbb{N}^{+}}\pi_{\alpha}(d)\cdot\left[I(P_{X},W^{\oplus d})+\Omega_{d}(\beta,P_{X},W)\right]-\frac{1}{\beta}\left(1-\pi_{\alpha}(0)\right).\label{eq: capacity lower bound}
\end{equation}
\end{thm}

\paragraph*{The main ideas of the proof}

The proof follows the main argument of \cite{shomorony2021dna,lenz2019upper}.
Thus, we first describe the argument and then emphasize where our
proof argument deviates from previous analysis. From Fano's inequality,
the rate of a reliable code is upper bounded by the mutual information
$I(X^{LM};Y^{LN})$, and so the main task is to upper bound the mutual
information of the, rather non-standard, DNA channel. This upper bound
should ideally match the lower bound of Theorem \ref{thm: achievable capacity},
which is comprised of two terms, to wit, $\sum_{d\in\mathbb{N}^{+}}\pi_{\alpha}(d)\cdot I(P_{X},W^{\oplus d})$
and $-\beta^{-1}(1-\pi_{\alpha}(0))$. If the decoder was aware of
$U^{N}$, say as side information, then standard arguments and Poissonization
bound the mutual information as 
\begin{equation}
I(X^{LM};Y^{LN},U^{N})=\underbrace{I(X^{LM};U^{N})}_{=0}+I(X^{LM};Y^{LN}\mid U^{N})\leq\sum_{d\in\mathbb{N}^{+}}\pi_{\alpha}(d)\cdot I(P_{X},W^{\oplus d}),\label{eq: converse (upper bound) for a known sampling vector}
\end{equation}
which is the first term of the lower bound. Therefore, the rate loss
term, $-\beta^{-1}(1-\pi_{\alpha}(0))$, is clearly related to the
lack of knowledge of $U^{N}$ by the decoder, or, loosely speaking,
the loss of order of the molecules in the DNA channel. Moreover, as
in the standard upper bound on the mutual information of a DMC, the
bound (\ref{eq: converse (upper bound) for a known sampling vector})
is achieved by choosing the $M$ molecules $\{X_{m}^{L}\}_{m\in[M]}$
to be i.i.d. (in fact, their $ML$ symbols are all i.i.d. too). If
one uses such independent molecules for the DNA channel, in which
$U^{N}$ is unknown, then the decoder must discern between $(X_{0}^{L},X_{1}^{L},\ldots,X_{M-1}^{L})$
and any other permutation of them. Thus, intuitively speaking, molecules
must contain some information on their index $m$.\footnote{Then, with this indexing information included, the molecules are not
identically distributed anymore.} Since there are $M!=e^{M\log M+O(M)}$ permutations, this information
exactly pertains to the rate loss term $-\beta^{-1}(1-\pi_{\alpha}(0))$.
This argument, however, is not complete on its own, since it is possible
that larger mutual information is achievable by statistically dependent
molecules. The key observation of \cite{shomorony2021dna}, which
was further developed in \cite{lenz2019upper}, is that one can maximize
over the statistical dependency between the molecules, and under some
conditions, independent molecules do maximize the mutual information
of the DNA channel. To intuitively demonstrate this phenomenon, we
consider the most simplistic case of $M=2$ input molecules, $N=2$
output molecules and $L=1$, while assuming that each of the two molecules
is sampled exactly once $S_{0}=S_{1}=1$ (so $\pi_{\alpha}(0)$ can
be set to $0$). On the one hand, independent molecules $X_{0},X_{1}\sim P_{X}$,
lead to the sum of mutual information terms, but since index information
must also be sent, the total rate is $2I(P_{X},W)-\beta^{-1}.$ On
the other hand, choosing fully dependent molecules $X_{0}=X_{1}$
reduces the mutual information to $I(P_{X},W^{\oplus2})$ (since the
two outputs can be considered the output of the binomial channel $W^{\oplus2}$
for input $X=X_{0}=X_{1}$), but trivially does not require index
information. Whenever the former rate is larger than the later, to
wit, 
\[
2I(P_{X},W)-\beta^{-1}>I(P_{X},W^{\oplus2}),
\]
or, equivalently $\beta>1/\mathsf{CID}(P_{X},W)$, the independent
inputs are optimal. The similarity of this condition to the ones appearing
in the excess term (\ref{eq: excess term in the capacity upper bound})
is not coincidental, and indeed the origin of this condition can be
traced to similar derivations. 

Naturally, the actual argument of \cite{shomorony2021dna,lenz2019upper}
is much more delicate, specifically regarding the statistical dependencies
between the $M$ molecules, and they rely on the Hamming distance
between either output molecules (in \cite{shomorony2021dna}) or input
molecules (in \cite{lenz2019upper}) to quantify this dependency.\footnote{Our argument will extend the \emph{input}-molecule based distance. }
If the Hamming distance between a pair of molecules, say, $X_{0}^{L}$
and $X_{1}^{L}$ is $\rho_{\text{H}}(X_{0}^{L},X_{1}^{L})=\gamma L$
with probability $1$, then the molecules are ``far'' apart, in
the sense that the mutual information they induce is approximately
as for independent inputs. Otherwise, they are ``close'' and the
mutual information they induce is approximately as for identical inputs.
The constant $\gamma>0$ is channel dependent, and for the $\mathsf{BSC}(w)$
considered in \cite{shomorony2021dna,lenz2019upper}, is given by
$\gamma=4w(1+o(1))$. 

Our proof argument mainly deviates from \cite{shomorony2021dna,lenz2019upper}
in the definition of the distance between molecules (Appendix  \ref{subsec:The-molecule-distance}),
which in general, is no longer the Hamming distance. It is defined
in terms of probabilities of \emph{conditional typical sets }\cite[Ch. 2]{csiszar2011information},
which are reviewed in Appendix  \ref{sec:A-brief-review}.\footnote{The notation in Appendix  \ref{subsec:The-molecule-distance} is slightly
different as the discussion there is general, and not necessarily
pertains only to the DNA channel.} By its construction, the conditional typical set ${\cal T}_{L}([W]\mid x_{0}^{L})$
of $x_{0}^{L}\in{\cal X}^{L}$ is a subset of ${\cal Y}^{L}$ for
which the random output $Y^{L}$ to the input $x_{0}^{L}$ (over the
channel $W^{L}$) belongs to with high probability (which tends to
$1$ as $L\to\infty$). Our distance definition is chosen so that
if $x_{1}^{L}$ is ``far'' from $x_{0}^{L}$ and $x_{1}^{L}$ is
the input to the channel $W^{L}$, then the conditional typical set
${\cal T}_{L}([W]\mid x_{0}^{L})$ no longer has high probability.
On the other hand, if $x_{1}^{L}$ is ``close'' to $x_{0}^{L}$
by our definition, then ${\cal T}_{L}([W]\mid x_{0}^{L})$ has high
probability (in fact, still possibly exponentially small, but with
a negligible exponent). Our definition of distance allows to sharply
characterize this property (Lemma \ref{lem: Probability of V-typical sets condition on a different vector}).
The key point, however, is that under our distance function, the distance
required to create a distinction between ``far'' and ``close''
molecules in terms of the resulting mutual information, is \emph{sub-linear}
in $L$. This is much smaller compared to the $\gamma L$ Hamming
distance of \cite{shomorony2021dna,lenz2019upper} (for the BSC case).
On top of that, the analysis of general sequencing channels, rather
than BSCs or symmetric channels in \cite{shomorony2021dna,lenz2019upper},
leads to various technical difficulties which our proof handles. In
the analysis of BSCs under Hamming distance between molecules in \cite{shomorony2021dna,lenz2019upper},
it holds in various parts of the proof that an i.i.d. uniform $P_{X}=(\frac{1}{2},\frac{1}{2})$
input distribution is extremal, which in turn reduces the analysis
to ``single-letter'' arguments. Here, under the distance function
we consider, this is not true in general (\emph{a priori} not even
for symmetric channels), and the arguments include analysis of probability
distributions over $L$-dimensional vectors (molecules). Thus, to
obtain the single-letter expression (\ref{eq: capacity lower bound})
two stages of ``single-letterization'' are required (from $ML$
to $L$ and from $L$ to $1$). Of course, for asymmetric channels,
even the final, single-letter bound (\ref{eq: capacity lower bound}),
is not necessarily maximized by the uniform input distribution. In
fact, \emph{a priori}, larger mutual information can be obtained by
assigning different input distributions to different molecules. These
are the main difficulties associated with using our distance function.\footnote{Beyond the technical issues associated with general sequencing channel,
which are handled, as usual, with method of types arguments.}

\paragraph*{Proof outline}

The proof of Theorem \ref{thm: converse capacity} appears in Appendix
\ref{sec:Converse:-Proof-of}. Since it is fairly complicated, we
next provide a proof outline. 

In Appendix  \ref{sec:A-brief-review}, we set notation conventions
and definitions of conditional typical sets, and briefly state their
defining property -- they asymptotically obtain high conditional
probability, and have cardinality given by the exponent of the conditional
entropy. 

In Appendix  \ref{subsec:The-molecule-distance}, we introduce our
distance function and its implications. First, we show its main defining
property -- the probability of a typical set ${\cal T}_{L}([W]\mid x_{0}^{L})$
when the channel input is $x_{1}^{L}$, under both cases of ``close''
and ``far'' $x_{0}^{L}$ and $x_{1}^{L}$. Second, we consider a
large set of molecules which are pairwise ``far'' apart, and assume
that each of these molecules is sequenced over a DMC, and that the
resulting output molecules are arbitrarily permuted. We show that
an observer of both the input and the output molecules, can gain information
on the permutation, in the sense that its equivocation given the input
and output is negligible compared to its unconditional entropy. This
result refines a similar result in \cite[proof of Lemma 3]{shomorony2021dna}.
Third, we estimate the mutual information for a pair of \emph{close}
input molecules under our distance definition. We show that this mutual
information behaves asymptotically as if the two input molecules are
identical, and thus strictly smaller compared to the mutual information
achieved by two independent input molecules (we refer to this as a
\emph{deficit} in the mutual information).

In Appendix  \ref{subsec:Structural-properties-capacity}, we show
that capacity-achieving codebooks can be assumed to have, without
loss of generality (w.l.o.g.), two simplifying structural properties.
Both these properties state that all the codewords can have the same
structure. The first one is that the codebook ${\cal C}_{M}$ is such
that all $\{x_{m}(j)\}_{j\in[|{\cal C}_{M}|]}$ have the same type
$P_{X,m}\in{\cal P}_{L}({\cal X})$ (but the type may change with
the molecule index $m$). The second one is related to the distances
between the molecules. Following \cite{shomorony2021dna,lenz2019upper},
we partition $[M]$ to two subsets for each codeword $x^{LM}(j)$.
In the first subset, the molecules are pairwise far, and each molecule
in the second subset has a close neighbor in the first subset. The
structural property shows that the subset can be the same for all
codewords in the codebook. These two structural properties allows
us to simplify the derivation in the next section. 

In Appendix  \ref{subsec:Upper-bound-on}, we upper bound the mutual
information. We follow the idea of \cite{lenz2019upper}, and consider
a genie-aided decoder which is capable of clustering its outputs.
That is, the decoder knows which output molecules are the result of
sequencing the same input molecule, but it does not know which input
molecule belongs to which cluster. Specifically, we may assume that
it knows $\Sigma(U^{N})=(\Sigma(U_{0}),\Sigma(U_{1}),\ldots,\Sigma(U_{N-1}))$
where $\Sigma\colon[M]\to[M]$ is a random permutation drawn from
the symmetric group $\mathfrak{S}_{M}$, which is unknown to the decoder.
There are $M$ such output clusters, which we denote by $\tilde{Y}^{LM}$
(some of them may be empty), and instead of upper bounding $I(X^{LM};Y^{LM})$
we upper bound $I(X^{LM};\tilde{Y}^{LM})$. The bounding of $I(X^{LM};\tilde{Y}^{LM})$
is done at three stages. 

At the first stage, we assume a fixed composition codebook, in which
all molecules have exactly the same type, that is $P_{X,m}=P_{X}$
for all $m\in[M]$, and condition on a fixed amplification vector
$Q^{N+1}=q^{N+1}$. We decompose 
\[
I(X^{LM};\tilde{Y}^{LM})\approx H(\tilde{Y}^{LM})-H(\tilde{Y}^{LM}\mid X^{LM},\tilde{U}^{M})+H(\tilde{U}^{M}\mid X^{LM},\tilde{Y}^{LM})-M\log M
\]
{[}see (\ref{eq: decomposition of mutual information}) for exact
statement{]}. As in \cite{shomorony2021dna,lenz2019upper}, the bound
is based on balancing between ``close'' and ``far'' molecules.
That is, using the structural property of the codebook from Appendix
 \ref{subsec:Structural-properties-capacity} we hypothesize that
the first subset has $M_{\rho}$ molecules which are all far apart,
and the second subset has a close molecule in the first. Then, we
utilize the results of Appendix  \ref{subsec:The-molecule-distance}
to bound these terms. Specifically, the term $H(\tilde{Y}^{LM})-H(\tilde{Y}^{LM}\mid X^{LM},\tilde{U}^{M})$
loosely represents the mutual information obtained $\tilde{U}^{M}$
is known,\footnote{Though note that the first entropy term is $H(\tilde{Y}^{LM})$ and
not $H(\tilde{Y}^{LM}\mid\tilde{U}^{M})$.} and can be upper bounded using the far/close property of the molecules,
and our characterization that close molecules lead to a deficit in
the mutual information. The term $H(\tilde{U}^{M}\mid X^{LM},\tilde{Y}^{LM})$
represents the equivocation of a permutation given the input and the
output of a permuting channel. For the first subset, in which molecules
are far, this equivocation can be bounded using the properties derived
in Appendix  \ref{subsec:The-molecule-distance}. At the second stage,
we still assume a fixed composition codebook, but bound the average
mutual information over $Q^{N+1}$. As in the proof of the lower bounds,
here the Poissonization effect of the multinomial is utilized. At
the third stage, we remove the fixed composition assumption, and allow
$P_{X,m}$ to vary with $m$, yet show that identical $P_{X,m}$ for
all $m\in[M]$ does not asymptotically limit the mutual information.
The order of these steps is crucial -- loosely speaking, an argument
based on less stages would lead to an upper bound on the mutual information
in which the input distribution can be optimized separately for any
given binomial channel order, to wit
\[
\sum_{d\in\mathbb{N}^{+}}\max_{P_{X}^{(d)}\in{\cal P}({\cal X})}\pi_{\alpha}(d)\cdot\left[I(P_{X}^{(d)},W^{\oplus d})+\Omega_{d}(\beta,P_{X}^{(d)},W)\right]-\frac{1}{\beta}\left(1-\pi_{\alpha}(0)\right).
\]
Such a bound is clearly loose since the encoder of the DNA channel
does not know $S_{m}$ -- how many times each molecule is sampled
by the channel. Finally, using this upper bound in the Fano's-inequality
based argument completes the proof. 

\paragraph*{The gap between the upper and the lower bound on the capacity}

The upper bound of Theorem \ref{thm: converse capacity} and the lower
bound of Theorem \ref{thm: achievable capacity} match in case $\Omega_{d}(\beta,P_{X},W)=0$
for all $d\in\mathbb{N}^{+}$, for the maximizing input distribution
of (\ref{eq: capacity lower bound}). Specifically, it holds that
there exists a critical value for the molecule length parameter $\beta_{\text{cr}}$
such that the capacity is known for all $\beta>\beta_{\text{cr}}$,
as follows:
\begin{cor}
Let 
\[
P_{X}^{*}(\alpha,\beta,W)\in\argmax_{_{P_{X}\in{\cal P}({\cal X})}}\sum_{d\in\mathbb{N}^{+}}\pi_{\alpha}(d)\cdot\left[I(P_{X},W^{\oplus d})+\Omega_{d}(\beta,P_{X},W)\right],
\]
and let 
\begin{equation}
\beta_{\text{\emph{cr}}}(\alpha,W)\dfn\min\left\{ \beta\colon\beta\geq\frac{2}{\mathsf{CID}(P_{X}^{*}(\alpha,\beta,W),W)}\right\} \label{eq: critical beta}
\end{equation}
Then, for all $\beta\geq\beta_{\text{\emph{cr}}}(\alpha,W)$ 
\begin{equation}
C(\mathsf{DNA})=\sum_{d\in\mathbb{N}^{+}}\pi_{\alpha}(d)\cdot I\left(P_{X}^{*}(\alpha,\beta_{\text{\emph{cr}}}(\alpha,W),W),W^{\oplus d})\right)-\frac{1}{\beta}\left(1-\pi_{\alpha}(0)\right).\label{eq: capacity no gap}
\end{equation}
\end{cor}
\begin{IEEEproof}
It is evident from (\ref{eq: excess term in the capacity upper bound})
that for tightness of the lower and upper bounds on capacity, it must
hold that 
\begin{equation}
\beta\geq\max_{d\in\mathbb{N}^{+}}\frac{2}{\mathsf{CID}(P_{X}^{*},W^{\oplus d})}=\frac{2}{\mathsf{CID}(P_{X}^{*},W)},\label{eq: condition on beta}
\end{equation}
where the equality follows since $\mathsf{CID}(P_{X},W^{\oplus d})$
is monotonic increasing in $d$ (See Corollary \ref{cor: common input deficit for binomial channels}).
It is also evident from (\ref{eq: excess term in the capacity upper bound})
that if (\ref{eq: condition on beta}) holds for the maximizer of
\begin{equation}
\max_{P_{X}\in{\cal P}({\cal X})}\sum_{d\in\mathbb{N}^{+}}\pi_{\alpha}(d)\cdot\left[I(P_{X},W^{\oplus d})+\Omega_{d}(\beta,P_{X},W)\right]\label{eq: maximization of capacity upper bound without the ordering term}
\end{equation}
then it holds for any larger $\beta$. Indeed, in (\ref{eq: excess term in the capacity upper bound}),
whenever $\beta$ is increased the set $\{P_{X}\colon\mathsf{CID}(P_{X},W^{\oplus d})\ge\frac{2}{\beta}\}\subset{\cal P}({\cal X})$
expands, while the excess-rate value at the other two regimes, to
wit $\frac{1}{\beta}$ and $-\mathsf{CID}(P_{X},W^{\oplus d})+\frac{2}{\beta}$
decreases. This implies that the value of (\ref{eq: maximization of capacity upper bound without the ordering term})
is fixed for all $\beta>\beta_{\text{cr}}(\alpha,W)$, which directly
leads to the capacity expression (\ref{eq: capacity no gap}). 
\end{IEEEproof}
\begin{rem}
Note that both sides of the inequality in the definition of the critical
value of $\beta$ in (\ref{eq: critical beta}) depend on $\beta$.
Thus, the critical value of $\beta$ is, in fact, a solution to a
fixed point equation. If we consider the simpler solution $P_{X}^{**}\in\argmax_{_{P_{X}\in{\cal P}({\cal X})}}\sum_{d\in\mathbb{N}^{+}}\pi_{\alpha}(d)\cdot I(P_{X},W^{\oplus d})$,
then it clearly holds that both $P_{X}^{*}$ and $P_{X}^{**}$ achieve
capacity for all $\beta>\beta_{\text{cr}}$, however, it does not
seem to hold, in general, that $\beta_{\text{cr}}(\alpha,W)$ equals
$\frac{2}{\mathsf{CID}(P_{X}^{**},W)}$. To see this, consider for
example, $\overline{d}=1$, that is, a single term in the sum of (\ref{eq: capacity lower bound}).
It may hold that $\beta>\frac{2}{\mathsf{CID}(P_{X}^{**},W)}$ but
there exists $\tilde{P}_{X}$ for which $\mathsf{CID}(\tilde{P}_{X},W^{\oplus d})<\frac{1}{\beta}$
and 
\[
I(\tilde{P}_{X},W)+\Omega_{1}(\beta,\tilde{P}_{X},W)>I(P_{X}^{**},W)+\Omega_{1}(\beta,P_{X}^{**},W).
\]
So $P_{X}^{**}$ does not attain the maximum of (\ref{eq: capacity lower bound})
for this $\beta$, and it is required to solve (\ref{eq: critical beta})
(with equality sign replacing inequality) in order to find the critical
$\beta$. 
\end{rem}

\paragraph*{Prospective refinement of the upper bound}

In general, the upper bound of Theorem \ref{thm: converse capacity}
and the lower bound of Theorem \ref{thm: achievable capacity} do
not match, and one may wonder if the upper bound of Theorem \ref{thm: converse capacity}
can be improved. We next discuss a possible method to refine the upper
bound of Theorem \ref{thm: converse capacity}. Recall that the CID
is defined by a common input to a \emph{pair} of channels, and indeed,
the proof of the upper bound is based on considering the loss in capacity
due to \emph{pairs} of ``close'' molecules, according to the defined
distance. It seems plausible that by considering the mutual information
loss of \emph{triplets} of molecules, which are ``close'' according
to a proper definition of scattering of triplets {[}defined in a way
that generalizes the distance between pairs of molecules in Appendix
 \ref{subsec:The-molecule-distance}, see (\ref{eq: Distance between molecules definition}){]}
would lead to more lenient constraints, and will show that the lower
and upper bounds match even when (\ref{eq: condition on beta}) does
not hold. In turn, this can be further generalized to quadruplets,
quintuplets, etc. of molecules, and offer further improvements of
the upper bound. It is conceivable, however, that the intricacy of
the details required in such a proof method greatly outweighs their
effectiveness in improving the bound. 

\paragraph*{The assumption on maximal log-likelihood ratio}

Our converse result requires that the maximal log-likelihood ratio
of the sequencing channel $\nu_{\text{min}}(W)$ is finite. The source
of this assumption is an application of the \emph{blowing-up lemma}
\cite[Ch. 5]{csiszar2011information} \cite{marton1986simple,marton1996bounding}\cite[Lemma 3.6.1]{raginsky2018concentration}
in the proof (see Appendix  \ref{subsec:Preliminaries:-Common-input},
proof of Lemma \ref{lem:entropy difference between the outputs of close vectors}).
This assumption precludes our result for being applicable for a binary
erasure channel (BEC) sequencing channel, for example. It is not obvious
that this is merely a technical assumption that can be removed. The
reason is that for channels with unbounded $\nu_{\text{min}}(W)$,
the ordering of the output molecules seems to be an easier task. For
example, in a BEC, even a disagreement in a single bit of a candidate
pair of input and output molecules reveals that this output molecule
is not a sequencing of the input molecule. As discussed, the implicit
necessity to order the molecules affects capacity. Thus, an upper
bound on the capacity for such channels remains an open problem.

\section{Modulo-Additive Sequencing Channels \label{sec:Modulo-additive-sequencing}}

In this section, we consider sequencing channels which are \emph{modulo-additive},
whose most notable special case is the BSC. For such channels, ${\cal X}={\cal Y}=[|{\cal X}|]$
and the input-output relationship for each symbol can be written as
\begin{equation}
Y=X\oplus Z,\label{eq: module additive channel}
\end{equation}
where $Z$ is independent of $X$ and $\oplus$ denotes addition modulo
$|{\cal X}|$. The channel transition function, $W(y\mid x)$, is
determined by the p.m.f. of $Z$. When ${\cal X}=\{0,1\}$ the modulo-additive
channel is the BSC with crossover probability $w\dfn\P[Z=1]$. The
modulo-additive channel is symmetric and so the unique capacity achieving
input distribution is uniform. Furthermore, per Prop. \ref{prop: symmetry of binomial extension}
(Appendix  \ref{subsec:The-binomial-extension of symmetric DMC is symmetric}),
its binomial extension is also symmetric. For modulo-additive channels,
the sufficient condition for the upper and lower bounds to coincide
is simplified as follows:
\begin{prop}
\label{prop: Critical beta for modulo additive channels}Let $P_{X}^{(\text{\emph{unif}})}$
be the uniform distribution over ${\cal X}$, and let 
\[
\beta_{\text{\emph{cr}}}^{(\text{\emph{unif}})}(\alpha,W)\dfn\frac{2}{\mathsf{CID}(P_{X}^{(\text{\emph{unif}})},W)}.
\]
Then, for all $\beta>\beta_{\text{\emph{cr}}}^{(\text{\emph{unif}})}(\alpha,W)$
the capacity of the DNA channel with modulo-additive sequencing channel
$W$ is given by
\begin{equation}
C(\mathsf{DNA})=\sum_{d\in\mathbb{N}^{+}}\pi_{\alpha}(d)\cdot I(P_{X}^{(\text{\emph{unif}})},W^{\oplus d})-\frac{1}{\beta}\left(1-\pi_{\alpha}(0)\right).\label{eq: capacity for modulo additive channels}
\end{equation}
\end{prop}
The proof of Prop. \ref{prop: Critical beta for modulo additive channels}
appears in Appendix \ref{subsec:Proof-of-Proposition}. We next consider
the BSC case, and compare our result with that of \cite[Thm. 1]{lenz2019upper}:
\begin{example}[Critical $\beta$ for BSCs]
Consider the BSC with crossover probability $w$. The results of
\cite[Thm. 1]{lenz2019upper} and \cite[Thm. 1]{lenz2020achieving}
combined show that under the condition $w<1/8$, the right-hand side
(r.h.s.) of (\ref{eq: capacity for modulo additive channels}) is
the capacity for all\footnote{Note that here we use nats rather than bits. Thus here $\beta=\frac{L}{\log_{e}M}$
whereas in \cite[Theorem 1]{lenz2019upper} the notation used is $\frac{1}{\beta}=\frac{L}{\log_{2}M}$.
The next condition takes this unit scaling and inverted definition
into consideration. } 
\[
\beta>\overline{\beta}_{\text{cr}}\dfn\frac{2}{\log2-h_{b}(4w)}.
\]
Similar restrictions on $(\beta,w)$ have appeared in \cite{shomorony2021dna}.
Here, Prop. \ref{prop: Critical beta for modulo additive channels}
implies that the r.h.s. of (\ref{eq: capacity for modulo additive channels})
is the capacity as long as 
\[
\beta\geq\beta_{\text{cr}}^{(\text{unif})}(\alpha,\mathsf{BSC}(w))=\frac{2}{\log2-h_{b}(w*w)}=\frac{2}{\log2-h_{b}\left(2w(1-w)\right)},
\]
where $\mathsf{CID}(P_{X}^{(\text{unif})},W)=\mathsf{CID}((1/2,1/2),\mathsf{BSC}(w))=\log2-h_{b}(w*w)=\log2-h_{b}(2w(1-w))$,
and where $*$ is the binary convolution operator, defined as $a*b\dfn a(1-b)+b(1-a)$
for $a,b\in[0,1]$. Note that there is no restriction on $w$, and
that $\beta_{\text{cr}}^{(\text{unif})}(\alpha,\mathsf{BSC}(w))$
is finite for any $w\in(0,\frac{1}{2})$ (though approaches $\infty$
as $w\uparrow\frac{1}{2}$). Fig. \ref{fig:Comparison-between-beta for BSC}
numerically compares the $\beta_{\text{cr}}^{(\text{unif})}(\alpha,\mathsf{BSC}(w))$
with $\overline{\beta}_{\text{cr}}$. 
\begin{figure}
\begin{centering}
\includegraphics[scale=0.25]{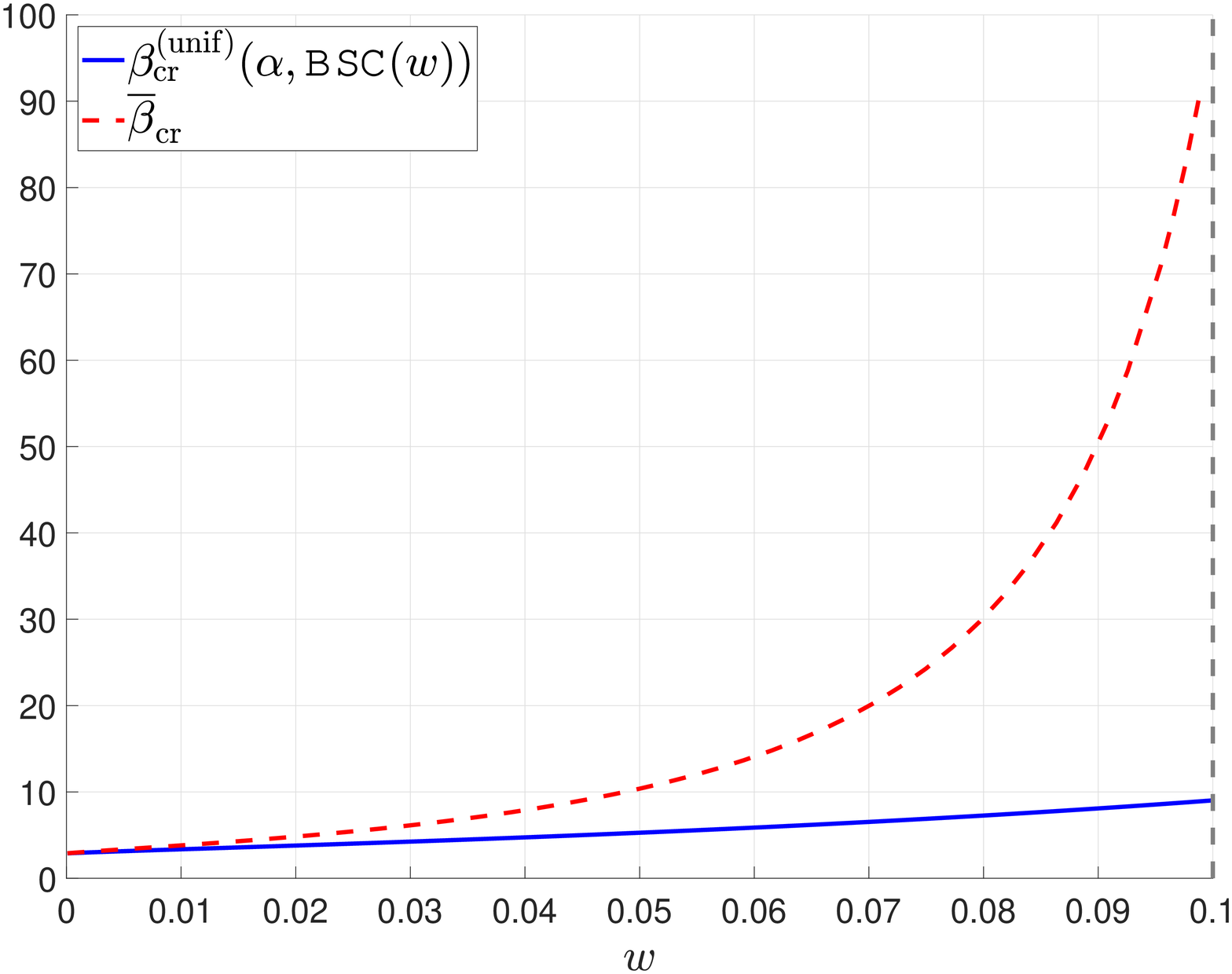}\includegraphics[scale=0.25]{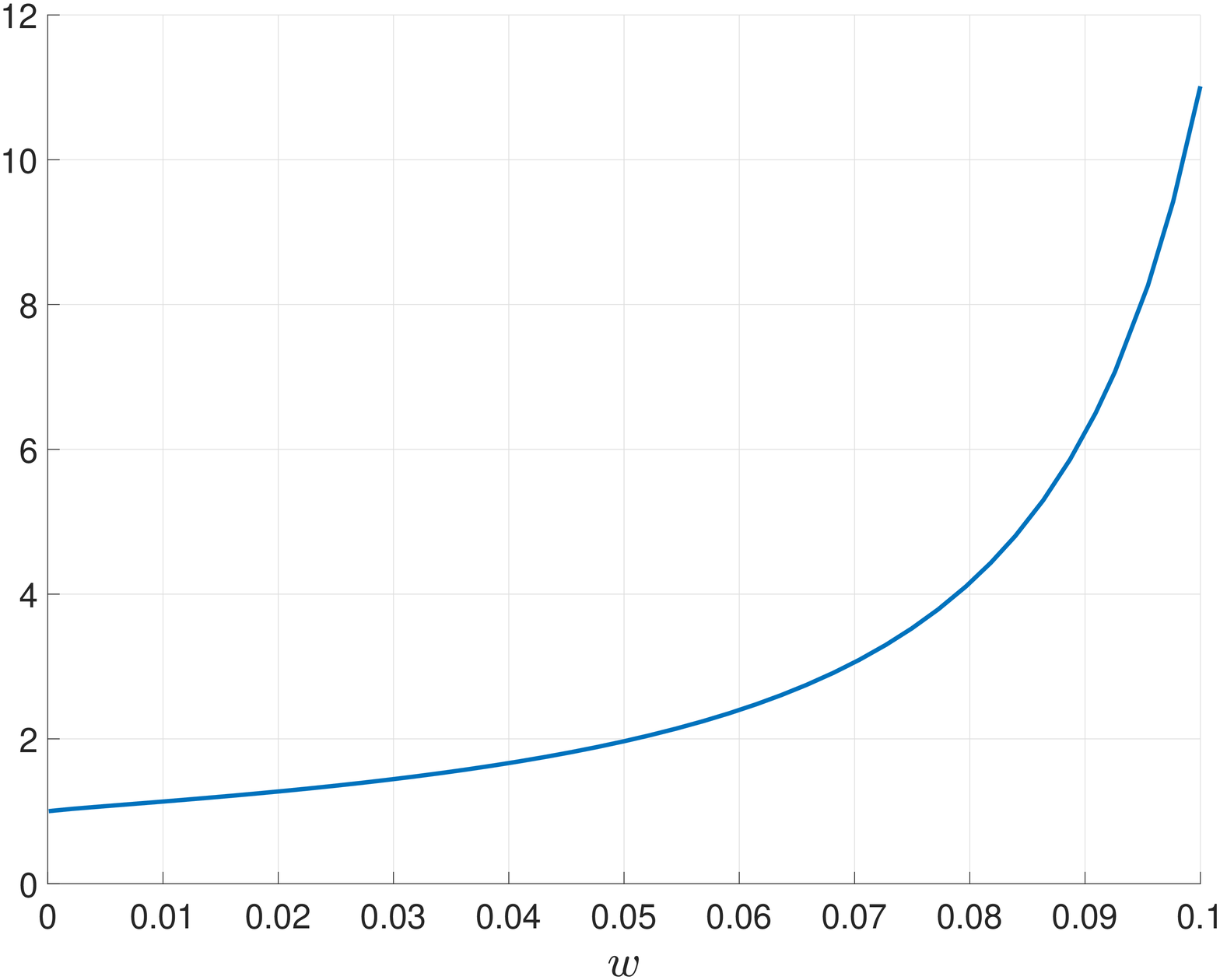}
\par\end{centering}
\caption{Comparison between \cite[Thm. 1]{lenz2019upper} and Prop. \ref{prop: Critical beta for modulo additive channels}
for BSC. Left -- $\beta_{\text{cr}}^{(\text{unif})}(\alpha,\mathsf{BSC}(w))$
and $\overline{\beta}_{\text{cr}}$ from \cite[Thm. 1]{lenz2019upper}.
Right -- the ratio $\overline{\beta}_{\text{cr}}/\beta_{\text{cr}}^{(\text{unif})}(\alpha,\mathsf{BSC}(w))$.
\label{fig:Comparison-between-beta for BSC}}

\end{figure}
\end{example}

\section{A Numerical Example \label{sec:A-numerical-example}}

\begin{figure}
\centering{}\includegraphics[scale=0.4]{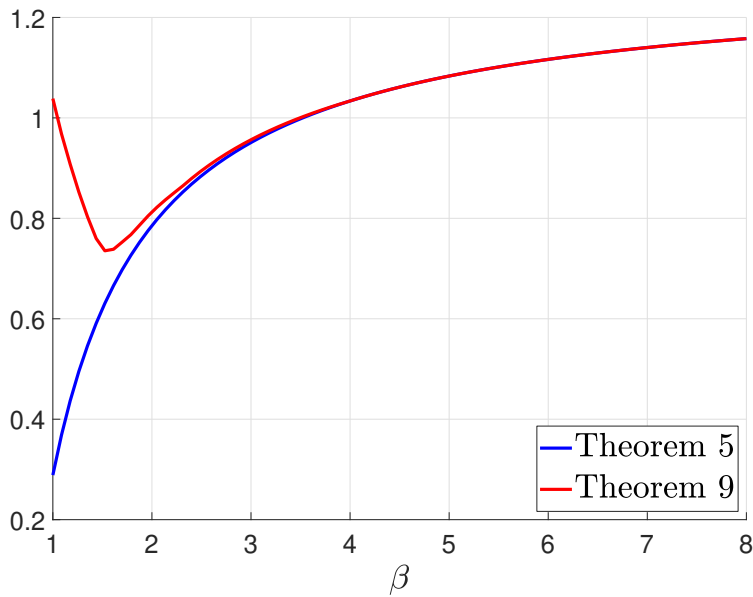}\caption{Upper and lower bounds on $C(\mathsf{DNA}(5,\beta,W_{0}))$ as a function
of $\beta$ (in nats).\label{fig: Capacity bounds example}}
\end{figure}

In general, all the bounds in this paper can be computed efficiently,
and a discussion on computational aspects of the bound appears in
Appendix \ref{sec:Computational-Aspects-of}. To exemplify our results,
we consider an alphabet ${\cal X}$ of size $|{\cal X}|=4$, which
is suitable to a practical DNA channel. For simplicity, we also set
$|{\cal Y}|=4$ and consider the sequencing DMC
\[
W_{0}(y\mid x)=\frac{1}{100}\cdot\left[\begin{array}{cccc}
94 & 2 & 2 & 2\\
2 & 70 & 25 & 3\\
3 & 2 & 85 & 10\\
10 & 5 & 5 & 80
\end{array}\right],
\]
which is an asymmetric channel. We assume $\alpha=5$ and truncate
our bounds with $\overline{d}=20$, so that the truncation error is
less than $10^{-6}$. The maximization over $P_{X}$ for the upper
and lower bounds is performed in two steps: First, a grid search over
the $(|{\cal X}|-1)$-dimensional simplex, with an accuracy of \textbf{$10^{-1}$
}(that is\textbf{ $P_{X}\in{\cal P}_{10}({\cal X})$) }and then refining
the result using Matlab's \texttt{fmincon} function (with its default
interior-point algorithm). The capacity lower and upper bounds for
various values of $\beta$ is plotted in Fig. \ref{fig: Capacity bounds example}.
For low values of $\beta$ (about $\beta\lesssim1.6$) the upper bound
is not monotonic increasing in $\beta$ (as might be expected from
the true capacity), and is most likely off at this regime. For larger
values, $1.6\lesssim\beta\lesssim3$ the upper bound is monotonic
increasing and is slightly above the lower bound. For $\beta\gtrsim3.5$
the distinction between the bounds is indiscernible. 

\begin{figure}
\centering{}\includegraphics[scale=0.4]{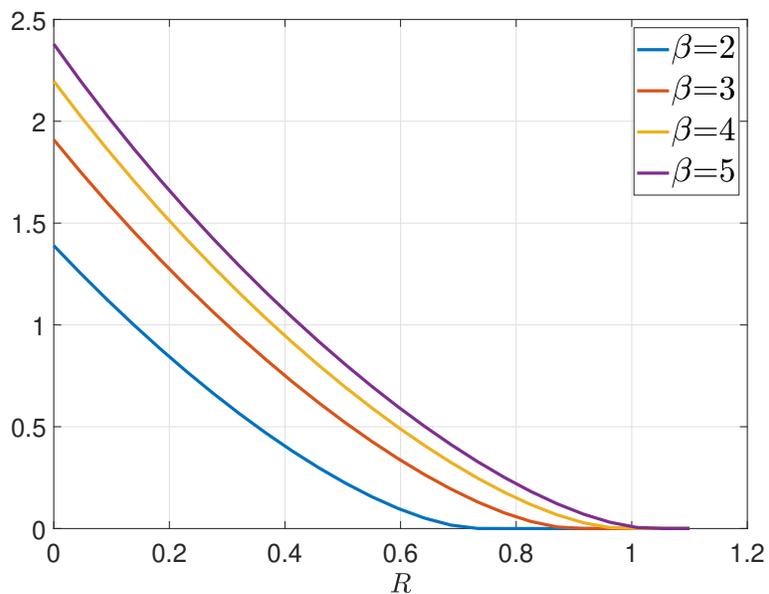}\caption{Lower bound on the reliability function $E^{*}(R,\mathsf{DNA}(5,\beta,W_{0}),\{M\})$
as a function of $R$ (in nats). \label{fig:Reliabilty function}}
\end{figure}

In Fig. \ref{fig:Reliabilty function} we plot the lower bound on
the reliability function, for $\alpha=5$, various values of $\beta$,
and using a truncation of $\overline{d}=10$. For simplicity of computation,
we do not optimize over $P_{X}$, but rather use the uniform $P_{X}$.
The advantage of using a larger $\beta$ on the reliability function
is easily observed. 


\section{Summary and Open Problems \label{sec:Summary-and-open}}

In this paper we have derived lower and upper bounds on the capacity
of the memoryless sequencing channel. In terms of lower bounds, we
show that the lower bound which was only known for BSC (or symmetric)
sequencing channels, and is restricted to a partial regime of the
molecule length parameter $\beta$ and the channel noise, holds in
fact for general DMCs and without any restrictions. We have shown
that this bound is achievable by a universal decoder, and obtained
asymptotic bounds on its error probability. This has revealed that
the DNA channel suffers from \emph{outage: }The random amplification
vector $Q^{N+1}$ is a measure of the quality of the channel, in terms
of the conditional mutual information it induces. When the mutual
information is larger than the required coding rate, the error probability
decays fast -- exponentially w.r.t. scaling $K_{M}=ML=\beta M\log M$.
However, the probability that $Q^{N+1}$ is such that the required
coding rate is not supported, the error probability is large.\footnote{At least, we do not have a non-trivial bound for it.}
Our bound on the probability of such events is exponential w.r.t.
scaling $K_{M}=M$ and so this is the dominant error event. In terms
of upper bounds (converse), we have refined the analysis of \cite{shomorony2021dna,lenz2019upper}
by introducing of a general distance function between molecules. This
resulted a tighter upper bound, which agrees with the lower bound
for a larger regime of $\beta$ and $W$. 

The following problems remain open for the memoryless sequencing channel:
\begin{enumerate}
\item Matching lower and upper bounds: It is tempting to conjecture that
the lower bound of Theorem \ref{thm: achievable capacity} is the
true capacity for any $\beta$ and $W$. This implies that the upper
bound of Theorem \ref{thm: converse capacity} is the one to be tightened.
However, as we have discussed after Theorem \ref{thm: converse capacity},
it is conceivable that we have pushed the capability of the converse
arguments of \cite{shomorony2021dna,lenz2019upper} to their maximum,
at least assuming reasonably complicated arguments.\footnote{Except perhaps, for the possibility of considering groups of molecules
larger than pairs -- as discussed after Theorem \ref{thm: converse capacity}.} Hence, it is of interest to find proof arguments of substantially
different nature that would be capable of establishing capacity in
the high-noise/low $\beta$ regime. 
\item Upper bounds on the reliability function: The lower bound derived
here (Theorem \ref{thm: achievable reliability function}) is based
on an analysis of the molecule duplicate vector $S^{M}\in[N]^{M}$.
In the proof of the lower bond, the multinomial distribution of $S^{M}$
-- for which the components $S_{m}$ are identically distributed
but statistically dependent -- is replaced by $\tilde{S}^{M}\in[N]^{M}$
which follows an i.i.d. Poisson distribution $\text{Pois}(\alpha)$.
This facilitates the analysis, but may not be tight. Specifically,
the analysis Lemma \ref{lem: Poissonization of outage event} is not
tight in general. As an illustrative example, in the Poisson model,
the probability that none of the molecules is sampled is given by
\[
\P\left[\cap_{m\in[M]}\tilde{S}_{m}=0\right]=(1-e^{-\alpha})^{M}=e^{-M\log(1-e^{-\alpha})}.
\]
This probability is exponential w.r.t. $M$ and thus affect the exponential
decay of the error probability at this scale. For example, $E^{*}(R)\leq\log(1-e^{-\alpha})$
because the error probability is large if none of the molecules have
been sampled. By contrast, in the true multinomial model $\P[\cap_{m\in[M]}S_{m}=0]=0$.
It is therefore of interest to find tight tails bounds on the multinomial
$S^{M}$, and utilize them to obtain upper bounds on the reliability
function. 
\item Tight finite blocklength bounds: Finite length bounds are of specific
interest in this problem, since the second-order terms in this problem
vanish very slow. In our arguments, the decay rate is either $O(\frac{1}{\log M})$
and occasionally or even $O(\frac{\log\log M}{\log M})$. This seems
to be an inherent aspect of the problem. For example, consider a basic
model in which $N=M$, each molecule is sampled exactly once (with
probability $1$), and the sequencing channel is noiseless. Then,
the capacity is the normalized logarithm of the possible number of
sequences. The stars and bars model (see footnote \ref{fn:The-stars-and-bars-model})
implies that it capacity is $\frac{1}{ML}\log{M+M^{\beta}-1 \choose M}$,
which by standard approximation of the binomial coefficient is $1-\frac{1}{\beta}+O(\frac{1}{\log M})$. 
\end{enumerate}

\section*{Acknowledgment}

The advice of Ido Tal regarding the computation of the bounds is acknowledged
with gratitude. 

\appendices{\numberwithin{equation}{section}}

\section{Proofs of Lower Bounds \label{sec: Achievability proofs} \label{sec:Achievability-proofs-}}

\begin{figure}
\begin{centering}
\includegraphics[scale=1.25]{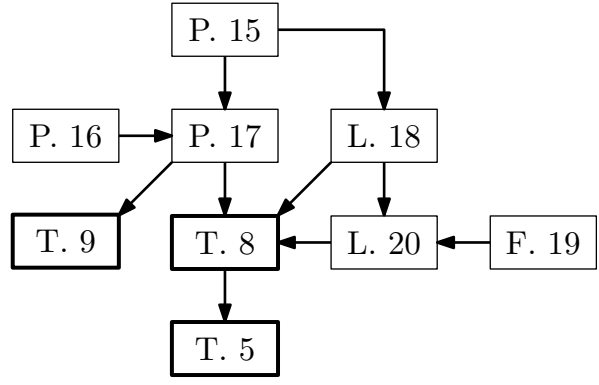}\caption{Roadmap for the proofs of the lower bounds (achievability). The used
abbreviations are Lemma (L.), Proposition (P.), Theorem (T.) and Fact
(F.). The result at the tail of an arrow is utilized to prove the
result pointed by its head. \label{fig:Roadmap-lower}}
\par\end{centering}
\end{figure}

In this section, we prove the lower (achievability) bounds. A roadmap
for the various results proved is provided in Fig. \ref{fig:Roadmap-lower}. 

Based on Theorem \ref{thm: achievable reliability function}, the
proof of Theorem \ref{thm: achievable capacity} is short and standard.
Thus, we provide it here before the proof of Theorem \ref{thm: achievable reliability function},
assuming the latter holds. 
\begin{IEEEproof}[Proof of Theorem \ref{thm: achievable capacity}]
 Let an arbitrary input distribution, $P_{X}\in{\cal P}({\cal X})$,
be given. By the strict positivity of the binary KL divergence $d_{b}(p\mid\mid q)>0$
for all $p\neq q$
\[
\sum_{d\in\mathbb{N}}\left(1-\sum_{i\in[d]}\theta_{i}\right)\cdot d_{b}\left(\frac{\theta_{d}}{1-\sum_{i\in[d]}\theta_{i}}\,\middle\vert\middle\vert\,\pi_{\alpha|\geq d}(d)\right)\geq0,
\]
and a simple calculation shows that equality holds if and only if
$\theta_{d}=\pi_{\alpha}(d)$ for all $d\in\mathbb{N}^{+}$. Let $R$
be given so that
\[
R<\sum_{d\in\mathbb{N}}\pi_{\alpha}(d)\cdot I(P_{X},W^{\oplus d})-\frac{1}{\beta}(1-\pi_{\alpha}(0)).
\]
Since $\{\pi_{\alpha}(d)\}$ does not belong to the feasible set (\ref{eq: minimization set for error exponent})
of the minimization (\ref{eq: error exponent for universal decoder}),
it holds that $E^{*}(R,\mathsf{DNA},\{M\})>0$. Hence, there exists
a sequence of codes-decoders $({\cal C}_{M},{\cal D}_{M})$ so that
the average error probability satisfies 
\[
\frac{1}{|{\cal C}_{M}|}\sum_{j\in[|{\cal C}_{M}|]}\pe({\cal C},{\cal D}\mid x^{LM}(j))\leq e^{-\Theta(M)}.
\]
A similar bound can be proved for the maximal error probability for
a codebook of rate slightly smaller, say $R-\delta$, for some $\delta>0$
(and all $M$ sufficiently large) by expurgating half of the codewords
from each codebook with the largest $\pe({\cal C},{\cal D}\mid x^{LM}(j))$.
Thus, $R-\delta$ is an achievable rate for any $\epsilon\in(0,1)$
and so the lower bound on capacity follows.
\end{IEEEproof}
The rest of this section is devoted to the proof of Theorem \ref{thm: achievable reliability function}.

\subsection{Preliminaries: Sampling Type Classes \label{subsec:Preliminaries:-Sampling-type}}

Recall that at the sampling stage, the pool of $L$ molecules $\{x_{m}^{L}\}_{m\in[M]}$
is sampled $N$ times, where $u^{N}\in[M]^{N}$, the molecule index
vector, is such that $u_{n}=m$ if $x_{m}^{L}$ is the sampled molecule
at the $n$th sampling trial. Recall also the definitions of the molecule
duplicate vector $s^{M}\in[N]^{M}$ and the amplification vector $q^{N+1}\in[M+1]^{N+1}$
from Sec. \ref{subsec:Formulation-of-the}. The triplet $(u^{N},s^{M},q^{N+1})$
can be described by the empirical count operator\emph{ }(\ref{eq: empirical count operator})
as $s^{M}=\mathscr{\mathscr{N}}(u^{N})$ and $q^{N+1}=\mathscr{\mathscr{N}}(s^{M})=\mathscr{\mathscr{N}}^{(2)}(u^{N})$,
and then, \emph{sampling type classes} can be defined as follows:
The \emph{molecule index type class} of $s^{M}$ is defined as
\[
\mathscr{T}_{s^{M}}\dfn\left\{ u^{N}\in[M]^{N}\colon\mathscr{\mathscr{N}}(u^{N})=s^{M}\right\} ,
\]
and the \emph{molecule duplicate type class} of a $q^{N+1}$ is defined
as
\[
\mathscr{T}_{q^{N+1}}\dfn\left\{ s^{M}\in[N]^{M}\colon\mathscr{\mathscr{N}}(s^{M})=q^{N+1}\right\} .
\]
In a similar fashion, we define the \emph{amplification type class}
as 
\[
\mathscr{T}_{q^{N+1}}^{(2)}\dfn\left\{ u^{N}\in[M]^{N}\colon\mathscr{\mathscr{N}}^{(2)}(u^{N})=q^{N+1}\right\} .
\]
Let 
\[
\mathscr{Q}(M,N)\dfn\left\{ q^{N+1}\in[M+1]^{N+1}\colon\sum_{d\in[N+1]}q_{d}=M,\;\sum_{d\in[N+1]}dq_{d}=N\right\} 
\]
be the set of all valid amplification vectors for $M$ input molecules
and $N$ output molecules. The next proposition provides a tight asymptotic
scaling of the sizes of the molecule index type class, the molecule
duplicate type class, and the amplification type class. This will
be used in the proofs of the lower bounds on capacity and reliability
function.
\begin{prop}
\label{prop: amplification type sizes estimates }For a coverage depth
$\alpha=\frac{N}{M}>0$:
\begin{enumerate}
\item The number of types is bounded as $|\mathscr{Q}(M,N)|\leq e^{[1+o(1)]\sqrt{\frac{2\pi^{2}}{3}M}}$. 
\item The size of the amplification type class is 
\[
|\mathscr{T}_{q^{N+1}}^{(2)}|=\exp\left[\left(1-\frac{q_{0}}{M}\right)\cdot M\log M+\epsilon_{M}\cdot M\right],
\]
where $|\epsilon_{M}|\leq\log\alpha^{2}e$.
\end{enumerate}
\end{prop}
\begin{IEEEproof}
~
\end{IEEEproof}
\begin{enumerate}
\item It is evident that each $q^{N+1}$ uniquely defines a restricted \emph{partition}
of $N$ to at most $M=\frac{N}{\alpha}$ summands (a bijection can
be defined between $q^{N+1}\in\mathscr{Q}(M,N)$ and a restricted
partition). For example, $q^{N+1}=(M-5,2,2,,\ldots0,\underbrace{1}_{q_{N-6}},\ldots,0,0)$
corresponds to the partition $N=1+1+2+2+(N-6)$. The number of restricted
partitions is asymptotically bounded as claimed by the celebrated
analysis of Hardy and Ramanujan, Uspensky, and Rademacher \cite[Ch. 5]{apostol2012modular}. 
\item Let $s^{M}$ and $q^{N+1}$ be such that $\mathscr{\mathscr{N}}(s^{M})=q^{N+1}$.
The exact value of $|\mathscr{T}_{q^{N+1}}|$ and $|\mathscr{T}_{s^{M}}|$
can be easily derived, and then $|\mathscr{T}_{q^{N+1}}^{(2)}|$ is
evaluated by $|\mathscr{T}_{q^{N+1}}^{(2)}|=|\mathscr{T}_{q^{N+1}}|\cdot|\mathscr{T}_{s^{M}}|$.
The size of $\mathscr{T}_{q^{N+1}}$ is the number of unique permutations
of the vector
\[
(\underbrace{0,0,\ldots0}_{q_{0}},\underbrace{1,1\ldots1}_{q_{1}},\ldots,\underbrace{d,d,\ldots,d}_{q_{d}},\ldots)\in[N+1]^{M}
\]
given by 
\[
|\mathscr{T}_{q^{N+1}}|={M \choose q_{0},q_{1},\ldots,q_{N}}=\frac{M!}{\prod_{d=0}^{N}q_{d}!}.
\]
It similarly holds that that size of $\mathscr{T}_{s^{M}}$ is 
\[
|\mathscr{T}_{s^{M}}|={N \choose s_{0},s_{1},\ldots,s_{M-1}}=\frac{N!}{\prod_{m=0}^{M-1}s_{m}!}=\frac{N!}{\prod_{d=0}^{N}(d!)^{q_{d}}},
\]
where the last equality holds by the definition $q^{N+1}=\mathscr{N}(s^{M})$.
To prove the claimed upper bound we first bound $|\mathscr{T}_{s^{M}}|$
by the number of ways to choose each of the $N$ entries of $u^{N}$
from a set of possible $M-q_{0}$ molecules (without taking into account
any other constraints on its empirical distribution). Thus
\begin{equation}
|\mathscr{T}_{s^{M}}|\leq N^{M-q_{0}}.\label{eq: upper bound on the sampling type class size}
\end{equation}
To bound the size of $\mathscr{T}_{q^{N+1}}$, let $\theta_{d}\dfn\frac{q_{d}}{M}$
so that $\theta=(\theta_{0},\ldots,\theta_{N})$ is a p.m.f., and
since $\sum_{d\in[N+1]}dq_{d}=N$ then $\sum_{d\in[N+1]}d\theta_{d}=\alpha$
must hold. Let $H(\theta)\dfn-\sum_{d\in[N+1]}\theta_{d}\log\theta_{d}$
be the entropy of $\theta$. Then, 
\begin{equation}
|\mathscr{T}_{q^{N+1}}|\trre[\leq,a]\exp\left[M\cdot H(\theta)\right]\trre[\leq,b]\exp\left[M\cdot\log\alpha e\right],\label{eq: samplimg type class upper bound max entropy}
\end{equation}
where $(a)$ is by the standard exponential upper bound of the multinomial
coefficient with the entropy being the exponent (just as bounding
the type class size in ordinary method of types \cite[Lemma 2.3]{csiszar2011information}),
and $(b)$ is by upper bounding $H(\theta)$ by the maximum entropy
under the mean constraint $\alpha$ \cite[Lemma 3.12 and Corollary 3.12]{csiszar2011information}
(it is easy to verify and well-known that the maximum entropy p.m.f.
in this case is geometric). Hence, $|\mathscr{T}_{q^{N+1}}|\leq e^{\Theta(M)}$.
Combining (\ref{eq: samplimg type class upper bound max entropy})
with (\ref{eq: upper bound on the sampling type class size}) and
$N=\alpha M$ results in
\[
|\mathscr{T}_{q^{N+1}}^{(2)}|=|\mathscr{T}_{q^{N+1}}|\cdot|\mathscr{T}_{s^{M}}|\leq e^{(M-q_{0})\log N+\Theta(M)}=e^{(M-q_{0})\log M+\Theta(M)}.
\]
To prove the claimed lower bound, let $\sigma_{m}\dfn\frac{s_{m}}{N}$,
so that $\sigma=(\sigma_{0},\ldots,\sigma_{M-1})$ is a p.m.f.. Here
we lower bound $|\mathscr{T}_{s^{M}}|$ by the exponent of the entropy
$H(\sigma)$, divided by a factor which will be shown to be $e^{\Theta(M)}$.
The proof is almost identical to the one used to lower bound the size
of the type class (a multinomial coefficient) in the standard method
of types: Considering $\sigma$ as the empirical type of $u\in[M]^{N}$
-- an $N$ dimensional vector from alphabet of size $M$ -- the
exact number of possible types is ${N+M-1 \choose M-1}$ (this follows
from the stars and bars model.\footnote{\label{fn:The-stars-and-bars-model}The stars and bars model: Let
$a,n,k\in\mathbb{N}^{+}$ such that $n\geq k$ and $n>ak$. The number
of ways to distribute $n$ stars into $k$ different bins (i.e., the
number of integer solutions $(x_{0},\ldots,x_{k-1})$ to $\sum_{i\in[k]}x_{i}=n$)
such that $x_{i}\geq a$ for all $i\in[k]$ is ${n-k(a-1)-1 \choose k-1}$.} See, also, e.g., \cite[Ex. 2.1]{csiszar2011information}). Then,
the standard estimate on a type class size \cite[Thm. 11.1.3]{cover2012elements}
\cite[Lemma 2.3]{csiszar2011information} implies that\footnote{In the stated results in both \cite[Thm. 11.1.3]{cover2012elements}
and \cite[Lemma 2.3]{csiszar2011information}, as well as in their
proofs, the pre-exponent is given by the inverse of $(N+1)^{M}$,
the latter being an \emph{upper} bound to the exact value ${N+M-1 \choose M-1}$.
Such a bound is useful in the standard method of types, since in that
analysis $N\to\infty$ while the alphabet size is $M=\Theta(1)$.
Here $M\to\infty$ and $N=\alpha M=\Theta(M)\to\infty$ and so a different
bound is required. Nonetheless, the proof itself carries over verbatim,
except for replacing the bound $(N+1)^{M}$ with the exact value ${N+M-1 \choose M-1}$.}
\begin{equation}
|\mathscr{T}_{s^{M}}|\geq\frac{1}{{N+M-1 \choose M-1}}\exp\left[N\cdot H(\sigma)\right].\label{eq: lower bound on s type size in terms of entropy}
\end{equation}
By the standard entropy bound on the binomial coefficient \textbf{
\[
{N+M-1 \choose M-1}\leq e^{(N+M-1)\cdot h_{b}(\frac{M}{N+M})}\leq e^{(\alpha+1)M\cdot h_{b}(\frac{1}{\alpha+1})}=e^{\Theta(M)},
\]
}and so continuing (\ref{eq: lower bound on s type size in terms of entropy})
\[
|\mathscr{T}_{s^{M}}|\geq\exp\left[N\cdot H(\sigma)-\Theta(M)\right]\geq\exp\left[N\cdot\min H(\zeta)-\Theta(M)\right],
\]
where the minimum is over p.m.f's $\zeta$ which have support of $M-q_{0}$.
We next show that the minimal entropy is obtained at 
\[
\zeta^{*}=\frac{1}{N}(\underbrace{0,0\ldots0}_{q_{0}},N-(M-q_{0}-1),\underbrace{1,1\ldots,1,1,1,1,\ldots1}_{M-q_{0}-1}).
\]
To see this, note that any other $\tilde{\zeta}$ which is supported
on exactly $M-q_{0}$ indices, and has entries which are integer multiples
of $1/N$ is \emph{majorized }by $\zeta^{*}$ \cite{marshall1979inequalities}.\emph{}\footnote{For any $a^{K}\in\mathbb{R}^{K}$, let $a_{\downarrow}^{K}\in\mathbb{R}^{K}$
be such that $\mathscr{N}(a^{K})=\mathscr{N}(a_{\downarrow}^{K})$
and the components of $a_{\downarrow}^{K}$ are sorted in descending
order. Then $a^{K}\in\mathbb{R}^{K}$ majorizes $b^{K}\in\mathbb{R}^{K}$
if $\sum_{i\in[k]}a_{\downarrow,i}\geq\sum_{i\in[k]}b_{\downarrow,i}$
for all $k\in[K]$.} It then follows from the Schur-concavity of the entropy function
that $H(\zeta^{*})\leq H(\tilde{\zeta})$ \cite[Ch. 3.D.1]{marshall1979inequalities}.
The resulting entropy is then 
\begin{align}
H(\zeta^{*}) & =\left[\frac{N-(M-q_{0}-1)}{N}\right]\log\left[\frac{N}{N-(M-q_{0}-1)}\right]+\frac{(M-q_{0}-1)}{N}\log N\\
 & \geq\frac{(M-q_{0}-1)}{N}\log N.
\end{align}
Thus, 
\begin{align}
|\mathscr{T}_{s^{M}}| & \geq\exp\left[N\cdot\frac{(M-q_{0}-1)}{N}\log N-\Theta(M)\right]\\
 & \geq\exp\left[(M-q_{0})\log M-\Theta(M)\right].
\end{align}
The proof is completed since $|\mathscr{T}_{q^{N+1}}^{(2)}|=|\mathscr{T}_{q^{N+1}}|\cdot|\mathscr{T}_{s^{M}}|\geq|\mathscr{T}_{s^{M}}|$
as $|\mathscr{T}_{q^{N+1}}|\geq1$. 
\end{enumerate}

\subsection{Preliminaries: The DNA Channel as a Mixture of Binomial Channels
\label{subsec:Preliminaries:-The-DNA}}

In this section, we develop an alternative representation for the
likelihood function of the DNA channel with a memoryless sequencing
channel. In the DNA channel, each symbol in the codeword is independently
sequenced a variable number of times, according to the number of times
the molecule it belongs to was sampled in the sampling stage. As a
result, the likelihood function of all symbols is a (random) mixture
of binomial channels. In this section, we make this property explicit
in the likelihood function. 

Let $x^{LM}$ be the input and $y^{LN}$ be the corresponding output
of the DNA channel. Assume, as usual, that the sampling step is such
that $U_{n}\sim\text{Uniform}[M]$, i.i.d., independent of $x^{LM}$.
The basic expression for the likelihood function starts from (\ref{eq: likelihood DNA channel})
and is further developed as follows: 
\begin{align}
{\cal L}\left[y^{LN}\mid x^{LM}\right] & =\sum_{u^{N}\in[M]^{N}}\P[U^{N}=u^{N}]\prod_{n\in[N]}W^{L}\left[y_{n}^{L}\mid x_{u_{n}}^{L}\right]\\
 & =\sum_{q^{N+1}\in\mathscr{Q}(M,N)}\sum_{s\in\mathscr{T}_{q^{N+1}}}\sum_{u\in\mathscr{T}_{s^{M}}}\P\left[U^{N}=u^{N}\right]\prod_{n=0}^{N}W^{L}\left[y_{n}^{L}\mid x_{u_{n}}^{L}\right]\\
 & =\sum_{q^{N+1}\in\mathscr{Q}(M,N)}\sum_{s\in\mathscr{T}_{q^{N+1}}}\P\left[\mathscr{T}_{s^{M}}\right]\sum_{u\in\mathscr{T}_{s^{M}}}\frac{1}{|\mathscr{T}_{s^{M}}|}\prod_{n=0}^{N}W^{L}\left[y_{n}^{L}\mid x_{u_{n}}^{L}\right]\\
 & =\sum_{q^{N+1}\in\mathscr{Q}(M,N)}\P\left[\mathscr{T}_{q^{N+1}}^{(2)}\right]\sum_{u\in\mathscr{T}_{q^{N+1}}^{(2)}}\frac{1}{|\mathscr{T}_{q^{N+1}}^{(2)}|}\prod_{n=0}^{N}W^{L}\left[y_{n}^{L}\mid x_{u_{n}}^{L}\right],\label{eq: derivation of likelihood function}
\end{align}
where $\P[\mathscr{T}_{q^{N+1}}^{(2)}]=|\mathscr{T}_{q^{N+1}}^{(2)}|\cdot\P[U^{N}=u^{N}]=|\mathscr{T}_{q^{N+1}}^{(2)}|/M^{N}$. 

Consider a fixed $(u^{N},s^{M},q^{N+1})$, where $u^{N}\in\mathscr{T}_{s^{M}}$,
$s^{M}\in\mathscr{T}_{q^{N+1}}$ and $q^{N+1}\in\mathscr{Q}(M,N)$.
We next present the conditional likelihood term, $\prod_{n=0}^{N-1}W^{L}[y_{n}^{L}\mid x_{u_{n}}^{L}]$,
in an alternative form, which better expresses the mixture-of-binomial
nature of this channel. Specifically, in this form, the symbols of
the molecules that were sampled the same number of times are grouped
together. Let $K=ML$ be the total number of symbols in the codeword
$x^{LM}$ and let $a^{K}\in{\cal A}^{K}$ be their concatenation in
an arbitrary fixed order. For concreteness, we choose the natural
ordering, that is, the one which satisfies $x_{0}^{L}=(a_{0},a_{1},\ldots,a_{L-1})$,
$x_{1}^{L}=(a_{L},a_{L+1},\ldots,a_{2L-1})$, etc. Given $(u^{N},s^{M},q^{N+1})$,
when $a_{k}$ belongs to a molecule $x_{m}$ which has been sampled
$d_{k}$ times, it has $d_{k}$ output symbols in $y^{LN}$. Let $b_{k}^{d_{k}}=(b_{k,0},\ldots,b_{k,d_{k}-1})\in{\cal Y}^{d_{k}}$
be those $d_{k}$ output symbols, ordered according to their order
in $y$, and let $(b^{*})^{K}=(b_{0}^{d_{0}},\ldots,b_{K-1}^{d_{K-1}})$.
It should be noted that $d_{k}$, the dimension of $b_{k}$ may be
different for each $k$, and depends on $(u^{N},s^{M},q^{N+1})$.
Furthermore, $d_{k}\in[N+1]$ since the maximal number of times a
single molecule can be sampled is $N$.\footnote{Which only occurs in the unlikely event that only a single molecule
has been sampled $N$ times.} 

With this interpretation, we now express the likelihood function in
terms of $W^{\oplus d}$, the $d$-order binomial extension of $W$
(Definition \ref{def: binomial channel}), Specifically, we summarize
the correspondence between $(x^{LM},y^{LN})$ and $(a^{K},b^{*K})$
(with $K=ML$) as follows: The molecule index vector $u^{N}$ transforms
an input-output pair $(x^{LM},y^{LN})$ to an equivalent input-output
pair $(a^{K},b^{*K})$ and a partition $\{{\cal K}_{d}(u^{N})\}_{d\in[N+1]}$
of $[K]=[ML]$ with $K_{d}=|{\cal K}_{d}(u)|=q_{d}L$ such that the
dimension of $b_{k}^{d_{k}}$, is $d_{k}=d$ if and only if $k\in{\cal K}_{d}(u^{N})$.
For brevity, we have omitted in this notation the explicit dependency
of $(a^{K},b^{*K})$ in $(x^{LM},y^{LN})$. It thus holds that
\begin{equation}
{\cal L}\left[y^{LN}\mid x^{LM},u^{N}\right]=\prod_{d=0}^{N}W^{\oplus d}\left[b_{{\cal K}_{d}(u^{N})}^{d}\mid a_{{\cal K}_{d}(u^{N})}\right],\label{eq: likelihood DNA as mixture of binomials conditioned on u}
\end{equation}
and combining this with (\ref{eq: derivation of likelihood function}),
leads to the likelihood form 
\begin{equation}
{\cal L}\left[y^{LN}\mid x^{LM}\right]=\sum_{q^{N+1}\in\mathscr{Q}(M,N)}\P\left[U^{N}\in\mathscr{T}_{q^{N+1}}^{(2)}\right]\sum_{u^{N}\in\mathscr{T}_{q^{N+1}}^{(2)}}\frac{1}{|\mathscr{T}_{q^{N+1}}^{(2)}|}\prod_{d=0}^{N}W^{\oplus d}\left[b_{{\cal K}_{d}(u^{N})}^{d}\mid a_{{\cal K}_{d}(u^{N})}\right].\label{eq: likelihood DNA as mixture of binomials}
\end{equation}
For brevity, we further simplify ${\cal K}_{d}\equiv{\cal K}_{d}(u^{N})$.
Furthermore, we note that the likelihood (\ref{eq: likelihood DNA as mixture of binomials conditioned on u}),
conditioned on $u^{N}$, is determined by the joint type in the $d$-order
binomial $a_{{\cal K}_{d}}\to b_{{\cal K}_{d}}^{d}$ channel given
by 
\[
\hat{P}^{d}(x^{LM},y^{LN};u^{N})\dfn\mathscr{P}(a_{{\cal K}_{d}},b_{{\cal K}_{d}}^{d})
\]
for all $d\in[N+1]$. 

\subsection{A Universal Decoder for the DNA Channel \label{subsec:A-universal-decoder}}

When the codewords are chosen under the uniform probability distribution,
the optimal decoding rule in terms of error probability is the (maximum
likelihood) ML\emph{ }decoding rule. However, it is also well known
that in random coding analysis of DMCs with a fixed composition ensemble,
the ML rule can be replaced by the MMI rule, without degrading the
random coding error exponent \cite[Ch. 10]{csiszar1998method} \cite[Thm. IV.1]{csiszar1998method}.
Here, however, we will analyze the i.i.d. random coding ensemble,
in which the $ML\cdot|{\cal C}|$ symbols in the codebook ${\cal C}=\{x^{LM}(j)\}$
are drawn i.i.d. from a distribution $P_{X}$. In this ensemble, the
codewords do not have fixed composition with probability $1$ and
so the standard MMI is not universal (see the discussion after the
proof of Prop. \ref{prop: random coding exponent condition on q}
in the next subsection for the reason to preferring this ensemble
of the fixed composition ensemble). Nonetheless, for this ensemble,
the random coding exponent of the ML rule can be achieved by a variant
of the MMI, which includes an additional penalty term. 

In this section, we first briefly describe the standard MMI rule and
then adapt it to DNA channel. We then modify the basic MMI with two
penalty terms: One is related to the use of the i.i.d. ensemble (instead
of the constant composition ensemble), and the other -- to the sampling
mechanism of the DNA channel. 

Specifically, let ${\cal A},{\cal B}$ be finite input and output
alphabets (respectively), and let $V\colon{\cal A}\to{\cal B}$ be
a DMC. Then, the normalized log-likelihood of an output sequence $b^{K}\in{\cal B}^{K}$,
conditioned on an input sequence $a^{K}\in{\cal A}^{K}$, is 
\[
\frac{1}{K}\log V^{K}(b^{K}\mid a^{K})=-H_{\mathscr{P}(a^{K},b^{K})}(B\mid A)-D\left(\mathscr{P}(a^{K},b^{K})\mid\mid\mathscr{P}(a^{K})\times V\right),
\]
where $\mathscr{P}(a^{K},b^{K})$ is the empirical joint distribution
of $(a^{K},b^{K})\in{\cal A}^{K}\times{\cal B}^{K}$. For a codebook
${\cal C}\subseteq{\cal A}^{K}$, and given an output vector $b^{K}\in{\cal B}^{K}$,
the ML decoding rule is then 
\[
\argmax_{a^{K}\in{\cal C}}\frac{1}{K}\log V^{K}(b^{K}\mid a^{K})=\argmin_{a^{K}\in{\cal C}}\left[H_{\mathscr{P}(a^{K},b^{K})}(B\mid A)+D\left(\mathscr{P}(a^{K},b^{K})\mid\mid\mathscr{P}(a^{K})\times V\right)\right].
\]
In universal decoding for DMCs, the decoding metric does not depend
on the channel $V$. A possible decoding rule replaces the normalized
log-likelihood with its maximized version over all DMCs from ${\cal A}\to{\cal B}$.
The resulting decoding rule is then 
\begin{equation}
\argmax_{a^{K}\in{\cal C}}\max_{V}\frac{1}{K}\log V^{K}(b^{K}\mid a^{K})=\argmax_{a^{K}\in{\cal C}}H_{\mathscr{P}(a^{K},b^{K})}(B\mid A).\label{eq: MMI as maximizing over channels}
\end{equation}
Equivalently, since $H_{\mathscr{P}(b^{K})}(B)$ only depends on the
output sequence, and thus is common to all codewords, the decoder
chooses the codeword which maximizes the empirical mutual information
$I_{\mathscr{P}(a^{K},b^{K})}(A;B).$ This decoder is called MMI,
and evidently it does not depend on the channel $V$. 

When condition on $u^{N}$, the likelihood function of the DNA channel
(\ref{eq: likelihood DNA as mixture of binomials conditioned on u})
corresponds to a mixture of binomial channels. The MMI can be generalized,
in a straightforward manner, to the mixture of binomial channels as
follows. Formally, consider a mixture of binomial channels, parameterized
by $(V,K,\{{\cal K}_{d}\}_{d\in[K+1]})$ (where ${\cal K}_{d}$ is
a partition of $[K]$). Given a codebook ${\cal C}\subseteq{\cal A}^{K}$,
an MMI decoder for this channel can be defined as 
\begin{equation}
\argmax_{a^{K}\in{\cal C}}\sum_{d\in\mathbb{N}}\frac{K_{d}}{K}\cdot I_{\mathscr{P}(a_{{\cal K}_{d}},b_{{\cal K}_{d}}^{d})}(A;B^{d}),\label{eq: MMI for mixture of binomial channels}
\end{equation}
which amounts to averaging of the empirical mutual information over
the possible orders $d$, while taking into account their proportion
in the $K$ symbols. 

Equipped with  these preliminaries, we next develop a universal decoder
for the DNA channel $\mathsf{DNA}=\{\alpha,\beta,W\}$. As said, the
maximum-likelihood rule $\argmax_{j\in[|{\cal C}|]}{\cal L}[y^{LN}\mid x^{LM}(j)]$
is the optimal decoder. Here, the main motivation for using the universal
decoder is mostly in order to facilitate the analysis (with the additional
benefit that the knowledge of the sequencing channel $W$ is not required).
It is evident from (\ref{eq: likelihood DNA as mixture of binomials})
that the likelihood is a mixture (over $u^{N}$) of a mixture (over
$d$) of binomial channels. Had $u^{N}$ been known to the decoder,
a possible universal decoder is the MMI for the mixture of binomial
channels stated in (\ref{eq: MMI for mixture of binomial channels}),
which in the DNA channel amounts to the metric
\[
\log\mathcal{L}(y^{LN}\mid x^{LM};u^{N})=\sum_{d\in[K+1]}q_{d}L\cdot I_{\hat{P}^{d}(x^{LM},y^{LN};u^{N})}(A;B^{d}),
\]
where $q^{N+1}=\mathscr{N}^{(2)}(u^{N})$. For the DNA channel, $u^{N}$
is unknown, and so we propose to further maximize this metric over
$u^{N}\in\mathscr{T}_{q^{N+1}}^{(2)}$, albeit with a penalty term
related to $|\mathscr{T}_{q^{N+1}}^{(2)}|$. From a technical perspective,
this penalty will compensate for a standard union bound that will
be used in the random coding analysis, as is common in related scenarios
(e.g., universal decoding for joint source-channel coding \cite{csiszar1980joint}).
Intuitively, this penalty will favor less the likelihood of $\{u^{N}\}$
which belong to a large amplification type class $\mathscr{T}_{q^{N+1}}^{(2)}$.
More precisely, the proposed penalty is the first order term in the
asymptotic expansion of the logarithm of the size of $|\mathscr{T}_{q^{N+1}}^{(2)}|$,
rather than its exact size, which according to Prop. \ref{prop: amplification type sizes estimates },
is 
\[
\log|\mathscr{T}_{q^{N+1}}^{(2)}|=(M-q_{0})\log M+O(M).
\]
Moreover, the maximization will not be over all possible $\{u^{N}\}$.
Instead, a finite $\overline{d}$ is set, and only $u^{N}$ for which
$q^{N+1}=\mathscr{N}^{(2)}(u^{N})\in\mathscr{Q}_{\overline{d}}(M,N)$
are feasible in the maximization, where 
\begin{equation}
\mathscr{Q}_{\overline{d}}(M,N)\dfn\left\{ q^{N+1}\in\mathscr{Q}(M,N)\colon q_{d}=0\;\forall d\geq\overline{d}\right\} .\label{eq: amplification vector restricted set}
\end{equation}
That is, the set of candidate $\{q^{N+1}\}$ does not include a single
molecule being sampled more than $\overline{d}$ times. This restriction
will be instrumental for the analysis. 

In addition to the penalty term $\log|\mathscr{T}_{q^{N+1}}^{(2)}|$,
we add another penalty term related to the choice of i.i.d. ensemble,
instead of the fixed composition ensemble. To this end, let $\hat{P}^{d}(x^{LM};u^{N})$
be the $x$-marginal of $\hat{P}^{d}(x^{LM},y^{LN};u^{N})$. Assuming
an input distribution $P_{X}$, this penalty term for the $d$ order
binomial channel, is the normalized asymptotic expansion of the probability
that the input to the $d$ order binomial channel has type $\hat{P}^{d}(x^{LM};u^{N})$,
which is given by $D(\hat{P}^{d}(x^{LM};u^{N})\mid\mid P_{X})$.

The result is that the universal decoder replaces log-likelihood $\log{\cal L}[y^{LN}\mid x^{LM}(j)]$
with the following universal metric: 
\begin{equation}
\log{\cal L}_{\text{u}}[y^{LN}\mid x^{LM}(j)]=\max_{q^{N+1}\in\mathscr{Q}_{\overline{d}}(M,N)}\max_{s\in\mathscr{T}_{q^{N+1}}}\max_{u\in\mathscr{T}_{s^{M}}}\lambda(y^{LN}\mid x^{LM}(j);q^{N+1},s^{M},u^{N}),\label{eq: universal decoder}
\end{equation}
with $\lambda(y^{LN}\mid x^{LM};u^{N})$ is defined as 
\begin{equation}
\lambda(y^{LN}\mid x^{LM};u^{N})\dfn-(M-q_{0})\log M+\sum_{d\in[N+1]}q_{d}L\cdot\left[D(\hat{P}^{d}(x^{LM};u^{N})\mid\mid P_{X})+I_{\hat{P}^{d}(x^{LM},y^{LN};u^{N})}(A;B^{d})\right],\label{eq: universal metric}
\end{equation}
where $q^{N+1}=\mathscr{N}^{(2)}(u^{N})$ in (\ref{eq: universal metric})
is assumed. As a final remark, we mention that constructing the decoder
on the basis of empirical types on a ``symbol-wise'' level is only
justified here since the sequencing channel $W^{L}$ is a product
of $L$ DMCs $W\colon{\cal X}\to{\cal Y}$.

\subsection{Random-Coding Error Probability Analysis for a Given $Q^{N+1}$ \label{subsec:Random-coding-error}}

We now turn to analyze the random coding error exponent of the proposed
universal decoder for the following random ensemble: The $ML\cdot|{\cal C}|$
symbols in the codebook ${\cal C}=\{x^{LM}(j)\}$ are drawn i.i.d.
from a distribution $P_{X}$. We evaluate the average error probability
conditioned on $Q^{N+1}=q^{N+1}$ and show that that if the rate does
not cross a prescribed threshold, then the average error probability
decays exponentially in $K_{M}=ML=\beta M\log M$. The proof for the
bounds on the reliability function of Theorems \ref{thm: achievable reliability function}
and \ref{thm: error exponent of universal decoder -- ideal amplification}
is concluded in the next subsections by averaging over $Q^{N+1}$.
As mentioned in the previous subsection, we set $\overline{d}\in\mathbb{N}^{+}$
and only consider $q^{N+1}\in\mathscr{Q}_{\overline{d}}(M,N)$, that
is $q^{N+1}$ for which $q_{d}=0$ for all $d\geq\overline{d}$. 

As we have seen, the likelihood function of the DNA channel is obtained
by a mixture of binomial channels. To facilitate the random coding
analysis, we first state a general result regarding exponential probability
bounds for such channels. Since our statements are for general DMCs,
we denote the blocklength by $K\in\mathbb{N}^{+}$, the input alphabet
by ${\cal A}$ and the output alphabet by ${\cal B}$ in order to
avoid confusion with the notation of the DNA channel. 

Let $V\colon{\cal A}\to{\cal B}$ be a DMC, and let $V^{\oplus d}\colon{\cal A}\to{\cal B}^{d}$
be its binomial extension. For a blocklength $K$, consider a \emph{mixture}
of $d$-order binomial channels $(V,K,\{{\cal K}_{d}\}_{d\in\mathbb{N}^{+}})$
where $\{{\cal K}_{d}\}_{d\in\mathbb{N}^{+}}$ is a partition of $[K]$.
Let $a^{K}$ (respectively $b^{K}$) be the input (respectively output)
to this channel, and assume that ${\cal K}_{d}\subseteq[K]$ are the
indices of inputs $a_{k}$ which are input to $V^{\oplus d}$. We
denote $a_{{\cal K}_{d}}=(a_{k})_{k\in{\cal K}_{d}}$ according to
increasing order in ${\cal K}_{d}$, and use a similar notation for
$b^{K}$. Note that ${\cal K}_{d}$ can be the empty set, but these
sets will simply be ignored in a natural way in the next derivations.
We thus assume that 
\begin{equation}
\P\left[B^{(*K)}=(b^{*})^{K}\mid A^{K}=a^{K}\right]=\prod_{d=0}^{\infty}V^{\oplus d}\left[b_{{\cal K}_{d}}^{d}\mid a_{{\cal K}_{d}}\right]=\prod_{d=0}^{\infty}\prod_{k\in{\cal K}_{d}}V^{\oplus d}\left[b_{k}^{d}\mid a_{k}\right]=\prod_{d=0}^{\infty}\prod_{k\in{\cal K}_{d}}\prod_{i=0}^{d-1}V\left[b_{k,i}\mid a_{k}\right].\label{eq: likelihood of binomial channel mixture of multiple orders}
\end{equation}
Note that in the last expression, $(b^{*})^{K}$ designates an output
vector whose components may have different dimensions (according to
the order $d$). The following is a generalization of well-known bounds:
\begin{prop}
\label{prop:Binomial channel types multiple  orders}Let $V\colon{\cal A}\to{\cal B}$
be a DMC and let $V^{\oplus d}$ be its $d$-order binomial extension.
Assume that $A^{K}\in{\cal A}^{K}$ has i.i.d. components distributed
according to $P_{A}$. Let $\{{\cal K}_{d}\}_{d\in\mathbb{N}^{+}}$
be a partition of $[K]$ and $K_{d}=|{\cal K}_{d}|$. Assume further
that $K_{d}=0$ for all $d\geq\overline{d}$. Then:
\begin{enumerate}
\item If $b_{{\cal K}_{d}}$ is fixed and has type $Q_{B^{d}}^{(d)}$ for
$d\in\mathbb{N}^{+}$, then for any given sequence of joint type $Q_{AB^{d}}^{(d)}$
(whose $B$-marginals all agree with $Q_{B}^{(d)}$)
\begin{multline}
\P\left[\bigcap_{d\in[\overline{d}]}(A_{{\cal K}_{d}},b_{{\cal K}_{d}}^{d})\in{\cal T}_{K_{d}}(Q_{AB^{d}}^{(d)})\right]\leq\\
\exp\left\{ -K\cdot\left[\sum_{d\in[K]}\frac{K_{d}}{K}\cdot\left(D(Q_{A}\mid\mid P_{A})+I_{Q^{(d)}}(A;B^{d})+O\left(\frac{|{\cal A}||{\cal B}|^{d}\log K_{d}}{K_{d}}\right)\right)\right]\right\} .\label{eq: exponential probability of joint class indepenent mixed}
\end{multline}
\item If $B_{k}\sim V^{\oplus d}(\cdot|A_{k})$ independently for all $k\in{\cal K}_{d}$
and all $d\in[K]$, that is $\P[B^{*K}=b^{*K}\mid A^{K}=a^{K}]$ is
as in (\ref{eq: likelihood of binomial channel mixture of multiple orders}),
then 
\begin{multline}
\P\left[\bigcap_{d\in[\overline{d}]}(A_{{\cal K}_{d}},B_{{\cal K}_{d}}^{d})\in{\cal T}_{K_{d}}^{(d)}(Q_{AB^{d}}^{(d)})\right]\leq\\
\exp\left\{ -K\cdot\left[\sum_{d\in[K]}\frac{K_{d}}{K}\cdot\left(D(Q_{A}^{(d)}\mid\mid P_{A})+D(Q_{B|A}^{(d)}\mid\mid V^{\oplus d}|Q_{A})+O\left(\frac{|{\cal A}||{\cal B}|^{d}\log K_{d}}{K_{d}}\right)\right)\right]\right\} .\label{eq: exponential probabiity of joint class over channel mixed}
\end{multline}
\end{enumerate}
\end{prop}
\begin{IEEEproof}
First assume that $K_{d}=K$ for some $d$. The proof then extends
in a straightforward manner the same results obtained by the standard
\emph{method of types }\cite{csiszar2011information,csiszar1998method}.
Specifically, by arguments as in \cite[Problem 2.3]{csiszar2011information}
for (\ref{eq: exponential probability of joint class indepenent mixed}),
and \cite[Lemma 2.6]{csiszar2011information} for (\ref{eq: exponential probabiity of joint class over channel mixed}).
The second-order term $O(|{\cal A}||{\cal B}|^{d}\cdot(\log K)/K)$
follows from the standard estimate \cite[Lemma 2.2]{csiszar2011information}
on the number of joint types for $V^{\oplus d}\colon{\cal A}\to{\cal B}^{d}$
\begin{equation}
|{\cal P}_{K}({\cal A}\times{\cal B}^{d})|\leq(K+1)^{|{\cal A}||{\cal B}|^{d}}=\exp\left[|{\cal A}||{\cal B}|^{d}\cdot\log(K+1)\right].\label{eq: number of types in a binomial channel}
\end{equation}
The bound for a general partition $\{{\cal K}_{d}\}$ then follows
from the statistical independence of the inputs to the various order-$d$
binomial channels. 
\end{IEEEproof}
We emphasize that since $K_{d}=0$ for all $d\geq\overline{d}$, the
second-order term in (\ref{eq: exponential probability of joint class indepenent mixed})
and (\ref{eq: exponential probabiity of joint class over channel mixed})
is \emph{uniformly} bounded as $O(|{\cal A}||{\cal B}|^{\overline{d}}\cdot(\log K)/K)$.
As common, (\ref{eq: exponential probability of joint class indepenent mixed})
will be used to analyze the probability that a competing codeword
has larger decoding metric than the true codeword (for a given output),
and (\ref{eq: exponential probabiity of joint class over channel mixed})
will be used to analyze the probability of observing a codeword and
its output under the random coding assumption and the channel's randomness.
We now have the following random coding bound conditioned on $Q^{N+1}=q^{N+1}$:
\begin{prop}
\label{prop: random coding exponent condition on q}Consider the DNA
channel $\mathsf{DNA}=(\alpha,\beta,W)$. Let $q^{N+1}\in\mathscr{Q}(M,N)$
be given. Suppose that the codebook ${\cal C}$ is drawn with i.i.d.
symbols and codewords from $P_{X}\in{\cal P}({\cal X})$. Let the
universal decoder ${\cal D}_{u}$ be such that $y^{LN}\in{\cal D}_{\text{u}}(j)$
if
\[
j\in\argmax_{\tilde{j}\in[|{\cal C}|]}\log\mathcal{L}_{\text{u}}(y^{LN}\mid x^{LM}(\tilde{j})).
\]
Then, 
\begin{multline}
\E_{{\cal C}}\left[\pe({\cal C},{\cal D}_{\text{u}}\mid x^{LM}(0)\text{ \emph{stored}},Q^{N+1}=q^{N+1})\right]\leq\\
\exp\left[-ML\cdot\left(E_{\overline{d}}(R,\mathsf{DNA}\mid q^{N+1})-O_{\overline{d}}\left(\frac{1}{\sqrt{M}\log M}\right)\right)\right],\label{eq:random coding exponent condition on q}
\end{multline}
where $\E_{{\cal C}}[\cdot]$ denotes expectation over the random
codebook, $\pe({\cal C},\mathsf{D_{\text{u}}}\mid x^{LM}(j),q^{N+1})$
is a conditional version of the error probability (\ref{eq: error proabability})
and
\begin{multline}
E_{\overline{d}}(R,\mathsf{DNA}\mid q^{N+1})\dfn\min_{\{Q_{AB^{d}}^{(d)}\}_{d\in[\overline{d}+1]}}\sum_{d\in[\overline{d}+1]}\frac{q_{d}}{M}\cdot\left(D(Q_{A}^{(d)}\mid\mid P_{X})+D(Q_{B^{d}|A}^{(d)}\mid\mid V^{\oplus d}\mid Q_{A})\right)\\
+\left[\sum_{d\in[\overline{d}+1]}\frac{q_{d}}{M}\cdot\left(D(Q_{A}^{(d)}\mid\mid P_{X})+I_{Q^{(d)}}(A;B^{d})\right)-\frac{1}{\beta}\left(1-\frac{q_{0}}{M}\right)-R\right]_{+}.\label{eq: conditional error exponent for universal decoder}
\end{multline}
\end{prop}
\begin{IEEEproof}
Since all $s^{M}\in\mathscr{T}_{q^{N+1}}$ are equiprobable, and all
$u^{N}\in\mathscr{T}_{q^{N+1}}^{(2)}$ are equiprobable too, symmetry
implies that we can condition on an arbitrary $s^{M}\in\mathscr{T}_{q^{N+1}}$
as well as an arbitrary $u^{N}\in\mathscr{T}_{s^{M}}=\mathscr{T}_{q^{N+1}}^{(2)}$.
We bound the average conditional error probability in the following
standard way, which comprises two steps. First, we condition on the
randomly chosen stored codeword, $x^{LM}(1)$, and the corresponding
channel output, $y^{LN}$, and compute the probability that a single
random codeword has a larger decoding metric than $\mathcal{L}_{\text{u}}(y^{LN}\mid x^{LM}(0))$,
and then we take a clipped union bound over $e^{MLR}-1$ competing
codewords. Second, we compute the average over $x^{LM}(0),y^{LN}$,
connected via the DNA channel. 

We begin with the first step. Assume that $x^{LM}(0)\equiv x^{LM}$
was stored and $y^{LN}$ was the output, and consider a competing
random codeword $x^{LM}(j)\equiv\tilde{X}^{LM}$ for some $j\in[|{\cal C}|]\backslash\{0\}$
(whose $ML$ symbols are $P_{X}$-i.i.d.). Denote temporarily, for
brevity, the event 
\[
{\cal E}\dfn\left\{ X^{LM}(1)=x^{LM},Y^{LN}=y^{LN},Q^{N+1}=q^{N+1},S^{M}=s^{M},U^{N}=u^{N}\right\} .
\]
The pairwise error probability is bounded as 
\begin{align}
 & \P\left[\log\mathcal{L}_{\text{u}}(y^{LN}\mid\tilde{X}^{LM})\geq\log\mathcal{L}_{\text{u}}(y^{LN}\mid x^{LM})\,\middle\vert\,{\cal E}\right]\nonumber \\
 & \trre[\leq,a]\P\left[\log\mathcal{L}_{\text{u}}(y^{LN}\mid\tilde{X}^{LM})\geq\lambda(y^{LN}\mid x^{LM};u^{N})\,\middle\vert\,{\cal E}\right]\\
 & =\P\left[\max_{\tilde{q}^{N+1}\in\mathscr{Q}_{\overline{d}}(M,N)}\max_{\tilde{u}^{N}\in\mathscr{T}_{\tilde{q^{N+1}}}^{(2)}}\lambda(y^{LN}\mid\tilde{X}^{LM};\tilde{u}^{N})\geq\lambda(y^{LN}\mid x^{LM};u^{N})\,\middle\vert\,{\cal E}\right]\\
 & \trre[\leq,b]\sum_{\tilde{q}^{N+1}\in\mathscr{Q}_{\overline{d}}(M,N)}\sum_{\tilde{u}^{N}\in\mathscr{T}_{\tilde{q^{N+1}}}^{(2)}}\P\left[\lambda(y^{LN}\mid\tilde{X}^{LM};\tilde{u}^{N})\geq\lambda(y^{LN}\mid x^{LM};u^{N})\,\middle\vert\,{\cal E}\right]\\
 & \trre[\leq,c]e^{O(\sqrt{M})}N^{M-\tilde{q}_{0}}\max_{\tilde{q}^{N+1}\in\mathscr{Q}_{\overline{d}}(M,N)}\max_{\tilde{u}\in\mathscr{T}_{\tilde{q^{N+1}}}^{(2)}}\cdot\P\left[\lambda(y^{LN}\mid\tilde{X}^{LM};\tilde{u}^{N})\geq\lambda(y^{LN}\mid x^{LM};u^{N})\,\middle\vert\,{\cal E}\right],\label{eq: pairwise error probability random coding universal}
\end{align}
where $(a)$ follows since from the definition of the universal metric
$\log\mathcal{L}_{\text{u}}(y^{LN}\mid x^{LM})$ in (\ref{eq: universal decoder})
\[
\log\mathcal{L}_{\text{u}}(y^{LN}\mid x^{LM})\geq\lambda(y^{LN}\mid x^{LM};u^{N})
\]
{[}using the possibly sub-optimal choice $u^{N}$ in (\ref{eq: universal decoder}){]},
$(b)$ follows from the union bound, and $(c)$ follows from Prop.
\ref{prop: amplification type sizes estimates }. We next evaluate
the inner probability for a specific choice of $\tilde{u}^{N}$. The
value of $\lambda(y^{LN}\mid\tilde{x}^{LM};\tilde{u}^{N})$ is equivalently
determined by $(\tilde{A}_{{\cal K}_{d}},b_{{\cal K}_{d}}^{d})_{d\in[\overline{d}+1]}$
corresponding to $(\tilde{u}^{N},\tilde{x}^{LM},y^{LN})$, where the
symbols $\tilde{A}^{K}\in{\cal A}^{K}$ are drawn i.i.d. according
to $P_{X}$. Letting $\{\tilde{Q}_{AB^{d}}^{(d)}\}_{d\in[\overline{d}+1]}$
be a collection of types with $\tilde{Q}_{B^{d}}^{(d)}=Q_{B^{d}}^{(d)}=\mathscr{P}(b_{{\cal K}_{d}}^{d})$,
Prop. \ref{prop:Binomial channel types multiple  orders} states that
\begin{align}
 & \P\left[\bigcap_{d\in[\overline{d}+1]\colon q_{d}\neq0}(\tilde{A}_{{\cal K}_{d}},b_{{\cal K}_{d}}^{d})\in{\cal T}_{K_{d}}(\tilde{Q}_{AB^{d}}^{(d)})\right]\nonumber \\
 & \leq\exp\left\{ -ML\cdot\left[\sum_{d\in[\overline{d}+1]\colon q_{d}\neq0}\frac{q_{d}}{M}\cdot\left(D(\tilde{Q}_{A}^{(d)}\mid\mid P_{X})+I_{\tilde{Q}^{(d)}}(A;B^{d})+O\left(\frac{|{\cal X}||{\cal Y}|^{d}\log Lq_{d}}{Lq_{d}}\right)\right)\right]\right\} \\
 & \leq\exp\left\{ -ML\cdot\left[\sum_{d\in[\overline{d}+1]\colon q_{d}\neq0}\frac{q_{d}}{M}\cdot\left(D(\tilde{Q}_{A}^{(d)}\mid\mid P_{X})+I_{\tilde{Q}^{(d)}}(A;B^{d})\right)+O_{\overline{d}}\left(\frac{\log(ML)}{ML}\right)\right]\right\} \label{eq: sepcific types probability for competing codeword}
\end{align}
where the last inequality follows from 
\begin{equation}
\sum_{d\in[\overline{d}+1]}\frac{q_{d}}{M}\cdot\frac{|{\cal X}||{\cal Y}|^{d}\log(Lq_{d})}{Lq_{d}}=\frac{1}{ML}\sum_{d\in[\overline{d}+1]}|{\cal X}||{\cal Y}|^{d}\log(Lq_{d})\leq\overline{d}|{\cal X}||{\cal Y}|^{\overline{d}}\frac{\log(ML)}{ML}=O_{\overline{d}}\left(\frac{\log(ML)}{ML}\right).\label{eq: second order term for the probability of a joint type class bound}
\end{equation}
Next, we note that $\lambda(y^{LN}\mid\tilde{x}^{LM};\tilde{u}^{N})$
is determined by the empirical types of $\mathscr{P}(\tilde{A}_{{\cal K}_{d}},b_{{\cal K}_{d}}^{d})=\hat{P}^{d}(\tilde{x}^{LM},y^{LN};\tilde{u}^{N})$,
for $d\in[\overline{d}+1]$. The number of possible types of $(\tilde{A}_{{\cal K}_{d}},b_{{\cal K}_{d}}^{d})$
is less than $(K_{d}+1)^{|{\cal X}||{\cal Y}|^{d}}=(q_{d}L+1)^{|{\cal X}||{\cal Y}|^{d}}=\exp\left[|{\cal X}||{\cal Y}|^{d}\cdot\log(q_{d}L+1)\right]$
{[}as in (\ref{eq: number of types in a binomial channel}){]}. Consequently,
the number of possible sets of types $\{\tilde{Q}_{AB^{d}}^{(d)}\}_{d\in[\overline{d}+1]}$
is upper bounded by 
\begin{equation}
\prod_{d=0}^{\overline{d}}\exp\left[|{\cal X}||{\cal Y}|^{d}\cdot\log(q_{d}L+1)\right]\leq\exp\left[\overline{d}|{\cal X}||{\cal Y}|^{\overline{d}}\cdot\log(ML+1)\right]\label{eq: number of type sets estimate}
\end{equation}
which is polynomial in $K=ML$ (for \emph{fixed} $\overline{d}$).
By the union bound over those set of types 
\begin{align}
 & \P\left[\lambda(y^{LN}\mid\tilde{X}^{LM};\tilde{u}^{N})\geq\lambda(y^{LN}\mid x^{LM};u^{N})\mid{\cal E}\right]\nonumber \\
 & \leq\sum_{\{\tilde{Q}_{AB^{d}}^{(d)}\}_{d\in[\overline{d}+1]}}\P\left[\bigcap_{d=0}^{\overline{d}}(\tilde{A}_{{\cal K}_{d}},b_{{\cal K}_{d}}^{d})\in{\cal T}_{K_{d}}^{(d)}(\tilde{Q}_{AB^{d}}^{(d)})\right],
\end{align}
where the sum is over $\{\tilde{Q}_{AB^{d}}^{(d)}\}_{d\in[\overline{d}+1]}$
such that $\tilde{Q}_{B^{d}}^{(d)}=Q_{B^{d}}^{(d)}$ and 
\begin{align}
 & -(M-\tilde{q}_{0})\log N+\sum_{d\in[\overline{d}+1]}\tilde{q}_{d}L\cdot\left(D(\tilde{Q}_{A}^{(d)}\mid\mid P_{X})+I_{\tilde{Q}_{AB}^{(d)}}(A;B^{d})\right)\nonumber \\
 & \geq\lambda(y^{LN}\mid x^{LM};u^{N})\nonumber \\
 & =-(M-q_{0})\log N+\sum_{d\in[\overline{d}+1]}q_{d}L\cdot\left(D(\hat{P}^{d}(x^{LM};u^{N})\mid\mid P_{X})+I_{\hat{P}^{d}(x^{LM},y^{LN};u^{N})}(A;B^{d})\right)
\end{align}
holds. By (\ref{eq: sepcific types probability for competing codeword}),
(\ref{eq: number of type sets estimate}), and the last display, we
obtain that 
\begin{align}
 & \P\left[\lambda(y^{LN}\mid\tilde{X}^{LM};\tilde{u}^{N})\geq\lambda(y^{LN}\mid x^{LM};u^{N})\mid{\cal E}\right]\nonumber \\
 & \leq e^{O(\log(ML))}\cdot\exp\left[-(M-\tilde{q}_{0})\log N+(M-q_{0})\log N\right]\nonumber \\
 & \hphantom{=}\times\exp\left\{ -ML\cdot\left[\sum_{d\in[\overline{d}+1]}\frac{q_{d}}{M}\left(D(\hat{P}^{d}(x^{LM};u^{N})\mid\mid P_{X})+I_{\hat{P}^{d}(x^{LM},y^{LN};u^{N})}(A;B^{d})\right)+O_{\overline{d}}\left(\frac{\log(ML)}{ML}\right)\right]\right\} .
\end{align}
Inserting this bound into (\ref{eq: pairwise error probability random coding universal}),
we then obtain
\begin{align}
 & \log\P\left[\mathcal{L}_{\text{u}}(y^{LN}\mid\tilde{X}^{LM})\geq\mathcal{L}_{\text{u}}(y^{LN}\mid x^{LM})\mid{\cal E}\right]\nonumber \\
 & \leq O(\sqrt{M})+(M-q_{0})\log N\nonumber \\
 & \hphantom{==}-ML\cdot\left[\sum_{d\in[\overline{d}+1]}\frac{q_{d}}{M}\cdot\left(D(\hat{P}^{d}(x^{LM};u^{N})\mid\mid P_{X})+I_{\hat{P}^{d}(x^{LM},y^{LN};u^{N})}(A;B^{d})\right)+O_{\overline{d}}\left(\frac{\log(ML)}{ML}\right)\right]\\
 & =-ML\cdot\left[\sum_{d\in[\overline{d}+1]}\frac{q_{d}}{M}\cdot\left(D(\hat{P}^{d}(x^{LM};u^{N})\mid\mid P_{X})+I_{\hat{P}^{d}(x^{LM},y^{LN};u^{N})}(A;B^{d})\right)-\frac{1}{\beta}\left(1-\frac{q_{0}}{M}\right)+O\left(\frac{1}{\sqrt{M}L}\right)\right],
\end{align}
where in $(a)$ we have used that $\alpha=\frac{N}{M}$ is constant\footnote{In fact, $\alpha=o(\log M)$ suffices.}
and $\beta=\frac{L}{\log M}$. This completes the bound on the pairwise
error probability for a single competing codeword. Letting now $\overline{\pe}({\cal C},{\cal D}_{\text{u}}\mid{\cal E})$
be the conditional average error probability over all $e^{MLR}-1$
competing codewords, the clipped union bound implies 
\begin{multline}
-\frac{1}{ML}\log\E_{{\cal C}}\left[\overline{\pe}({\cal C},{\cal D}_{\text{u}}\mid{\cal E})\right]\geq\\
\left[\sum_{d\in[\overline{d}+1]}\frac{q_{d}}{M}\cdot\left(D(\hat{P}^{d}(x^{LM};u^{N})\mid\mid P_{X})+I_{\hat{P}^{d}(x^{LM},y^{LN};u^{N})}(A;B^{d})\right)-\frac{1}{\beta}\left(1-\frac{q_{0}}{M}\right)-R\right]_{+}-O\left(\frac{1}{\sqrt{M}L}\right).\label{eq: competing codeword error probability}
\end{multline}
This completes the first step of the proof. 

We now move on to the second step in which we average over $(X^{LM}\equiv X^{LM}(1),Y^{LN})$.
In the first step, we have seen that the pairwise error probability
conditioned on $(X^{LM},Y^{LN})=(x^{LM},y^{LN})$ depends on $\sum_{d\in[\overline{d}+1]}\frac{q_{d}}{M}\cdot D(\hat{P}^{d}(x^{LM};u^{N})\mid\mid P_{X})+\sum_{d\in[\overline{d}+1]}\frac{q_{d}}{M}I_{\hat{P}^{d}(x^{LM},y^{LN};u^{N})}(A;B^{d})$,
which in turn depend on the types $\{\hat{P}^{d}(x^{LM},y^{LN};u^{N})\}_{d\in[\overline{d}+1]}$.
We next evaluate the probability distribution of this term, conditioned
on the given $q^{N+1}$, and an arbitrary representative $u^{N}\in\mathscr{T}_{q^{N+1}}^{(2)}$.
As in the first step, we denote $\hat{P}^{d}(x^{LM},y^{LN};u^{N})=Q_{AB^{d}}^{(d)}$.
Prop. \ref{prop:Binomial channel types multiple  orders} states that
in case the random output $B_{{\cal K}_{d}}^{d}$ is the result of
$A_{{\cal K}_{d}}$ passing in the channel $V^{\oplus d}$, it holds
that 
\begin{align}
 & \P\left[\bigcap_{d\in[\overline{d}+1]\colon q_{d}\neq0}(A_{{\cal K}_{d}},B_{{\cal K}_{d}})\in{\cal T}_{K_{d}}(Q_{AB^{d}}^{(d)})\right]\nonumber \\
 & \leq\exp\left\{ -ML\cdot\left[\sum_{d\in[\overline{d}+1]\colon q_{d}\neq0}\frac{q_{d}}{M}\cdot\left(D(Q_{A}^{(d)}\mid\mid P_{X})+D(Q_{B^{d}|A}^{(d)}\mid\mid V^{\oplus d}\mid Q_{A})\right)+O_{\overline{d}}\left(\frac{\log(ML)}{ML}\right)\right]\right\} ,\label{eq: sepcific types probability for true codeword}
\end{align}
where the second-order term is bounded as in (\ref{eq: second order term for the probability of a joint type class bound}).
Thus, for $s^{M}\in\mathscr{T}_{q^{N+1}}$ and $u^{N}\in\mathscr{T}_{s^{M}}$
\begin{align}
 & \E_{{\cal C}}\left[\pe({\cal C},{\cal D}_{\text{u}}\mid x^{LM}(1)\text{ stored},\;Q^{N+1}=q^{N+1})\right]\nonumber \\
 & =\E_{{\cal C}}\left[\pe({\cal C},{\cal D}_{\text{u}}\mid Q^{N+1}=q^{N+1},S^{M}=s^{M},U^{N}=u^{N})\right]\\
 & =\sum_{\{Q_{AB^{d}}^{(d)}\}_{d\in[\overline{d}]}}\P\left[\bigcap_{d=0}^{\overline{d}}(A_{{\cal K}_{d}},B_{{\cal K}_{d}}^{d})\in{\cal T}_{K_{d}}(Q_{AB^{d}}^{(d)})\right]\cdot\overline{\pe}({\cal C},{\cal D}_{\text{u}}\mid{\cal E}).
\end{align}
By (\ref{eq: competing codeword error probability}), (\ref{eq: sepcific types probability for true codeword})
and the bound (\ref{eq: number of type sets estimate}) on the total
number of sets of types $\{Q_{AB}^{(d)}\}_{d\in[\overline{d}+1]}$
(which from (\ref{eq: number of type sets estimate}) is polynomial
in $ML$ ) we obtain that 
\begin{alignat}{1}
 & -\frac{1}{ML}\log\E_{{\cal C}}\left[\pe({\cal C},{\cal D}\mid X^{LM}(1)\text{ stored},\;Q^{N+1}=q^{N+1})\right]\nonumber \\
 & \geq\min_{\{Q_{AB}^{(d)}\}}\sum_{d\in[\overline{d}+1]}\frac{q_{d}}{M}\cdot\left(D(Q_{A}^{(d)}\mid\mid P_{X})+D(Q_{B^{d}|A}^{(d)}\mid\mid V^{\oplus d}|Q_{A}^{(d)})\right)\nonumber \\
 & \hphantom{=}+\left[\sum_{d\in[\overline{d}+1]}\frac{q_{d}}{M}\cdot\left(D(Q_{A}^{(d)}\mid\mid P_{X})+I_{Q_{AB^{d}}^{(d)}}(A;B^{d})\right)-\frac{1}{\beta}\left(1-\frac{q_{0}}{M}\right)-R\right]_{+}-O_{\overline{d}}\left(\frac{1}{\sqrt{M}\log M}\right)\\
 & =E_{\overline{d}}(R,\mathsf{DNA}|q^{N+1})-O_{\overline{d}}\left(\frac{1}{\sqrt{M}\log M}\right).
\end{alignat}
This completes the second step, and the bound on the error probability
of the first codeword. 
\end{IEEEproof}
We pause for a few remarks:
\begin{enumerate}
\item The second-order term $O_{\overline{d}}(1/\sqrt{M}\log M)$ depends
only on $(\overline{d},{\cal X},{\cal Y})$ and thus is uniform in
$q^{N+1}$.
\item Furthermore, the $q^{N+1}$ being conditioned upon is not restricted
to $\mathscr{Q}_{\overline{d}}(M,N)$ but can have non-zero $q_{d}$
for $d>\overline{d}$. Nonetheless, the universal decoder ignores
the molecules which are sampled more than $\overline{d}$ times in
its metric computation, and in accordance, only $q_{d}$ for $d\in[\overline{d}+1]$
affect the error probability. The average error probability is then
determined by a finite number of types, say $\{\hat{P}^{d}(x^{LM},y^{LN};u^{N})\}_{d\in[\overline{d}+1]}$
for the true and competing codeword, and the total number of types
is uniformly bounded for all $M$ {[}see (\ref{eq: number of type sets estimate}){]}.
This is crucial to the proof as if $\overline{d}$ is not restricted,
and can be as large as $\Theta(N)=\Theta(M)$ the number of types
increases \emph{super-exponentially} with $M$ (rather than the standard
\emph{polynomial} increase).
\item The random coding ensemble is based on i.i.d. draws of symbols, which
can be compared with the simpler (and possibly better) constant composition
distribution over a type ${\cal T}_{ML}(P_{X})$. However, the random
sampling stage complicates the analysis of the latter. If codewords
are chosen to have a constant composition over all $ML$ symbols,
there is no guarantee that the specific inputs to the $d$-order binomial
channel will be of constant composition. If the codewords are chosen
to have a constant composition at each of the $M$ molecules, then
a random input to the $d$-order binomial channel is not distributed
uniformly over the type class of $P_{X}$. Indeed, if, for example,
$q_{d}=2$, that is, there are $2$ molecules which are input to the
$d$-order binomial channel, a random codeword drawn this way will
have constant composition $P_{X}$ at both its parts, and not just
as a whole.
\end{enumerate}

\subsection{Proof of Theorem \ref{thm: achievable reliability function} \label{subsec:The-proof-of error exponent}}

To prove Theorem \ref{thm: achievable reliability function}, we need
to average the conditional error probability derived in the previous
section (given $Q^{N+1}=q^{N+1}$) over the random sampling stage
of the decoder. To this end, we will need the following two lemmas.
The first lemma shows that the exponent of the probability that $Q^{N+1}$
belongs to some set ${\cal Q}$ is determined by the largest probability
of members in this set. 
\begin{lem}
\label{lem: outage set exponential charactarization}For any ${\cal Q}\subseteq\mathscr{Q}(M,N)$
\[
\P\left[Q^{N+1}\not\in{\cal Q}\right]=\exp\left[-M\cdot\left(\min_{q^{N+1}\not\in{\cal Q}}-\frac{1}{M}\log\P\left[Q^{N+1}=q^{N+1}\right]\right)+O(\sqrt{M})\right].
\]
\end{lem}
\begin{IEEEproof}
By Prop. \ref{prop: amplification type sizes estimates }, $|{\cal Q}|\leq e^{O(\sqrt{M})}$
and so by the union bound
\begin{align}
\max_{q^{N+1}\not\in{\cal Q}}\P\left[Q^{N+1}=q^{N+1}\right] & \leq\P\left[Q^{N+1}\not\in{\cal Q}\right]\\
 & =\sum_{q^{N+1}\not\in{\cal Q}}\P\left[Q^{N+1}=q^{N+1}\right]\leq|{\cal Q}|\cdot\max_{q^{N+1}\not\in{\cal Q}}\P\left[Q^{N+1}=q^{N+1}\right].
\end{align}
\end{IEEEproof}
We remark that if ${\cal Q}\subseteq\mathscr{Q}_{\overline{d}}(M,N)$
(as we will use next) then since $|\mathscr{Q}_{\overline{d}}(M,N)|\leq M^{\overline{d}}$
{[}see (\ref{eq: amplification vector restricted set}){]}, the second-order
term is actually $\overline{d}\log M$ rather than $O(\sqrt{M})$.
This is however, inconsequential to our analysis, and the bound of
Lemma \ref{lem: outage set exponential charactarization} nonetheless
holds even when $\overline{d}$ is not restricted.

We next turn to evaluate $\frac{1}{M}\log\P[Q^{N+1}=q^{N+1}]$ required
for the bound of Lemma \ref{lem: outage set exponential charactarization}.
The main complication is that $Q^{N+1}=\mathscr{N}(S^{M})$ is the
empirical count of $S^{M}$, but $S^{M}$ is distributed according
to the multinomial distribution, and so its components are statistically
dependent. We next use the ``Poissonization of the multinomial''
effect to evaluate this bound. Recall that $S^{M}\sim\text{Multinomial}(N;(\frac{1}{M},\frac{1}{M},\ldots\frac{1}{M}\})$
where $N=M\alpha$ is fixed, and the following fact:
\begin{fact}[Poissonization of the multinomial distribution]
\label{fact: Poissonization of the multinomial distribution}Let
$\tilde{N}\sim\text{\emph{Pois}}(\lambda)$, and let $\tilde{S}^{M}$
be a random vector such that $\tilde{S}^{M}\sim\text{\emph{Multinomial}}(\tilde{N},(p_{0},\ldots p_{M-1}))$
conditioned on $\tilde{N}$, where $\sum_{m\in[M]}p_{m}=1$ and $p_{m}>0$.
Then, $\{\tilde{S}_{m}\}_{m\in[M]}$ are statistically independent
and $\tilde{S}_{m}\sim\text{\emph{Pois}}(p_{m}\lambda)$ (unconditioned
on $\tilde{N}$). 
\end{fact}
Fact \ref{fact: Poissonization of the multinomial distribution} can
be verified by spelling out the conditional p.m.f. of $\tilde{S}^{M}$
conditioned on $\tilde{N}$ (e.g. \cite[Thm. 5.6]{mitzenmacher2017probability}).
The following then is similar to \cite[Corollary 5.9]{mitzenmacher2017probability}:
\begin{lem}
\label{lem: Poissonization of outage event}Let $S^{M}\sim\text{\emph{Multinomial}}(N,(\frac{1}{M},\ldots,\frac{1}{M}))$
with $N=\alpha M$ and let $Q^{N+1}=\mathscr{N}(S^{M})$. Further
let $\tilde{S}_{m}\sim\text{\emph{Pois}}(\frac{N}{M})=\text{\emph{Pois}}(\alpha)$,
i.i.d. for $m\in[M]$, and let $\tilde{Q}^{N+1}=\mathscr{N}(\tilde{S}^{M})$.\textbf{
}Then, 
\[
\P\left[Q^{N+1}=q^{N+1}\right]\leq\sqrt{e^{2}\alpha M}\cdot\P\left[\tilde{Q}^{N+1}=q^{N+1}\right].
\]
\end{lem}
\begin{IEEEproof}[Proof of Lemma \ref{lem: Poissonization of outage event}]
Let $\tilde{N}=\sum_{m\in[M]}\tilde{S}_{m}$ so that $\tilde{N}\sim\text{Pois}(N)=\text{Pois}(\alpha M)$.
By Stirling's bound $n!\leq e\cdot n^{n+1/2}e^{-n}$, and so 
\[
\P[\tilde{N}=N]=\frac{N^{N}e^{-N}}{N!}\geq\frac{1}{e\sqrt{N}}.
\]
Then, by Poissonization (Fact \ref{fact: Poissonization of the multinomial distribution})
\begin{align}
\P\left[Q^{N+1}=q^{N+1}\right] & =\P\left[\tilde{Q}^{N+1}=q^{N+1}\mid\tilde{N}=N\right]\\
 & =\frac{\P\left[\tilde{Q}^{N+1}=q^{N+1},\;\tilde{N}=N\right]}{\P\left[\tilde{N}=N\right]}\\
 & \leq\frac{\P\left[\tilde{Q}^{N+1}=q^{N+1}\right]}{\P\left[\tilde{N}=N\right]}\\
 & \leq e\sqrt{N}\cdot\P\left[\tilde{Q}^{N+1}=q^{N+1}\right].
\end{align}
This completes the proof of the upper bound.
\end{IEEEproof}
We now prove Theorem \ref{thm: achievable reliability function},
which bounds the error probability for the sampling stage of the DNA
channel. 
\begin{IEEEproof}[Proof of Theorem \ref{thm: achievable reliability function}]
Let $\theta_{d}\dfn q_{d}/M$. We first rewrite the bound of Prop.
\ref{prop: random coding exponent condition on q}, with a slight
abuse of notation obtained by replacing $q_{d}$ with $\theta_{d}$,
as 
\[
\E_{{\cal C}}\left[\pe({\cal C},{\cal D}_{\text{u}}\mid X^{LM}(0)\text{ stored},\theta)\right]\leq\exp\left[-ML\cdot\left(E_{\overline{d}}(R,\mathsf{DNA}\mid\theta)-\tau_{M}\right)\right],
\]
where $\tau_{M}=O_{\overline{d}}\left(\frac{1}{\sqrt{M}\log M}\right)$
and
\begin{align}
E_{\overline{d}}(R,\mathsf{DNA}\mid\theta) & \dfn\min_{\{Q^{(d)}\}_{d\in[\overline{d}+1]}}\sum_{d\in[\overline{d}+1]}\theta_{d}\cdot\left(D(Q_{A}^{(d)}\mid\mid P_{X})+D(Q_{B^{d}|A}^{(d)}\mid\mid V^{\oplus d}|Q_{A})\right)\nonumber \\
 & \hphantom{==}+\left[\sum_{d\in[\overline{d}+1]}\theta_{d}\cdot\left(D(Q_{A}^{(d)}\mid\mid P_{X})+I_{Q^{(d)}}(A;B^{d})\right)-\frac{1}{\beta}\left(1-\theta_{0}\right)-R\right]_{+}.
\end{align}
Let 
\[
\Gamma_{\overline{d}}(\theta)\dfn\sum_{d\in[\overline{d}+1]}\theta_{d}\cdot I(P_{X},W^{\oplus d})-\frac{1}{\beta}\left(1-\theta_{0}\right),
\]
and let $\rho>0$ be arbitrary. Since the KL divergence $D(P\mid\mid Q)>0$
for all $P\neq Q$ (strictly positive) it can be easily verified that
if $\Gamma_{\overline{d}}(\theta)\geq R+\rho$ then $E_{\overline{d}}(R,\mathsf{DNA}\mid\theta)>0$.
Consider the set 
\[
\Theta_{\overline{d}}(R,\rho)\dfn\left\{ \theta\colon\theta_{d}\geq0,\;\forall d\in[\overline{d}],\sum_{d\in[\overline{d}+1]}\theta_{d}\leq1,\;\Gamma_{\overline{d}}(\theta)\geq R+\rho\right\} ,
\]
which, loosely speaking, is comprised of $\theta$ for which the conditional
``capacity'' $\Gamma_{\overline{d}}(\theta)$ is larger than the
rate $R+\rho$. Since $\Theta_{\overline{d}}(R,\rho)$ is compact,
and since $E_{\overline{d}}(R,\mathsf{DNA}\mid\theta)$ is a continuous
function of $\theta$, it holds that 
\[
\epsilon(\rho)\dfn\min_{\theta\in\Theta_{\overline{d}}(R,\rho)}E_{\overline{d}}(R,\mathsf{DNA}\mid\theta)>0.
\]
Note that since the minimization defining $\epsilon(\rho)$ is over
$\theta$, rather than over $q^{N+1}$, $\epsilon(\rho)$ does not
depend on $M$, and the last inequality is strict. Now, taking $M$
to be sufficiently large so that $\tau_{M}\leq\epsilon(\rho)/2$ it
holds that if $\frac{q^{N+1}}{M}=\theta\in\Theta_{\overline{d}}(\rho)$
then 
\[
\E_{{\cal C}}\left[\pe({\cal C},{\cal D}_{\text{u}}\mid x^{LM}(0)\text{ stored},\theta)\right]=e^{-ML\cdot\epsilon(\rho)/2}.
\]
Based on this bound, the conditional error probability bound of Prop.
\ref{prop: random coding exponent condition on q} can be averaged
over $Q^{N+1}$ as follows:
\begin{align}
 & \E\left[\E_{{\cal C}}\left[\pe({\cal C},{\cal D}_{\text{u}}\mid x^{LM}(0)\text{ stored},Q^{N+1})\right]\right]\nonumber \\
 & =\sum_{q^{N+1}\in\mathscr{Q}(M,N)\colon\frac{q^{N+1}}{M}\in\Theta_{\overline{d}}(R,\rho)}\P\left[Q^{N+1}=q^{N+1}\right]\cdot\E_{{\cal C}}\left[\pe({\cal C},{\cal D}_{\text{u}}\mid x^{LM}(0)\text{ stored},Q^{N+1}=q^{N+1})\right]\nonumber \\
 & \hphantom{=}+\sum_{q^{N+1}\in\mathscr{Q}(M,N)\colon\frac{q^{N+1}}{M}\not\in\Theta_{\overline{d}}(R,\rho)}\P\left[Q^{N+1}=q^{N+1}\right]\cdot\E_{{\cal C}}\left[\pe({\cal C},{\cal D}_{\text{u}}\mid x^{LM}(0)\text{ stored},Q^{N+1}=q^{N+1})\right]\\
 & \leq\P\left[\frac{Q^{N+1}}{M}\not\in\Theta_{\overline{d}}(R,\rho)\right]+e^{-ML\cdot\epsilon(\rho)/2}\\
 & \trre[=,a]\exp\left[-M\cdot\left(\min_{\frac{q^{N+1}}{M}\not\in\Theta_{\overline{d}}(R,\rho)}-\frac{1}{M}\log\P[Q^{N+1}=q^{N+1}]\right)+O(\sqrt{M})\right]+e^{-ML\cdot\epsilon(\rho)/2}\\
 & \trre[=,b]\exp\left[-M\cdot\left(\min_{\frac{q^{N+1}}{M}\not\in\Theta_{\overline{d}}(R,\rho)}-\frac{1}{M}\log\P\left[\tilde{Q}^{N+1}=q^{N+1}\right]+O_{\alpha}\left(\frac{1}{\sqrt{M}}\right)\right)\right]+e^{-ML\cdot\epsilon(\rho)/2}\\
 & \trre[\leq,c]2\exp\left[-M\cdot\left(\min_{\frac{q^{N+1}}{M}\not\in\Theta_{\overline{d}}(R,\rho)}-\frac{1}{M}\log\left[\tilde{Q}^{N+1}=q^{N+1}\right]+O_{\alpha}\left(\frac{1}{\sqrt{M}}\right)\right)\right],\label{eq: averaging random coding bound w.r.t q}
\end{align}
where $(a)$ is using Lemma \ref{lem: outage set exponential charactarization},
$(b)$ is using Lemma \ref{lem: Poissonization of outage event},
where $\tilde{Q}^{N+1}=\mathscr{N}(\tilde{S}^{M})$ and $\tilde{S}_{m}\sim\text{Pois}(\alpha)$
are independent, $m\in[M]$, and $(c)$ holds for all $M\geq M_{0}(\rho)$
for some $M_{0}(\mathsf{DNA},\overline{d},\rho)\in\mathbb{N}^{+}$. 

We next evaluate $\min_{\frac{q^{N+1}}{M}\not\in\Theta_{\overline{d}}(R,\rho)}\frac{1}{M}\log\P[\tilde{Q}^{N+1}=q^{N+1}].$
We begin by considering the distribution of a fixed $q_{d}$ and then
the joint distribution of $\{q_{d}\}_{d\in[\overline{d}]}$. Since
$\tilde{S}^{M}$ has independent components, it holds for any specific
$d\in[\overline{d}]$ that $\tilde{Q}_{d}=\sum_{m\in[M]}\I\{\tilde{S}_{m}=d\}\sim\text{Binomial}(M,\pi_{\alpha}(d))$
where $\pi_{\alpha}(d)=\frac{\alpha^{d}e^{-\alpha}}{d!}$ is the Poisson
p.m.f.. By large deviations of the Binomial distribution (as in the
method of types for Bernoulli vectors \cite[Lemma 2.3]{csiszar2011information})
\[
\P\left[Q_{d}=\theta_{d}M\right]=\exp\left\{ -M\cdot\left[d_{b}\left(\theta_{d}\mid\mid\pi_{\alpha}(d)\right)-O\left(\frac{\log M}{M}\right)\right]\right\} .
\]
The joint distribution of $\{Q_{d}\}_{d\in[\overline{d}+1]}$ is determined
by the distribution of $Q_{d}$ conditioned on $\{Q_{0},\ldots,Q_{d-1}\}$
for all $d\in[\overline{d}+1]$. For such $d$, the total number of
molecules for which $\tilde{S}_{m}\leq d-1$ is $\gamma_{d}\dfn\sum_{i\in[d]}Q_{i}$.
By symmetry, the identity of those $\gamma_{d}$ molecules in $[M]$
is immaterial, and so we further condition, w.l.o.g. that these are
the last $\gamma_{d}$ molecules of $\tilde{S}^{M}$, that is $\tilde{S}_{M-\gamma_{d}},\ldots,\tilde{S}_{M-1}$.
Now, the p.m.f. of $\tilde{S}_{m}$ for $m\in[M-\gamma_{d}]$ conditioned
on this event is 
\begin{align}
\P\left[\tilde{S}_{m}=i\mid\tilde{S}_{m}\geq d\right] & =\frac{\pi_{\alpha}(i)}{\P\left[\tilde{S}_{m}\geq d\right]}\\
 & =\frac{\pi_{\alpha}(i)}{1-\sum_{i'\in[d]}\pi_{\alpha}(i')}\\
 & \dfn\pi_{\alpha|\geq d}(i),
\end{align}
and the $\tilde{S}_{m}$ are independent, where $\pi_{\alpha|\geq d}(i)$
is the Poisson hazard probability {[}as defined in (\ref{eq: Poisson hazard probability}){]}.
Hence, $\tilde{Q}_{d}\sim\text{Binomial}(M-\gamma_{d},\pi_{\alpha|\geq d}(d))$.
By large deviations of the Binomial distribution (as in the method
of types \cite[Lemma 2.3]{csiszar2011information})
\begin{align}
 & \P\left[\tilde{Q}_{d}=\theta_{d}M\mid\{\tilde{Q}_{i}=\theta_{i}M\}_{i\in[d]}\right]\nonumber \\
 & =\exp\left\{ -M\left(1-\frac{\gamma_{d}}{M}\right)\cdot\left[d_{b}\left(\frac{\theta_{d}}{1-\gamma_{d}/M}\,\middle\vert\middle\vert\,\pi_{\alpha|\geq d}(d)\right)-O\left(\frac{\log M}{M}\right)\right]\right\} \\
 & =\exp\left[-M\left(1-\sum_{i\in[d]}\theta_{i}\right)\cdot d_{b}\left(\frac{\theta_{d}}{(1-\sum_{i\in[d]}\theta_{i})}\,\middle\vert\middle\vert\,\pi_{\alpha|\geq d}(d)\right)-O\left(\frac{\log M}{M}\right)\right].
\end{align}
Thus, for $\{\theta_{d}M\}$ integers
\[
-\frac{1}{M}\log\P\left[\cap_{d\in[\overline{d}]}\{\tilde{Q}_{d}=\theta_{d}M\}\right]=\sum_{d\in[\overline{d}+1]}\left(1-\sum_{i\in[d]}\theta_{i}\right)\cdot d_{b}\left(\frac{\theta_{d}}{1-\sum_{i\in[d]}\theta_{i}}\,\middle\vert\middle\vert\,\pi_{\alpha|\geq d}(d)\right)+O\left(\frac{\log M}{M}\right).
\]
Using this estimate in (\ref{eq: averaging random coding bound w.r.t q})
we obtain 
\begin{align}
 & -\frac{1}{M}\log\E_{{\cal C}}\left[\pe({\cal C},{\cal D}_{\text{u}}\mid x^{LM}(0))\right]\nonumber \\
 & \geq\min_{\theta\not\in\Theta_{\overline{d}}(R,\rho)}\sum_{d\in[\overline{d}+1]}\left(1-\sum_{i\in[d]}\theta_{i}\right)\cdot d_{b}\left(\frac{\theta_{d}}{1-\sum_{i\in[d]}\theta_{i}}\,\middle\vert\middle\vert\,\pi_{\alpha|\geq d}(d)\right)+O_{\alpha}\left(\frac{1}{\sqrt{M}}\right).
\end{align}
From the linearity of expectation and symmetry, 
\[
\E_{{\cal C}}\left[\frac{1}{|{\cal C}|}\sum_{j\in[|{\cal C}|]}\pe({\cal C},{\cal D}\mid X^{LM}(j))\right]=\E_{{\cal C}}\left[\pe({\cal C},{\cal D}\mid X^{LM}(0))\text{ stored}\right],
\]
and so the same bounds hold for the average error probability. Finally,
we take $\rho\downarrow0$ and then $\overline{d}\uparrow\infty$.
The bound on the average error probability implies a suitable bound
to on the maximal error probability by a standard expurgation argument. 
\end{IEEEproof}

\subsection{Proof of Theorem \ref{thm: error exponent of universal decoder -- ideal amplification}
\label{subsec:The-proof-of theore ideal sampling exponent}}

We next prove Theorem \ref{thm: error exponent of universal decoder -- ideal amplification}
which bounds the error probability for idealized version of the sampling
stage of the DNA channel. 
\begin{IEEEproof}[Proof of Theorem \ref{thm: error exponent of universal decoder -- ideal amplification}]
 The result is a simple corollary to Prop. \ref{prop: random coding exponent condition on q},
obtained by setting $q_{d}=M\cdot\I[d=\alpha]$ in the conditioning
event. For any finite $\overline{d}>\alpha$, the resulting reliability
function, w.r.t. scaling $ML$, is given by 
\[
E_{\overline{d}}(R,\mathsf{DNA}\mid q)=\min_{Q_{AB^{\alpha}}}\left(D(Q_{A}\mid\mid P_{X})+D(Q_{B^{\alpha}|A}\mid\mid V^{\oplus\alpha}|Q_{A})\right)+\left[D(Q_{A}\mid\mid P_{X})+I_{Q}(A;B^{\alpha})-\frac{1}{\beta}-R\right]_{+}.
\]
For ideal sampling $q_{d}=M\cdot\I[d=\alpha]$ with probability $1$,
and so no further averaging is required to obtain the bound on the
error probability.
\end{IEEEproof}

\section{Proof of Theorem \ref{thm: converse capacity} \label{sec:Converse:-Proof-of}}

\begin{figure}
\centering{}\includegraphics[scale=1.25]{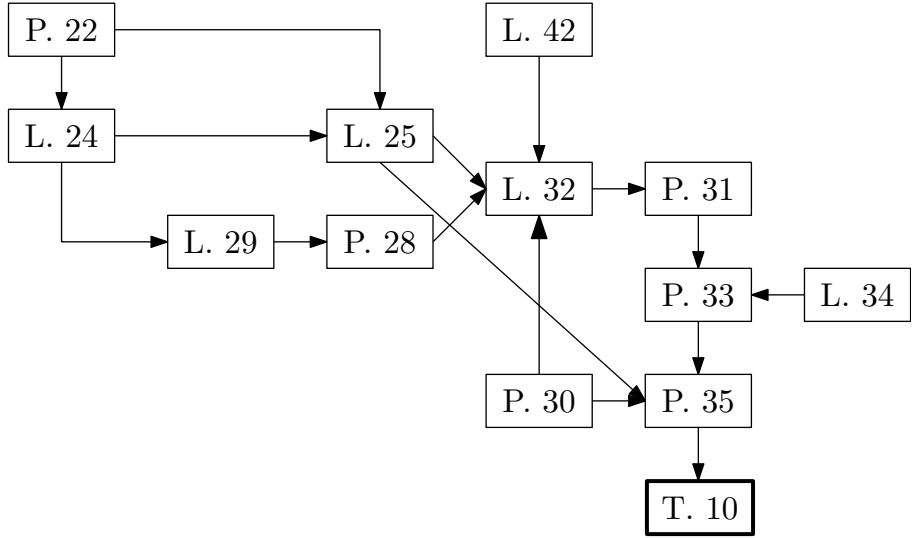}\caption{Roadmap for the proofs of the upper bound (converse). The used abbreviations
and arrow interpretation are as in Fig. \ref{fig:Roadmap-lower}.
\label{fig:Roadmap-upper}.}
\end{figure}

In this section, we prove Theorem \ref{thm: converse capacity}. A
roadmap for the results required to prove it appears in Fig. \ref{fig:Roadmap-upper}.
 Recall the definition of the CID in (\ref{eq: definition common input deficit}),
and define
\begin{equation}
J_{d}(P_{X},W,\beta)\dfn\left(I(P_{X},W^{\oplus d})-\left[\mathsf{CID}(P_{X},W^{\oplus d})-\frac{1}{\beta}\right]_{+}+\left[\mathsf{CID}(P_{X},W^{\oplus d})-\frac{2}{\beta}\right]_{+}\right).\label{eq: definition of J for a channel, input distribution and beta}
\end{equation}

The main argument of the proof is standard, and given as follows:
\begin{IEEEproof}[Proof of Theorem \ref{thm: converse capacity}]
Assume that ${\cal C}_{M}$ is an $(M,\epsilon)$ codebook of size
$e^{MLR}$. Fano's inequality implies that 
\begin{equation}
R\leq\frac{1}{ML}+\epsilon+\frac{I(X^{LM};Y^{LN})}{ML}.\label{eq: use of Fano's inequality}
\end{equation}
The rest of the section is devoted to bounding $I(X^{LM};Y^{LN})$.
Its final result Prop. \ref{prop:converse genie aided avearge general composition},
which states that 
\[
\frac{1}{ML}I(X^{LM};Y^{LN})\leq\max_{P_{X}}\sum_{d\in\mathbb{N}}\pi_{\alpha}(d)\cdot J_{d}(P_{X},W,\beta)+o(1).
\]
Taking $\epsilon\downarrow0$, and replacing $J_{d}(P_{X},W,\beta)$
with the excess term $\Omega_{d}(P_{X},W,\beta)$ from (\ref{eq: excess term in the capacity upper bound})
by simple algebraic manipulations completes the proof of the upper
bound on the capacity.
\end{IEEEproof}

\subsection{Typical Sets\label{sec:A-brief-review}}

In this short section we set basic definitions of conditional typical
sets \cite[Ch. 2]{csiszar2011information}, and briefly review some
standard results. Let $V\colon{\cal A}\to{\cal B}$ be a DMC for the
finite alphabets ${\cal A}$ and ${\cal B}$, and let $K\in\mathbb{N}^{+}$
be the blocklength. 

For a given $Q_{A}\in{\cal P}_{K}({\cal A})$ and $\delta>0$ let
\begin{multline}
[V\mid Q_{A}]_{\delta}\dfn\left\{ V\in{\cal P}_{K}({\cal B}\mid Q_{A})\colon\;Q_{A}(a)\cdot\left|\tilde{V}(b\mid a)-V(b\mid a)\right|\leq\delta,\;\forall(a,b)\in{\cal A}\times{\cal B}\right\} \\
\cap\left\{ \tilde{V}(\cdot\mid a)\ll V(\cdot\mid a),\;\forall a\in\supp(Q_{A})\right\} \label{eq: conditional types distance}
\end{multline}
be a set of conditional types which are $\delta$-close to $V$ in
a total variation sense, uniformly over ${\cal A}$. In what follows
whenever $Q_{A}$ in the definition (\ref{eq: conditional types distance})
can be understood from the context, it will be omitted from the notation,
as in $[V]_{\delta}$. With this definition, the \emph{$V$-typical
set }is defined as 
\begin{equation}
{\cal T}_{K}([V]_{\delta}\mid a^{K})\dfn\cup_{\tilde{V}\in[V]_{\delta}}{\cal T}_{K}(\tilde{V}\mid a^{K}),\label{eq: V-typical set of a DMC}
\end{equation}
where $Q_{A}=\mathscr{P}(a^{K})$ is implicit in the notation. Let
$B^{K}\sim V^{K}(\cdot\mid a^{K})$ where $V^{K}$ is the $K$th product
of $V$. For brevity, we next denote 
\[
V\left({\cal T}_{K}([V]_{\delta}\mid a^{K})\mid a^{K}\right)\equiv\P\left[B^{K}\in{\cal T}_{K}([V]_{\delta}\mid a^{K})\mid a^{K}\right].
\]

\begin{prop}
\label{prop:Typical-set}(Typical sets \cite[Variation of Lemma 2.12 and Lemma 2.13]{csiszar2011information}):
Let $\delta>0$ be given. Then, 
\begin{equation}
V\left({\cal T}_{K}([V]_{\delta}\mid a^{K})\mid a^{K}\right)\geq1-2|{\cal A}||{\cal B}|e^{-K\delta^{2}}\label{eq: typical set high probability}
\end{equation}
and 
\begin{equation}
\exp\left[K\cdot\left(H_{\mathscr{P}(a^{k})\times V}(B\mid A)-\mu\right)\right]\leq\left|{\cal T}_{K}([V]_{\delta}\mid a^{k})\right|=\exp\left[K\cdot\left(H_{\mathscr{P}(a^{k})\times V}(B\mid A)+\mu\right)\right]\label{eq: size of V-typical set}
\end{equation}
where 
\begin{equation}
\mu\equiv\mu_{K}\equiv|{\cal A}||{\cal B}|\left(\delta\log\frac{1}{\delta}+\frac{\log(K+1)}{K}\right).\label{eq: definition of mu}
\end{equation}
\end{prop}
Prop. \ref{prop:Typical-set} implies that $V[{\cal T}_{K}([V]_{\delta}\mid a^{K})\mid a^{K}]\to1$
as $K\to\infty$ as long as $\delta\equiv\delta_{K}$ satisfies the
\emph{delta-convention} \cite[Convention 2.11]{csiszar2011information},
that is -- $\delta_{K}=o(1)$, $\delta_{K}=\omega(\frac{1}{\sqrt{K}})$.
The constant in the order terms depend only on $|{\cal A}|,|{\cal B}|$.
Henceforth, we assume that the delta-convention is satisfied, that
$\mu_{K}$ is as in (\ref{eq: definition of mu}) and that $\delta_{K}$
is chosen to accommodate the largest alphabet sizes possible. For
example, we will consider channels ${\cal A}\to{\cal B}^{d}$ where
$d\leq\overline{d}$ for some bounded $\overline{d}$, and accordingly,
$\delta_{K}$ is chosen to satisfy the delta convention for $|{\cal A}|,|{\cal B}|^{d}$.
To lighten the notation, we will mostly write $\delta\equiv\delta_{K}$. 
\begin{rem}
\label{rem: From conditional types to types}When $|{\cal A}|=1$,
that is, the input is deterministic, Prop. \ref{prop:Typical-set}
and the discussion between them provide bounds on the probability
and the size of typical sets. Following standard notation, we will
omit in this case the conditioning variable, and use, ${\cal T}_{K}([Q_{A\overline{A}}]_{\delta_{K}})$,
for example, for the typical set of $Q_{A\overline{A}}.$\textbf{ }
\end{rem}

\subsection{The Molecule Distance Function and its Implications \label{subsec:The-molecule-distance}}

As was stated in Prop. \ref{prop:Typical-set}, the $V$-typical set
${\cal T}_{K}([V]_{\delta}\mid a^{K})\subseteq{\cal B}^{K}$ has high
probability when the output is drawn as $B^{K}\sim V^{K}(\cdot\mid a^{K})$.
We next explore the probability of the same $V$-typical set conditioned
on $a^{K}$, when $B^{K}\sim V^{K}(\cdot\mid\overline{a}^{K})$, that
is, with a different channel input $\overline{a}^{K}$. We define
a proper distance between $a^{K}$ and $\overline{a}^{K}$, and show
that if $a^{K}$ and $\overline{a}^{K}$ are ``far'' according to
this distance, then this probability is exponentially small. On the
other hand, we also show that if they are ``close'', then this probability
is still exponentially small, albeit with an arbitrarily small exponent. 

The key issue is a proper choice of the distance measure, which is
of course channel dependent. Specifically, for a joint type $Q_{A\overline{A}}\in{\cal P}_{K}({\cal A}\times{\cal A})$
and a DMC $V\colon{\cal A}\to{\cal B}$, we define the distance function

\begin{equation}
\Delta(Q_{A\overline{A}},V)\dfn H_{Q_{A}\times V}(B\mid A)-\max_{Q_{B\mid A\overline{A}}\colon Q_{B|A}\in[V\mid Q_{A}]_{\delta},\;Q_{B|\overline{A}}\in[V\mid Q_{\overline{A}}]_{\delta}}H_{Q}(B\mid A,\overline{A}).\label{eq: Distance between molecules definition}
\end{equation}
Let $Q_{A\overline{A}B}^{*}$ be the maximizer in (\ref{eq: Distance between molecules definition}).
Since by definition $Q_{B|A}^{*}\in[V\mid Q_{A}]_{\delta}$ then its
conditional entropy satisfies $H_{Q^{*}}(B\mid A)\approx H_{Q_{A}\times V}(B\mid A)$,
that is, close to the conditional entropy induced by the channel $V$
itself. Since conditioning reduces entropy $H_{Q^{*}}(B\mid A,\overline{A})\lesssim H_{Q_{A}\times V}(B\mid A)$,
and (approximate) equality is achieved when $Q_{\overline{A}|A}$
is the identity, noiseless, channel (this is however not the only
case). In this case $\Delta(Q_{A\overline{A}},V)\approx0$. Thus,
if $Q_{\overline{A}|A}$ is close to the identity, the distance is
close to zero. 

In what follows, we will consider a pair of inputs $a^{K}$ and $\overline{a}^{K}$
to be ``close'' if $\Delta(\mathscr{P}(a^{K},\overline{a}^{K}),V)\leq\underline{\rho}_{K}$
where 
\begin{equation}
\underline{\rho}_{K}\dfn2\left[\delta_{K}^{2}+\mu_{K}+|{\cal A}||{\cal B}|\delta_{K}\log\frac{1}{\delta_{K}}+\frac{3|{\cal A}||{\cal B}|\log K}{K}\right],\label{eq: minimal value for rho}
\end{equation}
and ``far'' otherwise. It should be stressed that by the delta-convention,
$\underline{\rho}_{K}=o(1)$, and thus even a ``far'' pair of inputs
may have in fact, normalized distance which tends to zero, albeit
with a slow enough decay rate. 

In the next three subsections we consider various implications of
the distance function.

\subsubsection{Probability of a Typical Set Conditioned on Close v.s. Far Inputs}
\begin{lem}
\label{lem: Probability of V-typical sets condition on a different vector}Let
$\delta_{K}$ satisfy the delta-convention be given and $\rho_{K}$
be given such that $\rho_{K}\geq\underline{\rho}_{K}$. Further, let
$a^{K},\overline{a}^{K}\in{\cal A}^{K}$ be given for some $K\in\mathbb{N}^{+}$.
Then, there exists $K_{0}$ (which only depends on $|{\cal A}|,|{\cal B}|$)
such that for all $K\geq K_{0}$ it holds that 
\begin{equation}
V\left({\cal T}_{K}([V]_{\delta_{K}}\mid\overline{a}^{K})\mid a^{K}\right)\le e^{-K\rho_{K}/2}\label{eq: V-typical set of a different input vector upper bound}
\end{equation}
if $\Delta(\mathscr{P}(a^{K},\overline{a}^{K}),V)\geq\rho_{K}$ and
\begin{equation}
V\left({\cal T}_{K}([V]_{\delta_{K}}\mid\overline{a}^{K})\mid a^{K}\right)\geq e^{-2K\rho_{K}}\label{eq: V-typical set of a different input vector lower bound}
\end{equation}
if $\Delta(\mathscr{P}(a^{K},\overline{a}^{K}),V)<\rho_{K}$. 
\end{lem}
\begin{IEEEproof}
For brevity, we denote $\delta\equiv\delta_{K}$, $\mu\equiv\mu_{K}$
and $\rho\equiv\rho_{K}$. We first prove (\ref{eq: V-typical set of a different input vector upper bound}),
and to this end, we upper bound the size of the intersection of a
pair of $V$-shells given by $\tilde{V}\in[V\mid\mathscr{P}(a^{K})]_{\delta}$
and $\overline{V}\in[V\mid\mathscr{P}(\overline{a}^{K})]_{\delta}$.
By the method of types 
\begin{align}
\left|{\cal T}_{K}(\tilde{V}\mid a^{K})\cap{\cal T}_{K}(\overline{V}\mid\overline{a}^{K})\right| & =\sum_{Q_{B\mid A\overline{A}}}\left|{\cal T}_{K}(Q_{B\mid A\overline{A}}\mid a^{K},\overline{a}^{K})\right|\\
 & \trre[\leq,a]\sum_{Q_{B\mid A\overline{A}}}\exp\left[K\cdot H_{\mathscr{P}(a^{K},\overline{a}^{K})\times Q_{B\mid A\overline{A}}}(B\mid A,\overline{A})\right]\\
 & \leq\left|{\cal P}_{K}({\cal B}\mid\mathscr{P}(a^{K},\overline{a}^{K}))\right|\cdot\max_{Q_{B\mid A\overline{A}}}\exp\left[K\cdot H_{Q}(B\mid A,\overline{A})\right]\\
 & \trre[\leq,b]\left|{\cal P}_{K}({\cal B}\mid\mathscr{P}(a^{K},\overline{a}^{K}))\right|\cdot\exp\left[K\cdot\left(H_{\mathscr{P}(a^{K})\times V}(B\mid A)-\rho\right)\right],\label{eq: intersection of V-shells size upper bound}
\end{align}
where the first two summations and the following maximization are
over 
\[
\left\{ Q_{B\mid A\overline{A}}\in{\cal P}_{K}({\cal B}\mid\mathscr{P}(a^{K},\overline{a}^{K}))\colon\;Q_{B|A}=\tilde{V},\;Q_{B|\overline{A}}=\overline{V}\right\} ,
\]
$(a)$ follows by a generalization of size of a $V$-shell in \cite[Lemma 2.5]{csiszar2011information}
(for this specific result, see \cite[Problem 2.10]{csiszar2011information}),
and $(b)$ follows from the definition of $\Delta(Q_{A\overline{A}},V)$
in (\ref{eq: Distance between molecules definition}), and the assumption
$\Delta(Q_{A\overline{A}},V)\geq\rho$. 

Based on (\ref{eq: intersection of V-shells size upper bound}), we
evaluate the $V$-probability of intersection of a pair of typical
sets:
\begin{align}
 & V\left({\cal T}_{K}([V]_{\delta}\mid a^{K})\cap{\cal T}_{K}([V]_{\delta}\mid\overline{a}^{K})\mid a^{K}\right)\nonumber \\
 & =V\left(\bigcup_{\tilde{V}\in[V\mid\mathscr{P}(a^{K})]_{\delta},\;\overline{V}\in[V\mid\mathscr{P}(\overline{a}^{K})]_{\delta}}{\cal T}_{K}(\tilde{V}\mid a^{K})\cap{\cal T}_{K}(\overline{V}\mid\overline{a}^{K})\,\middle\vert\,a^{K}\right)\\
 & \trre[\leq,a]\sum_{\tilde{V}\in[V\mid\mathscr{P}(a^{K})]_{\delta},\;\overline{V}\in[V\mid\mathscr{P}(\overline{a}^{K})]_{\delta}}V\left({\cal T}_{K}(\tilde{V}\mid a^{K})\cap{\cal T}_{K}(\overline{V}\mid\overline{a}^{K})\,\middle\vert\,a^{K}\right)\\
 & \trre[=,b]\sum_{\tilde{V}\in[V\mid\mathscr{P}(a^{K})]_{\delta},\;\overline{V}\in[V\mid\mathscr{P}(\overline{a}^{K})]_{\delta}}\frac{\left|{\cal T}_{K}(\tilde{V}\mid a^{K})\cap{\cal T}_{K}(\overline{V}\mid\overline{a}^{K})\right|}{\left|{\cal T}_{K}(\tilde{V}\mid a^{K})\right|}V\left({\cal T}_{K}(\tilde{V}\mid a^{K})\,\middle\vert\,a^{K}\right)\\
 & \trre[\leq,c]\left|{\cal P}_{K}({\cal A}\times{\cal A}\times{\cal B})\right|^{2}\sum_{\tilde{V}\in[V\mid\mathscr{P}(a^{K})]_{\delta}}\frac{\exp\left[K\cdot\left(H_{\mathscr{P}(a^{K})\times V}(B\mid A)-\rho\right)\right]}{\left|{\cal T}_{K}(\tilde{V}\mid a^{K})\right|}V\left({\cal T}_{K}(\tilde{V}\mid a^{K})\,\middle\vert\,a^{K}\right)\\
 & \trre[\leq,d]\left|{\cal P}_{K}({\cal A}\times{\cal A}\times{\cal B})\right|^{2}\exp\left[K\cdot\left(\text{\ensuremath{\mu}}-\rho+|{\cal A}||{\cal B}|\delta\log\frac{1}{\delta}+\frac{|{\cal A}||{\cal B}|\log K}{K}\right)\right]\sum_{\tilde{V}\in[V\mid\mathscr{P}(a^{K})]_{\delta}}V\left({\cal T}_{K}(\tilde{V}\mid a^{K})\,\middle\vert\,a^{K}\right)\\
 & \trre[\leq,e]\exp\left[K\cdot\left(\text{\ensuremath{\mu}}-\rho+|{\cal A}||{\cal B}|\delta\log\frac{1}{\delta}+\frac{3|{\cal A}||{\cal B}|\log K}{K}\right)\right]V\left({\cal T}_{K}([V]_{\delta}\mid a^{K})\,\middle\vert\,a^{K}\right)\\
 & \trre[\leq,f]e^{-K\rho/2},\label{eq: probability of intersection of V-typical sets upper bound}
\end{align}
where $(a)$ follows from the union bound, $(b)$ follows since the
$V$-probabilities of vectors which belong to the same $V$-shell
${\cal T}_{K}(\tilde{V}\mid a^{K})$ are equal, $(c)$ follows from
(\ref{eq: intersection of V-shells size upper bound}), $(d)$ follows
since from the method-of-types bound on the size of a $V$-shell \cite[Lemma 2.5]{csiszar2011information}
\[
\left|{\cal T}_{K}(\tilde{V}\mid a^{K})\right|\geq\exp\left[K\cdot\left(H_{\mathscr{P}(a^{K})\times V}(B\mid A)-\frac{|{\cal A}||{\cal B}|\log K}{K}\right)\right],
\]
and since as was shown in \cite[p. 22, proof of Lemma 2.13]{csiszar2011information},
if $\tilde{V}\in[V\mid\mathscr{P}(a^{K})]_{\delta}$, then 
\begin{equation}
\left|H_{\mathscr{P}(a^{K})\times\tilde{V}}(B\mid A)-H_{\mathscr{P}(a^{K})\times V}(B\mid A)\right|\leq|{\cal A}||{\cal B}|\delta\log\frac{1}{\delta},\label{eq: continuity of conditional entropy for shells which are typical}
\end{equation}
$(e)$ follows again from type counting 
\[
|{\cal P}_{K}({\cal A}\times{\cal A}\times{\cal B})|\leq(K+1)^{|{\cal A}|^{2}|{\cal B}|}=\exp\left[|{\cal A}|^{2}|{\cal B}|\cdot\log(K+1)\right]
\]
(\cite[Lemma 2.5]{csiszar2011information}), and $(f)$ from the assumption
on $\rho$. Hence, 
\begin{align}
 & V\left({\cal T}_{K}([V]_{\delta}\mid\overline{a}^{K})\mid a^{K}\right)\nonumber \\
 & =V\left({\cal T}_{K}([V]_{\delta}\mid\overline{a}^{K})\cap{\cal T}_{K}([V]_{\delta}\mid a^{K})\mid a^{K}\right)+V\left({\cal T}_{K}([V]_{\delta}\mid\overline{a}^{K})\cap{\cal T}_{K}^{c}([V]_{\delta}\mid a^{K})\mid a^{K}\right)\\
 & \leq V\left({\cal T}_{K}([V]_{\delta}\mid\overline{a}^{K})\cap{\cal T}_{K}([V]_{\delta}\mid a^{K})\mid a^{K}\right)+V\left({\cal T}_{K}^{c}([V]_{\delta}\mid a^{K})\mid a^{K}\right)\\
 & \trre[\leq,a]e^{-K\rho/2}+e^{-K\delta^{2}}\leq e^{-K\rho/2},
\end{align}
where $(a)$ follows from (\ref{eq: probability of intersection of V-typical sets upper bound})
and Prop. \ref{prop:Typical-set} and $(b)$ from the assumption $\rho>2\delta^{2}$
{[}see (\ref{eq: minimal value for rho}){]}. 

The reverse statement (\ref{eq: V-typical set of a different input vector lower bound})
is proved analogously, and so only the main steps are given (with
fewer explanations). Analogously to (\ref{eq: intersection of V-shells size upper bound})
it holds that 
\begin{align}
\left|{\cal T}_{K}(\tilde{V}\mid a^{K})\cap{\cal T}_{K}(\overline{V}\mid\overline{a}^{K})\right| & =\sum_{Q_{B\mid A\overline{A}}}\left|{\cal T}_{K}(Q_{B\mid A\overline{A}}\mid a^{K},\overline{a}^{K})\right|\\
 & \geq\left|{\cal P}_{K}({\cal B}\mid\mathscr{P}(a^{K},\overline{a}^{K}))\right|^{-1}\sum_{Q_{B\mid A\overline{A}}}\exp\left[K\cdot H_{\mathscr{P}(a^{K},\overline{a}^{K})\times Q_{B\mid A\overline{A}}}(B\mid A,\overline{A})\right]\\
 & \geq\left|{\cal P}_{K}({\cal B}\mid\mathscr{P}(a^{K},\overline{a}^{K}))\right|^{-2}\cdot\max_{Q_{B\mid A\overline{A}}}\exp\left[K\cdot H_{\mathscr{P}(a^{K},\overline{a}^{K})\times Q_{B\mid A\overline{A}}}(B\mid A,\overline{A})\right]\\
 & >\left|{\cal P}_{K}({\cal B}\mid\mathscr{P}(a^{K},\overline{a}^{K}))\right|^{-2}\cdot\exp\left[K\cdot\left(H_{\mathscr{P}(a^{K})\times V}(B\mid A)-\rho\right)\right].\label{eq: intersection of V-shells size lower bound}
\end{align}
with the maximization over the same set as in (\ref{eq: intersection of V-shells size upper bound}).
Then, analogously to (\ref{eq: probability of intersection of V-typical sets upper bound})
\begin{align}
 & V\left({\cal T}_{K}([V]_{\delta}\mid a^{K})\cap{\cal T}_{K}([V]_{\delta}\mid\overline{a}^{K})\mid a^{K}\right)\nonumber \\
 & =V\left(\bigcup_{\tilde{V}\in[V\mid\mathscr{P}(a^{K})]_{\delta},\;\overline{V}\in[V\mid\mathscr{P}(\overline{a}^{K})]_{\delta}}{\cal T}_{K}(\tilde{V}\mid a^{K})\cap{\cal T}_{K}(\overline{V}\mid\overline{a}^{K})\,\middle\vert\,a^{K}\right)\\
 & \geq\max_{\tilde{V}\in[V\mid\mathscr{P}(a^{K})]_{\delta},\;\overline{V}\in[V\mid\mathscr{P}(\overline{a}^{K})]_{\delta}}V\left({\cal T}_{K}(\tilde{V}\mid a^{K})\cap{\cal T}_{K}(\overline{V}\mid\overline{a}^{K})\,\middle\vert\,a^{K}\right)\\
 & =\max_{\tilde{V}\in[V\mid\mathscr{P}(a^{K})]_{\delta},\;\overline{V}\in[V\mid\mathscr{P}(\overline{a}^{K})]_{\delta}}\frac{\left|{\cal T}_{K}(\tilde{V}\mid a^{K})\cap{\cal T}_{K}(\overline{V}\mid\overline{a}^{K})\right|}{\left|{\cal T}_{K}(\tilde{V}\mid a^{K})\right|}V\left({\cal T}_{K}(\tilde{V}\mid a^{K})\,\middle\vert\,a^{K}\right)\\
 & \trre[\geq,c]\left|{\cal P}_{K}({\cal A}\times{\cal A}\times{\cal B})\right|^{-2}\max_{\tilde{V}\in[V\mid\mathscr{P}(a^{K})]_{\delta}}\frac{\exp\left[K\cdot\left(H_{\mathscr{P}(a^{K})\times V}(B\mid A)-\rho\right)\right]}{\left|{\cal T}_{K}(\tilde{V}\mid a^{K})\right|}V\left({\cal T}_{K}(\tilde{V}\mid a^{K})\,\middle\vert\,a^{K}\right)\\
 & \trre[\geq,d]\left|{\cal P}_{K}({\cal A}\times{\cal A}\times{\cal B})\right|^{-2}\exp\left[K\cdot\left(-\rho-|{\cal A}||{\cal B}|\delta\log\frac{1}{\delta}\right)\right]\max_{\tilde{V}\in[V\mid\mathscr{P}(a^{K})]_{\delta}}V\left({\cal T}_{K}(\tilde{V}\mid a^{K})\mid a^{K}\right)\\
 & \geq\left|{\cal P}_{K}({\cal A}\times{\cal A}\times{\cal B})\right|^{-3}\exp\left[K\cdot\left(-\rho-|{\cal A}||{\cal B}|\delta\log\frac{1}{\delta}\right)\right]\sum_{\tilde{V}\in[V\mid\mathscr{P}(a^{K})]_{\delta}}V\left({\cal T}_{K}(\tilde{V}\mid a^{K})\mid a^{K}\right)\\
 & \geq\exp\left[K\cdot\left(-\rho-|{\cal A}||{\cal B}|\delta\log\frac{1}{\delta}-\frac{3|{\cal A}||{\cal B}|\log K}{K}\right)\right](1-e^{-K\delta^{2}})\\
 & \geq e^{-2K\rho},\label{eq: probability of intersection of V-typical sets lower bound}
\end{align}
for all $K$ large enough. 
\end{IEEEproof}

\subsubsection{Equivocation of a Permutation for Far Inputs}

Even in a simplified DNA channel, in which $N=M$ and any molecule
in the codeword is sampled exactly once ($S_{m}=1$ for all $m\in[M])$,
the decoder still faces an uncertainty regarding the \emph{order }in
which the molecules were sequenced (that is, the\emph{ }molecule index
vector \emph{$U^{N}$}).\emph{ }In this section, we show that if the
input molecules are sufficiently far apart, then the decoder can infer
some information regarding their order. To this end, we measure closeness
according to the given the distance definition in (\ref{eq: Distance between molecules definition}),
and utilize the characterization of typical set probability in Lemma
\ref{lem: Probability of V-typical sets condition on a different vector}.
As said, the result refines \cite[proof of Lemma 3]{shomorony2021dna}.

We thus next focus on the following scenario, which involves a ``large''
number $T\in\mathbb{N}^{+}$ of an \emph{ordered sequence of }vectors
$a^{KT}=(a_{0}^{K},a_{1}^{K},\ldots,a_{T-1}^{K})$ so that $a_{t}^{K}\in{\cal A}^{K}$
for all $t\in[T]$, and a DMC $V\colon{\cal A}\to{\cal B}$.\footnote{Here, $K$ plays the role of the length of the molecule $L$, and
$T$ the number of molecules in a codeword $M$. However, when the
result of the next Lemma \ref{lem: decoding permutation lemma} will
be utilized to prove the upper bound on the capacity, we will set
$K=L$ and $T$ to be roughly $M$. Hence the different notation. } The order of the $T$ vectors is permuted by a permutation $\Sigma\colon[T]\to[T]$
chosen uniformly at random from the symmetric group $\mathfrak{S}_{T}$
to obtain $(a_{\Sigma(0)}^{K},a_{\Sigma(1)}^{K},\ldots,a_{\Sigma(T-1)}^{K})$.
Then, each $a_{\Sigma(t)}^{K}$, $t\in[T]$ is input to the channel
$V$ according to order (a total of $KT$ uses) and the result is
$B^{KT}=(B_{0}^{K},B_{1}^{K},\ldots,B_{T-1}^{K})$ where $B_{t}^{K}\in{\cal B}^{K}$
and by the description above $B_{t}^{K}\sim V^{K}(\cdot\mid a_{\Sigma(t)}^{K})$.
The observer of the channel output is assumed to be aware of the input
vectors, but not the permutation $\Sigma$. We next quantify the ability
of the observer to re-order the output and essentially guess the permutation,
in terms of its equivocation. Clearly, if the channel $V$ is noiseless,
and all $a_{t}^{K}$ are different from one another, then perfect
ordering is possible, with probability $1$, and so $H(\Sigma\mid B^{KT})=0$.
On the other hand, if the channel is extremely noisy (the output is
independent of the input), or if $a_{t}^{K}$ are all equal, then
no information on the permutation is revealed by the channel to the
observer. In that case 
\[
H(\Sigma\mid B^{KT})=\log\left|\mathfrak{S}_{T}\right|=\log\left|T!\right|=T\log T-T+O(\log T),
\]
where the last equality is by Stirling's approximation. The next lemma
concerns an intermediate situation. It assumes that $a_{t_{1}}^{K},a_{t_{2}}^{K}$
are ``far'' according to the definition of $\Delta(\mathscr{P}(a_{t_{1}}^{K},a_{t_{2}}^{K}),V)$
in (\ref{eq: Distance between molecules definition}), for any $t_{1},t_{2}\in[T]$,
$t_{1}\neq t_{2}$, and that the channel can be noisy. In this case,
it is shown that the entropy is $o(T\log T)$, that is, negligible
compared to the first-order term $T\log T$. Importantly, this is
shown for an exponential number of vectors, which scales as $T\geq e^{\tau_{0}K}$
for some $\tau_{0}>0$. 
\begin{lem}
\label{lem: decoding permutation lemma} Let $T_{K}$ be such that
$T_{K}\geq e^{K\tau_{0}}$ for some $\tau_{0}>0$ and all large enough
$K$. Let $V\colon{\cal A}\to{\cal B}$ be a DMC, and let $\rho_{K}$
be such that $\underline{\rho}_{K}\leq\rho_{K}<\tau_{0}$ for all
large enough $K$ (where $\underline{\rho}_{K}$ is as defined in
(\ref{eq: minimal value for rho})). Further let $(a_{0}^{K},a_{1}^{K},\ldots,a_{T-1}^{K})\subset({\cal A}^{K})^{T}$
be such that any pair of vectors is ``far'', to wit
\[
\Delta(\mathscr{P}(a_{t_{1}}^{K},a_{t_{2}}^{K})),V)\geq\underline{\rho}_{K}
\]
for any $t_{1},t_{2}\in[T]$ with $t_{1}\neq t_{2}$. Moreover, let
$\Sigma\sim\text{Uniform}(\mathfrak{S}_{T})$ be a uniformly random
permutation, and let $B_{t}^{K}\sim V^{K}(\cdot\mid a_{\Sigma(t)}^{K})$
independently for all $t\in[T]$. Then, there exists $K_{0}$ such
that for all $K\geq K_{0}$
\[
H(\Sigma\mid B^{KT})=o(T\log T).
\]
\end{lem}
\begin{IEEEproof}
We denote $T=e^{\tau_{K}K}$ for $K\in\mathbb{N}^{+}$ such that according
to the assumption of the lemma $\tau_{K}\geq\tau_{0}$ for all large
enough $K$. From symmetry we may assume that $\Sigma$ was drawn
to be the identity permutation $\Sigma(t)=t$, and thus we (implicitly)
assume in the rest of the proof that $B_{t}^{K}\sim V^{K}(\cdot\mid a_{t}^{K})$,
independently for all $t\in[T]$. Let $\delta_{K}>0$ be given which
defines the $V$-typical set ${\cal T}_{K}([V]_{\delta_{K}}\mid a_{t}^{K})$.

When $a_{t}^{K}$ is fed into the channel $V$ there are two possible
``bad'' events to consider:
\begin{enumerate}
\item \emph{A-typical} \emph{output}: The channel output $B_{t}^{K}$ is
not $V$-typical conditioned on its input $a_{t}^{K}$, namely the
event whose indicator is 
\[
e_{t}\dfn\I\left\{ B_{t}^{K}\notin{\cal T}_{K}([V]_{\delta}\mid a_{t}^{K})\right\} .
\]
\item \emph{Large ambiguity output}: For some $t'\neq t$, $B_{t}^{K}$
belongs to a large number of $V$-typical sets for inputs which are
\emph{not} $a_{t}^{K}$, namely the event whose indicator is 
\[
f_{t}\dfn\I\left\{ \sum_{t'\in[T]\backslash\{t\}}\I\{B_{t}^{K}\in{\cal T}_{K}([V]_{\delta}\mid a_{t'}^{K})\}\geq T_{0}\right\} ,
\]
where $T_{0}>0$ is a threshold that will be chosen later. 
\end{enumerate}
We next analyze the events related to $\{e_{t}\}_{t\in[T]}$ and $\{f_{t}\}_{t\in[T]}$.
We begin by analyzing $\{e_{t}\}_{t\in[T]}$. By Prop. \ref{prop:Typical-set}
$\E[e_{t}]\leq e^{-K\delta_{K}^{2}}$. We assume that $K$ is large
so that $\delta_{K}$ is sufficiently small so that $\tau_{K}>\tau_{0}>\delta_{K}^{2}$.
Thus, the expected of total number of atypical outputs is $\E[\sum_{t\in[T]}e_{t}]\leq e^{K(\tau_{K}-\delta_{K}^{2})}=o(T)$
as $T=e^{K\tau_{K}}$ and $\delta_{K}$ satisfies the delta-convention.
We now let 
\[
{\cal E}\dfn\left\{ \sum_{t\in[T]}e_{t}\geq e^{K(\tau_{K}-\delta_{K}^{2}/2)}\right\} 
\]
be the event that the total number of atypical events exceeds its
expectation by a factor of more than $e^{K\delta_{K}^{2}/2}$. Then,
since $e_{t}$ are independent events, Chernoff's bound implies that
\begin{align}
\P\left[{\cal E}\right] & \leq\exp\left[-e^{K\tau_{K}}\cdot d_{b}\left(e^{-K\delta_{K}^{2}/2}\mid\mid\E[e_{t}]\right)\right]\\
 & \trre[\leq,a]\exp\left[-e^{K\tau_{K}}\cdot e^{-K\delta_{K}^{2}/2}\left(\log\frac{e^{-K\delta_{K}^{2}/2}}{\E[e_{t}]}-1\right)\right]\\
 & \leq\exp\left[-e^{K(\tau_{K}-\delta_{K}^{2}/2)}\left(K\delta_{K}^{2}/2-1\right)\right],
\end{align}
where $(a)$ follows from the bound $d_{b}(p_{1}\mid\mid p_{2})>p_{1}(\log\frac{p_{1}}{p_{2}}-1)$
for $p_{1},p_{2}\in[0,1]$ \cite[Sec. 6.3]{merhav2010statistical}.
Since $\delta_{K}$ satisfies the delta-convention, it holds that
$K\delta_{K}^{2}=\omega(1)$ and hence the probability of ${\cal E}$
decays double-exponentially with $K$.

Next, we analyze the probability of the events $\{f_{t}\}_{t\in[T]}$,
in a similar fashion to the previous analysis. For a single alternative
vector $a_{t'}^{K}$ with $t'\neq t$ it holds from (\ref{eq: V-typical set of a different input vector upper bound})
(Lemma \ref{lem: Probability of V-typical sets condition on a different vector})
\[
\P\left[B_{t}^{K}\in{\cal T}_{K}([V]_{\delta}\mid a_{t'}^{K})\right]=V\left({\cal T}_{K}([V]_{\delta}\mid a_{t'}^{K})\mid a_{t}^{K}\right)\leq e^{-K\rho_{K}/2},
\]
using the assumption that $a_{t}^{K}$ and $a_{t'}^{K}$ are far,
that is $\Delta(\mathscr{P}(a_{t}^{K},a_{t'}^{K}),V)\geq\rho_{K}$
for $t\neq t'$, and for all $K$ large enough. We further assume
that $K$ is large so that $\tau_{K}>\tau_{0}>\rho_{K}$. Thus, by
linearity of expectation
\[
\E\left[\sum_{t'\in[T]\backslash\{t\}}\I\{B_{t}^{K}\in{\cal T}_{K}([V]_{\delta}\mid a_{t'}^{K})\}\right]\leq e^{K(\tau_{K}-\rho_{K}/2)},
\]
and then by Markov's inequality
\[
\E[f_{t}]=\P\left[\sum_{t'\in[T]\backslash\{t\}}\I\{B_{t}^{K}\in{\cal T}_{K}([V]_{\delta}\mid a_{t'}^{K})\}>T_{0}\right]\leq\frac{e^{K(\tau_{K}-\rho_{K}/2)}}{T_{0}}.
\]
Thus, if we choose $T_{0}=e^{K(\tau_{K}-\rho_{K}/4)}$ then the expected
of total number of ``large-ambiguity'' outputs is 
\[
\E\left[\sum_{t\in[T]}f_{t}\right]\leq\frac{e^{K(2\tau_{K}-\rho_{K}/2)}}{e^{K(\tau_{K}-\rho_{K}/4)}}=e^{K(\tau_{K}-\rho_{K}/4)}=o(T)
\]
 (as $T=e^{K\tau_{K}}$, and $\rho_{K}>2\delta_{K}^{2}=\omega(\frac{1}{K})$).
We now let 
\[
{\cal F}\dfn\left\{ \sum_{t\in[T]}f_{t}\geq e^{K(\tau_{K}-\rho_{K}/8)}\right\} 
\]
be the event that the total number of atypical events exceeds its
expectation by a factor of more than $e^{K\rho_{K}/8}$. Since $f_{t}$
are independent events, Chernoff's bound implies that 
\begin{align}
\P\left[{\cal F}\right] & \leq\exp\left[-e^{K\tau_{K}}\cdot d_{b}\left(e^{-K\rho_{K}/8}\mid\mid\E[f_{t}]\right)\right]\\
 & \trre[\leq,a]\exp\left[-e^{K\tau_{K}}\cdot e^{-K\rho_{K}/8}\left(\log\frac{e^{-K\delta_{K}^{2}/8}}{\E[f_{t}]}-1\right)\right]\\
 & \leq\exp\left[-e^{K(\tau_{K}-\rho_{K}/8)}\left(K\rho_{K}/8-1\right)\right],
\end{align}
where $(a)$ follows from the bound $d_{b}(p_{1}\mid\mid p_{2})>p_{1}(\log\frac{p_{1}}{p_{2}}-1)$.
By the assumption of the lemma, $\rho_{K}\geq\underline{\rho}_{K}$
defined in (\ref{eq: minimal value for rho}) and so, specifically,
$\rho_{K}\geq2\delta_{K}^{2}$. By the delta-convention $\delta_{K}^{2}=\omega(\frac{1}{K})$
and hence the probability of ${\cal F}$ decays double-exponentially
with $K$. 

We thus conclude that ${\cal G}={\cal E}\cup{\cal F}$ decays double-exponentially
with $K$, as by the union bound
\[
\P[{\cal G}]\leq\P[{\cal {\cal E}}]+\P[{\cal F}]\leq\exp\left[-e^{\Theta(K)}\right].
\]
The equivocation of the permutation conditioned on $B^{KT}$ is now
upper bounded as 
\[
H(\Sigma\mid B^{KT})\leq H(\Sigma,{\cal G}\mid B^{KT})=h_{b}(\P[{\cal G}])+\P[{\cal G}]\cdot H(\Sigma\mid B^{KT},{\cal G})+\P[{\cal G}^{c}]\cdot H(\Sigma\mid B^{KT},{\cal G}^{c}).
\]
We next bound each term separately. For the first term $h_{b}(\P[{\cal G}])\leq\log2$,
and for the second term, due to the double-exponential decay of the
probability of ${\cal G}$
\begin{align}
\P[{\cal G}]\cdot H(\Sigma\mid B^{KT},{\cal G}) & \leq\exp\left[-e^{\Theta(K)}\right]\log(T!)\\
 & \leq\exp\left[-e^{\Theta(K)}\right]T\log T\\
 & =\exp\left[-e^{\Theta(\log T)}\right]T\log T\\
 & =o(T\log T).
\end{align}
To complete the proof of the lemma, it suffices to show that the third
term is $o(T\log T)$. To this end, we bound the equivocation under
the event that ${\cal G}^{c}={\cal E}^{c}\cap{\cal F}^{c}$ as follows.
Under this event, there are at most $T_{1}\dfn e^{K(\tau_{K}-\delta_{K}^{2}/2)}+e^{K(\tau_{K}-\rho_{K}/8)}=o(e^{K\tau_{K}})=o(T)$
output sequences $B_{t}^{K}$ which are either ``a-typical'' to
their input or have ``large-ambiguity'' as they are typical to many
inputs. We say that an output $B_{t}^{K}$ is ``bad'' if either
of the two occurs, that is, if either $e_{t}=1$ or $f_{t}=1$ (and
``good'' otherwise). Note, however, that observer, which is aware
of both $(a_{t}^{K})_{t\in[T]}$ and $(B_{t}^{K})_{t\in[T]}$ (but
not of the permutation $\Sigma$) cannot determine if $B_{t}^{K}$
is bad or not (that would require the knowledge that $\Sigma$ is
the identity permutation and that it is $a_{t}^{K}$ which was the
channel input that resulted the output $B_{t}^{K}$). Nonetheless,
we next upper bound the total number of permutations which have non-zero
probability conditioned on ${\cal G}^{c}$, and then further upper
bound the conditional entropy by the logarithm of the number of such
permutations. 

Consider the set of permutations which are constructed as follows: 
\begin{enumerate}
\item The number of bad indices is chosen. There are at most $T_{1}$ possibilities
for this choice. We denote the number of bad indices by $\tilde{T}_{1}$.
\item Given a choice of a total of exactly $\tilde{T}_{1}$ bad indices,
each index $t\in[T]$ is labeled as bad or good. There are at most
${T \choose \tilde{T}_{1}}\leq{T \choose T_{1}}$ possibilities for
this choice. 
\item For any $t$ currently labeled as good find, at most $T_{0}=e^{K(\tau_{K}-\rho_{K}/4)}=o(T)$
arbitrary indices $t'\in[T]$ such that $B_{t}^{K}\in{\cal T}_{K}([V]_{\delta}\mid a_{t'}^{K})$.
There are at most $T^{T_{0}}$ possibilities for this choice, and
there are at most $T$ total good indices. Thus, the number of choices
for the permutation mapping to the good output indices is at most
$T\cdot T^{T_{0}}$. 
\item Given the association of the $T-\tilde{T}_{1}$ good indices, choose
the $\tilde{T}_{1}$ remaining values of the permutation. There are
at most $\tilde{T}_{1}!\leq T_{1}!\leq T_{1}\log T_{1}$ possibilities
for this choice.
\end{enumerate}
It is easy to verify that conditioned on ${\cal G}^{c}$, the true
permutation belongs to one of this possibilities constructed above,
with probability $1$. Thus, the entropy of the permutation is bounded
by the logarithm of its support. Hence, from the above count of possible
permutations
\begin{align}
\P[{\cal G}^{c}]\cdot H(\Sigma\mid B^{KT},{\cal G}^{c}) & \leq1\cdot H(\Sigma\mid B^{KT},{\cal G}^{c})\\
 & \leq\log\left[T_{1}\cdot{T \choose T_{1}}\cdot T^{T_{0}+1}\cdot T_{1}!\right]\\
 & \trre[\leq,a]\log T_{1}+Th_{b}\left(\frac{T_{1}}{T}\right)+(T_{0}+1)\log T+T_{1}\log T_{1}\\
 & \trre[\leq,b]\log T_{1}+T\cdot\log2+(T_{0}+1)\log T+T_{1}\log T_{1}\\
 & \trre[=,c]o(T\log T),
\end{align}
where $(a)$ follows from the entropy bound on the binomial coefficient
${n \choose k}\leq e^{nh_{b}(k/n)}$ for $n\geq k$ integers, and
$(b)$ follows from $h_{b}(\epsilon)\leq\log2$, and $(c)$ follows
since $T_{0}=o(T)$ and $T_{1}=o(T)$. 
\end{IEEEproof}

\subsubsection{Common Input Mutual Information Deficit for Close Inputs\label{subsec:Preliminaries:-Common-input}}

A possible way of overcoming the ordering problem in the DNA channel
mentioned in the previous section, is to send codewords in which many
of identical molecules, or even just similar. In the extreme case,
when all input molecules $x_{m}^{L}$ are equal, then the ordering
is of course immaterial. However, sending the same input in a memoryless
channel reduces the output entropy, and thus the achievable mutual
information. In this section, we focus on such mutual information
for a pair of molecules. We show that if the molecules are ``close''
according to the distance (\ref{eq: Distance between molecules definition}),
then, up to lower order terms, the resulting mutual information is
as if the two inputs are \emph{identical}, which, in turn, is strictly
less than the mutual information achieved for a pair of independent
inputs. As the discussion is general we use general notation, not
necessarily the one of the DNA channel. 

\begin{figure}
\centering{}\includegraphics{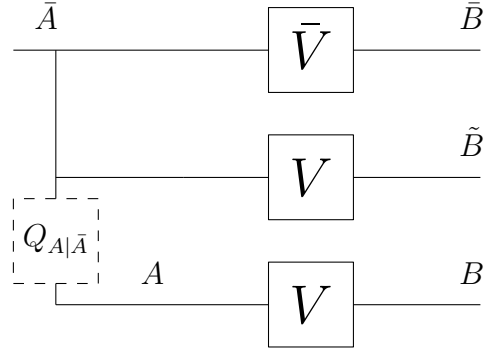}\caption{Channel model for a pair of ``close'' inputs \label{fig:Channel-model for close channel inputs}}
\end{figure}

Specifically, we focus on the scenario depicted in Fig. \ref{fig:Channel-model for close channel inputs},
where we assume throughout this section that $Q_{A}=Q_{\overline{A}}$.

Consider the pair of DMCs $\overline{V}$ and the lower $V$ in Fig.
\ref{fig:Channel-model for close channel inputs}, and suppose that
$Q_{A\mid\overline{A}}$ is a DMC too. Clearly, to maximize $I(A,\overline{A};B,\overline{B})$
the channel $Q_{A\mid\overline{A}}$ should be chosen so that $A$
and $\overline{A}$ are\emph{ independent} -- the resulting mutual
information is the sum of the mutual-information in each of the channels.
At the other extreme, if $Q_{A\mid\overline{A}}$ is restricted to
be a noiseless channel, $A=\overline{A}$, and this \emph{common}
input in fact minimizes the mutual information $I(A,\overline{A};B,\overline{B})$.
In this case, $B,\overline{B}$ are two conditionally independent
observations of the same input (as in the binomial channel of Definition
\ref{def: binomial channel}). We may thus define the CID of a \emph{pair}
of channels $V,\overline{V}$ and input distribution $Q_{A}$ as\emph{
}
\begin{equation}
\mathsf{CID}(Q_{\overline{A}},V,\overline{V})\dfn I_{Q_{\overline{A}}\times V}(\overline{A};\tilde{B})+I_{Q_{A}\times\overline{V}}(\overline{A};\overline{B})-I_{Q_{\overline{A}\tilde{B}\overline{B}}}(\overline{A};\overline{B},\tilde{B}),\label{eq: CID two channels}
\end{equation}
where the first two terms represent the total mutual information obtained
by independent inputs to $V,\overline{V}$ (with the same input distribution
$Q_{\overline{A}}$), and the last term represents the total mutual
information obtained by a common input to $V$ and $\overline{V}$
given by $A=\overline{A}$. The definition (\ref{eq: CID two channels})
slightly generalizes the definition of the CID for a single channel
$V$ in (\ref{eq: definition common input deficit}), where with a
slight abuse of notation, it agrees with the convention that $\mathsf{CID}(Q_{\overline{A}},V,V)\equiv\mathsf{CID}(Q_{\overline{A}},V)$. 

The following proposition provides a useful characterization of $\mathsf{CID}(Q_{\overline{A}},V,\overline{V})$. 
\begin{prop}
\label{prop: common input deficit} Let $V\colon{\cal A}\to{\cal B}$
and $\overline{V}\colon{\cal A}\to\overline{{\cal B}}$ be a pair
of DMCs, let $Q_{\overline{A}}\in{\cal P}({\cal A})$ and let $\overline{B}\sim\overline{V}(\cdot\mid\overline{A})$
and $\tilde{B}\sim V(\cdot\mid\overline{A})$. Then, $\mathsf{CID}(Q_{\overline{A}},V,\overline{V})=I(\overline{B};\tilde{B})\geq0$
and equality holds if and only if either $V$ or $\overline{V}$ are
completely noisy. 
\end{prop}
\begin{IEEEproof}
Under the given distributions (see Fig. \ref{fig:Channel-model for close channel inputs}),
Markovity implies that 
\begin{align}
I(\overline{A};\overline{B},\tilde{B}) & =I(\overline{A};\overline{B})+I(\overline{A};\tilde{B}\mid\overline{B})\\
 & =I(\overline{A};\overline{B})+H(\tilde{B}\mid\overline{B})-H(\tilde{B}\mid\overline{A}).
\end{align}
Then, by standard identities 
\[
\mathsf{CID}(Q_{\overline{A}},V,\overline{V})=H(\tilde{B})-H(\tilde{B}\mid\overline{B})=I(\overline{B};\tilde{B})\geq0
\]
Since $\overline{B}-\overline{A}-\tilde{B}$ holds, $I(\overline{B};\tilde{B})=0$
if and only if $\overline{B}$ and $\tilde{B}$ are independent, which
occurs if and only if either $I(\overline{A};\overline{B})=0$ or
$I(\overline{A};\tilde{B})=0$. 
\end{IEEEproof}
\begin{cor}
\label{cor: common input deficit for binomial channels}Let $V\colon{\cal A}\to{\cal B}$
be a DMC which is not completely noisy for $Q_{\overline{A}}$ (i.e.,
$I(Q_{\overline{A}},V)>0$) and let $V^{\oplus d_{1}}$ and $V^{\oplus d_{2}}$
be its binomial extensions for some $d_{1},d_{2}\geq1$. Then, 
\[
\mathsf{CID}(Q_{\overline{A}},V^{\oplus d_{1}},V^{\oplus d_{2}})\geq\mathsf{CID}(Q_{\overline{A}},V).
\]
\end{cor}
This property follows by applying the data processing theorem to the
mutual information representation of the CID in Prop. \ref{prop: common input deficit}. 

For the DNA channel, we will encounter the case in which $Q_{A\mid\overline{A}}$
is such that the inputs $A$ and $\overline{A}$ are ``close'' (in
terms of the distance $\Delta(Q_{A\overline{A}},V)$ defined in the
previous section), but not exactly identical. As the next proposition
shows, for small enough distance $\Delta(Q_{A\overline{A}},V)$, the
resulting mutual information asymptotically behaves as the mutual
information achieved by a common input, and thus it is \emph{strictly
lower }compared to independent inputs, by an amount of $\mathsf{CID}(Q_{\overline{A}},V,\overline{V})$
(plus an additional negligible term).
\begin{prop}
\label{prop: mutual information given close inputs}Let $V\colon{\cal A}\to{\cal B}$
and $\overline{V}\colon{\cal A}\to\overline{{\cal B}}$ be a pair
of discrete channels such that $V$ satisfies 
\[
\nu_{\text{\emph{min}}}(V)\dfn\max_{a\in{\cal A},\;b\in{\cal B}}\log\frac{1}{V(b\mid a)}<\infty.
\]
Further let $Q_{A\overline{A}}$ be such that $Q_{A}=Q_{\overline{A}}$
and $\Delta(Q_{A\overline{A}},V)\leq\underline{\rho}_{K}$ where $\underline{\rho}_{K}$
is as defined in (\ref{eq: minimal value for rho}). 

If $A^{K},\overline{A}^{K}\in{\cal A}^{K}$ are random inputs such
that $\mathscr{P}(A^{K},\overline{A}^{K})=Q_{A\overline{A}}$ with
probability $1$, and $B^{K}\sim V^{K}(\cdot\mid A^{K})$ and $\overline{B}^{K}\sim\overline{V}^{K}(\cdot\mid\overline{A}^{K})$
where $B^{K}\in{\cal B}^{K}$ and $\overline{B}^{K}\in\overline{{\cal B}}^{K}$,
then,
\begin{align}
I(A^{K},\overline{A}^{K};B^{K},\overline{B}^{K}) & \leq K\cdot\left[I_{Q}(\overline{A};\overline{B},\tilde{B})+o_{\nu_{\text{\emph{min}}}(V)}(1)\right]\\
 & =K\cdot\left[I_{Q_{A}\times V}(\overline{A};\tilde{B})+I_{Q_{\overline{A}}\times\overline{V}}(\overline{A};\overline{B})-\mathsf{CID}(Q_{\overline{A}},V,\overline{V})+o_{\nu_{\text{\emph{min}}}}(1)\right],
\end{align}
where $Q_{B|A}=V$ and $Q_{\overline{B}|B\overline{A}}=Q_{\overline{B}|\overline{A}}=\overline{V}$. 
\end{prop}
To prove Prop. \ref{prop: mutual information given close inputs}
we will need the following lemma, which is concerned with a single-letter
joint output entropy:
\begin{lem}
\label{lem:entropy difference between the outputs of close vectors}Assume
that $V,\overline{V},Q_{A\overline{A}}$ satisfy the conditions of
Prop. \ref{prop: mutual information given close inputs}, with $\Delta(Q_{A\overline{A}},V)\leq\rho$.
If $(A,\overline{A})\sim Q_{A\overline{A}}$, $\overline{B}\sim\overline{V}(\cdot\mid\overline{A})$,
$B\sim V(\cdot\mid A)$ and $\tilde{B}\sim V(\cdot\mid\overline{A})$
(as in Fig. \ref{fig:Channel-model for close channel inputs}) then,
\begin{equation}
H(B,\overline{B})\leq H(\tilde{B},\overline{B})+O(\rho).\label{eq: entropy difference due to small input perturbation}
\end{equation}
\end{lem}
\begin{IEEEproof}
We prove the claim by showing that if $\Delta(Q_{A\overline{A}},V)=\Delta(Q_{\overline{A}A},V)\leq\rho\to0$\footnote{The equality follows from the assumption $Q_{A}=Q_{\overline{A}}$.}
then $|H(B,\overline{B})-H(\tilde{B},\overline{B})|\to0$. Let $P_{A\overline{A}\tilde{B}\overline{B}}$
and $P_{A\overline{A}B\overline{B}}$ be the distributions of the
relevant random variables. The entropy difference is upper bounded
by the total variation between the corresponding distributions as
\cite[Lemma 2.7]{csiszar2011information}\footnote{The logarithmic term is not sharp, and can be removed by using the
refined bound \cite[Problem 3.10]{csiszar2011information}. This is
however inconsequential for the proof.}
\begin{equation}
\left|H(B,\overline{B})-H(\tilde{B},\overline{B})\right|\leq-\left|P_{\tilde{B}\overline{B}}-P_{B\overline{B}}\right|\cdot\log\frac{\left|P_{\tilde{B}\overline{B}}-P_{B\overline{B}}\right|}{|{\cal B}|^{2}}.\label{eq: application of entropy continuity for output entropies}
\end{equation}
In turn, the total variation is bounded by Pinsker's inequality (e.g.
\cite[Problem 3.18]{csiszar2011information}) 
\begin{equation}
\left|P_{\tilde{B}\overline{B}}-P_{B\overline{B}}\right|\leq\sqrt{2\log2\cdot D(P_{\tilde{B}\overline{B}}\mid\mid P_{B\overline{B}})},\label{eq: application of Pinsker for output distributions}
\end{equation}
and so it suffices to show that $D(P_{\tilde{B}\overline{B}}\mid\mid P_{B\overline{B}})\to0$
as $\rho\to0$. While this is property pertains to single-letter distributions
of $(\overline{B},\tilde{B})$ and $(B,\overline{B})$, it will be
convenient to prove it by considering $K$ dimensional vectors $A^{K},\overline{A}^{K}\in{\cal A}^{K}$,
$B^{K},\tilde{B}^{K}\in{\cal B}^{K}$ and $\overline{B}^{K}\in\overline{{\cal B}}^{K}$
whose marginals $(A_{k},\overline{A}_{k},B_{k},\tilde{B}_{k},\overline{B}_{k})$
are distributed i.i.d. according to $P_{A\overline{A}B\tilde{B}\overline{B}}$.
We further set $\delta_{K}$ to satisfy the delta-convention, and
let $\rho\equiv\rho_{K}=\underline{\rho}_{K}$ where $\underline{\rho}_{K}$
is as defined in (\ref{eq: minimal value for rho}) and satisfies
$\underline{\rho}_{K}=o(1)$. Then, the proof of the lemma is completed
by showing that $D(P_{\tilde{B}^{K}\overline{B}^{K}}\mid\mid P_{B^{K}\overline{B}^{K}})=K\cdot D(P_{\tilde{B}\overline{B}}\mid\mid P_{B\overline{B}})=o(K)$,
substituting it to (\ref{eq: application of Pinsker for output distributions})
and the resulting bound in (\ref{eq: application of entropy continuity for output entropies}).

Let $P_{A^{K}\overline{A}^{K}\tilde{B}^{K}\overline{B}^{K}}$ and
$P_{A^{K}\overline{A}^{K}B^{K}\overline{B}^{K}}$ be the resulting
$K$ dimensional (product) distributions of the relevant random variables.
To complete the proof it is thus required to upper bound the KL divergence
$D(P_{\tilde{B}^{K}\overline{B}^{K}}\mid\mid P_{B^{K}\overline{B}^{K}})$,
and to this end, we first prove several properties. 

\uline{Property 1:} Recall that it is assumed that $\Delta(Q_{A\overline{A}},V)\leq\rho_{K}$.
We show that a similar bound holds for joint distributions of $\breve{Q}_{A\overline{A}}\in[Q_{A\overline{A}}]_{\delta_{K}}$
which are typical to the memoryless distribution $Q_{A\overline{A}}$,
to wit, satisfy 
\[
|\breve{Q}_{A\overline{A}}(a,\overline{a})-Q_{A\overline{A}}(a,\overline{a})|\leq\delta_{K},\quad\forall(a,\overline{a})\in{\cal A}\times\overline{{\cal A}}.
\]
Clearly in this case $\breve{Q}_{A}\in[Q_{A}]_{\delta_{K}\cdot|{\cal A}|}$
also holds (see, e.g., \cite[Lemma 2.10]{csiszar2011information}).
The conditional entropy is known to be continuous in the input distribution,
and so, as in \cite[p. 22, proof of Lemma 2.13]{csiszar2011information},
\begin{equation}
\left|H_{Q_{A}\times V}(B\mid A)-H_{\breve{Q}_{A}\times V}(B\mid A)\right|\leq\delta_{K}|{\cal A}|\cdot\log|{\cal B}|,\label{eq: continuity of conditional entropy in input distribution}
\end{equation}
and, similarly, 
\[
\left|H_{Q_{A\overline{A}}\times Q_{B\mid A\overline{A}}}(B\mid A,\overline{A})-H_{\breve{Q}_{A\overline{A}}\times Q_{B\mid A\overline{A}}}(B\mid A,\overline{A})\right|\leq\delta_{K}\log|{\cal B}|
\]
for any $Q_{B\mid A\overline{A}}\in{\cal P}({\cal B}\mid{\cal A}\times{\cal A})$.
Hence, by the assumption $\Delta(Q_{A\overline{A}},V)\leq\rho_{K}$
{[}see the distance definition in (\ref{eq: Distance between molecules definition}){]}
\begin{align}
\Delta(\breve{Q}_{A\overline{A}},V) & =\Delta(Q_{A\overline{A}},V)+\Delta(\breve{Q}_{A\overline{A}},V)-\Delta(Q_{A\overline{A}},V)\\
 & \leq\Delta(Q_{A\overline{A}},V)+\left|H_{Q_{A}\times V}(B\mid A)-H_{\breve{Q}_{A}\times V}(B\mid A)\right|\nonumber \\
 & \hphantom{-}+\max_{Q_{B\mid A\overline{A}}}\left|H_{Q_{A\overline{A}}\times Q_{B\mid A\overline{A}}}(B\mid A,\overline{A})-H_{\breve{Q}_{A\overline{A}}\times Q_{B\mid A\overline{A}}}(B\mid A,\overline{A})\right|\\
 & \leq\rho_{K}+\delta_{K}(|{\cal A}|+1)\log|{\cal B}|\dfn\rho_{K}'.\label{eq: distance for a joint type which is typically close to close joint type}
\end{align}
By the delta-convention, $\rho_{K}'=o(1)$. The bound (\ref{eq: distance for a joint type which is typically close to close joint type})
is the first required property. 

\uline{Property 2:} We relate the typical sets ${\cal T}_{K}([V]_{\delta_{K}}\mid a^{K})\subseteq{\cal B}^{K}$
and ${\cal T}_{K}([V]_{\delta_{K}}\mid\overline{a}^{K})$ for $(a^{K},\overline{a}^{K})\in{\cal T}_{K}([Q_{A\overline{A}}]_{\delta_{K}})$
which are ``close'' according to the $\Delta$-distance, and specifically,
according to Property 1 satisfy that $\Delta(\mathscr{P}(a^{K},\overline{a}^{K}),V)\leq\rho_{K}'$.
To this end, note that Lemma \ref{lem: Probability of V-typical sets condition on a different vector}
(with the role of $a^{K}$ and $\overline{a}^{K}$ switched) implies
that 
\[
V^{K}\left({\cal T}_{K}([V]_{\delta}\mid a^{K})\mid\overline{a}^{K}\right)\geq e^{-2K\rho_{K}'},
\]
where $\rho_{K}'=o(1)$ holds. Denote the \emph{Hamming neighborhood}
of a set ${\cal S}_{K}\subset{\cal B}^{K}$ by 
\begin{equation}
\Gamma^{\ell}{\cal S}_{K}\dfn\left\{ \tilde{b}^{K}\in{\cal B}^{K}\colon\min_{b^{K}\in{\cal S}_{K}}\rho_{\text{H}}(\tilde{b}^{K},b^{K})\leq\ell\right\} ,\label{eq: Hamming neighborhood}
\end{equation}
where $\rho_{\text{H}}(\cdot,\cdot)$ is the Hamming distance. By
the non-asymptotic \emph{blowing-up lemma} \cite[Lemma 3.6.1]{raginsky2018concentration}
(see also \cite[Lemma 5.4]{csiszar2011information} \cite{marton1986simple,marton1996bounding})
\begin{equation}
V^{K}\left(\Gamma^{\ell_{K}}{\cal T}_{K}([V]_{\delta}\mid a^{K})\mid\overline{a}^{K}\right)\geq1-\frac{1}{K^{2}},\label{eq: blown-up set as high probability}
\end{equation}
where $\frac{\ell_{K}}{K}=\sqrt{\rho_{K}'}+\sqrt{\frac{\log K}{K}}=o(1)$.\footnote{With the choice $\alpha=1$ in \cite[Eq. 3.6.5]{raginsky2018concentration}
which leads in the notation there to $\eta_{K}=\frac{1}{K^{2}}$.} The probability bound of (\ref{eq: blown-up set as high probability})
is the second required property. 

\uline{Property 3:} If $\tilde{b}^{K}\in\Gamma^{\ell_{K}}{\cal T}_{K}([V]_{\delta}\mid a^{K})$
then there exists $\breve{b}\in{\cal T}_{K}([V]_{\delta}\mid a^{K})$
such that $\rho_{\text{H}}(\tilde{b}^{K},\breve{b}^{K})\leq\ell_{K}$.
It then follows from the definition of $\nu_{\text{min}}(V)$ that
\[
\frac{V^{K}\left(\tilde{b}^{K}\mid a^{K}\right)}{V^{K}\left(\breve{b}^{K}\mid a^{K}\right)}\geq e^{-\ell_{K}\nu_{\text{min}}}.
\]
This is the third required property. 

\uline{Property 4:} We upper bound the ratio $\frac{V^{K}(\tilde{b}^{K}\mid\overline{a}^{K})}{V^{K}(\tilde{b}^{K}\mid a^{K})}$
for $\tilde{b}^{K}\in{\cal T}_{K}([V]_{\delta_{K}}\mid\overline{a}^{K})\cap\Gamma^{\ell_{K}}{\cal T}_{K}([V]_{\delta_{K}}\mid a^{K})$.
First, we upper bound $V(\tilde{b}^{K}\mid\overline{a}^{K})$ for
$\tilde{b}^{K}\in{\cal T}_{K}([V]_{\delta_{K}}\mid\overline{a}^{K})$.
There exists $\check{V}\in[V\mid Q_{\overline{A}}]_{\delta}\subset{\cal P}_{K}({\cal B}\mid Q_{\overline{A}})$
such that $\tilde{b}^{K}\in{\cal T}_{K}(\check{V}\mid\overline{a}^{K})$
and so
\begin{align}
V^{K}(\tilde{b}^{K}\mid\overline{a}^{K}) & =\exp\left[-K\cdot\left(D(\check{V}\mid\mid V\mid Q_{\overline{A}})+H_{Q_{\overline{A}}\times\check{V}}(B\mid\overline{A})\right)\right]\\
 & \leq\exp\left[-K\cdot\left(-|{\cal A}||{\cal B}|\delta_{K}\log\delta_{K}+H_{Q_{\overline{A}}\times V}(B\mid\overline{A})\right)\right],\label{eq: upper bound on limelihood}
\end{align}
where the equality follows from the standard identity \cite[Lemma 2.6]{csiszar2011information},
and the inequality follows from the continuity of the conditional
entropy in the channel that determines the relation between the variables
\cite[Proof of Lemma 2.13]{csiszar2011information}, as well as the
non-negativity of the KL divergence. Second, we lower bound $V^{K}(\tilde{b}^{K}\mid a^{K})$
for $\tilde{b}^{K}\in\Gamma^{\ell_{K}}{\cal T}_{K}([V]_{\delta}\mid a^{K})$.
By Property 3 there exists $\check{V}\in{\cal P}_{K}({\cal B}\mid Q_{A})$
and $\breve{b}^{K}\in{\cal T}_{K}(\check{V}\mid a^{K})\subset{\cal T}_{K}([V]_{\delta_{K}}\mid a^{K})$
such that
\begin{equation}
V^{K}(\tilde{b}^{K}\mid a^{K})\geq e^{-\ell_{K}\nu_{\text{min}}}\cdot V^{K}(\breve{b}^{K}\mid a^{K})\geq\exp\left[-K\cdot\left(\frac{\ell_{K}}{K}\nu_{\text{min}}+\nu_{\text{min}}\delta_{K}|{\cal A}||{\cal B}|+H_{Q_{A}\times V}(B\mid A)\right)\right],\label{eq: lower bound on likelihood}
\end{equation}
where the right-most inequality follows from 
\begin{align}
\frac{1}{K}\log V^{K}(\breve{b}^{K}\mid a^{K}) & =\sum_{a\in\supp(Q_{A})}Q_{A}(a)\sum_{b\in\supp(V(\cdot\mid a))}\check{V}(b\mid a)\cdot\log V(b\mid a)\\
 & \leq\sum_{a\in\supp(Q_{A})}Q_{A}(a)\sum_{b\in\supp(V(\cdot\mid a))}V(b\mid a)\cdot\log V(b\mid a)\nonumber \\
 & \hphantom{=}+\sum_{a\in\supp(Q_{A})}Q_{A}(a)\sum_{b\in\supp(V(\cdot\mid a))}\left|\check{V}(b\mid a)-V(b\mid a)\right|\cdot\log\frac{1}{V(b\mid a)}\\
 & \leq\sum_{a\in\supp(Q_{A})}Q_{A}(a)\sum_{b\in\supp(V(\cdot\mid a))}V(b\mid a)\cdot\log V(b\mid a)+\nu_{\text{min}}\delta_{K}|{\cal A}||{\cal B}|\\
 & =-H_{Q_{A}\times V}(B\mid A)+\nu_{\text{min}}\delta_{K}|{\cal A}||{\cal B}|.
\end{align}
Combining (\ref{eq: upper bound on limelihood}) and (\ref{eq: lower bound on likelihood}),
it holds that if $\tilde{b}^{K}\in{\cal T}_{K}([V]\mid\overline{a}^{K})\cap\Gamma^{\ell_{K}}{\cal T}_{K}([V]_{\delta_{K}}\mid a^{K})$,
and $Q_{A}=Q_{\overline{A}}$ then 
\[
\frac{V^{K}(\tilde{b}^{K}\mid\overline{a}^{K})}{V^{K}(\tilde{b}^{K}\mid a^{K})}\leq\exp\left[K\cdot\left(\frac{\ell_{K}}{K}\nu_{\text{min}}+\nu_{\text{min}}\delta_{K}|{\cal A}||{\cal B}|-|{\cal A}||{\cal B}|\delta_{K}\log\delta_{K}\right)\right].
\]
This is the fourth required property. 

With the above four properties, we now bound the required KL divergence,
as follows:
\begin{align}
 & D(P_{\tilde{B}^{K}\overline{B}^{K}}\mid\mid P_{B^{K}\overline{B}^{K}})\nonumber \\
 & \trre[\leq,a]D(P_{\tilde{B}^{K}\overline{B}^{K}\mid A^{K}\overline{A}^{K}}\mid\mid P_{B^{K}\overline{B}^{K}\mid A^{K}\overline{A}^{K}}\mid P_{A^{K}\overline{A}^{K}})\\
 & \trre[\leq,b]\sum_{(a^{K},\overline{a}^{K})\in{\cal T}_{K}([Q_{A\overline{A}}])}P_{A^{K}\overline{A}^{K}}(a^{K},\overline{a}^{K})\cdot D\left(P_{\tilde{B}^{K}\overline{B}^{K}\mid A^{K}\overline{A}^{K}}(\cdot\mid a^{K},\overline{a}^{K})\mid\mid P_{B^{K}\overline{B}^{K}\mid A^{K}\overline{A}^{K}}(\cdot\mid a^{K},\overline{a}^{K})\right)+e^{-K\delta_{K}^{2}}\\
 & \trre[=,c]\sum_{(a^{K},\overline{a}^{K})\in{\cal T}_{K}([Q_{A\overline{A}}])}P_{A^{K}\overline{A}^{K}}(a^{K},\overline{a}^{K})\times\nonumber \\
 & \hphantom{==}\sum_{\tilde{b}\in{\cal B}^{K}}\sum_{\overline{b}\in{\cal B}^{K}}V^{K}(\tilde{b}\mid\overline{a}^{K})\cdot\overline{V}^{K}(\overline{b}\mid\overline{a}^{K})\log\frac{V^{K}(\tilde{b}\mid\overline{a}^{K})\cdot\overline{V}^{K}(\overline{b}\mid\overline{a}^{K})}{V^{K}(\tilde{b}\mid a^{K})\cdot V^{K}(\overline{b}\mid\overline{a}^{K})}\\
 & =\sum_{(a^{K},\overline{a}^{K})\in{\cal T}_{K}([Q_{A\overline{A}}])}P_{A^{K}\overline{A}^{K}}(a^{K},\overline{a}^{K})\cdot\sum_{\tilde{b}\in{\cal B}^{K}}V^{K}(\tilde{b}^{K}\mid\overline{a}^{K})\log\frac{V^{K}(\tilde{b}^{K}\mid\overline{a}^{K})}{V^{K}(b^{K}\mid a^{K})}+e^{-K\delta_{K}^{2}},\label{eq: KL divergence bound between two type of output pairs}
\end{align}
where $(a)$ follows from the convexity of the KL divergence \cite[Thm. 2.7.2]{cover2012elements},
$(b)$ follows since $\P[(a^{K},\overline{a}^{K})\in{\cal T}_{K}^{c}([Q_{A\overline{A}}])]\leq e^{-K\delta_{K}^{2}}$
under $P_{A^{K}\overline{A}^{K}}$, and $(c)$ follows from (see Fig.
\ref{fig:Channel-model for close channel inputs}) 
\[
P_{\tilde{B}^{K}\overline{B}^{K}\mid A^{K}\overline{A}^{K}}(\tilde{b}^{K},\overline{b}^{K}\mid a^{K},\overline{a}^{K})=V^{K}(\tilde{b}^{K}\mid\overline{a}^{K})\cdot\overline{V}^{K}(\overline{b}^{K}\mid\overline{a}^{K})
\]
 and 
\[
P_{B^{K}\overline{B}^{K}\mid A^{K}\overline{A}^{K}}(\tilde{b}^{K},\overline{b}^{K}\mid a^{K},\overline{a}^{K})=V(\tilde{b}^{K}\mid a^{K})\cdot\overline{V}^{K}(\overline{b}^{K}\mid\overline{a}^{K}).
\]
We next bound the KL divergence in (\ref{eq: KL divergence bound between two type of output pairs})
for an arbitrary $(a^{K},\overline{a}^{K})\in{\cal T}_{K}([Q_{A\overline{A}}])$.
The idea is that under the measure $V^{K}(\cdot\mid\overline{a}^{K})$,
both the typical set ${\cal T}_{K}([V]\mid\overline{a}^{K})$ and
the blown-up set $\Gamma^{\ell_{K}}{\cal T}_{K}([V]_{\delta}\mid a^{K})$
have high probability (larger than $1-e^{-K\delta_{K}^{2}}$ and $1-1/K^{2}$,
respectively). Thus their intersection also has high probability (larger
than $1-e^{-K\delta_{K}^{2}}-1/K^{2}$). Using this observation we
may bound the conditional KL divergence as follows: 
\begin{align}
 & \sum_{\tilde{b}^{K}\in{\cal B}^{K}}V^{K}(\tilde{b}^{K}\mid\overline{a}^{K})\log\frac{V^{K}(\tilde{b}^{K}\mid\overline{a}^{K})}{V^{K}(\tilde{b}^{K}\mid a^{K})}\nonumber \\
 & \trre[\leq,a]\sum_{\tilde{b}^{K}\in{\cal T}_{K}([V]\mid\overline{a}^{K})\cap\Gamma^{\ell_{K}}{\cal T}_{K}([V]_{\delta}\mid a^{K})}V^{K}(\tilde{b}\mid\overline{a}^{K})\log\frac{V^{K}(\tilde{b}\mid\overline{a}^{K})}{V^{K}(\tilde{b}\mid a^{K})}+\left(e^{-K\delta_{K}^{2}}+\frac{1}{K^{2}}\right)\cdot K\nu_{\text{min}}\\
 & \trre[\leq,b]\left[\frac{\ell_{K}}{K}\nu_{\text{min}}+\nu_{\text{min}}\delta_{K}|{\cal A}||{\cal B}|-|{\cal A}||{\cal B}|\delta_{K}\log\delta_{K}+\left(e^{-K\delta_{K}^{2}}+\frac{1}{K^{2}}\right)\nu_{\text{min}}\right]\cdot K\\
 & \trre[=,c]o(K),
\end{align}
where $(a)$ follows since 
\begin{equation}
\frac{1}{K}\log\frac{V^{K}(\tilde{b}^{K}\mid\overline{a}^{K})}{V^{K}(\tilde{b}^{K}\mid a^{K})}\leq\max_{a,\overline{a}\in\supp(Q_{A}),\tilde{b}\in{\cal B}}\log\frac{V(\tilde{b}\mid\overline{a})}{V(\tilde{b}\mid a)}\leq\nu_{\text{min}},\label{eq: maximal likelihood ratio}
\end{equation}
$(b)$ follows from Property 4, and $(c)$ follows since $\delta_{K}$
satisfies the delta convention, and $\ell_{K}=o(K)$. Inserting this
bound back to (\ref{eq: KL divergence bound between two type of output pairs})
we obtain that $D(P_{\tilde{B}^{K}\overline{B}^{K}}\mid\mid P_{B^{K}\overline{B}^{K}})=o(K)$,
as was required to be proved in order to complete the proof of the
lemma.
\end{IEEEproof}
We now prove Prop. \ref{prop: mutual information given close inputs}. 
\begin{IEEEproof}[Proof of Prop. \ref{prop: mutual information given close inputs}]
Let $P_{A^{K}\overline{A}^{K}}$ denote the joint probability distribution
of $(A^{K},\overline{A}^{K})$ and let us explicitly denote the dependence
of the mutual information on this distribution as $I(A^{K},\overline{A}^{K};B^{K},\overline{B}^{K})\equiv I(P_{A^{K}\overline{A}^{K}},(V^{K},\overline{V}^{K}))$.
Under the assumption of the proposition, $P_{A^{K}\overline{A}^{K}}$
is supported on (a subset of) ${\cal T}_{K}(Q_{A\overline{A}})$.
For a permutation $\sigma\colon[K]\to[K]$ from the symmetric group
$\mathfrak{S}_{K}$, let 
\[
P_{A^{K}\overline{A}^{K}}^{(\sigma)}\left((a_{0},\ldots,a_{K-1}),(\overline{a}_{0},\ldots,\overline{a}_{K-1})\right)\dfn P_{A^{K}\overline{A}^{K}}\left((a_{\sigma(0)},\ldots,a_{\sigma(K-1)}),(\overline{a}_{\sigma(0)},\ldots,\overline{a}_{\sigma(K-1)})\right).
\]
Then, $P_{A^{K}\overline{A}^{K}}^{(\sigma)}$ is also supported on
a subset of ${\cal T}_{K}(Q_{A\overline{A}})$. Since $B^{K}\sim V^{K}(\cdot\mid A^{K})$
and $\overline{B}^{K}\sim\overline{V}^{K}(\cdot\mid\overline{A}^{K})$
and the channels $V^{K},\overline{V}^{K}$ are memoryless, then 
\begin{equation}
I(P_{A^{K}\overline{A}^{K}},(V^{K},\overline{V}^{K}))=I(P_{A^{K}\overline{A}^{K}}^{(\sigma)},(V^{K},\overline{V}^{K}))\label{eq: equality of mutual information under permutation}
\end{equation}
for any $\sigma\in\mathfrak{S}_{K}$. Letting $\overline{P}_{A^{K}\overline{A}^{K}}\dfn\frac{1}{|\mathfrak{S}_{K}|}\sum_{\sigma\in\mathfrak{S}_{K}}P_{A^{K}\overline{A}^{K}}^{(\sigma)}$
denote a uniform averaging over all permutations in $\mathfrak{S}_{K}$,
the concavity of the mutual information in its input distribution
and (\ref{eq: equality of mutual information under permutation})
imply 
\[
I\left(P_{A^{K}\overline{A}^{K}},(V^{K},\overline{V}^{K})\right)\leq I\left(\overline{P}_{A^{K}\overline{A}^{K}},(V^{K},\overline{V}^{K})\right).
\]
To bound the mutual information for $\overline{P}_{A^{K}\overline{A}^{K}}$
we note that under this distribution, $(A^{K},\overline{A}^{K})\sim\text{Uniform}[{\cal T}_{K}(Q_{A\overline{A}})]$,
and so the distribution of a marginal pair is $(A_{k},\overline{A}_{k})\sim Q_{A\overline{A}}$
for any $k\in[K]$. By the standard bound on mutual information for
the memoryless channel $(A,\overline{A})\to(B,\overline{B})$ it holds
that 
\begin{equation}
I\left(P_{A^{K}\overline{A}^{K}},(V^{K},\overline{V}^{K})\right)\leq K\cdot I_{Q_{A\overline{A}}\times Q_{B\mid A}\times Q_{\overline{B}|\overline{A}}}(A,\overline{A};B,\overline{B}),\label{eq: single-letter bound on MI}
\end{equation}
where $Q_{B|A}=V$ and $Q_{\overline{B}|\overline{A}}=\overline{V}$.
We next bound the single-letter expression $I(A,\overline{A};B,\overline{B})$
where the subscript defining the distribution $Q$ is omitted for
brevity. From Markovity (see Fig. \ref{fig:Channel-model for close channel inputs})
\begin{align}
I(A;B,\overline{B}\mid\overline{A}) & =H(B,\overline{B}\mid\overline{A})-H(B,\overline{B}\mid A,\overline{A})\\
 & =H(\overline{B}\mid\overline{A})+H(B\mid\overline{B},\overline{A})-H(\overline{B}\mid A,\overline{A})-H(B\mid\overline{B},A,\overline{A})\\
 & =H(\overline{B}\mid\overline{A})+H(B\mid\overline{A})-H(\overline{B}\mid\overline{A})-H(B\mid A)\\
 & =H(B\mid\overline{A})-H(B\mid A).\label{eq: an identiy for MI}
\end{align}
Then,
\begin{align}
I(A,\overline{A};B,\overline{B}) & =I(\overline{A};B,\overline{B})+I(A;B,\overline{B}\mid\overline{A})\\
 & =H(B,\overline{B})-H(B,\overline{B}\mid\overline{A})+I(A;B,\overline{B}\mid\overline{A})\\
 & =H(B,\overline{B})-H(\overline{B}\mid\overline{A})-H(B\mid\overline{A},\overline{B})+I(A;B,\overline{B}\mid\overline{A})\\
\text{} & \trre[=,a]H(B,\overline{B})-H(\overline{B}\mid\overline{A})-H(B\mid\overline{A})+I(A;B,\overline{B}\mid\overline{A})\\
 & \trre[=,b]H(B,\overline{B})-H(\overline{B}\mid\overline{A})-H(B\mid A)\\
 & \trre[=,c]H(B,\overline{B})-H(\overline{B}\mid\overline{A})-H(\tilde{B}\mid\overline{A})\\
 & \trre[\leq,d]H(\tilde{B},\overline{B})-H(\overline{B}\mid\overline{A})-H(\tilde{B}\mid\overline{A})+o(\rho)\\
 & \trre[=,a]H(\overline{B})+H(\tilde{B}\mid\overline{B})-H(\overline{B}\mid\overline{A})-H(\tilde{B}\mid\overline{A},\overline{B})+o(\rho)\\
 & =I(\overline{A};\overline{B})+I(\overline{A};\tilde{B}\mid\overline{B})+o(\rho)\\
 & =I(\overline{A};\overline{B},\tilde{B})+o(\rho),
\end{align}
where $(a)$ follows from Markovity (see Fig. \ref{fig:Channel-model for close channel inputs}),
$(b)$ follows from (\ref{eq: an identiy for MI}), $(c)$ follows
since the fact that $Q_{A}=Q_{\overline{A}}$, and $(d)$ follows
from Lemma \ref{lem:entropy difference between the outputs of close vectors}.
Plugging this bound into (\ref{eq: single-letter bound on MI}) leads
to the required result. 
\end{IEEEproof}

\subsection{Structural Properties of Capacity Achieving Codebooks \label{subsec:Structural-properties-capacity}}

In this section, we return to the DNA channel formulation, and show
that if a rate $R$ is achievable by a sequence of codebooks ${\cal C}_{M}$
(and appropriate decoders), then it is also achievable by a sequence
of codebooks which have convenient structural properties. 

The first structural property regards the \emph{minimum-distance maximal
sets} of the codewords, which we define next. Recall the distance
definition of $\Delta(Q_{A\overline{A}},V)$ in (\ref{eq: Distance between molecules definition})
which we now utilize with $K=L$. For any given codeword $x^{LM}(j)$
in a codebook ${\cal C}_{M}$, we define the \emph{minimum-distance
maximal set} ${\cal M}_{\rho}(j)\subseteq[M]$ as the largest subset
of $[M]$ such that 
\[
\Delta(\mathscr{P}(x_{m_{1}}^{L}(j),x_{m_{2}}^{L}(j)),W^{\oplus d})\geq\rho
\]
for all $m_{1},m_{2}\in{\cal M}_{\rho}(j)$, and so $x_{m_{1}}^{L}(j)$
and $x_{m_{2}}^{L}(j)$ are ``far'' according to the distance $\Delta$.
Furthermore, by the maximal property of this set, it is clear that
if $m\notin{\cal M}_{\rho}(j)$, then there exists $m'\in{\cal M}_{\rho}(j)$
such that $x_{m}^{L}(j)$ and $x_{m'}^{L}(j)$ are ``close''. Thus
the size of ${\cal M}_{\rho}(j)$ is a measure of the scattering of
the molecules of the codeword $x^{LM}(j)$. In principle, the index
set ${\cal M}_{\rho}(j)$ can be different for each codeword $j\in[|{\cal C}_{M}|]$.
However, as we show a sequence of capacity achieving codebooks can
be found for which ${\cal M}_{\rho}(j)=[M_{\rho}]$, that is, identical
to all the codewords $j\in[|{\cal C}_{M}|]$ in ${\cal C_{M}}$. 

The second structural property is that the molecules of the codeword
in a specific index $m\in[M]$ can have a constant composition $P_{X,m}\in{\cal P}_{L}({\cal X})$,
again, for all $j\in[|{\cal C}_{M}|]$. 

These properties are summarized in the following proposition. 
\begin{prop}
\label{prop: reduction to constant composition}Suppose that the rate
$R$ is achievable for the DNA channel $\mathsf{DNA}=\{\alpha,\beta,W\}$.
Then $R$ is also achievable by a sequence of codebooks $\{{\cal C}_{M}\}_{M\in\mathbb{N}^{+}}$
for which:
\begin{enumerate}
\item There exists $M_{\rho}$ such that ${\cal M}_{\rho}(j)=[M_{\rho}]$
for all $j\in[|{\cal C}_{M}|]$.\textbf{ }
\item There exists $\{P_{X,m}\}_{m\in[M]}\subseteq{\cal P}_{L}({\cal X})$
(with $L=\beta\log M$) such that $\mathscr{P}(x_{m}^{L}(j))=P_{X,m}$
for all $j\in[|{\cal C}_{M}|]$.
\end{enumerate}
\end{prop}
\begin{IEEEproof}
The result is a simple consequence of the number of possibilities
for sets ${\cal M}_{\rho}$, and for the number of types, as well
as the DNA channel property that the order of the molecules does not
affect the error probability of the codebook. We will establish these
properties in two stages. In each stage we find a suitable sub-codebook
which satisfies the property and has negligible rate loss, and, trivially,
lower error probability (under optimal decoding). 

Let a codebook ${\cal C}_{M}$ of size $e^{ML(R-\delta)}$ be given.
First, the number of possible sets ${\cal M}_{\rho}(j)\subseteq[M]$
is at most $2^{M}=e^{o(ML)}$ since $L=\beta\log M$. Thus, for all
$M$ large enough, there must exist a sub-code ${\cal C}_{M}^{*}$
of size larger than 
\[
|{\cal C}_{M}^{*}|\geq\frac{e^{ML(R-\delta)}}{2^{M}}\geq\exp\left[ML\left(R-\frac{\delta}{2}\right)\right]
\]
such that ${\cal M}_{\rho}(j)$ is identical to all codewords. We
denote its size by $M_{\rho}$. By re-ordering the molecules for each
codeword, we may assure that ${\cal M}_{\rho}(j)=[M_{\rho}]$. Second,
the number of possible types for each molecule $x_{m}^{L}(j)\in{\cal X}^{L}$
is less than $(L+1)^{|{\cal X}|}$ and so the number of sequences
of $M$ ordered types for a codeword is less than 
\[
(L+1)^{|{\cal X}|M}=\exp\left[ML\cdot\frac{|{\cal X}|M\cdot\log(L+1)}{ML}\right]=e^{o(ML)}.
\]
A proper sub-code can be found similarly to the first stage.
\end{IEEEproof}
We will henceforth assume that the codebooks $\{{\cal C}_{M}\}$ satisfy
the structural properties of Prop. \ref{prop: reduction to constant composition}. 

\subsection{Upper Bound on the Mutual Information\label{subsec:Upper-bound-on}}

\begin{figure}
\begin{centering}
\includegraphics[scale=1.5]{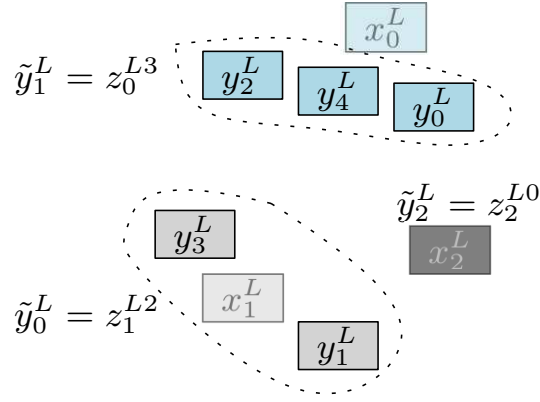}
\par\end{centering}
\caption{Illustration of permuted output clusters for $M=3$ and $N=5$. Here,
$z_{0}^{L3}=(y_{0}^{L},y_{2}^{L},y_{4}^{L})$, $z_{1}^{L2}=(y_{1}^{L},y_{3}^{L})$
and $z_{2}^{L0}=()$ (is empty), while $\tilde{u}^{3}=(1,0,2)$. \label{fig:Illustration-of-permuted-output-clusters}}

\end{figure}

In this section, we derive an upper bound on $I(X^{LM};Y^{LN})$,
and so complete the proof of the upper bound on the capacity. To this
end, we consider a genie-aided decoder proposed in \cite{shomorony2021dna,lenz2019upper},
and specifically, we assume that the genie-aided decoder has a modified
output $\tilde{Y}^{LM}$ instead of $Y^{LN}$, as we next define.
Let $Z_{m}^{LS_{m}}$ be the \emph{output clusters} of the molecule
$X_{m}^{L}$, that is 
\[
Z_{m}^{LS_{m}}\dfn\left\{ Y_{n}^{L}\colon U_{n}=m\right\} 
\]
is the set of outputs which are the result of the sequencing of the
$m$th molecule. Note $Z_{m}^{LS_{m}}\in{\cal Y}^{LS_{m}}$, that
is, $Z_{m}$ has a variable number of $LS_{m}$ symbols from ${\cal Y}$,
where $S_{m}$ is the number of times molecule $m$ have been sampled.
Thus, whenever $S_{m}=d$, we may consider $Z_{m}$ to be the $L$-tuple
output of $L$ memoryless uses of the binomial channel $W^{\oplus d}$
, for the input molecule $X_{m}^{L}$. So, we equivalently consider
$Z_{m}^{LS_{m}}$ to have $L$ of symbols from ${\cal Y}^{d}$, and
slightly change the notation to $Z_{m}^{L}.$ We then let $Z^{ML}=(Z_{0}^{L},\ldots,Z_{M-1}^{L})$
(where each $Z_{m}$ might have a different alphabet ${\cal Y}^{S_{m}}$).
As discussed in \cite{shomorony2021dna,lenz2019upper}, a genie-aided
decoder whose output is $Z^{ML}$ is too strong, and would lead to
a loose upper bound. So, we only assume that the decoder knows a random
permutation of $Z^{ML}$. In other words, the decoder knows the outputs
clusters, but does not know which input molecule is the source of
which cluster. Concretely, let $\tilde{U}^{M}\colon[M]\to[M]$ be
a permutation drawn uniformly at random from the symmetric group $\mathfrak{S}_{M}$,
and let 
\[
\tilde{Y}_{m}^{L}=Z_{\tilde{U}_{m}}^{L}\in{\cal Y}^{S_{\tilde{U}_{m}}L}
\]
for all $m\in[M]$, and then consider the vector $\tilde{Y}^{LM}=(\tilde{Y}_{0}^{L},\ldots,\tilde{Y}_{M-1}^{L})$.
This is illustrated in Fig. \ref{fig:Illustration-of-permuted-output-clusters}.

We assume that the output of the genie-aided channel is $\tilde{Y}^{LM}$
rather than $Y^{LN}$. Note that this also implies that the genie-aided
decoder knows the amplification vector $Q^{N+1}$, that is, how many
molecules have been sampled $d$ times for $d\in[N+1]$. This assumption
essentially transforms the DNA channel as follows: On the original
DNA channel with the original decoder, the $M$ input molecules $\{X_{m}^{L}\}_{m\in[M]}$
are transformed into output molecules $\{Y_{n}^{L}\}_{n\in[N]}$,
where each $Y_{n}^{L}$ is the result of $X_{U_{n}}^{L}$ sequenced
by the DMC $W^{L}$. In the modified channel, there are only $M$
outputs $\tilde{Y}^{LM}$, and each output $\tilde{Y}_{m}$ is the
result of $X_{\tilde{U}_{m}}^{L}$ sequenced by the binomial channel,
$W^{\oplus d}$, where $d=S_{\tilde{U}_{m}}$, and ${\cal Y}^{d}$
is the alphabet size of $\tilde{Y}_{m}^{L}$. In other words, while
the original DNA channel had $N$ homogeneous outputs -- each is
a sequencing of a single molecule, the modified channel has $M$ possibly
non-homogeneous outputs -- each is a (possibly) multiple sequencing
of the same molecule. 

Evidently, given $\tilde{Y}^{LM}$, the genie-aided decoder can generate
an output which is distributed as $Y^{LN}$ by separating its clusters
and performing a random permutation. Hence, by the data processing
theorem 
\begin{equation}
I(X^{LM};Y^{LN})\leq I(X^{LM};\tilde{Y}^{LM}).\label{eq: data processing bound for channel and modified channel output}
\end{equation}
Thus in what follows we concentrate on upper bounding $I(X^{LM};\tilde{Y}^{LM})$.
Moreover, since the genie-aided decoder, in fact, is aware of $Q^{N+1}$,
we will focus on bounding $I(X^{LM};\tilde{Y}^{LM}\mid Q^{N+1})$
{[}which also satisfies a data processing bound as in (\ref{eq: data processing bound for channel and modified channel output}){]}.
The error probability in the DNA channel depends on $U^{N}$ only
via $Q^{N+1}$. Thus $Q^{N+1}$ can be considered a random state of
the channel. Since $Q_{d}$ is the number of molecules which have
been sampled $d$ times, or, otherwise stated, were input to a $d$-order
binomial channel, $Q^{N+1}$ represents the distribution of channel
``quality'' for the $M$ molecules (with $d$ being larger means
that the channel is less noisy). We thus upper bound the mutual information
$I(X^{LM};\tilde{Y}^{LM}\mid Q^{N+1})$ in three steps:
\begin{enumerate}
\item Appendix  \ref{subsec:Fixed-composition-molecules conditional q}:
Conditioned on a fixed given amplification vector $Q^{N+1}=q^{N+1}$,
and assuming a fixed composition codebook ${\cal C}_{M}$, namely,
$\mathscr{P}(x_{m}^{L}(j))=P_{X}$ for all $m\in[M]$, and for all
$j\in[{\cal C}_{M}]$ (per the structural property of Prop. \ref{prop: reduction to constant composition}).
\item Appendix  \ref{subsec:Fixed-composition-molecules}: Averaged over
the distribution of $Q^{N+1}$, while still assuming a fixed composition
codebook.
\item Appendix  \ref{subsec:General-composition-molecules}: Averaged over
the distribution of $Q^{N+1}$, for a general composition codebook.
Per the structural property of Prop. \ref{prop: reduction to constant composition},
it can still be assumed that $\mathscr{P}(x_{m}^{L}(j_{1}))=\mathscr{P}(x_{m}^{L}(j_{2}))$
for any $j_{1},j_{2}\in[{\cal C}_{M}]$ and any $m\in[M]$. 
\end{enumerate}
The order in which we develop the upper bound on the mutual information
is not arbitrary, or just chosen for the convenience of the analysis.
Allowing a general composition codebook at an earlier stage of the
derivation would lead to an upper bound in which a different input
distribution $P_{X,d}$ can be chosen for input molecules which are
sampled $d$ times. Clearly, such a bound will not be generally tight
in a DNA channel, in which the encoder does not how many times each
molecule is sampled.

\subsubsection{Fixed Composition Molecules and Given $Q^{N+1}$ \label{subsec:Fixed-composition-molecules conditional q}}

We next upper bound the mutual information for a DNA channel with
genie-aided decoder. We will need a slight generalization of the original
model, which pertains to the molecule scaling length. For the DNA
channel, there are $M$ molecules, each of length $L=\beta\log M$.
Here we still assume that $L=\beta\log M$, yet the total number of
molecules is $\theta_{M}M$ for a general $\theta_{M}\in(0,1]$, which
may even decay to zero, yet sub-polynomially.
\begin{prop}
\label{prop:converse genie aided conditional}Let $\overline{M}=\theta M$
for some $\theta\equiv\theta_{M}=\frac{1}{M^{o(1)}}\in(0,1]$, and
assume a DNA codebook with molecules of length $L=\beta\log M$ and
a total of $\overline{M}$ molecules per codeword. Let $\overline{d}\in\mathbb{N}^{+}$
and $q^{N+1}\in\mathscr{Q}(M,N)$ be given, and assume that
\[
\nu_{\text{min}}(W)\dfn\max_{x\in{\cal X},\;y\in{\cal Y}}\log\frac{1}{W(y\mid x)}<\infty.
\]
Further assume that $\mathscr{P}(X_{m}^{L})=P_{X}$ with probability
$1$. Then,
\begin{equation}
\frac{1}{\overline{M}L}I(X^{L\overline{M}};\tilde{Y}^{L\overline{M}}\mid Q^{N+1}=q^{N+1})\leq\sum_{d\in[\overline{d}]}\frac{q_{d}}{\overline{M}}\cdot J_{d}(P_{X},W,\beta)+\left(1-\sum_{d\in[\overline{d}]}\frac{q_{d}}{\overline{M}}\right)\log|{\cal X}|+o_{\overline{d},|{\cal X}|}(1),\label{eq: upper bound on fixed composition conditional q capacity}
\end{equation}
where $J_{d}(P_{X},W,\beta)$ is as defined in (\ref{eq: definition of J for a channel, input distribution and beta}).
\end{prop}
To prove Prop. \ref{prop:converse genie aided conditional}, we need
the following lemma which bounds $I(X^{LM};\tilde{Y}^{LM}\mid Q^{N+1}=q^{N+1})$
in case all molecules have been sampled the same number of times $d$,
that is, under the event, $S_{m}=d$ for all $m\in[M]$, and so the
total number of output molecules is $dM$. Note that in the DNA channel
$N=\alpha M$ and so $d=\alpha$ in this event, yet here we consider
a general $d$. 
\begin{lem}
\label{lem:mutual information upper bound for codebook}Let $\overline{M}=\theta M$
for some $\theta\equiv\theta_{M}=\frac{1}{M^{o(1)}}\in(0,1]$, and
assume a DNA codebook with molecules of length $L=\beta\log M$ and
a total of $\overline{M}$ molecules per codeword. Assume that the
codebook ${\cal C}_{\overline{M}}$ is such that $\mathscr{P}(x_{m}(j))=P_{X}$
for all $m\in[\overline{M}]$ and $j\in[|{\cal C}_{\overline{M}}|]$
and that the sequencing channel satisfies $\nu_{\text{min}}(W)<\infty$.
Let $q^{N+1}$ be given such that $q_{d}=M$ for some $d\in[N+1]$
(that is $S_{m}=d$ for all $m\in[M]$). 

If $X^{L\overline{M}}$ is uniformly distributed over ${\cal C}_{\overline{M}}$
then 
\begin{multline}
\frac{1}{\overline{M}L}I(X^{L\overline{M}};\tilde{Y}^{L\overline{M}}\mid Q^{N+1}=q^{N+1})\\
\leq\begin{cases}
I(P_{X},W^{\oplus d})-\frac{1}{\beta}+o(1), & \mathsf{CID}(P_{X},W^{\oplus d})\ge\frac{2}{\beta}\\
I(P_{X},W^{\oplus d})-\mathsf{CID}(P_{X},W^{\oplus d})+\frac{1}{\beta}+o(1), & \frac{1}{\beta}\leq\mathsf{CID}(P_{X},W^{\oplus d})<\frac{2}{\beta}\\
I(P_{X},W^{\oplus d}), & \mathsf{CID}(P_{X},W^{\oplus d})<\frac{1}{\beta}
\end{cases}.
\end{multline}
\end{lem}
We remark that the following proof utilizes Lemma \ref{lem: mutual information difference between iid and fixed composition inputs}
from Appendix \ref{subsec:Mutual-information-of-DMC with i.i.d. vs fixed composition},
which shows that the mutual information obtained by $K$ DMC uses
is roughly the same when the input distribution is i.i.d. and when
the input has a fixed composition over the type class induced by the
input distribution. This result may be of independent interest. 
\begin{IEEEproof}
To simplify the notation, we remove throughout the conditioning on
$Q^{N+1}=q^{N+1}$, and simply write $M$ instead of $\overline{M}=\theta M$.
The exact location in which this matters will be highlighted. 

We begin by modifying the input distribution. Recall that the distribution
of the input $X^{LM}$ is 
\[
p_{{\cal C}_{M}}(x^{LM})=\frac{1}{|{\cal C}_{M}|}\sum_{j\in[|{\cal C}_{M}|]}\I\left\{ x^{LM}(j)=x^{LM}\right\} .
\]
Let us denote $x_{m}^{L}(j)=(x_{m,0}(j),\ldots,x_{m,L-1}(j))$ so
that $x_{m,\ell}(j)\in{\cal X}$ is the $\ell$th symbol in the $m$th
molecule of the codeword $x^{LM}(j)$. Let $x_{m,\sigma}(j)=(x_{m,\sigma(0)}(j),\ldots,x_{m,\sigma(L-1)}(j))$,
where $\sigma\colon[L]\to[L]$ is a permutation from the symmetric
group $\mathfrak{S}_{L}$. Let ${\cal C}_{M,\sigma}$ denote the codebook
obtained by applying the same permutation $\sigma$ to all molecules,
and for all codewords. Further let us temporarily denote the mutual
information, $I(X^{LM};\tilde{Y}^{LM})$, with an input distribution
$P_{{\cal C}_{M}}$ as $I(P_{{\cal C}_{M}})$. Then, since the sequencing
channel is memoryless $I(P_{{\cal C}_{M}})=I(P_{{\cal C}_{M},\sigma})$
for any permutation $\sigma\in\mathfrak{S}_{L}$. Let 
\begin{equation}
\overline{P}=\frac{1}{|\mathfrak{S}_{L}|}\sum_{\sigma\in\mathfrak{S}_{L}}P_{{\cal C}_{M},\sigma}\label{eq: a mixture of permuted distributions}
\end{equation}
denote the uniform mixture of $P_{{\cal C}_{M},\sigma}$. By concavity
of the mutual information in its input distribution, it holds that
$I(\overline{P})\geq I(P_{{\cal C}_{M}}).$ Thus we may upper bound
$I(\overline{P})=I(X^{LM};\tilde{Y}^{LM})$ under the input distribution
$\overline{P}$. Note that under $\overline{P}$ the marginal distribution
of any molecule $X_{m}^{L}$ is uniform over the type class ${\cal T}_{L}(P_{X})$.
Other than that, the order of the molecules is unchanged, and the
closeness relations between molecules (according to the distance $\Delta$)
are unaltered.

Recall that $\tilde{U}^{M}$ is the random permutation which maps
$X_{m}^{L}$ to $\tilde{Y}_{\tilde{U}_{m}^{-1}}$. It holds that 
\begin{align}
I(X^{LM};\tilde{Y}^{LM}) & =H(\tilde{Y}^{LM})-H(\tilde{Y}^{LM}\mid X^{LM})\\
 & =H(\tilde{Y}^{LM})-H(\tilde{Y}^{LM}\mid X^{LM},\tilde{U}^{M})-H(\tilde{Y}^{LM}\mid X^{LM})+H(\tilde{Y}^{LM}\mid X^{LM},\tilde{U}^{M})\\
 & =H(\tilde{Y}^{LM})-H(\tilde{Y}^{LM}\mid X^{LM},\tilde{U}^{M})-I(\tilde{Y}^{LM};\tilde{U}^{M}\mid X^{LM})\\
 & =H(\tilde{Y}^{LM})-H(\tilde{Y}^{LM}\mid X^{LM},\tilde{U}^{M})-H(\tilde{U}^{M}\mid X^{LM})+H(\tilde{U}^{M}\mid X^{LM},\tilde{Y}^{LM})\\
 & =H(\tilde{Y}^{LM})-H(\tilde{Y}^{LM}\mid X^{LM},\tilde{U}^{M})-M\log M+H(\tilde{U}^{M}\mid X^{LM},\tilde{Y}^{LM})+O(M),\label{eq: decomposition of mutual information}
\end{align}
where the last equality follows since the permutation $\tilde{U}^{M}$
is distributed uniformly over the symmetric group $\mathfrak{S}_{M}$,
independently of $X^{LM}$, and since by Stirling's approximation
$H(\tilde{U}^{M})=\log M!\geq M\log M+O(M)$. 

We continue by bounding $H(\tilde{U}^{M}\mid X^{LM},\tilde{Y}^{LM})$
and to this end we recall the distance $\Delta(Q_{A\overline{A}},V)$
defined in (\ref{eq: Distance between molecules definition}), and
set $\rho_{L}\equiv\underline{\rho}_{L}$ obtained by setting $K=L$
in (\ref{eq: minimal value for rho}), and taking $\delta_{L}$ which
satisfies the delta-convention for alphabet as large as ${\cal X}\times{\cal Y}^{d}$.
In essence, we will utilize here Lemma \ref{lem: decoding permutation lemma},
which implies that if a set of far molecules is randomly permuted
and then sequenced, the permutation can be ``decoded'' by observing
the inputs and the outputs in the sense that the conditional entropy
of the permutation is greatly reduced conditioned on the input and
the output. Recall that according to Prop. \ref{prop: reduction to constant composition},
we may assume that there exists $M_{\rho}\in[M]$ such that 
\[
\Delta\left(\mathscr{P}(x_{m_{1}}^{L}(j),x_{m_{2}}^{L}(j)),W^{\oplus d}\right)\geq\rho_{L}
\]
for all $m_{1},m_{2}\in[M_{\rho}]$ $m_{1}\neq m_{2}$, and that for
any $m_{3}\not\in[M_{\rho}]$ there exists $m_{4}\in[M_{\rho}]$ such
that\footnote{Note that since $\mathscr{P}(x_{m_{3}}(j))=\mathscr{P}(x_{m_{4}}(j))$
the distance $\Delta$ is symmetric.}
\[
\Delta\left(\mathscr{P}(x_{m_{3}}^{L}(j),x_{m_{4}}^{L}(j)),W^{\oplus d}\right)\leq\rho_{L}.
\]
Thus, $\{x_{m}^{L}(j)\}_{m\in[M_{\rho}]}$ is a minimum-distance maximal
set for any $j\in[|{\cal C}_{M}|]$. Let $G^{M}\in\{0,1\}^{M}$ be
such that $G_{m}=\I\{\tilde{U}_{m}\not\in[M_{\rho}]\}$, that is,
$G_{m}$ indicates whether $\tilde{Y}_{m}$ is the output of a molecule
which does not belong to the minimum-distance maximal set. Given $G^{M}$
the decoder can partition each of $X^{LM}$, $\tilde{Y}^{LM}$ and
$\tilde{U}^{M}$ into two parts. Intuitively, the knowledge of $G^{m}$
allows to separate the input, output and channel permutation to ones
which pertain to the maximal set and their complement. Concretely,
first $X_{[M_{\rho}]}=(X_{0}^{L},\ldots,X_{M_{\rho}-1}^{L})$ and
$X_{[M_{\rho}]^{c}}=(X_{M_{\rho}}^{L},\ldots,X_{M-1}^{L})$ are, respectively,
the molecules which belong to the maximal set, and the remaining molecules.
Second, in accordance, $\tilde{Y}_{(0)}=(\tilde{Y}_{m}^{L}\colon G{}_{m}=0)$
and $\tilde{Y}_{(1)}=(\tilde{Y}_{m}^{L}\colon G{}_{m}=1)$, the outputs
clusters which are the result of sequencing $X_{[M_{\rho}]}$ and
$X_{[M_{\rho}]^{c}}$, respectively, and $\tilde{U}_{(0)}=(U_{m}:G_{m}=0)$
and $\tilde{U}_{(1)}=(U_{m}:G_{m}=1)$. It holds that $\tilde{U}_{(0)}$
(respectively $\tilde{U}_{(1)}$) is distributed uniformly over a
group isomorphic to the symmetric group $\mathfrak{S}_{M_{\rho}}$
(respectively $\mathfrak{S}_{[M]\backslash[M_{\rho}]})$. 

Let $\eta\in(0,1)$ be a given constant (which does not depend on
$M$). Then, for all sufficiently large $M$\footnote{The minimal value of $M$ required for this to hold only depends on
$|{\cal X}|,|{\cal Y}|$, $\overline{d}$ and $\eta$. } 
\begin{align}
H(\tilde{U}^{M}\mid X^{LM},\tilde{Y}^{LM}) & \leq H(\tilde{U}^{M},G^{M}\mid X^{LM},\tilde{Y}^{LM})\nonumber \\
 & =H(G^{M}\mid X^{LM},\tilde{Y}^{LM})+H(\tilde{U}^{M}\mid G^{M},X^{LM},\tilde{Y}^{LM})\\
 & \leq H(G^{M})+H(\tilde{U}_{(1)}\mid G^{M},X^{LM},\tilde{Y}^{LM})+H(\tilde{U}_{(0)}\mid G^{M},X^{LM},\tilde{Y}^{LM})\\
 & \trre[\leq,a]M\log2+(M-M_{\rho})\log M+O(M)+H(\tilde{U}_{(0)}\mid G^{M},X^{LM},\tilde{Y}^{LM})\\
 & \trre[\leq,b]O(M)+(M-M_{\rho})\log M+H(\tilde{U}_{(0)}\mid X_{[M_{\rho}]},\tilde{Y}_{(0)})\\
 & \trre[\leq,c]O(M)+(M-M_{\rho})\log M+o(M_{\rho}\log M_{\rho})\\
 & \leq O(M)+(M-M_{\rho})\log M+o(M\log M)+\eta M\log M,\label{eq: bound on permutation entropy}
\end{align}
where:
\begin{itemize}
\item $(a)$ follows since $H(G^{M})\leq\sum_{m\in[M]}H(G_{m})\leq M\log2$,
and since $\tilde{U}_{(1)}$ has at most $(M-M_{\rho})!$ outcomes
and so Stirling's bound implies $H(\tilde{U}_{(1)}\mid G^{M},X^{LM},\tilde{Y}^{LM})\leq\log(M-M_{\rho})!\leq(M-M_{\rho})\log M+O(M)$. 
\item $(b)$ follows since conditioning reduces entropy. 
\item $(c)$ is justified as follows. If $M_{\rho}\geq\eta M$ then we utilize
Lemma \ref{lem: decoding permutation lemma}. Indeed, $\tilde{U}_{(0)}$
is a uniform random permutation, $X_{[M_{\rho}]}$ is a set of ``far''
inputs, and $\tilde{Y}_{(0)}$ is the set of corresponding permuted
outputs over the channel $W^{\oplus d}$. Specifically, we set therein
$K=L=\beta\log M$. However, at this point we recall that $M$ is
a simplified notation for $\theta_{M}M$, and so we set $T=\eta\theta M\geq\exp[L(\frac{1}{\beta}-\frac{1}{L}\log(\frac{1}{\eta\theta_{M}}))]$.
It then holds that $\tau_{L}$ therein satisfies $\tau_{L}\geq\frac{1}{2\beta}=\tau_{0}$
for all $L$ large enough (or $M$) since $\eta$ is a constant and
$\theta_{M}=\omega(\frac{1}{L})$. Thus, the claim of Lemma \ref{lem: decoding permutation lemma}
applies here and results 
\[
H(\tilde{U}_{(0)}\mid X_{[M_{\rho}]},\tilde{Y}_{(0)})=o(M_{\rho}\log M_{\rho}).
\]
This is the only place in the proof in which it matters that $\theta$
may not equal $1$. Otherwise, if $M_{\rho}\leq\eta M$ then we bound
by Stirling's bound $H(\tilde{U}_{(0)}\mid X_{[M_{\rho}]},\tilde{Y}_{(0)})\leq\eta M\log(\eta M)+O(M)=\eta M\log M+O(M)$.
The upper bound in $(c)$ replaces the maximum of both cases ($M_{\rho}\lessgtr\eta M$)
by a sum. 
\end{itemize}
We continue with the analysis of (\ref{eq: decomposition of mutual information}),
and next turn to upper bound $H(\tilde{Y}^{LM})$. To this end, let
$\overline{G}^{M}\in[M]$ be the following refinement of $G^{M}$.
If $G_{m}=0$ then $\overline{G}_{m}=0$ too. Otherwise, if $G_{m}=1$
then 
\[
\overline{G}_{m}=\min\left\{ \overline{m}:G_{\overline{m}}=0,\;\Delta(\mathscr{P}(X_{\tilde{U}_{\overline{m}}}^{L},X_{\tilde{U}_{m}}^{L}),W^{\oplus d})\leq\rho\right\} ,
\]
that is $\overline{G}_{m}$, indicates which is the output cluster
$\tilde{Y}_{\overline{G}_{m}}^{L}$ so that $\tilde{Y}_{m}^{L}$ and
$\tilde{Y}_{\overline{G}_{m}}^{L}$ are the outputs of ``close''
input molecules, where $\tilde{Y}_{\overline{G}_{m}}^{L}$ is the
output of a molecule in the minimum-distance maximal set. In addition,
let $F_{m}$ be an arbitrary member of ${\cal P}_{L}({\cal X}\times{\cal X})$
if $G_{m}=0$ and $F_{m}=\mathscr{P}(X_{\tilde{U}_{\overline{G}_{m}}}^{L},X_{\tilde{U}_{m}}^{L})$
if $G_{m}=1$, that is, in case $\tilde{Y}_{m}^{L}$ is not a cluster
which is the result of sequencing a molecule in the maximal set, $F_{m}$
indicates the joint type its input molecule and the ``close'' molecule
in the maximal set. Note that $F_{m}\in{\cal P}_{L}({\cal X}\times{\cal X})$.
Then, 
\begin{align}
H(\tilde{Y}^{LM}) & \leq H(\tilde{Y}^{LM},\overline{G}^{M},F^{M})\\
 & =H(\overline{G}^{M},F^{M})+H(\tilde{Y}^{LM}\mid\overline{G}^{M},F^{M})\\
 & \leq O(M\log(\log M)))+(M-M_{\rho})\log M+H(\tilde{Y}^{LM}\mid\overline{G}^{M},F^{M}),\label{eq: first bound on output entropy}
\end{align}
where in the last inequality $H(\overline{G}^{M},F^{M})$ is bounded
as follows: There are ${M \choose M_{\rho}}\leq2^{M}$ ways to choose
$G^{M}$, that is, the set of output clusters $\tilde{Y}_{(0)}$ which
are the result of molecules in $X_{[M_{\rho}]}$. This can be specified
with $O(M)$ nats. Then, for each of the $M-M_{\rho}$ indices in
which $G_{m}=1$, it requires $\log M_{\rho}$ nats to specify the
cluster index $\overline{m}$, and 
\[
\log|{\cal P}_{L}({\cal X}\times{\cal X})|\leq\log\left(L+1\right)^{|{\cal X}|^{2}}=O(\log(\log M))
\]
to specify the joint type (as follows from standard type counting,
e.g., \cite[Lemma 2.2]{csiszar2011information}). 

Continuing (\ref{eq: first bound on output entropy}), we next bound
$H(\tilde{Y}^{LM}\mid\overline{G}^{M},F^{M})$ as 
\begin{align}
 & H(\tilde{Y}^{LM}\mid\overline{G}^{M},F^{M})\nonumber \\
 & =H(\tilde{Y}_{(0)}\mid\overline{G}^{M},F^{M})+H(\tilde{Y}_{(1)}\mid\tilde{Y}_{(0)},\overline{G}^{M},F^{M})\\
 & \leq H(\tilde{Y}_{(0)}\mid\overline{G}^{M})+H(\tilde{Y}_{(1)}\mid\tilde{Y}_{(0)},\overline{G}^{M},F^{M})\\
 & \trre[\leq,a]\sum_{m\in[M]\colon G_{m}=0}H(\tilde{Y}_{m}^{L})+\sum_{m\in[M]\colon G_{m}=1}H(\tilde{Y}_{m}^{L}\mid\tilde{Y}_{\overline{G}_{m}}^{L},F_{m})\\
 & =\sum_{m\in[M]\colon G_{m}=0}H(\tilde{Y}_{m}^{L})+\sum_{m\in[M]\colon G_{m}=1}H(\tilde{Y}_{m}^{L}\mid\tilde{Y}_{\overline{G}_{m}}^{L},F_{m})+H(\tilde{Y}_{\overline{G}_{m}}^{L}\mid F_{m})-H(\tilde{Y}_{\overline{G}_{m}}^{L}\mid F_{m})\\
 & \trre[=,b]\sum_{m\in[M]\colon G_{m}=0}H(\tilde{Y}_{m}^{L})+\sum_{m\in[M]\colon G_{m}=1}H(\tilde{Y}_{m}^{L},\tilde{Y}_{\overline{G}_{m}}^{L}\mid F_{m})-H(\tilde{Y}_{\overline{G}_{m}}^{L}),\label{eq: output entropy bound conditioned on side information}
\end{align}
where $(a)$ follows from the standard independence bound on entropy
bound $H(A,B)\leq H(A)+H(B)$ and since conditioning reduces entropy,
and $(b)$ follows since $\tilde{Y}_{\overline{G}_{m}}^{L}$ is independent
of $F_{m}$ {[}under $\overline{P}$ defined in (\ref{eq: a mixture of permuted distributions}){]}.
Next, we combine this bound with the value of $H(\tilde{Y}^{LM}\mid X^{LM},\tilde{U}^{M})$
which is the last entropy term in the decomposition of $I(X^{LM};\tilde{Y}^{LM})$
in (\ref{eq: decomposition of mutual information}). For this term,
we simply write:
\begin{align}
H(\tilde{Y}^{LM}\mid X^{LM},\tilde{U}^{M}) & =\sum_{m\in[M]}H(\tilde{Y}_{m}\mid X_{\tilde{U}_{m}}^{L})\\
 & =\sum_{m\in[M]\colon G_{m}=0}H(\tilde{Y}_{m}\mid X_{\tilde{U}_{m}}^{L})+\sum_{m\in[M]\colon G_{m}=1}H(\tilde{Y}_{m}\mid X_{\tilde{U}_{m}}^{L}).
\end{align}
Then, combining this expression with (\ref{eq: output entropy bound conditioned on side information})
\begin{align}
 & H(\tilde{Y}^{LM}\mid\overline{G}^{M},F^{M})-H(\tilde{Y}^{LM}\mid X^{LM},\tilde{U}^{M})\nonumber \\
 & =\sum_{m\in[M]\colon G_{m}=0}H(\tilde{Y}_{m}^{L})-H(\tilde{Y}_{m}^{L}\mid X_{\tilde{U}_{m}}^{L})\nonumber \\
 & \hphantom{=}+\sum_{m\in[M]\colon G_{m}=1}H(\tilde{Y}_{m}^{L},\tilde{Y}_{\overline{G}_{m}}^{L}\mid F_{m})-\sum_{m\in[M]\colon G_{m}=1}H(\tilde{Y}_{m}^{L}\mid X_{\tilde{U}_{m}}^{L})-\sum_{m\in[M]\colon G_{m}=1}H(\tilde{Y}_{\overline{G}_{m}}^{L}\mid X_{\tilde{U}_{\overline{G}_{m}}}^{L})\nonumber \\
 & \hphantom{=}-\sum_{m\in[M]\colon G_{m}=1}H(\tilde{Y}_{\overline{G}_{m}}^{L})+\sum_{m\in[M]\colon G_{m}=1}H(\tilde{Y}_{\overline{G}_{m}}^{L}\mid X_{\tilde{U}_{\overline{G}_{m}}}^{L})\\
 & =\sum_{m\in[M]\colon G_{m}=0}I(X_{\tilde{U}_{m}}^{L};\tilde{Y}_{m}^{L})+\sum_{m\in[M]\colon G_{m}=1}I(X_{\tilde{U}_{m}}^{L},X_{\tilde{U}_{\overline{G}_{m}}}^{L};\tilde{Y}_{m}^{L},\tilde{Y}_{\overline{G}_{m}}^{L}\mid F_{m})-I(X_{\tilde{U}_{\overline{G}_{m}}}^{L};\tilde{Y}_{\overline{G}_{m}}^{L})\\
 & \trre[=,a](2M_{\rho}-M)\cdot L\cdot I(P_{X},W^{\oplus d})+o(ML)+\sum_{m\in[M]\colon G_{m}=1}I(X_{\tilde{U}_{m}}^{L},X_{\tilde{U}_{\overline{G}_{m}}}^{L};\tilde{Y}_{m}^{L},\tilde{Y}_{\overline{G}_{m}}^{L}\mid F_{m})\\
 & \trre[\leq,b](2M_{\rho}-M)\cdot L\cdot I(P_{X},W^{\oplus d})+2(M-M_{\rho})L\cdot I(P_{X},W^{\oplus d})-L(M-M_{\rho})\cdot\mathsf{CID}(P_{X},W^{\oplus d})+o(ML)\\
 & =ML\cdot I(P_{X},W^{\oplus d})-L(M-M_{\rho})\cdot\mathsf{CID}(P_{X},W^{\oplus d})+o(ML),\label{eq: bound on the mutual entreopy of output given SI minus the channel entropy}
\end{align}
where:
\begin{itemize}
\item $(a)$ holds since under the assumed input distribution to the channel
$\overline{P}$ it holds that $X_{m}^{L}\sim\text{Uniform}[{\cal T}_{L}(P_{X})]$,
and by Lemma \ref{lem: mutual information difference between iid and fixed composition inputs}
from Appendix \ref{subsec:Mutual-information-of-DMC with i.i.d. vs fixed composition}
the mutual information in this case is $L\cdot I(P_{X},W^{\oplus d})+o(L)$
(where $I(P_{X},W^{\oplus d})$ is the single-letter mutual information
in the DMC $W^{\oplus d}$ with input distribution $P_{X}$). 
\item $(b)$ follows from Prop. \ref{prop: mutual information given close inputs},
where here we utilized the fact that conditioned on $F_{m}$, the
input distribution is supported on the type indicated by $F_{m}$
with probability $1$, say $Q_{X\overline{X}}$, and this joint type
satisfies $\Delta(Q_{X\overline{X}},W^{\oplus d})\leq\rho\equiv\rho_{L}=o(L)$. 
\end{itemize}
Combining (\ref{eq: bound on the mutual entreopy of output given SI minus the channel entropy})
with (\ref{eq: first bound on output entropy}) we obtain the bound
\begin{multline}
H(\tilde{Y}^{LM})-H(\tilde{Y}^{LM}\mid X^{LM},\tilde{U}^{M})\leq ML\cdot I(P_{X},W^{\oplus d})-L(M-M_{\rho})\cdot\mathsf{CID}(P_{X},W^{\oplus d})\\
+(M-M_{\rho})\log M+o(ML).\label{eq: final bound on mutual information given ordering}
\end{multline}
Further combining the bounds (\ref{eq: bound on permutation entropy})
and (\ref{eq: final bound on mutual information given ordering})
on the entropy terms in (\ref{eq: decomposition of mutual information})
and setting, $L=\beta\log M$ we obtain
\begin{equation}
I(X^{LM};\tilde{Y}^{LM})\leq ML\left[I(P_{X},W^{\oplus d})-\left(1-\frac{M_{\rho}}{M}\right)\mathsf{CID}(P_{X},W^{\oplus d})+\left(1-\frac{2M_{\rho}}{M}\right)\frac{1}{\beta}+\frac{\eta}{\beta}+o(1)\right].\label{eq: first bound on MI for outputs of fixed sizes}
\end{equation}
Since $M_{\rho}$ is arbitrary, a general bound must take the worst
case over $M_{\rho}\in[M]$. The term in the last bound which depends
on $M_{\rho}$ is given by
\[
\frac{M_{\rho}}{M}\left[\mathsf{CID}(P_{X},W^{\oplus d})-\frac{2}{\beta}\right]
\]
and if $\mathsf{CID}(P_{X},W^{\oplus d})\geq\frac{2}{\beta}$ then
this term is maximized by $M_{\rho}=M$ and otherwise by $M_{\rho}=0$.
Inserting this into (\ref{eq: first bound on MI for outputs of fixed sizes}),
and taking $\eta\downarrow0$ we obtain 
\[
\frac{I(X^{LM};\tilde{Y}^{LM})}{ML}\leq\begin{cases}
I(P_{X},W^{\oplus d})-\frac{1}{\beta}+o(1), & \mathsf{CID}(P_{X},W^{\oplus d})\ge\frac{2}{\beta}\\
I(P_{X},W^{\oplus d})-\mathsf{CID}(P_{X},W^{\oplus d})+\frac{1}{\beta}+o(1), & \mathsf{CID}(P_{X},W^{\oplus d})<\frac{2}{\beta}
\end{cases}.
\]
The second case above will be used when $\mathsf{CID}(P_{X},W^{\oplus d})>\frac{1}{\beta}$.
Otherwise, the trivial bound $I(X^{LM};\tilde{Y}^{LM})\leq ML\cdot I(P_{X}\times W^{\oplus d})$
-- which can be easily obtained by analyzing a decoder which knows
the molecule order $\tilde{U}^{M}$ -- is better. Summarizing all
three cases and utilizing the definition of $J_{d}(P_{X},W,\beta)$
in (\ref{eq: definition of J for a channel, input distribution and beta})
completes the proof. 
\end{IEEEproof}
We can now prove Prop. \ref{prop:converse genie aided conditional}:
\begin{IEEEproof}[Proof of Prop. \ref{prop:converse genie aided conditional}]
To simplify the notation we again use $M$ instead of $\theta_{M}M$.
Recall the molecule duplicate vector $S^{M}$ where $S_{m}$ is the
number of times that molecule $X_{m}^{L}$ have been sampled, and
that $Q^{N+1}=\mathscr{N}(S^{M})$. Then,
\begin{align}
 & I(X^{LM};\tilde{Y}^{LM}\mid Q^{N+1}=q^{N+1})\nonumber \\
 & \leq I(X^{LM};\tilde{Y}^{LM},S^{M}\mid Q^{N+1}=q^{N+1})\\
 & =I(X^{LM};S^{M}\mid Q^{N+1}=q^{N+1})+I(X^{LM};\tilde{Y}^{LM}\mid S^{M},Q^{N+1}=q^{N+1})\\
 & =I(X^{LM};\tilde{Y}^{LM}\mid S^{M},Q^{N+1}=q^{N+1}),
\end{align}
where the right-most equality follows since $X^{LM}$ and $S^{M}$
are independent given $Q^{N+1}$ and so $I(X^{LM};S^{M}\mid Q^{N+1}=q^{N+1})=0$.
We next bound the last mutual information for an arbitrary $s^{M}$
for which $\mathscr{N}(s^{M})=q^{N+1}$. 

Let ${\cal M}_{d}\dfn\{m\in[M]\colon s_{m}=d\}$ denote the set of
molecule indices which have been sampled $d$ times (where $|{\cal M}_{d}|=q_{d}$).
Conditioned on $S^{M}=s^{M}$ (in essence, assuming that the decoder
knows $S^{M}=s^{M}$), the decoder can partition any codeword $X^{LM}$
to sets $X_{{\cal M}_{d}}=(X_{m}^{L})_{m\in{\cal M}_{d}}$ for $d\in[N+1]$,
and, in accordance, partition the entire codebook ${\cal C}_{M}$
to shorter codebooks ${\cal C}_{{\cal M}_{d}}$. Since the number
of $L$-tuples in $\tilde{Y}_{m}^{L}$ is trivially known to the decoder,
it knows $\overline{{\cal M}}_{d}\dfn\{m\in[M]\colon\tilde{Y}_{m}^{L}\in({\cal Y}^{d})^{L}\}$
and so it can similarly partition $\tilde{Y}_{m}^{L}$ to the corresponding
sets $\tilde{Y}_{\overline{{\cal M}}_{d}}$, so that the $q_{d}$
clusters in $\tilde{Y}_{\overline{{\cal M}}_{d}}$ are the result
of sequencing $X_{{\cal M}_{d}}$ in a random order (uniform over
all permutations). Then, 
\begin{align}
I(X^{LM};\tilde{Y}^{LM}\mid S^{M}=s^{M}) & =I\left(X_{{\cal M}_{0}},X_{{\cal M}_{1}},\ldots,X_{{\cal M}_{N}};\tilde{Y}_{{\cal \overline{M}}_{0}},\ldots,\tilde{Y}_{\overline{{\cal M}}_{N}}\mid S^{M}=s^{M}\right)\\
 & \trre[=,a]H\left(\tilde{Y}_{{\cal \overline{M}}_{0}},\ldots,\tilde{Y}_{\overline{{\cal M}}_{N}}\mid S^{M}=s^{M}\right)-\sum_{d\in[N+1]}H\left(\tilde{Y}_{\overline{{\cal M}}_{d}}\mid X_{{\cal M}_{d}},S^{M}=s^{M}\right)\\
 & \trre[\leq,b]\sum_{d\in[N+1]}H\left(\tilde{Y}_{\overline{{\cal M}}_{d}}\right)-\sum_{d\in[N+1]}H\left(\tilde{Y}_{\overline{{\cal M}}_{0}}\mid X_{{\cal M}_{d}}\right)\\
 & =\sum_{d\in[N+1]}I(X_{{\cal M}_{d}};\tilde{Y}_{\overline{{\cal M}}_{d}}),\label{eq: upper bound of mutual information by partition inputs and outputs to set of those sampled d times}
\end{align}
where $(a)$ follows since conditioned on $X_{{\cal M}_{d}}$, $\tilde{Y}_{\overline{{\cal M}}_{d}}$
is independent of all other variables, and $(b)$ follows from the
standard independence bound on entropy $H(A,B)\leq H(A)+H(B)$, and
since conditioning reduces entropy. Each mutual information term $I(X_{{\cal M}_{d}};\tilde{Y}_{\overline{{\cal M}}_{d}})$
in the sum (\ref{eq: upper bound of mutual information by partition inputs and outputs to set of those sampled d times})
corresponds to a channel $x_{{\cal M}_{d}}\to\tilde{y}_{{\cal M}_{d}}$
in which each molecule is sampled exactly $d$ times, and so Lemma
\ref{lem:mutual information upper bound for codebook} can be utilized.
Specifically, first, for $d=0$ (the erased molecules) $I(x_{{\cal M}_{0}};\tilde{y}_{{\cal M}_{0}})=0$.
Second, for any $d\geq\overline{d}$ (molecules which are ``over-sampled'')
we bound
\[
\sum_{d\in[N+1]\backslash[\overline{d}]}I(X_{{\cal M}_{d}};\tilde{Y}_{\overline{{\cal M}}_{d}})\leq\sum_{d\in[N+1]\backslash[\overline{d}]}H(x_{{\cal M}_{d}})\leq\left(M-\sum_{d\in[\overline{d}]}q_{d}\right)L\log|{\cal X}|.
\]
Third, for $d\in[\overline{d}]$ with $d>0$ we have two sub-cases.
Let $\underline{\theta}>0$ be arbitrary.
\begin{itemize}
\item If $q_{d}\leq\underline{\theta}M$ then we bound, 
\[
I(X_{{\cal M}_{d}};\tilde{Y}_{\overline{{\cal M}}_{d}})\leq\underline{\theta}ML\log|{\cal X}|
\]
\item If $q_{d}>\underline{\theta}M$ then we use Lemma \ref{lem:mutual information upper bound for codebook},
which can be succinctly written as 
\[
I(X_{{\cal M}_{d}};\tilde{Y}_{\overline{{\cal M}}_{d}})\leq q_{d}\cdot\left\{ I_{P_{X}\times W^{\oplus d}}(X;Y)-\left[\mathsf{CID}(P_{X},W^{\oplus d})-\frac{1}{\beta}\right]_{+}+\left[\mathsf{CID}(P_{X},W^{\oplus d})-\frac{2}{\beta}\right]_{+}+o(1)\right\} .
\]
Note that $\underline{\theta}M$ is fixed, and so Lemma \ref{lem:mutual information upper bound for codebook}
holds even if $M$ is replaced with $\theta_{M}M$. 
\end{itemize}
Summing up all mutual information bounds for all $d$ and substituting
in (\ref{eq: upper bound of mutual information by partition inputs and outputs to set of those sampled d times}),
and then taking $\underline{\theta}\downarrow0$ and using the definition
of $J_{d}(P_{X},W,\beta)$ in (\ref{eq: definition of J for a channel, input distribution and beta})
establishes the claimed bound. 
\end{IEEEproof}

\subsubsection{Fixed Composition Molecules and Random $Q^{N+1}$ \label{subsec:Fixed-composition-molecules}}

In this section, we remove the conditioning on a given amplification
vector $Q^{N+1}=q^{N+1}$, and prove an upper bound on the average
mutual information for a random $Q^{N+1}$. 
\begin{prop}
\label{prop:converse genie aided avearge}Consider the same assumptions
as in Prop. \ref{prop:converse genie aided conditional}. Then,
\begin{equation}
\frac{1}{\overline{M}L}I(X^{L\overline{M}};\tilde{Y}^{L\overline{M}}\mid Q^{N+1})\leq\sum_{d\in\mathbb{N}}\pi_{\alpha}(d)\cdot J_{d}(P_{X},W,\beta)+o(1),\label{eq: upper bound on fixed composition average q capacity}
\end{equation}
where $J_{d}(P_{X},W,\beta)$ is as defined in (\ref{eq: definition of J for a channel, input distribution and beta}),
and $\pi_{\alpha}(d)\dfn\frac{\alpha^{d}e^{-\alpha}}{d!}$ Poisson
p.m.f. with parameter $\alpha$. 
\end{prop}
To prove this proposition, we will need a notion of a typical set
for amplification vectors, defined as follows:
\[
\mathscr{A}_{\overline{d}}(\chi)\dfn\left\{ q^{N+1}\in\mathscr{Q}(M,N)\colon\max_{d\in[\overline{d}]}\left|\frac{q_{d}}{M}-\pi_{\alpha}(d)\right|\leq\chi\right\} ,
\]
and the next lemma which estimates the tolerance $\chi$ required
for $\mathscr{A}_{\overline{d}}(\chi)$ to have probability $\psi$.
\begin{lem}[Typical sampling set]
\label{lem: Typical amplification set}Let $\overline{d}\in[N+1]$
be given, let $S^{M}\sim\text{\emph{Multinomial}}(N,(\frac{1}{M},\ldots,\frac{1}{M}))$,
let $Q^{N+1}=\mathscr{N}(S^{M})$ be its empirical distribution, and
let $\chi(\psi)$ satisfy 
\[
\P\left[Q^{N+1}\in\mathscr{A}_{\overline{d}}\left(\chi(\psi)\right)\right]=\psi.
\]
Then, $\chi(\psi)=O_{\overline{d},\psi}(\frac{1}{\sqrt{M}})$. 
\end{lem}
\begin{IEEEproof}
By the union bound
\[
\P\left[Q^{N+1}\in\mathscr{A}_{\overline{d}}^{c}(\chi)\right]\leq\overline{d}\cdot\P\left[\left|\frac{Q_{d}}{M}-\pi_{\alpha}(d)\right|>\chi\right].
\]
Recall that $\frac{Q_{d}}{M}$ is the average of $M$ indicator random
variables 
\[
\frac{Q_{d}}{M}=\frac{1}{M}\sum_{m\in[M]}\I\{S_{m}=d\},
\]
and while these variables are not statistically independent, as we
shall next show, the correlation between them is low, and so the probability
of the deviation of $\frac{Q_{d}}{M}$ from its mean can be bounded
by Chebyshev's inequality. The mean is $\E\left[Q_{d}/M\right]=\P[S_{m}=d]$,
and the second moment is 
\begin{align}
\E\left[\left(\frac{Q_{d}}{M}\right)^{2}\right] & =\frac{1}{M^{2}}\sum_{m\in[M]}\sum_{m'\in[M]}\P\left[\{S_{m}=d\}\cap\{S_{m'}=d\}\right]\\
 & =\frac{\P[S_{0}=d]}{M}+\frac{1}{M^{2}}\sum_{m,m'\in[M]\colon m'\neq m}\P\left[\{S_{m}=d\}\cap\{S_{m'}=d\}\right]\\
 & =\frac{\P[S_{0}=d]}{M}+\frac{M^{2}-M}{M^{2}}\cdot\P\left[\{S_{m}=d\}\cap\{S_{m'}=d\}\right].\label{eq: second moment of normalized Qd}
\end{align}
Now, for $m'\neq m$ it holds that $S_{m}\sim\text{Binomial}(N,\frac{1}{M})$
and $S_{m'}\mid S_{m}=d\sim\text{Binomial}(N-d,\frac{1}{M-1})$.\footnote{This is because $U^{N}\sim\text{Uniform}[M]\backslash\{m\}$ for in
the indices $n\in[N]$ for which $U_{n}\neq m$.} So, 
\[
\P\left[S_{m}=d\right]={N \choose d}\left(\frac{1}{M}\right)^{d}\left(1-\frac{1}{M}\right)^{N-d}
\]
and
\[
\P\left[S_{m'}=d\mid S_{m}=d\right]={N-d \choose d}\left(\frac{1}{M-1}\right)^{d}\left(1-\frac{1}{M-1}\right)^{N-2d}.
\]
So, as long as $d\in[\overline{d}]$ and $\overline{d}=\Theta(1)$
\begin{align}
\frac{\P\left[S_{m'}=d\mid S_{m}=d\right]}{\P\left[S_{m}=d\right]} & =\frac{{N-d \choose d}\left(\frac{M-1}{M-2}\right)^{2d}\left(\frac{M(M-2)}{(M-1)^{2}}\right)^{N}}{{N \choose d}}\\
 & \trre[\leq,a]\frac{{N-d \choose d}\left(\frac{M-1}{M-2}\right)^{2d}}{{N \choose d}}\\
 & =\frac{[(N-d)!]^{2}}{(N-2d)!N!}\left(\frac{M-1}{M-2}\right)^{2d}\\
 & =\frac{(N-d)\cdots(N-2d+1)}{N(N-1)\cdots(N-d+1)}\left(\frac{M-1}{M-2}\right)^{2d}\\
 & =\left(1-\frac{d}{N}\right)\left(1-\frac{d}{N-1}\right)\cdot\left(1-\frac{d}{N-d+1}\right)\left(1+\frac{1}{M-2}\right)^{2d}\\
 & \trre[=,b]1+O_{\overline{d}}\left(\frac{1}{M}\right),
\end{align}
where $(a)$ follows since $M(M-2)/(M-1)^{2}\leq1$, and $(b)$ follows
since $d\in[\overline{d}]$. So, 
\begin{align}
\P\left[\{S_{m}=d\}\cap\{S_{m'}=d\}\right] & =\P\left[S_{m}=d\right]\P\left[S_{m'}=d\mid S_{m}=d\right]\\
 & \leq\P^{2}\left[S_{m}=d\right]\left(1+O_{\overline{d}}\left(\frac{1}{M}\right)\right).
\end{align}
Hence, from (\ref{eq: second moment of normalized Qd})
\begin{align}
\E\left[\left(\frac{Q_{d}}{M}\right)^{2}\right] & \leq\frac{\P[S_{0}=d]}{M}+\frac{M^{2}-M}{M^{2}}\cdot\P^{2}\left[S_{m}=d\right]\left(1+O_{\overline{d}}\left(\frac{1}{M}\right)\right)\\
 & \leq\frac{1}{M}+\P^{2}\left[S_{m}=d\right]\left(1+O_{\overline{d}}\left(\frac{1}{M}\right)\right)\\
 & =\P^{2}\left[S_{m}=d\right]+O_{\overline{d}}\left(\frac{1}{M}\right),
\end{align}
and then
\[
\V\left[\frac{Q_{d}}{M}\right]=\E\left[\left(\frac{Q_{d}}{M}\right)^{2}\right]-\E^{2}\left[\frac{Q_{d}}{M}\right]=O_{\overline{d}}\left(\frac{1}{M}\right).
\]
We thus denote $\V\left[\frac{Q_{d}}{M}\right]\dfn\frac{c_{\overline{d}}}{M}$
where $c_{\overline{d}}$ depends only on $\overline{d}$. By Chebyshev's
inequality 
\[
\P\left[\left|\frac{Q_{d}}{M}-\E\left[\frac{Q_{d}}{M}\right]\right|>\chi\right]=\P\left[\left|\frac{Q_{d}}{M}-\P\left[S_{m}=d\right]\right|>\chi\right]\leq\frac{c_{\overline{d}}}{M\chi^{2}},
\]
and by the union bound 
\[
\P\left[\max_{d\in[\overline{d}]}\left|\frac{Q_{d}}{M}-\P\left[S_{m}=d\right]\right|>\chi\right]\leq\frac{\overline{d}c_{\overline{d}}}{M\chi^{2}}.
\]
Let $\tilde{S}_{0}\sim\text{Pois}(\alpha)$. Since $S_{m}\sim\text{Binomial}(N,\frac{1}{M})$
and $N\cdot\frac{1}{M}=\alpha$, a constant, the binomial distribution
tends to a Poisson distribution. The approximation error is due to
Hodges and Le-Cam (e.g., \cite[Sec. 3.6.1]{durrett2019probability})
\[
\left|\P[S_{0}=d]-\P[\tilde{S}_{0}=d]\right|\leq\frac{N}{M^{2}}=\frac{\alpha}{M}.
\]
Since $\P[\tilde{S}_{0}=d]=\pi_{\alpha}(d)$ (by definition)
\begin{align}
 & \P\left[\max_{d\in[\overline{d}]}\left|\frac{Q_{d}}{M}-\pi_{\alpha}(d)\right|>\chi\right]\nonumber \\
 & =\P\left[\max_{d\in[\overline{d}]}\left|\frac{Q_{d}}{M}-\P[S_{0}=d]+\P[S_{0}=d]-\P\left[\tilde{S}_{0}=d\right]\right|>\chi\right]\\
 & \leq\P\left[\max_{d\in[\overline{d}]}\left|\frac{Q_{d}}{M}-\P[S_{0}=d]\right|+\left|\P[S_{0}=d]-\P\left[\tilde{S}_{0}=d\right]\right|>\chi\right]\\
 & \leq\P\left[\max_{d\in[\overline{d}]}\left|\frac{Q_{d}}{M}-\P[S_{0}=d]\right|+\left|\P[S_{0}=d]-\P\left[\tilde{S}_{0}=d\right]\right|>\chi-\frac{\alpha}{M}\right]\\
 & \leq\frac{\overline{d}c_{\overline{d}}}{M\left(\chi-\frac{\alpha}{M}\right)^{2}}.
\end{align}
Thus for $\P[Q^{N+1}\in\mathscr{A}_{\overline{d}}\left(\chi(\psi)\right)]\geq\psi$
it suffices to require $\chi=\frac{\alpha}{M}+\sqrt{\frac{\overline{d}c_{\overline{d}}}{\psi M}}=O_{\overline{d},\psi}(\frac{1}{\sqrt{M}})$. 
\end{IEEEproof}
We are now ready to prove an upper bound on the mutual information
for random $Q^{N+1}$.
\begin{IEEEproof}[Proof of Prop. \ref{prop:converse genie aided avearge}]
We again use $M$ instead of $\overline{M}=\theta_{M}M$ to lighten
notation. The next proof holds verbatim for any $\theta_{M}=\frac{1}{M^{o(1)}}$. 

Let arbitrary $\psi\in(0,1)$ and $\eta>0$ be given, and choose $\overline{d}$
such that$\sum_{d\in[\overline{d}]}\pi_{\alpha}(d)\geq1-\eta$ (which
is always possible since $\sum_{d\in\mathbb{N}}\pi_{\alpha}(d)=1$,
$\pi_{\alpha}(d)\geq0$). Then, 
\begin{align}
I(X^{LM};\tilde{Y}^{LM}\mid Q^{N+1}) & =\P\left[Q^{N+1}\in\mathscr{A}_{\overline{d}}(\chi(\psi))\right]\cdot I\left(X^{LM};\tilde{Y}^{LM}\mid Q^{N+1}\in\mathscr{A}_{\overline{d}}(\chi(\psi))\right)\nonumber \\
 & \hphantom{=}+\P\left[Q^{N+1}\in\mathscr{A}_{\overline{d}}^{c}(\chi(\psi))\right]\cdot I\left(X^{LM};\tilde{Y}^{LM}\mid Q^{N+1}\in\mathscr{A}_{\overline{d}}^{c}(\chi(\psi))\right)\\
 & \leq I\left(X^{LM};\tilde{Y}^{LM}\mid Q^{N+1}\in\mathscr{A}_{\overline{d}}(\chi(\psi))\right)+(1-\psi)ML\log|{\cal X}|,
\end{align}
since it holds for any $q^{N+1}\in\mathscr{Q}(M,N)$ that 
\[
I(X^{LM};\tilde{Y}^{LM}\mid Q^{N+1}=q^{N+1})\leq H(X^{LM})\leq ML\log|{\cal X}|.
\]
We next bound $I(X^{LM};\tilde{Y}^{LM}\mid Q^{N+1})$ for an arbitrary
$q^{N+1}\in\mathscr{A}_{\overline{d}}(\chi(\psi))$. By Lemma \ref{lem: Typical amplification set},
for any $q^{N+1}\in\mathscr{A}_{\overline{d}}(\chi(\psi))$ it holds
\[
\left|\frac{q_{d}}{M}-\pi_{\alpha}(d)\right|\leq O_{\overline{d},\psi}\left(\frac{1}{\sqrt{M}}\right)
\]
for all $d\in[\overline{d}]$. Thus, $\frac{q_{d}}{M}\leq\pi_{\alpha}(d)+O_{\overline{d},\psi}\left(\frac{1}{\sqrt{M}}\right)$
for $d\in[\overline{d}]$ and also 
\[
1-\sum_{d\in[\overline{d}]}\frac{q_{d}}{M}\leq1-\sum_{d\in[\overline{d}]}\pi_{\alpha}(d)+O_{\overline{d},\psi}\left(\frac{1}{\sqrt{M}}\right)\leq\eta+O_{\overline{d},\psi}\left(\frac{1}{\sqrt{M}}\right).
\]
So Prop. \ref{prop:converse genie aided conditional} implies that
\[
I(X^{LM};\tilde{Y}^{LM}\mid Q^{N+1}=q^{N+1})\leq\sum_{d\in[\overline{d}]}\pi_{\alpha}(d)\cdot J_{d}(P_{X},W,\beta)+\eta\cdot\log|{\cal X}|+o_{\overline{d},|{\cal X}|,\alpha}(1),
\]
where $J_{d}(P_{X},W,\beta)$ is as defined in (\ref{eq: definition of J for a channel, input distribution and beta}).
Taking $M\to\infty$ and then accordingly $\overline{d}\uparrow\infty$,
and $\eta\downarrow0$, $\psi\uparrow1$, proves the claim of the
proposition.
\end{IEEEproof}

\subsubsection{General Composition Molecules\label{subsec:General-composition-molecules}}

In the previous section, we have bounded the mutual information between
the input $X^{LM}$ and the output $\tilde{Y}^{LM}$, conditioned
on $Q^{N+1}$, and assuming that the input molecules all have the
same fixed composition. In this section, we remove the fixed composition
assumption.
\begin{prop}
\label{prop:converse genie aided avearge general composition}Assume
that the sequencing channel is such that $\nu_{\text{min}}(W)<\infty$.
Then,
\begin{equation}
\frac{1}{ML}I(X^{LM};\tilde{Y}^{LM})\leq\max_{P_{X}}\sum_{d\in\mathbb{N}}\pi_{\alpha}(d)\cdot J_{d}(P_{X},W,\beta)+o(1),\label{eq: upper bound on general composition average q capacity}
\end{equation}
where $J_{d}(P_{X},W,\beta)$ is as defined in (\ref{eq: definition of J for a channel, input distribution and beta}),
and $\pi_{\alpha}(d)\dfn\frac{\alpha^{d}e^{-\alpha}}{d!}$ Poisson
p.m.f. with parameter $\alpha$.
\end{prop}
\begin{IEEEproof}
By Prop. \ref{prop: reduction to constant composition}, it can be
assumed that there exists $\{P_{X,m}\}_{m\in[M]}\subseteq{\cal P}_{L}({\cal X})$
(with $L=\beta\log M$) such that $\mathscr{P}(x_{m}(j))=P_{X,m}$
for all $j\in[|{\cal C}_{M}|]$. That is, the composition is fixed
for each molecule index $m\in[M]$ over all codewords, but may change
from one index to the another. Let $T^{M}\in[|{\cal P}_{L}({\cal X})|]^{M}$
be such that $T_{m}=\mathscr{P}(x_{\tilde{U}_{m}}(j))$, that is $T_{m}$
states the molecule type $P_{X}\in{\cal P}_{L}({\cal X})$ from which
the outputs in $\tilde{Y}_{m}^{L}$ were sequenced. 

We bound the mutual information as follows: \textbf{
\begin{align}
I(X^{LM};\tilde{Y}^{LM}) & \leq I(X^{LM};\tilde{Y}^{LM},T^{M})\\
 & =I(X^{LM};T^{M})+I(X^{LM};\tilde{Y}^{LM}\mid T^{M})\\
 & \leq H(T^{M})+I(X^{LM};\tilde{Y}^{LM}\mid T^{M})\\
 & =o(ML)+I(X^{LM};\tilde{Y}^{LM}\mid T^{M}),\label{eq: first bound on mutual information for multiple types}
\end{align}
}where the last equality follows since $|{\cal P}_{L}({\cal X})|\leq(L+1)^{|{\cal X}|}$
and so 
\[
H(T^{M})\leq\sum_{m\in[M]}H(T_{m})\leq M\log\left[(L+1)^{|{\cal X}|}\right]=O\left(M\log(\log M)\right)=o(ML).
\]
We continue by bounding $I(X^{LM};\tilde{Y}^{LM}\mid T^{M})$. Recall
that in the proof of Prop. \ref{prop:converse genie aided conditional},
the assumption that the decoder knows $S^{M}$ allowed it to partition
$X^{LM}$ to subsets of molecules $X_{{\cal M}_{d}}$ which have been
sampled $d$ times (therein, we have used ${\cal M}_{d}=\{m\in[M]\colon S_{m}=d\}$),
and coupled them with the set of output clusters which contain $d$
molecules, namely $\tilde{Y}_{\overline{{\cal M}}_{d}}$. Then the
mutual information was upper bounded by $\sum_{d}I(X_{\overline{{\cal M}}_{d}};Y_{\overline{{\cal M}}_{d}})$.
The same strategy is taken here only with the molecule type replaces
the number of samples $d$. To this end, let $\overline{{\cal M}}_{P_{X}}=\{m\in[M]\colon\mathscr{P}(X_{\tilde{U}_{m}}^{L})=P_{X}\}$
where we recall that $\tilde{U}_{m}$ is such that $\tilde{Y}_{m}^{L}=Z_{\tilde{U}_{m}}^{L}=\{Y_{n}^{L}\colon U_{n}=m\}$.
We then partition $\tilde{Y}$ to sub-vectors $Y_{\overline{{\cal M}}_{P_{X}}}=(Y_{m}^{L})_{m\in{\cal \overline{M}}_{P_{X}}}$
for $P_{X}\in{\cal P}_{L}({\cal X})$. In addition, we set $X_{{\cal M}_{P_{X}}}$
to be the set of $X_{m}^{L}$ such that $\mathscr{P}(X_{m}^{L})=P_{X}$
(by the structural assumption on the codebook, the indices of this
set are the same for all codewords, and thus known to the decoder).
Then, as in (\ref{eq: upper bound of mutual information by partition inputs and outputs to set of those sampled d times})
\begin{equation}
I(X^{LM};\tilde{Y}^{LM}\mid T^{M})\leq\sum_{P_{X}\in{\cal P}_{L}({\cal X})}I(X_{{\cal M}_{P_{X}}};\tilde{Y}_{\overline{{\cal M}}_{P_{X}}}).\label{eq: mutual infomration bound conditioned on types}
\end{equation}
Let $\eta>0$ be given, and denote for brevity $\omega_{L}=|{\cal P}_{L}({\cal X})|$
and $M_{P_{X}}=|{\cal M}_{P_{X}}|$. We then split the possible types
${\cal P}_{L}({\cal X})$ to those with $\frac{M_{P_{X}}}{M}<\frac{\eta}{\omega_{L}}$
and their complement, and bound as in (\ref{eq: mutual infomration bound conditioned on types})
\begin{align}
I(X^{LM};\tilde{Y}^{LM}\mid T^{M}) & \leq\sum_{P_{X}\in{\cal P}_{L}({\cal X})\colon\frac{M_{P_{X}}}{M}<\frac{\eta}{\omega_{L}}}I(X_{{\cal M}_{P_{X}}};\tilde{Y}_{\overline{{\cal M}}_{P_{X}}})+\sum_{P_{X}\in{\cal P}_{L}({\cal X})\colon\frac{M_{P_{X}}}{M}\geq\frac{\eta}{\omega_{L}}}I(X_{{\cal M}_{P_{X}}};\tilde{Y}_{\overline{{\cal M}}_{P_{X}}})\\
 & \leq\eta ML\log|{\cal X}|+\sum_{P_{X}\in{\cal P}_{L}({\cal X})\colon\frac{M_{P_{X}}}{M}\geq\frac{\eta}{\omega_{L}}}I(X_{{\cal M}_{P_{X}}};\tilde{Y}_{\overline{{\cal M}}_{P_{X}}}),\label{eq: bound on mutual for many molecule types and less molecules types}
\end{align}
where the inequality follows from the (generous) bound $I(X_{{\cal M}_{P_{X}}};\tilde{Y}_{\overline{{\cal M}}_{P_{X}}})\leq M_{P_{X}}L\log|{\cal X}|$
and since the number of total terms in both sums is $\omega_{L}$.
We continue by bounding $I(X_{{\cal M}_{P_{X}}};\tilde{Y}_{\overline{{\cal M}}_{P_{X}}})$
for the case $\frac{M_{P_{X}}}{M}\geq\frac{\eta}{\omega_{L}}$. Let
$Q_{P_{X}}$ be the restriction of the amplification vector $Q^{N+1}$
to the molecules which have type $P_{X}$. Then, 
\begin{align}
I(X_{{\cal M}_{P_{X}}};\tilde{Y}_{\overline{{\cal M}}_{P_{X}}}) & \leq I(X_{{\cal M}_{P_{X}}};\tilde{Y}_{\overline{{\cal M}}_{P_{X}}},Q_{P_{X}})\\
 & =I(X_{{\cal M}_{P_{X}}};Q_{P_{X}})+I(X_{{\cal M}_{P_{X}}};\tilde{Y}_{\overline{{\cal M}}_{P_{X}}}\mid Q_{P_{X}})\\
 & \trre[=,a]I(X_{{\cal M}_{P_{X}}};\tilde{Y}_{\overline{{\cal M}}_{P_{X}}}\mid Q_{P_{X}})\\
 & \trre[\leq,b]M_{P_{X}}L\sum_{d\in\mathbb{N}}\pi_{\alpha}(d)\cdot J_{d}(P_{X},W,\beta)+o(ML)\\
 & \leq M_{P_{X}}L\cdot\max_{P_{X}\in{\cal P}({\cal X})}\sum_{d\in\mathbb{N}}\pi_{\alpha}(d)\cdot J_{d}(P_{X},W,\beta)+o(ML),\label{eq: bounds on MI for specfic type within many types}
\end{align}
where $(a)$ follows since $X_{{\cal M}_{P_{X}}}$ and $Q_{P_{X}}$
are independent, and $(b)$ follows from Prop. \ref{prop:converse genie aided avearge},
yet requires the following justification: Prop. (\ref{prop:converse genie aided avearge})
requires that the codewords have $\theta_{M}M$ molecules, each of
length $L=\beta\log M$. Here, we apply this proposition to the case
in which there are $M_{P_{X}}\geq\frac{\eta M}{\omega_{L}}\geq\frac{\eta M}{(L+1)^{|{\cal X}|}}=\Theta\left(\frac{M}{\log^{|{\cal X}|}M}\right)$
and so $\theta=\Theta\left(\frac{1}{\log^{|{\cal X}|}M}\right)$ which
decreases sub-polynomially, as required. It should be emphasized,
that that the asymptotic terms in Lemma \ref{lem: decoding permutation lemma}
-- which the proof Prop. (\ref{prop:converse genie aided avearge})
hinges on -- do not depend on the input distribution of the molecules
$P_{X}$. 

Inserting the bounds (\ref{eq: bounds on MI for specfic type within many types})
into (\ref{eq: bound on mutual for many molecule types and less molecules types})
and then to (\ref{eq: first bound on mutual information for multiple types}),
and taking $\eta\downarrow0$ leads to the claim.
\end{IEEEproof}

\section{Auxiliary Results and Proofs \label{sec:Auxiliary-Results}}

\subsection{Symmetry of Binomial Extensions of Symmetric DMCs \label{subsec:The-binomial-extension of symmetric DMC is symmetric}}

In this section, we explore when symmetry of a DMC $V\colon{\cal A}\to{\cal B}$
is preserved for its binomial extension $V^{\oplus d}\colon{\cal A}\to{\cal B}^{d}$
(Definition \ref{def: binomial channel}). To this end, it will be
convenient to identify a DMC $V\colon{\cal A}\to{\cal B}$ with a
matrix of $|{\cal A}|$ rows (for input letters) and $|{\cal B}|$
columns (for output letters). Thus, the $a$th row of $V$ contains
the values of $V(\cdot\mid a)$, and so sums to $1$. We will denote
the sub-matrix of $V$ of dimension $|{\cal A}|\times|{\cal B}_{0}|$
comprised of the columns pertaining to ${\cal B}_{0}$ by $V_{\mid{\cal B}_{0}}$.
We begin with the following definition:
\begin{defn}
A matrix $U$ is said to be \emph{doubly-permutation }if each row
of $U$ is a permutation of each other row, and each column of $U$
(if more than $1$) is a permutation of each other column.
\end{defn}
We recall the definition of a symmetric channel, in a strong sense,
and in Gallager's sense. 
\begin{defn}
\label{def: symmetric DMC}A DMC $V\colon{\cal A}\to{\cal B}$ is
said to be \emph{symmetric }if its channel matrix is doubly-permutation.
A DMC $V\colon{\cal A}\to{\cal B}$ is said to be \emph{symmetric
in Gallager's sense }\cite[p. 94]{gallager1968information} if there
exists a (disjoint) partition of ${\cal B}=\bigcup_{i}{\cal B}_{i}$
such that $V_{\mid{\cal B}_{i}}$ is doubly-permutation for all $i$.
A DMC is called \emph{modulo-additive} if ${\cal A}={\cal B}$ and
the rows of $V$ are cyclic shifts of one another. 
\end{defn}
As discussed in Sec. \ref{sec:Modulo-additive-sequencing}, modulo-additive
channels are obtained by the relation $B=A\oplus C$ where $C$ is
independent of the input $A$ and $\oplus$ is addition modulo $|{\cal A}|$
{[}see (\ref{eq: module additive channel}){]}. 

In this appendix we establish two results regarding the symmetry of
$V^{\oplus d}$. First, we consider the special case of modulo-additive
channels, and show:
\begin{prop}
\label{prop: binomial extension of modulo additive channels}Let $V\colon{\cal A}\to{\cal B}$
be a modulo-additive DMC. Then $V^{\oplus d}$ is symmetric in Gallager's
sense for all $d\in\mathbb{N}^{+}$. 
\end{prop}
For general symmetric channels, symmetry does not always hold for
$V^{\oplus d}$ (see counterexamples at the end of this section).
Nonetheless, we prove that, specifically, for DMCs with $4$ letter
input and output alphabets, which are especially relevant for the
DNA channel, such symmetry does hold: 
\begin{prop}
\label{prop: symmetry of binomial extension}Let $V\colon{\cal A}\to{\cal B}$
be a DMC and let $V^{\oplus d}$ be its $d$-order binomial extension.
Assume that $|{\cal A}|\leq4$ and that $|{\cal B}|\leq|{\cal A}|$.
If $V$ is symmetric in Gallager's sense then so is $V^{\oplus d}$
for all $d\in\mathbb{N}^{+}$. 
\end{prop}
We begin with the proof of Prop. \ref{prop: binomial extension of modulo additive channels}.
\begin{IEEEproof}[Proof of Prop. \ref{prop: binomial extension of modulo additive channels}]
 Assume w.l.o.g. that ${\cal A}=[A]$ where $A=|{\cal A}|$, and
let an arbitrary $d\in\mathbb{N}^{+}$ be given. We will partition
the output letters of ${\cal {\cal B}}^{d}=[A]^{d}$ to subsets $\{{\cal B}_{i}^{d}\}$
such that $V_{\mid{\cal B}_{i}^{d}}$ is doubly-permutation for all
$i$. Let $(a_{0},\ldots,a_{d-1})\in[A]^{d}$ be an arbitrary output
letter of $V^{\oplus d}$. We construct from the output letter $(a_{0},\ldots,a_{d-1})$
the subset of at most $d$ output letters 
\begin{equation}
{\cal B}_{0}^{d}=\left\{ \left(a_{0},\ldots,a_{d-1}\right),\left(a_{0}\oplus1,\ldots,a_{d-1}\oplus1\right),\ldots,\left(a_{0}\oplus(A-1),\ldots,a_{d-1}\oplus(A-1)\right)\right\} \label{eq: construction of subset for module additive}
\end{equation}
(if an output letter appears more than once in this construction then
we take just one instance of it). Next, we take an output letter in
$[A]^{d}\backslash{\cal B}_{0}^{d}$ and construct ${\cal B}_{1}^{d}$
in the same manner as in (\ref{eq: construction of subset for module additive}).
We continue constructing sets in this manner until $\bigcup_{i}\{{\cal B}_{i}^{d}\}=[A]^{d}$.
Furthermore, it is readily follows from the group property of the
set $[A]$ with the modulo addition $\oplus$ implies that ${\cal B}_{i_{1}}^{d}\cap{\cal B}_{i_{2}}^{d}$
is empty. Thus $\{{\cal B}_{i}^{d}\}$ is a partition of the output
alphabet $[A]^{d}$. Now, assume that ${\cal B}_{i}^{d}$ was generated,
as in (\ref{eq: construction of subset for module additive}), from
the output letter $(a_{0},\ldots,a_{d-1})$. Then, 
\begin{align}
V_{\mid{\cal B}_{i}^{d}} & =\left[\begin{array}{cccc}
\prod_{i=0}^{d-1}V(a_{i}\mid0) & \prod_{i=0}^{d-1}V(a_{i}\oplus1\mid0) & \cdots & \prod_{i=0}^{d-1}V(a_{i}\oplus(A-1)\mid0)\\
\prod_{i=0}^{d-1}V(a_{i}\mid1) & \prod_{i=0}^{d-1}V(a_{i}\oplus1\mid1) & \cdots & \prod_{i=0}^{d-1}V(a_{i}\oplus(A-1)\mid1)\\
\vdots & \ddots & \ddots & \vdots\\
\prod_{i=0}^{d-1}V(a_{i}\mid A-1) & \prod_{i=0}^{d-1}V(a_{i}\mid A-2) & \cdots & \prod_{i=0}^{d-1}V(a_{i}\oplus(A-1)\mid A-1)
\end{array}\right]\\
 & =\left[\begin{array}{cccc}
\prod_{i=0}^{d-1}V(a_{i}\mid0) & \prod_{i=0}^{d-1}V(a_{i}\mid A-1) & \cdots & \prod_{i=0}^{d-1}V(a_{i}\mid1)\\
\prod_{i=0}^{d-1}V(a_{i}\mid1) & \prod_{i=0}^{d-1}V(a_{i}\mid0) & \cdots & \prod_{i=0}^{d-1}V(a_{i}\mid2)\\
\vdots & \ddots & \ddots & \vdots\\
\prod_{i=0}^{d-1}V(a_{i}\mid A-1) & \prod_{i=0}^{d-1}V(a_{i}\mid A-2) & \cdots & \prod_{i=0}^{d-1}V(a_{i}\mid0)
\end{array}\right],
\end{align}
where the second equality follows since $V$ is a modulo-additive
channel. It is evident that $V_{\mid{\cal B}_{i}^{d}}$ is a circulant
matrix, and hence doubly-permutation. As this is true for any subset
${\cal B}_{i}^{d}$, it holds that $V^{\oplus d}$ is symmetric in
Gallager's sense. 
\end{IEEEproof}
We now turn to the proof of Prop. \ref{prop: symmetry of binomial extension}.
The proof essentially exhausts all possible DMCs which satisfy the
condition of the proposition. The first step is to identify all doubly-permutation
matrices of up to $4$ rows. In order to identify only matrices which
are essentially different, we classify matrices $U_{1}$ and $U_{2}$
as \emph{equivalent} if $U_{2}$ can be obtained from $U_{1}$ by
row permutations followed by column permutations, and denote this
relation by $U_{1}\equiv U_{2}$. We say that a doubly-permutation
matrix $U$ is an \emph{atom} if its columns cannot be partitioned
into subsets $\{{\cal B}_{i}\}$ such that $V_{\mid{\cal B}_{i}}$
are doubly-permutation for all $i$. 
\begin{claim}[Taxonomy of small doubly-permutation atoms]
\label{claim: atoms of size less than 4}Let $U$ be a doubly-permutation
matrix of $r$ rows and $c$ columns. Then, any doubly-permutation
matrix atom of $r=2$ is equivalent to either 
\[
U_{(2,1)}\dfn\left[\begin{array}{c}
p_{0}\\
p_{0}
\end{array}\right],U_{(2,2)}\dfn\left[\begin{array}{cc}
p_{0} & p_{1}\\
p_{1} & p_{0}
\end{array}\right]
\]
for some $p_{0},p_{1}$; of $r=3$ is equivalent to either one of
the following matrices
\[
U_{(3,1)}\dfn\left[\begin{array}{c}
p_{0}\\
p_{0}\\
p_{0}
\end{array}\right],U_{(3,3)}\dfn\left[\begin{array}{ccc}
p_{0} & p_{1} & p_{2}\\
p_{1} & p_{2} & p_{0}\\
p_{2} & p_{0} & p_{1}
\end{array}\right]
\]
for some $p_{0},p_{1},p_{2}$; and of $r=4$ is equivalent to either
one of the following matrices
\[
U_{(4,1)}\dfn\left[\begin{array}{c}
p_{0}\\
p_{0}\\
p_{0}\\
p_{0}
\end{array}\right],U_{(4,2)}\dfn\left[\begin{array}{cc}
p_{0} & p_{1}\\
p_{0} & p_{1}\\
p_{1} & p_{0}\\
p_{1} & p_{0}
\end{array}\right],
\]
\[
U_{(4,4)}^{(\text{A})}\dfn\left[\begin{array}{cccc}
p_{0} & p_{1} & p_{2} & p_{3}\\
p_{1} & p_{0} & p_{3} & p_{2}\\
p_{2} & p_{3} & p_{0} & p_{1}\\
p_{3} & p_{2} & p_{1} & p_{0}
\end{array}\right],U_{(4,4)}^{(\text{B})}\dfn\left[\begin{array}{cccc}
p_{0} & p_{1} & p_{2} & p_{3}\\
p_{1} & p_{0} & p_{3} & p_{2}\\
p_{2} & p_{3} & p_{1} & p_{0}\\
p_{3} & p_{2} & p_{0} & p_{1}
\end{array}\right],
\]
\[
U_{(4,4)}^{(\text{C})}\dfn\left[\begin{array}{cccc}
p_{0} & p_{1} & p_{2} & p_{3}\\
p_{1} & p_{2} & p_{3} & p_{0}\\
p_{2} & p_{3} & p_{0} & p_{1}\\
p_{3} & p_{0} & p_{1} & p_{2}
\end{array}\right],U_{(4,4)}^{(\text{D})}\dfn\left[\begin{array}{cccc}
p_{0} & p_{1} & p_{2} & p_{3}\\
p_{1} & p_{3} & p_{0} & p_{2}\\
p_{2} & p_{0} & p_{3} & p_{1}\\
p_{3} & p_{2} & p_{1} & p_{0}
\end{array}\right].
\]
for some $p_{0},p_{1},p_{2},p_{3}$.
\end{claim}
\begin{IEEEproof}
All these matrices can be found either an automated or a manual exhaustive
search procedure, in which all doubly-permutation atom $U$ will be
constructed. Few general properties may aid this search:
\begin{itemize}
\item If all rows of $U$ are equal then it is not an atom, and thus henceforth
we ignore these matrices.
\item If $c=1$ then the matrix $U=[p_{0},p_{0},\ldots]^{T}$ is doubly-permutation
and an atom for any $r\geq1$. Thus, we next only consider $c>1$. 
\item $c\leq r$ must hold. Indeed, assume by contradiction that $c>r$
then as we have assumed that all entries of the first row are unique,
at least one of these elements does not appear in the first column.
Since the columns are permutation of one another, that element does
not appear in any of the other columns too, a contradiction. 
\end{itemize}
In what follows, we will refer to the requirement that $U$ has rows
(respectively columns) which are permutations of one another as the
row-permutation \emph{property} (respectively column-permutation \emph{property}).
In addition, we will refer to the fact that permuting the rows (respectively
columns) leads to an equivalent matrix by row-permutation \emph{equivalence}
(respectively column-permutation \emph{equivalence}). If both row-
and column-permutation equivalence are used, we will refer to this
as permutation equivalence. 

We first consider the case in which the first row of $U$ has \emph{unique}
entries. Under this assumption, it further holds that:
\begin{itemize}
\item If $c=r$, it can be assumed by permutation equivalence that the first
row of $U$, say $(p_{0},p_{1},\ldots,p_{r-1})$ is the transpose
of its first column. Indeed, if this is not case, and the first column
of $U$ is not comprised of the values $\{p_{0},p_{1},\ldots,p_{r-1}\}$
-- one time each -- then there is some $p_{i}$ which does not appear
in the first column, and by the column-permutation property, not anywhere
in $U$, which is a contradiction. 
\end{itemize}
We now go over all possible dimensions:
\begin{casenv}
\item $r=2$. The only $2\times2$ atom is clearly $U_{(2,2)}$. 
\item $r=3$ and $c=2$. By the row-permutation property any doubly-permutation
matrix is equivalent to $\left[\begin{array}{cc}
p_{0} & p_{1}\\
p_{1} & p_{0}\\
p_{0} & p_{1}
\end{array}\right]$. By the column-permutation property $p_{0}=p_{1}$ must hold, and
so the resulting matrix is not an atom. 
\item $r=3$ and $c=3$. We may first set the first row of $U$ to $(p_{0},p_{1},p_{2})$
and the first column to its transpose. From this point, there is only
a single way to complete the matrix to assure the doubly-permutation
property, as in $U_{(3,3)}$. 
\item $r=4$ and $c=2$. By the row-permutation property, the rows of $U$
must be either $(p_{0},p_{1})$ or $(p_{1},p_{0})$. The only way
to satisfy the column-permutation property is by having two rows of
$(p_{0},p_{1})$ and two rows of $(p_{1},p_{0})$, as in $U_{(4,2)}$.
\item $r=4$ and $c=3$. Let the first row of $U$ be $(p_{0},p_{1},p_{2})$.
If we set $U(2,1)=p_{0}$, then we can complete the second row either
with $(p_{1},p_{2})$ or $(p_{2},p_{1})$. Assuming the former for
example, by the column-permutation property, we must complete $U$
as 
\[
\left[\begin{array}{ccc}
p_{0} & p_{1} & p_{2}\\
p_{0} & p_{1} & p_{2}\\
* & p_{0} & p_{0}\\
* & p_{0} & p_{0}
\end{array}\right],
\]
which cannot satisfy the row-permutation property. Thus we cannot
set $U(2,1)=p_{0}$, given the first row. Suppose that given the first
row, we set $U(2,1)\neq p_{0}$, which can be taken w.l.o.g as $U(2,1)=p_{1}$.
Then, we may complete the second row either with $(p_{0},p_{2})$
or $(p_{2},p_{0})$. Suppose that we set the former $(p_{0},p_{2})$,
then by the column-permutation property, we must complete $U$ as
\[
\left[\begin{array}{ccc}
p_{0} & p_{1} & p_{2}\\
p_{1} & p_{0} & p_{2}\\
p_{2} & p_{2} & *\\
p_{2} & p_{2} & *
\end{array}\right],
\]
which cannot satisfy the row-permutation property. Suppose that we
set the latter $(p_{2},p_{0})$. Then, by the column-permutation property
we can set $U(3,1)=p_{2}$. Then, by row-permutation property, we
must complete the third row in either of these two ways: 
\[
\left[\begin{array}{ccc}
p_{0} & p_{1} & p_{2}\\
p_{1} & p_{2} & p_{0}\\
p_{2} & p_{0} & p_{1}\\
* & * & *
\end{array}\right],\left[\begin{array}{ccc}
p_{0} & p_{1} & p_{2}\\
p_{1} & p_{2} & p_{0}\\
p_{2} & p_{1} & p_{0}\\
* & * & *
\end{array}\right].
\]
Both evidently cannot be completed to be doubly-permutation. Thus,
there is no $4\times3$ doubly-permutation matrix. 
\item $r=4$ and $c=4$. Let the first row of $U$ be $(p_{0},p_{1},p_{2},p_{3})$
and set its first column to its transpose. To satisfy the column-permutation
property, there are $3$ ways to complete the second row $(p_{0},p_{3},p_{2})$,
$(p_{2},p_{3},p_{0})$ $(p_{3},p_{0},p_{2})$ (otherwise, there is
a column with two identical values). Then, using the row- and column-permutation
properties, we may complete the matrix in either of the following
ways:
\[
\left[\begin{array}{cccc}
p_{0} & p_{1} & p_{2} & p_{3}\\
p_{1} & p_{0} & p_{3} & p_{2}\\
p_{2} & * & * & *\\
p_{3} & * & * & *
\end{array}\right]\to\left[\begin{array}{cccc}
p_{0} & p_{1} & p_{2} & p_{3}\\
p_{1} & p_{0} & p_{3} & p_{2}\\
p_{2} & p_{3} & * & *\\
p_{3} & p_{2} & * & *
\end{array}\right]\to U_{(4,4)}^{(\text{A})},U_{(4,4)}^{(\text{B})}.
\]
\[
\left[\begin{array}{cccc}
p_{0} & p_{1} & p_{2} & p_{3}\\
p_{1} & p_{2} & p_{3} & p_{0}\\
p_{2} & * & * & *\\
p_{3} & * & * & *
\end{array}\right]\to\left[\begin{array}{cccc}
p_{0} & p_{1} & p_{2} & p_{3}\\
p_{1} & p_{2} & p_{3} & p_{0}\\
p_{2} & p_{3} & * & *\\
p_{3} & p_{0} & * & *
\end{array}\right]\to\left[\begin{array}{cccc}
p_{0} & p_{1} & p_{2} & p_{3}\\
p_{1} & p_{2} & p_{3} & p_{0}\\
p_{2} & p_{3} & p_{0} & *\\
p_{3} & p_{0} & p_{1} & *
\end{array}\right]\to U_{(4,4)}^{(\text{C})}
\]
\[
\left[\begin{array}{cccc}
p_{0} & p_{1} & p_{2} & p_{3}\\
p_{1} & p_{3} & p_{0} & p_{2}\\
p_{2} & * & * & *\\
p_{3} & * & * & *
\end{array}\right]\to\left[\begin{array}{cccc}
p_{0} & p_{1} & p_{2} & p_{3}\\
p_{1} & p_{3} & p_{0} & p_{2}\\
p_{2} & p_{0} & * & *\\
p_{3} & p_{2} & * & *
\end{array}\right]\to\left[\begin{array}{cccc}
p_{0} & p_{1} & p_{2} & p_{3}\\
p_{1} & p_{3} & p_{0} & p_{2}\\
p_{2} & p_{0} & p_{3} & *\\
p_{3} & p_{2} & p_{1} & *
\end{array}\right]\to U_{(4,4)}^{(\text{D})}.
\]
\end{casenv}
This exhausts all possible cases with unique first row entries. Now
suppose the first row of $U$ does not have unique entries. If $c=2$
then this cannot lead to an atom, and so we consider only $c=3$ and
$c=4$.
\begin{casenv}
\item $r=3$ and $c=3$. By column-permutation equivalence, we may set the
first row of $U$ to $(p_{0},p_{0},p_{1})$ where $p_{0}\neq p_{1}$.
Since the first column must contain $p_{1}$ we may then set $U(2,1)=p_{1}$.
This leads to a matrix which is equivalent to $U_{(3,3)}$. 
\item $r=4$ and $c=3$. By column-permutation equivalence, we may first
set the first row of $U$ to $(p_{0},p_{0},p_{1})$ where $p_{0}\neq p_{1}$.
Since the first column must contain $p_{1}$ we may then set $U(2,1)=p_{1}$.
By the row-permutation property
\[
\left[\begin{array}{ccc}
p_{0} & p_{0} & p_{1}\\
p_{1} & * & *\\
* & * & *\\
* & * & *
\end{array}\right]\to\left[\begin{array}{ccc}
p_{0} & p_{0} & p_{1}\\
p_{1} & p_{0} & p_{0}\\
* & * & *\\
* & * & *
\end{array}\right].
\]
Suppose we next set $U(3,1)=p_{0}$. Then attempting to complete the
third row with $(p_{0},p_{1})$ violates the column-permutation property.
Completing the third row with $(p_{1},p_{0})$ leads to 
\[
\left[\begin{array}{ccc}
p_{0} & p_{0} & p_{1}\\
p_{1} & p_{0} & p_{0}\\
p_{0} & p_{1} & p_{0}\\
* & * & *
\end{array}\right],
\]
and attempting to complete the fourth row with any permutation of
$(p_{0},p_{0},p_{1})$ violates the column-permutation property. Thus,
$U(3,1)=p_{0}$ is impossible, and so suppose we set $U(3,1)=p_{1}$.
Then, by the row-permutation property, the third row is completed
with $(p_{0},p_{0})$ 
\[
\left[\begin{array}{ccc}
p_{0} & p_{0} & p_{1}\\
p_{1} & p_{0} & p_{0}\\
p_{1} & p_{0} & p_{0}\\
* & * & *
\end{array}\right],
\]
which violates the column-permutation property. Thus, there are no
atoms with $r=4$ and $c=3$, even with non-unique first row entries. 
\item $r=4$ and $c=4$. There are $3$ sub-cases:
\begin{casenv}
\item By column-permutation equivalence we may assume that the first row
is $(p_{0},p_{0},p_{0},p_{1})$ where $p_{0}\neq p_{1}$. We may set
$U(2,1)=p_{1}$ and then by row-permutation property, we must complete
the second row with $(p_{0},p_{0},p_{0})$. Then, setting $U(3,1)=p_{1}$
violates the column-permutation property, and so we set $U(3,1)=p_{0}$.
Then, there are two possible choices $U(3,2)=p_{0}$ and $U(3,2)=p_{1}$.
After completing the matrix while preserving column- and row- permutation
properties, the resulting matrix is equivalent to $U_{(4,4)}^{(\text{C})}$
. 
\item By column-permutation equivalence we may assume that the first row
is $(p_{0},p_{0},p_{1},p_{2})$ where $\{p_{0},p_{1},p_{2}\}$ are
unique. We may then set $U(2,1)=p_{1}$ and $U(3,1)=p_{2}$. Suppose
we set $U(2,2)=p_{0}$. By column- and row-permutation properties
\begin{align*}
\left[\begin{array}{cccc}
p_{0} & p_{0} & p_{1} & p_{2}\\
p_{1} & p_{0} & * & *\\
p_{2} & * & * & *\\
* & * & * & *
\end{array}\right] & \to\left[\begin{array}{cccc}
p_{0} & p_{0} & p_{1} & p_{2}\\
p_{1} & p_{0} & * & *\\
p_{2} & * & * & *\\
p_{0} & * & * & *
\end{array}\right]\to\left[\begin{array}{cccc}
p_{0} & p_{0} & p_{1} & p_{2}\\
p_{1} & p_{0} & p_{2} & p_{0}\\
p_{2} & * & * & *\\
p_{0} & * & * & *
\end{array}\right]\\
 & \to\left[\begin{array}{cccc}
p_{0} & p_{0} & p_{1} & p_{2}\\
p_{1} & p_{0} & p_{2} & p_{0}\\
p_{2} & p_{1} & * & *\\
p_{0} & p_{2} & * & *
\end{array}\right]\to\left[\begin{array}{cccc}
p_{0} & p_{0} & p_{1} & p_{2}\\
p_{1} & p_{0} & p_{2} & p_{0}\\
p_{2} & p_{1} & p_{0} & p_{0}\\
p_{0} & p_{2} & p_{0} & p_{1}
\end{array}\right],
\end{align*}
which is equivalent to $U'''_{(4,4)}.$ Next, suppose we set $U(2,2)=p_{2}$.
By the By row- and column-permutation properties 
\[
\left[\begin{array}{cccc}
p_{0} & p_{0} & p_{1} & p_{2}\\
p_{1} & p_{2} & * & *\\
p_{2} & * & * & *\\
* & * & * & *
\end{array}\right]\to\left[\begin{array}{cccc}
p_{0} & p_{0} & p_{1} & p_{2}\\
p_{1} & p_{2} & p_{0} & p_{0}\\
p_{2} & * & * & *\\
* & * & * & *
\end{array}\right].
\]
If we set $U(3,2)=p_{1}$ then this leads to 
\[
\left[\begin{array}{cccc}
p_{0} & p_{0} & p_{1} & p_{2}\\
p_{1} & p_{2} & p_{0} & p_{0}\\
p_{2} & p_{1} & p_{0} & p_{0}\\
* & * & * & *
\end{array}\right]\to\left[\begin{array}{cccc}
p_{0} & p_{0} & p_{1} & p_{2}\\
p_{1} & p_{2} & p_{0} & p_{0}\\
p_{2} & p_{1} & p_{0} & p_{0}\\
p_{0} & p_{0} & p_{2} & p_{1}
\end{array}\right],
\]
which is equivalent to $U'_{(4,4)}$. If we set $U(3,2)=p_{0}$ then
we may only complete the third row either with $(p_{0},p_{1})$ as
otherwise the column-permutation property is violated. This leads
to 
\[
\left[\begin{array}{cccc}
p_{0} & p_{0} & p_{1} & p_{2}\\
p_{1} & p_{2} & p_{0} & p_{0}\\
p_{2} & p_{0} & p_{0} & p_{1}\\
* & * & * & *
\end{array}\right]\to\left[\begin{array}{cccc}
p_{0} & p_{0} & p_{1} & p_{2}\\
p_{1} & p_{2} & p_{0} & p_{0}\\
p_{2} & p_{0} & p_{0} & p_{1}\\
p_{0} & p_{1} & p_{2} & p_{0}
\end{array}\right],
\]
which is equivalent to $U'''_{(4,4)}$. 
\item By column-permutation equivalence we may assume that the first row
is $(p_{0},p_{0},p_{1},p_{1})$ where $p_{0}\neq p_{1}$. We may then
set $U(2,1)=p_{1}$. Suppose we next set $U(2,2)=p_{0}$. Then we
complete the second row with either $(p_{0},p_{1})$ or $(p_{1},p_{0})$.
Assuming the former $(p_{0},p_{1})$, the row- and column- properties
lead to two possibilities
\[
\left[\begin{array}{cccc}
p_{0} & p_{0} & p_{1} & p_{1}\\
p_{1} & p_{0} & p_{0} & p_{1}\\
* & * & * & *\\
* & * & * & *
\end{array}\right]\to\left[\begin{array}{cccc}
p_{0} & p_{0} & p_{1} & p_{1}\\
p_{1} & p_{0} & p_{0} & p_{1}\\
* & p_{1} & * & p_{0}\\
* & p_{1} & * & p_{0}
\end{array}\right]\to\begin{cases}
\left[\begin{array}{cccc}
p_{0} & p_{0} & p_{1} & p_{1}\\
p_{1} & p_{0} & p_{0} & p_{1}\\
p_{1} & p_{1} & p_{0} & p_{0}\\
p_{0} & p_{1} & p_{1} & p_{0}
\end{array}\right]\\
\left[\begin{array}{cccc}
p_{0} & p_{0} & p_{1} & p_{1}\\
p_{1} & p_{0} & p_{0} & p_{1}\\
p_{0} & p_{1} & p_{1} & p_{0}\\
p_{1} & p_{1} & p_{0} & p_{0}
\end{array}\right]
\end{cases},
\]
which are both equivalent to $U'''_{(4,4)}$. Complete the second
row with $(p_{1},p_{0})$ leads to a similar result. Now, suppose
we set $U(2,2)=p_{1}$. By row-permutation property, we may complete
the second row with $(p_{0},p_{0})$. There are four ways to set $U(3,1)$
and $U(3,2)$. By symmetry, we may consider $(p_{0},p_{0})$ and $(p_{1},p_{0})$.
By row- and column- permutation properties, the first possibility
leads to 
\[
\left[\begin{array}{cccc}
p_{0} & p_{0} & p_{1} & p_{1}\\
p_{1} & p_{1} & p_{0} & p_{0}\\
p_{0} & p_{0} & * & *\\
* & * & * & *
\end{array}\right]\to\left[\begin{array}{cccc}
p_{0} & p_{0} & p_{1} & p_{1}\\
p_{1} & p_{1} & p_{0} & p_{0}\\
p_{0} & p_{0} & p_{1} & p_{1}\\
* & * & * & *
\end{array}\right]\to\left[\begin{array}{cccc}
p_{0} & p_{0} & p_{1} & p_{1}\\
p_{1} & p_{1} & p_{0} & p_{0}\\
p_{0} & p_{0} & p_{1} & p_{1}\\
p_{1} & p_{1} & p_{0} & p_{0}
\end{array}\right],
\]
which is not an atom (the first subset is the first and third columns
and the second one is its complement). Similarly, the second possibility
leads to 
\[
\left[\begin{array}{cccc}
p_{0} & p_{0} & p_{1} & p_{1}\\
p_{1} & p_{1} & p_{0} & p_{0}\\
p_{1} & p_{0} & * & *\\
* & * & * & *
\end{array}\right]\to\left[\begin{array}{cccc}
p_{0} & p_{0} & p_{1} & p_{1}\\
p_{1} & p_{1} & p_{0} & p_{0}\\
p_{1} & p_{0} & * & *\\
p_{0} & p_{1} & * & *
\end{array}\right]\to\begin{cases}
\left[\begin{array}{cccc}
p_{0} & p_{0} & p_{1} & p_{1}\\
p_{1} & p_{1} & p_{0} & p_{0}\\
p_{1} & p_{0} & p_{1} & p_{0}\\
p_{0} & p_{1} & p_{0} & p_{1}
\end{array}\right]\\
\left[\begin{array}{cccc}
p_{0} & p_{0} & p_{1} & p_{1}\\
p_{1} & p_{1} & p_{0} & p_{0}\\
p_{1} & p_{0} & p_{0} & p_{1}\\
p_{0} & p_{1} & p_{1} & p_{0}
\end{array}\right]
\end{cases},
\]
where the first is not an atom, and the second is equivalent to $U'''_{(4,4)}$. 
\end{casenv}
\end{casenv}
This exhausts all possible cases with non-unique first row entries,
and completes the proof. 
\end{IEEEproof}
Now, consider a DMC $V\colon{\cal A}\to{\cal B}$ which is symmetric
in Gallager's sense and for which $|{\cal A}|\leq4$ and $|{\cal B}|\leq|{\cal A}|$.
Claim \ref{claim: atoms of size less than 4} restricts the number
of such channels.
\begin{claim}
\label{claim:Gallager symmetric channels of input size less than 4}Let
$V\colon{\cal A}\to{\cal B}$ be a DMC with $|{\cal A}|\leq4$ and
$|{\cal B}|\leq|{\cal A}|$ which is symmetric in Gallager's sense,
and does not have equal rows.\footnote{If two rows are equal they can be combined into a single input letter,
thus effectively reduce $|{\cal A}|$. } Then, 
\begin{itemize}
\item If $|{\cal A}|=2$ then $V\equiv U_{(2,2)}$. 
\item If $|{\cal A}|=3$ then $V\equiv U_{(3,3)}$. 
\item If $|{\cal A}|=4$ then $V$ is equivalent to either one of $U_{(4,4)}^{(\text{A})},U_{(4,4)}^{(\text{B})},U_{(4,4)}^{(\text{C})},U_{(4,4)}^{(\text{D})}$
or to 
\[
U_{(4,4)}^{(\text{E})}\dfn\left[\begin{array}{cccc}
p_{0} & p_{1} & p_{2} & p_{3}\\
p_{0} & p_{1} & p_{3} & p_{2}\\
p_{1} & p_{0} & p_{2} & p_{3}\\
p_{1} & p_{0} & p_{3} & p_{2}
\end{array}\right].
\]
\end{itemize}
\end{claim}
\begin{IEEEproof}
The case $|{\cal B}|=1$, and the partition of ${\cal B}$ to $|{\cal B}|$
subsets of size $1$ leads to identical rows (and zero capacity) and
thus can be ignored. Then, the only cases for $|{\cal A}|=2$ (respectively
$|{\cal A}|=3$) are $|{\cal B}|=2$ (respectively $|{\cal B}|=3)$
and the claim follows directly from Claim \ref{claim: atoms of size less than 4},
since there are no non-trivial ways to partition ${\cal B}$. For
$|{\cal A}|=4$, we may consider either $|{\cal B}|=2$, $|{\cal B}|=3$
and $|{\cal B}|=4$. For $|{\cal B}|=2$, $V\equiv U_{(4,2)}$ which
leads to identical rows, and thus ignored. For $|{\cal B}|=3$ the
only symmetric channel in Gallager's sense is 
\[
\left[U_{(4,1)},U_{(4,2)}\right]=\left[\begin{array}{ccc}
p_{0} & p_{1} & p_{2}\\
p_{0} & p_{1} & p_{2}\\
p_{0} & p_{2} & p_{1}\\
p_{0} & p_{2} & p_{1}
\end{array}\right],
\]
which again has identical rows, and thus can be ignored (it effectively
creates a channel of size $2\times3$ for which we make no claims).
For $|{\cal B}|=4$ we may partition the output alphabet to subsets
of sizes $(1,1,1,1),(2,1,1),(2,2)$ and $(4,0)$. In the case of no
partition $(4,0)$, Claim \ref{claim: atoms of size less than 4}
directly implies that $V$ is equivalent to either $U_{(4,4)}^{(\text{A})}$,
$U_{(4,4)}^{(\text{B})},$ $U_{(4,4)}^{(\text{C})},$ or $U_{(4,4)}^{(\text{D})}$.
The case $(1,1,1,1)$ can be ignored, as discussed above. Next, the
case $(2,1,1)$ leads to identical rows and thus can be ignored. It
remains to consider the case $(2,2)$. In this case $V$ is equivalent
to 
\[
\left[U_{(4,2)},\Pi[U_{(4,2)}]\right]
\]
for some row permutation $\Pi$. All such channels with non-equal
rows are equivalent to $U_{(4,4)}^{(\text{E})}$. 
\end{IEEEproof}

We now turn to binomial extension of channels. In principle, the output
alphabet of the binomial channel $V^{\oplus d}$ is ${\cal B}^{d}$.
However, since $V^{\oplus d}(b^{d}\mid a)$ only depends on $a$ and
the composition of $b^{d}$, we may assume that the output alphabet
of $V^{\oplus d}$ is the set of all compositions in ${\cal B}^{d}$.
Specifically, we change the output alphabet to 
\[
\tilde{{\cal B}}^{(d)}\dfn\left\{ \delta^{|{\cal B}|}\in[d+1]^{|{\cal B}|}\colon\delta_{b}\in\mathbb{N}^{+},\;\sum_{b\in{\cal B}}\delta_{b}=d\right\} ,
\]
where with this representation $\delta^{|{\cal B}|}=\mathscr{N}(b^{d})$,
that is $\delta_{b}$ is the number of times $b\in{\cal B}$ appears
in the composition, such that the probability of this output alphabet
under $a$ is 
\[
V^{\oplus d}(\delta^{|{\cal B}|}\mid a)={d \choose \delta_{0},\delta_{1},\ldots,\delta_{|{\cal B}|-1}}\cdot\prod_{b\in{\cal B}}V^{\delta_{b}}(b\mid a).
\]

\begin{IEEEproof}[Proof of Prop. \ref{prop: symmetry of binomial extension}]
By Claim \ref{claim:Gallager symmetric channels of input size less than 4},
it suffices to verify the statement of Prop. \ref{prop: symmetry of binomial extension}
for the symmetric channels $U_{(2,2)}$, $U_{(3,3)}$ and $U_{(4,4)}^{(\text{A})},U_{(4,4)}^{(\text{B})},U_{(4,4)}^{(\text{C})},U_{(4,4)}^{(\text{D})},U_{(4,4)}^{(E)}$,
while assuming $p_{0},p_{1},p_{2},p_{3}$ are all unique. In fact,
$U_{(2,2)}$ (BSC), $U_{(3,3)}$ and $U_{(4,4)}^{(\text{C})}$ are
modulo-additive channels, for which optimality of uniform input holds
for general size input (output) alphabet in Sec. \ref{sec:Modulo-additive-sequencing}.

To show that $V^{\oplus d}$ is symmetric in Gallager's sense, we
need to prove that $\tilde{{\cal B}}^{(d)}$ can be partitioned into
subsets, such that the columns and rows of the channel transition
matrix restricted to each of the subsets is doubly-permutation.
\begin{casenv}
\item $V\equiv U_{(2,2)}$. Considering $\{(\delta_{0},\delta_{1}),(\delta_{1},\delta_{0})\}\in\tilde{{\cal B}}^{(d)}$
we observe that 
\[
U_{(2,2)\mid\{(\delta_{0},\delta_{1}),(\delta_{1},\delta_{0})\}}={d \choose \delta_{0},\delta_{1}}\cdot\left[\begin{array}{cc}
p_{0}^{\delta_{0}}p_{1}^{\delta_{1}} & p_{0}^{\delta_{1}}p_{1}^{\delta_{0}}\\
p^{\delta_{0}} & p_{0}^{\delta_{1}}p_{1}^{\delta_{0}}
\end{array}\right],
\]
which is doubly-permutation. Note that this subset may be degenerated,
that is, only contain a single element $\delta_{0}=\delta_{1}=d/2$,
but the claim holds for this case too. This is true for all next cases
too. 
\item $V\equiv U_{(3,3)}$. Considering the subsets $\{(\delta_{0},\delta_{1},\delta_{2}),(\delta_{1},\delta_{2},\delta_{0}),(\delta_{2},\delta_{0},\delta_{1})\}\in\tilde{{\cal B}}^{(d)}$
we observe that 
\begin{multline*}
U_{(3,3)\mid\{(\delta_{0},\delta_{1},\delta_{2}),(\delta_{1},\delta_{2},\delta_{0}),(\delta_{2},\delta_{0},\delta_{1})\}}={d \choose \delta_{0},\delta_{1},\delta_{2}}\times\\
\left[\begin{array}{ccc}
p_{0}^{\delta_{0}}\cdot p_{1}^{\delta_{1}}\cdot p_{2}^{\delta_{2}}\quad & p_{0}^{\delta_{2}}\cdot p_{1}^{\delta_{0}}\cdot p_{2}^{\delta_{1}}\quad & p_{2}^{\delta_{0}}\cdot p_{0}^{\delta_{1}}\cdot p_{1}^{\delta_{2}}\\
p_{1}^{\delta_{0}}\cdot p_{2}^{\delta_{1}}\cdot p_{0}^{\delta_{2}}\quad & p_{1}^{\delta_{2}}\cdot p_{2}^{\delta_{0}}\cdot p_{0}^{\delta_{1}}\quad & p_{2}^{\delta_{0}}\cdot p_{0}^{\delta_{1}}\cdot p_{1}^{\delta_{2}}\\
p_{2}^{\delta_{0}}\cdot p_{0}^{\delta_{1}}\cdot p_{1}^{\delta_{2}}\quad & p_{2}^{\delta_{2}}\cdot p_{0}^{\delta_{0}}\cdot p_{1}^{\delta_{1}}\quad & p_{2}^{\delta_{0}}\cdot p_{0}^{\delta_{1}}\cdot p_{1}^{\delta_{2}}
\end{array}\right],
\end{multline*}
which is doubly-permutation. 
\item $V\equiv U_{(4,4)}^{(\text{A})}$. Considering the subsets
\[
\left\{ (\delta_{0},\delta_{1},\delta_{2},\delta_{3}),(\delta_{1},\delta_{0},\delta_{3},\delta_{2}),(\delta_{2},\delta_{3},\delta_{0},\delta_{1}),(\delta_{3},\delta_{2},\delta_{1},\delta_{0})\right\} 
\]
(the vectors are isomorphic to the rows of $U_{(4,4)}^{(\text{A})}$),
we observe that
\[
U_{(4,4)}^{(\text{A})}=\left[\begin{array}{cccc}
p_{0} & p_{1} & p_{2} & p_{3}\\
p_{1} & p_{0} & p_{3} & p_{2}\\
p_{2} & p_{3} & p_{0} & p_{1}\\
p_{3} & p_{2} & p_{1} & p_{0}
\end{array}\right],
\]
and that
\begin{multline*}
U_{(4,4)\mid(\delta_{0},\delta_{1},\delta_{2},\delta_{3}),(\delta_{1},\delta_{1},\delta_{3},\delta_{2}),(\delta_{2},\delta_{3},\delta_{0},\delta_{1}),(\delta_{3},\delta_{2},\delta_{1},\delta_{1})}^{(\text{A})}={d \choose \delta_{0},\delta_{1},\delta_{2},\delta_{3}}\times\\
\left[\begin{array}{cccc}
p_{0}^{\delta_{0}}\cdot p_{1}^{\delta_{1}}\cdot p_{2}^{\delta_{2}}\cdot p_{3}^{\delta_{3}}\quad & p_{0}^{\delta_{1}}\cdot p_{1}^{\delta_{0}}\cdot p_{2}^{\delta_{3}}\cdot p_{3}^{\delta_{2}}\quad & p_{0}^{\delta_{2}}\cdot p_{1}^{\delta_{3}}\cdot p_{2}^{\delta_{0}}\cdot p_{3}^{\delta_{1}}\quad & p_{0}^{\delta_{3}}\cdot p_{1}^{\delta_{2}}\cdot p_{2}^{\delta_{1}}\cdot p_{3}^{\delta_{0}}\\
p_{1}^{\delta_{0}}\cdot p_{0}^{\delta_{1}}\cdot p_{3}^{\delta_{2}}\cdot p_{2}^{\delta_{3}}\quad & p_{1}^{\delta_{1}}\cdot p_{0}^{\delta_{0}}\cdot p_{3}^{\delta_{3}}\cdot p_{2}^{\delta_{2}}\quad & p_{1}^{\delta_{2}}\cdot p_{0}^{\delta_{3}}\cdot p_{3}^{\delta_{0}}\cdot p_{2}^{\delta_{1}}\quad & p_{1}^{\delta_{3}}\cdot p_{0}^{\delta_{2}}\cdot p_{3}^{\delta_{1}}\cdot p_{2}^{\delta_{0}}\\
p_{2}^{\delta_{0}}\cdot p_{3}^{\delta_{1}}\cdot p_{0}^{\delta_{2}}\cdot p_{1}^{\delta_{3}}\quad & p_{2}^{\delta_{1}}\cdot p_{3}^{\delta_{0}}\cdot p_{0}^{\delta_{3}}\cdot p_{1}^{\delta_{2}}\quad & p_{2}^{\delta_{2}}\cdot p_{3}^{\delta_{3}}\cdot p_{0}^{\delta_{0}}\cdot p_{1}^{\delta_{1}}\quad & p_{2}^{\delta_{3}}\cdot p_{3}^{\delta_{2}}\cdot p_{0}^{\delta_{1}}\cdot p_{1}^{\delta_{0}}\\
p_{3}^{\delta_{0}}\cdot p_{2}^{\delta_{1}}\cdot p_{1}^{\delta_{2}}\cdot p_{0}^{\delta_{3}}\quad & p_{3}^{\delta_{1}}\cdot p_{2}^{\delta_{0}}\cdot p_{1}^{\delta_{3}}\cdot p_{0}^{\delta_{2}}\quad & p_{3}^{\delta_{2}}\cdot p_{2}^{\delta_{3}}\cdot p_{1}^{\delta_{0}}\cdot p_{0}^{\delta_{1}}\quad & p_{3}^{\delta_{3}}\cdot p_{2}^{\delta_{2}}\cdot p_{1}^{\delta_{1}}\cdot p_{0}^{\delta_{0}}
\end{array}\right],
\end{multline*}
which is doubly-permutation, and equivalent to $U_{(4,4)}^{(\text{A})}$
\item $V\equiv U_{(4,4)}^{(\text{B})}$ and $V\equiv U_{(4,4)}^{(\text{C})}$
$V\equiv U_{(4,4)}^{(\text{D})}$ and $V\equiv U_{(4,4)}^{(\text{E})}$
can be verified similarly to the previous case. 
\end{casenv}
\end{IEEEproof}

\paragraph*{Counterexamples}

We conclude this section we two counterexamples. The channel 
\[
W_{1}=\frac{1}{15}\cdot\left[\begin{array}{ccccc}
1 & 2 & 3 & 4 & 5\\
4 & 3 & 2 & 5 & 1\\
2 & 5 & 1 & 3 & 4\\
3 & 4 & 5 & 1 & 2\\
5 & 1 & 4 & 2 & 3
\end{array}\right]
\]
is symmetric, yet $W_{1}^{\oplus2}$ is not symmetric, not even in
Gallager's sense. The channel\footnote{The channel matrix is a juxtaposition of two symmetric channels of
sizes $|{\cal X}|=|{\cal Y}|=4$.} 
\[
W_{2}=\frac{1}{20}\cdot\left[\begin{array}{cccccccc}
1 & 2 & 3 & 4 & 1 & 2 & 3 & 4\\
2 & 1 & 4 & 3 & 2 & 1 & 4 & 3\\
3 & 4 & 1 & 2 & 3 & 4 & 2 & 1\\
4 & 3 & 2 & 1 & 4 & 3 & 1 & 2
\end{array}\right]
\]
is symmetric in Gallager's sense, yet $W_{2}^{\oplus2}$ is not symmetric
in Gallager's sense. Furthermore, the capacity-achieving input distribution
of both $W_{1}^{\oplus2}$ and $W_{2}^{\oplus2}$ is not uniform (though
very close to being uniform, with deviation on the order of $10^{-3}$). 

\subsection{Proof of Proposition \ref{prop: Critical beta for modulo additive channels}
\label{subsec:Proof-of-Proposition}}
\begin{IEEEproof}[Proof of Prop. \ref{prop: Critical beta for modulo additive channels}]
We first loosen the upper bound of Theorem \ref{thm: converse capacity}
by upper bounding 
\begin{align}
 & I(P_{X},W^{\oplus d})+\Omega_{d}(\beta,P_{X},W)\nonumber \\
 & =\begin{cases}
I(P_{X},W^{\oplus d})+\frac{1}{\beta}, & \mathsf{CID}(P_{X},W^{\oplus d})<\frac{1}{\beta}\\
I(P_{X},W^{\oplus d})-\mathsf{CID}(P_{X},W^{\oplus d})+\frac{2}{\beta}, & \frac{1}{\beta}\leq\mathsf{CID}(P_{X},W^{\oplus d})<\frac{2}{\beta}\\
I(P_{X},W^{\oplus d}), & \mathsf{CID}(P_{X},W^{\oplus d})\ge\frac{2}{\beta}
\end{cases}\\
 & =\left[\left(I(P_{X},W^{\oplus d})+\frac{1}{\beta}\right)\wedge\left((P_{X},W^{\oplus d})-\mathsf{CID}(P_{X},W^{\oplus d})+\frac{2}{\beta}\right)\right]\vee\left[I(P_{X},W^{\oplus d})\right]\\
 & \trre[=,a]\left[I(P_{X},W^{\oplus d})+\frac{1}{\beta}\right]\wedge\left[\left(I(P_{X},W^{\oplus d})-\mathsf{CID}(P_{X},W^{\oplus d})+\frac{2}{\beta}\right)\vee\left(I(P_{X},W^{\oplus d})\right)\right]\\
 & \leq\left(I(P_{X},W^{\oplus d})-\mathsf{CID}(P_{X},W^{\oplus d})+\frac{2}{\beta}\right)\vee I(P_{X},W^{\oplus d}),\label{eq: loosening of Omega}
\end{align}
where $(a)$ follows from the distributive law for minima and maxima
$(c_{1}\vee c_{2})\wedge c_{3}=(c_{1}\vee c_{3})\wedge(c_{2}\vee c_{3})$
for $c_{1},c_{2},c_{3}\in\mathbb{R}$ and since $1/\beta>0$. 

We next show that for any $d\in\mathbb{N}^{+}$, each of the two terms
in the maximization of (\ref{eq: loosening of Omega}) is individually
maximized by the uniform distribution $P_{X}^{(\text{unif})}$. First,
$I(P_{X},W^{\oplus d})$ is the mutual information of a symmetric
channel (Prop.\textbf{ }\ref{prop: binomial extension of modulo additive channels}),
and so clearly maximized by the uniform distribution $P_{X}^{(\text{unif})}$.
Second, we let $(X,Y,\overline{Y})\in{\cal X}\times{\cal Y}^{2}$
be such that 
\[
\P[X=x,Y^{d}=y^{d},\overline{Y}^{d}=\overline{y}^{d}]=P_{X}(x)\cdot W^{\oplus d}(y^{d}\mid x)W^{\oplus d}(\overline{y}^{d}\mid x)
\]
for some $d\in\mathbb{N}^{+}$. Then,
\begin{align}
I(P_{X},W^{\oplus d})-\mathsf{CID}(P_{X},W^{\oplus d}) & =I(X,Y^{d})-2I(X;Y^{d})+I(X;Y^{d},\overline{Y}^{d})\\
 & =I(X;Y^{d},\overline{Y}^{d})-I(X;Y^{d})\\
 & =I(X;\overline{Y}^{d}\mid Y^{d})\\
 & =H(\overline{Y}^{d}\mid Y^{d})-H(\overline{Y}^{d}\mid X,Y^{d})\\
 & \trre[=,a]H(\overline{Y}^{d}\mid Y^{d})-H(\overline{Y}^{d}\mid X)\\
 & \trre[=,b]H(\overline{Y}^{d}\mid Y^{d})-H(\overline{Y}^{d}\mid X=x),
\end{align}
where $(a)$ follows from Markovity, and $(b)$ holds for any $x\in{\cal X}$
and follows from the fact that $W$ is a modulo-additive channel,
for which the conditional output entropy $H(\overline{Y}^{d}\mid X=x)$
does not depend on $x\in{\cal X}$. Indeed, if the noise p.m.f. in
the modulo-additive channel is $\P[Z=z]=w_{z}$ for $z\in{\cal X}$,
then the multiset of ${\cal X}^{d}$ possible values of $\P[Y^{d}=y^{d}\mid X=x]$
is given by
\[
\left\{ \prod_{i=0}^{d-1}w_{(y_{i}-x)}\right\} _{y^{d}\in{\cal X}^{d}}=\left\{ \prod_{i=0}^{d-1}w_{z_{i}}\right\} _{z^{d}\in{\cal X}^{d}}
\]
for any $x\in{\cal X}$. Thus, the $P_{X}$-maximizer of $I(P_{X},W^{\oplus d})-\mathsf{CID}(P_{X},W^{\oplus d})+\frac{2}{\beta}$
over $P_{X}$ is the maximizer of $H(\overline{Y}^{d}\mid Y^{d})$.
We next show that $H(\overline{Y}^{d}\mid Y^{d})$ is maximized by
the uniform distribution. To this end, we let $G(\omega,t)\dfn t\cdot H(\frac{\omega}{t})$
be the \emph{perspective function }of the entropy function, defined
for $(\omega,t)\in[0,1]^{\infty}\times\mathbb{R}^{+}$ such that $\frac{1}{t}\sum\omega_{i}=1$
and $\omega_{i}/t\geq0$. Then, 
\begin{align}
H(\overline{Y}^{d}\mid Y^{d}) & =\sum_{y^{d}\in{\cal X}^{d}}P_{Y^{d}}(y^{d})\cdot H\left(P_{\overline{Y}^{d}\mid Y^{d}}(\cdot\mid y^{d})\right)\\
 & =\sum_{y^{d}\in{\cal X}^{d}}P_{Y^{d}}(y^{d})\cdot H\left(\frac{P_{Y^{d}\overline{Y}^{d}}(y^{d},\cdot)}{P_{Y^{d}}(y^{d})}\right)\\
 & =\sum_{y^{d}\in{\cal X}^{d}}P_{Y^{d}}(y^{d})\cdot H\left(\frac{\sum_{x\in{\cal X}}P_{XY^{d}\overline{Y}^{d}}(x,y^{d},\cdot)}{P_{Y^{d}}(y^{d})}\right)\\
 & =\sum_{y^{d}\in{\cal X}^{d}}P_{Y^{d}}(y^{d})\cdot H\left(\frac{\sum_{x\in{\cal X}}P_{X}(x)\cdot P_{Y^{d}\overline{Y}^{d}\mid X}(\cdot,y^{d}\mid x)}{P_{Y^{d}}(y^{d})}\right)\\
 & \trre[=,a]\sum_{y^{d}\in{\cal X}^{d}}P_{Y^{d}}(y^{d})\cdot H\left(\frac{\sum_{x\in{\cal X}}P_{X}(x)\cdot P_{Y^{d}\mid X}(y^{d}\mid x)P_{\overline{Y}^{d}\mid X}(\cdot,\mid x)}{P_{Y^{d}}(y^{d})}\right)\\
 & =\sum_{y^{d}\in{\cal X}^{d}}\left(\sum_{x\in{\cal X}}P_{X}(x)\cdot P_{Y^{d}\mid X}(y^{d}\mid x)\right)\cdot H\left(\frac{\sum_{x\in{\cal X}}P_{X}(x)\cdot P_{Y^{d}\mid X}(y^{d}\mid x)P_{\overline{Y}^{d}\mid X}(\cdot,\mid x)}{\sum_{x\in{\cal X}}P_{X}(x)\cdot P_{Y^{d}\mid X}(y^{d}\mid x)}\right)\\
 & =\sum_{y^{d}\in{\cal X}^{d}}G\left(\sum_{x\in{\cal X}}P_{X}(x)\cdot P_{Y^{d}\mid X}(y^{d}\mid x)P_{\overline{Y}^{d}\mid X}(\cdot,\mid x),\;\sum_{x\in{\cal X}}P_{X}(x)\cdot P_{Y^{d}\mid X}(y^{d}\mid x)\right),\label{eq: conditional entropy as perspective function}
\end{align}
where $(a)$ follows from Markovity. Now since the entropy function
is concave, then so is $(\omega,t)\to G(\omega,t)$ \cite[Sec. 3.2.6]{boyd2004convex}.
Furthermore, both the arguments of $G$ in (\ref{eq: conditional entropy as perspective function}),
to wit,
\[
\sum_{x\in{\cal X}}P_{X}(x)\cdot P_{Y^{d}\mid X}(y^{d}\mid x)P_{\overline{Y}^{d}\mid X}(\cdot,\mid x)
\]
and 
\[
\sum_{x\in{\cal X}}P_{X}(x)\cdot P_{Y^{d}\mid X}(y^{d}\mid x)
\]
are affine mappings of $\{P_{X}(x)\}_{x\in{\cal X}}$, and composition
with an affine mapping preserves concavity \cite[Sec. 3.2.2]{boyd2004convex}.
Thus, $H(\overline{Y}^{d}\mid Y^{d})$ is a concave function of $P_{X}$.
For any $z\in{\cal X}$, let $P_{X}^{(z)}$ be the distribution defined
via 
\[
P_{X}^{(z)}(x)=P_{X}(x+z).
\]
Due to the symmetry in the modulo-additive channel, the entropy $H(\overline{Y}^{d}\mid Y^{d})$
is the same for any choice $P_{X}^{(z)}(x)$ for any $z\in{\cal X}$.\textbf{
}By concavity of $H(\overline{Y}^{d}\mid Y^{d})$ in $P_{X}$, the
distribution $\overline{P}_{X}=\frac{1}{|{\cal X}|}\sum_{z\in{\cal X}}P_{X}^{(z)}$
has larger $H(\overline{Y}^{d}\mid Y^{d})$ than any for $P_{X}^{(z)}(x)$.
However, $\overline{P}_{X}$ is clearly the uniform distribution over
${\cal X}$. 

We thus have shown that both terms in the maximization of (\ref{eq: loosening of Omega})
are maximized by the uniform distribution, for all $d\in\mathbb{N}^{+}$.
Now, if $-\mathsf{CID}(P_{X}^{(\text{unif})},W^{\oplus d})+\frac{2}{\beta}<0$
for all $d\in\mathbb{N}^{+}$ then the resulting (loosened) upper
bound on the capacity, obtained by replacing $I(P_{X},W^{\oplus d})+\Omega_{d}(\beta,P_{X},W)$
with their upper bound in (\ref{eq: loosening of Omega}), is given
by

\begin{align}
 & \max_{P_{X}\in{\cal P}({\cal X})}\sum_{d\in\mathbb{N}^{+}}\pi_{\alpha}(d)\cdot\left[\left(I(P_{X},W^{\oplus d})-\mathsf{CID}(P_{X},W^{\oplus d})+\frac{2}{\beta}\right)\vee\left(I(P_{X},W^{\oplus d})\right)\right]-\frac{1}{\beta}\left(1-\pi_{\alpha}(0)\right)\nonumber \\
 & \leq\sum_{d\in\mathbb{N}^{+}}\pi_{\alpha}(d)\cdot\left[\max_{P_{X}\in{\cal P}({\cal X})}\left(I(P_{X},W^{\oplus d})-\mathsf{CID}(P_{X},W^{\oplus d})+\frac{2}{\beta}\right)\vee\left(\max_{P_{X}\in{\cal P}({\cal X})}I(P_{X},W^{\oplus d})\right)\right]-\frac{1}{\beta}\left(1-\pi_{\alpha}(0)\right)\\
 & =\sum_{d\in\mathbb{N}^{+}}\pi_{\alpha}(d)\cdot\left[\left(I(P_{X}^{(\text{unif})},W^{\oplus d})-\mathsf{CID}(P_{X}^{(\text{unif})},W^{\oplus d})+\frac{2}{\beta}\right)\vee I(P_{X}^{(\text{unif})},W^{\oplus d})\right]-\frac{1}{\beta}\left(1-\pi_{\alpha}(0)\right)\\
 & =\sum_{d\in\mathbb{N}^{+}}\pi_{\alpha}(d)\cdot I(P_{X}^{(\text{unif})},W^{\oplus d})-\frac{1}{\beta}\left(1-\pi_{\alpha}(0)\right),
\end{align}
which matches the lower bound. Since $\mathsf{CID}(P_{X}^{(\text{unif})},W^{\oplus d})$
is monotonic increasing (Corollary \ref{cor: common input deficit for binomial channels})
then this condition is satisfied if $\beta\geq2/\mathsf{CID}(P_{X}^{(\text{unif})},W)$,
as claimed. 
\end{IEEEproof}

\subsection{Mutual Information of $K$ DMC Uses with i.i.d. v.s. Fixed Composition
Inputs \label{subsec:Mutual-information-of-DMC with i.i.d. vs fixed composition}}

In this appendix, we consider $K$ independent uses of a DMC $V\colon{\cal A}\to{\cal B}$,
so the input is $A^{K}$ and the output is $B^{K}\sim V^{K}(\cdot\mid A^{K})$.
In this channel, the maximal mutual information is obtained by choosing
the input $A_{k}$ to be i.i.d.. For the next lemma, we compare this
maximal mutual information to that obtained by the $K$-dimensional
mutual information obtained for $A^{K}\sim\text{Uniform}[{\cal T}_{K}(P_{A})]$,
that is, a uniform distribution over the type class ${\cal T}_{K}(P_{A})$
(where we insist $P_{A}\in{\cal {\cal P}}_{K}({\cal A})$). The next
lemma shows that the difference in mutual information is $o(K)$. 
\begin{lem}
\label{lem: mutual information difference between iid and fixed composition inputs}Let
a DMC $V\colon{\cal A}\to{\cal B}$, $K\in\mathbb{N}^{+}$ and $P_{A}\in{\cal P}_{K}({\cal A})$
be given. Let $A^{K}$ be distributed i.i.d. so that $A_{k}\sim P_{A}$
and let $\tilde{A}^{K}\sim\text{Uniform}[{\cal T}_{K}(P_{A})]$. Let
$B^{K}\sim V^{K}(\cdot\mid A^{K})$ and $\tilde{B}^{K}\sim V^{K}(\cdot\mid\tilde{A}^{K})$.
Then, 
\begin{equation}
0\leq I(A^{K};B^{K})-I(\tilde{A}^{K};\tilde{B}^{K})=O(\sqrt{K}\cdot\log K)=o(K).\label{eq: mutual information difference iid and fixed composition}
\end{equation}
The constant involved in the asymptotic order term only depends on
$|{\cal A}|,|{\cal B}|$. 
\end{lem}
\begin{IEEEproof}
The non-negativity of the mutual information difference in (\ref{eq: mutual information difference iid and fixed composition})
stems from the fact that memoryless input distributions maximize mutual
information for memoryless channels. We turn to prove the upper bound.
To this end, write
\begin{equation}
I(A^{K};B^{K})-I(\tilde{A}^{K};\tilde{B}^{K})=H(A^{K})-H(\tilde{A}^{K})+H(B^{K})-H(\tilde{B}^{K})-H(A^{K},B^{K})+H(\tilde{A}^{K},\tilde{B}^{K}).\label{eq: mutual information as three enropies}
\end{equation}
We upper bound each of the three entropy differences above. The first
entropy difference can be easily bounded 
\[
H(A^{K})-H(\tilde{A}^{K})=K\cdot H(P_{A})-\log|{\cal T}_{K}(P_{A})|\leq|{\cal A}|\cdot\log(K+1),
\]
since $\tilde{A}^{K}$ is distributed uniformly over ${\cal T}_{K}(P_{A})$
and using $|{\cal T}_{K}(P_{A})|\geq(K+1)^{-|{\cal A}|}\cdot e^{K\cdot H(P_{A})}$
(from \cite[Lemma 2.5]{csiszar2011information}). We next turn to
bound the \emph{third} entropy difference, to wit, $H(A^{K},B^{K})-H(\tilde{A}^{K},\tilde{B}^{K})$.
To bound this entropy difference we first bound the KL divergence
between the distributions $P_{\tilde{A}^{K}\tilde{B}^{K}}$ and $P_{A^{K}B^{K}}$
of $\tilde{A}^{K},\tilde{B}^{K}$ (respectively $A^{K},B^{K}$). Then,
we use Marton's transportation inequality \cite{marton1986simple,marton1996bounding}
to bound Ornstein's $\overline{d}$-distance between these distributions.
Finally, we use \cite[Prop. 8]{polyanskiy2016wasserstein} to bound
the entropy difference using the bound on the Ornstein's $\overline{d}$-distance.
We begin by bounding the KL divergence:
\begin{align}
D(P_{\tilde{A}^{K}\tilde{B}}\mid\mid P_{A^{K}B^{K}}) & =D(P_{\tilde{A}^{K}}\times V^{K}\mid\mid P_{A^{K}}\times V^{K})\\
 & =D(P_{\tilde{A}^{K}}\mid\mid P_{A^{K}})\\
 & =\sum_{a^{K}\in{\cal A}^{K}}P_{\tilde{A}^{K}}(a^{K})\log\frac{P_{\tilde{A}^{K}}(a^{K})}{P_{A^{K}}(a^{K})}\\
 & =\sum_{a^{K}\in{\cal T}_{K}(P_{A})}\frac{1}{|{\cal T}_{K}(P_{A})|}\log\frac{1/|{\cal T}_{K}(P_{A})|}{P_{A^{K}}(a^{K})}\\
 & \trre[=,a]\log\frac{1}{P_{A^{K}}({\cal T}_{K}(P_{A}))}\\
 & \trre[\leq,b]|{\cal A}|\cdot\log(K+1),\label{eq: bound on the divergence between fixed composition and iid inputs}
\end{align}
where $(a)$ follows since $P_{A^{K}}(a^{K})$ is identical to all
$a^{K}\in{\cal T}_{K}(P_{A})$, and $(b)$ follows since $P_{A^{K}}({\cal T}_{K}(P_{A}))\geq(K+1)^{-|{\cal A}|}$
\cite[Proof of Lemma 2.3]{csiszar2011information}. Now, let Ornstein's
$\overline{d}$-distance between distributions over $({\cal A}\times{\cal B})^{K}$
be\footnote{The $\overline{d}$ distance function denoted here is not related
to the constant $\overline{d}$ used in the proofs.}
\[
\overline{d}(P_{\tilde{A}^{K}\tilde{B}},P_{A^{K}B^{K}})\dfn\frac{1}{K}\inf\E\left[\rho_{H}(\tilde{A}^{K}\tilde{B}^{K},A^{K}B^{K})\right],
\]
where for $(a^{K},b^{K}),(\tilde{a}^{K},\tilde{b}^{K})\in({\cal A}\times{\cal B})^{K}$
\[
\rho_{H}\left((\tilde{a}^{K},\tilde{b}^{K}),(a^{K},b^{K})\right)=\sum_{k\in[K]}\I\left\{ (a_{k},b_{k})\neq(\tilde{a}_{k},\tilde{b}_{k})\right\} 
\]
is the Hamming distance, and the infimum is taken over all couplings
$P_{\tilde{A}\tilde{B}^{K}A^{K}B^{K}}$ of $P_{\tilde{A}\tilde{B}^{K}}$
and $P_{A^{K}B^{K}}$. Now, since $P_{A^{K}B^{K}}=\prod_{k\in[K]}P_{A_{k}}\times P_{B_{k}}$
is a memoryless distribution, Marton's transportation inequality implies
that 
\begin{equation}
\overline{d}(P_{\tilde{A}^{K}\tilde{B}},P_{A^{K}B^{K}})\leq\sqrt{\frac{D(P_{\tilde{A}^{K}\tilde{B}}\mid\mid P_{A^{K}B^{K}})}{2K}}\leq\sqrt{\frac{|{\cal A}|\cdot\log(K+1)}{2K},}\label{eq: bound on the d-bar distance}
\end{equation}
where the right inequality is from (\ref{eq: bound on the divergence between fixed composition and iid inputs}).
As was stated in \cite[Prop. 8]{polyanskiy2016wasserstein},\footnote{This is a generalization to $K>1$ of \cite{audenaert2006sharp,zhang2007estimating}.}
the $\overline{d}$-distance controls entropy difference, in the sense
that for any pair of distributions $P_{1},P_{2}\in{\cal P}({\cal A}^{K})$,
\begin{equation}
\left|H(P_{1})-H(P_{2})\right|\leq\overline{d}(P_{1},P_{2})\cdot K\cdot\log\left(|{\cal A}|-1\right)+h_{b}\left(\overline{d}(P_{1},P_{2})\right).\label{eq: entropy difference via d-bar}
\end{equation}
Thus, for all $K$ large enough so that $\overline{d}(P_{\tilde{A}^{K}\tilde{B}},P_{A^{K}B^{K}})<1/2$
it holds that 
\begin{align}
 & \left|H(A^{K},B^{K})-H(\tilde{A}^{K},\tilde{B}^{K})\right|\nonumber \\
 & \trre[\leq,a]K\cdot\left[\overline{d}(P_{\tilde{A}^{K}\tilde{B}},P_{A^{K}B^{K}})\cdot\left(\log|{\cal A}\times{\cal B}|-1\right)+h_{b}\left(\overline{d}(P_{\tilde{A}^{K}\tilde{B}},P_{A^{K}B^{K}})\right)\right]\\
 & \trre[\leq,b]K\cdot\left[\overline{d}(P_{\tilde{A}^{K}\tilde{B}},P_{A^{K}B^{K}})\cdot\left(\log|{\cal A}\times{\cal B}|-1\right)+2\overline{d}(P_{\tilde{A}^{K}\tilde{B}},P_{A^{K}B^{K}})\log\frac{1}{\overline{d}(P_{\tilde{A}^{K}\tilde{B}},P_{A^{K}B^{K}})}\right]\\
 & \trre[=,c]O(\sqrt{K}\cdot\log K)=o(K),
\end{align}
where $(a)$ follows from (\ref{eq: entropy difference via d-bar}),
$(b)$ follows from $h_{b}(t)\leq-2t\log t$ for all $t\leq1/2$,
and $(c)$ follows from (\ref{eq: bound on the d-bar distance}).
This completes the bound on the third entropy-difference term in (\ref{eq: mutual information as three enropies}).
The bound of the second entropy-difference term $H(B^{K})-H(\tilde{B}^{K})$
is similar to the third one since by the lumping property of the KL
divergence and (\ref{eq: bound on the divergence between fixed composition and iid inputs})
\[
D(P_{\tilde{B}}\mid\mid P_{B^{K}})\leq D(P_{\tilde{A}^{K}\tilde{B}}\mid\mid P_{A^{K}B^{K}})\leq|{\cal A}|\cdot\log(K+1).
\]
Combining the bounds on the three entropy-difference terms and inserting
to (\ref{eq: mutual information as three enropies}) completes the
proof. 
\end{IEEEproof}

\section{Computational Aspects of the Bounds \label{sec:Computational-Aspects-of}}

\paragraph*{Capacity lower bound}

In general, the infinite sum in (\ref{eq: lower bound on capacity})
can be truncated to a finite sum ending at $\overline{d}<\infty$,
so that if $\sum_{d\in\mathbb{N}^{+}\backslash[\overline{d}]}^{\infty}\pi_{\alpha}(d)\le\epsilon$
then the loss in the capacity bound is at most $\epsilon\log|{\cal X}|$.
Regarding the maximization over $P_{X}$, since $I(P_{X},W^{\oplus d})$
is a concave function of $P_{X}$, then so is the capacity lower bound.
It should be noted that while $\overline{d}$ should be chosen finite,
it does not have to be small. Naively, the output alphabet size of
$W^{\oplus d}$ is indeed $|{\cal Y}|^{d}$, however, since $W^{\oplus d}(y^{d}\mid x)$
only depends on $x$ and the type $\mathscr{P}(y^{d})$, it holds
that if the output letters $y^{d},\overline{y}^{d}\in{\cal Y}^{d}$
of $W^{\oplus d}$ satisfy $\mathscr{P}(y^{d})=\mathscr{P}(\overline{y}^{d})$,
then $W(y^{d}\mid x)=W(\overline{y}^{d}\mid x)$ is the same for all
$x\in{\cal X}$, and the two letters can be merged to a single output
letter, without changing the mutual information. For example, if $W$
is a BSC with crossover probability $w$ and $d=3$, then $W^{\oplus3}$
has the output letters $\{000,100,110,111\}$. In general, the number
of merged output letters is simply the number of output types $|{\cal P}_{d}({\cal Y})|={d+|{\cal Y}|-1 \choose |{\cal Y}|-1}\leq(d+1)^{|{\cal Y}|}$
(by \cite{csiszar2011information}, or by the stars and bars model,
see footnote \ref{fn:The-stars-and-bars-model}), which is only polynomial
in $d$. 

\paragraph*{Reliability function lower bound}

As for capacity, a computationally feasible bound can be obtained
by restricting the infinite sums in (\ref{eq: error exponent for universal decoder})
and (\ref{eq: minimization set for error exponent}) to (even different)
finite value $\overline{d}$, and replacing $\sum_{d\in\mathbb{N}}\theta_{d}=1$
with $\sum_{d\in[\overline{d}]}\theta_{d}\leq1$ (this can be deduced
from the Proof of Theorem \ref{thm: achievable reliability function}).
Assuming such a truncation, the minimization over $\{\theta_{d}\}_{d\in[\overline{d}]}$
is a convex optimization problem. To see this, let $\mu=1-\sum_{i\in[d]}\theta_{i}$
and consider the $d$th term in the sum, i.e., $\mu\cdot d_{b}(\frac{\theta_{d}}{\mu}\mid\mid\pi_{\alpha|\geq d}(d)).$
Since $\theta\to d_{b}(\theta\mid\mid\pi)$ is convex in $\theta$,
its perspective function $(\mu,\theta)\to\mu\cdot d_{b}(\frac{\theta}{\mu}\mid\mid\pi)$
is jointly convex in $(\mu,\theta)$ for $\frac{\theta}{\mu}\in[0,1]$
and $\mu>0$ \cite[Sec. 3.2.6]{boyd2004convex}. A composition of
a linear function $\mu=1-\sum_{i\in[d]}\theta_{i}$ with a convex
function results a convex function. Thus any of the $\overline{d}$
terms in the (truncated) sum of (\ref{eq: error exponent for universal decoder})
is convex, and so is their sum. The next step is maximization over
$P_{X}$, which can be simply performed by a grid search over the
$(|{\cal X}|-1)$-dimensional simplex ${\cal P}({\cal X})$. In principle,
it can be shown that maximization-minimization optimization problem
is concave-convex over $P_{X}\in{\cal P}({\cal X})$ and $\{\theta_{d}\}_{d\in\overline{d}}\in{\cal P}_{\text{sub}}([\overline{d}])$,
where ${\cal P}_{\text{sub}}([\overline{d}])\dfn\sum_{d\in[\overline{d}]}\theta_{d}\leq1$,
by using strong Lagrange duality \cite[Sec. 5.2.3]{boyd2004convex}
to cast (\ref{eq: error exponent for universal decoder}) as a Lagrange
optimization problem. However, since there are two constraints in
(\ref{eq: minimization set for error exponent}) (beyond the interval
constraints $\theta_{d}\in(0,1]$), the resulting optimization problem
requires searching over Lagrange multipliers $(\lambda_{0},\lambda_{1})\in(\mathbb{R}^{+})^{2}$,
and solving a concave-convex maximization-minimization optimization
problem for each (which can nonetheless be efficiently solved, e.g.
\cite[Sec. 5.2]{bubeck2014convex}). However, while this extra complication
compared to a simple grid search over ${\cal P}({\cal X})$ is feasible,
it does not seem to be computationally beneficial here, since for
the DNA channel $|{\cal X}|=4$ is typical, and thus the grid search
is merely three-dimensional.

\paragraph*{Capacity upper bound}

First, as for the lower bound on capacity, the infinite sum in (\ref{eq: capacity lower bound})
can be truncated to a finite value $\overline{d}$, so that if $\sum_{d\in\mathbb{N}^{+}\backslash\overline{d}}\pi_{\alpha}(d)\le\epsilon$
then the loss in capacity bound is at most $\epsilon\log|{\cal X}|$.
Assuming that this approximation has been made, the next computational
step is the maximization over $P_{X}$. While we cannot propose any
general computationally effective algorithm for this task, in a practical
DNA channel $|{\cal X}|=4$, and so, the maximization problem is only
three-dimensional, and it can be easily solved by standard global
optimization methods. It should be noted, however, that the computation
of the upper bound is more complex than the lower bound. Specifically,
for a given $d$, the computation of the upper bound requires computing
$\mathsf{CID}(P_{X},W^{\oplus d})$, which, in turn requires computing
the mutual information $I(P_{X},W^{\oplus(2d)})$ which has larger
output alphabet size then than $I(P_{X},W^{\oplus d})$. As said,
however, the effective output size of $W^{\oplus d}$ increase rather
slowly as $(d+1)^{|{\cal Y}|}$. 

\bibliographystyle{plain}
\bibliography{DNA_storage}

\begin{thebibliography}{10}

\bibitem{antkowiak2020low}
P.~L. Antkowiak, J.~Lietard, M.~Z. Darestani, M.~M. Somoza, W.~J. Stark,
  R.~Heckel, and R.~N. Grass.
\newblock Low cost {DNA} data storage using photolithographic synthesis and
  advanced information reconstruction and error correction.
\newblock {\em Nature communications}, 11(1):1--10, 2020.

\bibitem{apostol2012modular}
T.~M. Apostol.
\newblock {\em Modular Functions and Dirichlet Series in Number Theory}.
\newblock Graduate Texts in Mathematics. Springer New York, 2012.

\bibitem{audenaert2006sharp}
K.~M.~R. Audenaert.
\newblock A sharp {F}annes-type inequality for the von {N}eumann entropy.
\newblock {\em arXiv preprint quant-ph/0610146}, 2006.

\bibitem{bornholt2016dna}
J.~Bornholt, R.~Lopez, D.~M. Carmean, L.~Ceze, G.~Seelig, and K.~Strauss.
\newblock A {DNA}-based archival storage system.
\newblock In {\em Proceedings of the Twenty-First International Conference on
  Architectural Support for Programming Languages and Operating Systems}, pages
  637--649, 2016.

\bibitem{boyd2004convex}
S.~P. Boyd and L.~Vandenberghe.
\newblock {\em Convex Optimization}.
\newblock Cambridge University Press, 2004.

\bibitem{bubeck2014convex}
S.~Bubeck.
\newblock Convex optimization: {A}lgorithms and complexity.
\newblock {\em arXiv preprint arXiv:1405.4980}, 2014.

\bibitem{church2012next}
G.~M. Church, Y.~Gao, and S.~Kosuri.
\newblock Next-generation digital information storage in {DNA}.
\newblock {\em Science}, 337(6102):1628--1628, 2012.

\bibitem{cover2012elements}
T.~M. Cover and J.~A. Thomas.
\newblock {\em Elements of Information Theory}.
\newblock Wiley-Interscience, Hoboken, NJ, USA, 2006.

\bibitem{csiszar1980joint}
I.~Csisz{\'a}r.
\newblock Joint source-channel error exponent.
\newblock {\em Problems of Control and Information Theory}, 9(5):315--327,
  1980.

\bibitem{csiszar1998method}
I.~Csisz{\'a}r.
\newblock The method of types.
\newblock {\em IEEE Transactions on Information Theory}, 44(6):2505--2523,
  1998.

\bibitem{csiszar2011information}
I.~Csisz{\'a}r and J.~K{\"o}rner.
\newblock {\em Information Theory: {C}oding Theorems for Discrete Memoryless
  Systems}.
\newblock Cambridge University Press, Cambridge, U.K., 2011.

\bibitem{csiszar1977new}
I.~Csisz{\'a}r, J.~K\"{o}rner, and K.~Marton.
\newblock A new look at the error exponent of discrete memoryless channels.
\newblock In {\em IEEE International Symposium on Information Theory}, 1977.
\newblock unpublished.

\bibitem{durrett2019probability}
R.~Durrett.
\newblock {\em Probability: {T}heory and examples}, volume~49.
\newblock Cambridge University Press, 2019.

\bibitem{erlich2017dna}
Y.~Erlich and D.~Zielinski.
\newblock {DNA} fountain enables a robust and efficient storage architecture.
\newblock {\em Science}, 355(6328):950--954, 2017.

\bibitem{gabrys2015asymmetric}
R.~Gabrys, H.~M. Kiah, and O.~Milenkovic.
\newblock Asymmetric {L}ee distance codes: {N}ew bounds and constructions.
\newblock In {\em 2015 IEEE Information Theory Workshop}, pages 1--5. IEEE,
  2015.

\bibitem{gallager1968information}
R.~G. Gallager.
\newblock {\em Information Theory and Reliable Communication}.
\newblock John Wiley and Sons, 1968.

\bibitem{goldman2013towards}
N.~Goldman, S.~Bertone, P.and~Chen, C.~Dessimoz, E.~M. LeProust, B.~Sipos, and
  E.~Birney.
\newblock Towards practical, high-capacity, low-maintenance information storage
  in synthesized {DNA}.
\newblock {\em Nature}, 494(7435):77--80, 2013.

\bibitem{goppa1975nonprobabilistic}
V.~D. Goppa.
\newblock Nonprobabilistic mutual information without memory.
\newblock {\em Problems of Control and Information Theory}, 4(2):97--102, 1975.

\bibitem{grass2015robust}
R.~N. Grass, R.~Heckel, M.~Puddu, Daniela Paunescu, and Wendelin~J. S.
\newblock Robust chemical preservation of digital information on {DNA} in
  silica with error-correcting codes.
\newblock {\em Angewandte Chemie International Edition}, 54(8):2552--2555,
  2015.

\bibitem{heckel2019characterization}
R.~Heckel, G.~Mikutis, and R.~N. Grass.
\newblock A characterization of the {DNA} data storage channel.
\newblock {\em Scientific reports}, 9(1):1--12, 2019.

\bibitem{kiah2016codes}
H.~M. Kiah, G.~J. Puleo, and O.~Milenkovic.
\newblock Codes for {DNA} sequence profiles.
\newblock {\em IEEE Transactions on Information Theory}, 62(6):3125--3146,
  2016.

\bibitem{kovavcevic2018codes}
M.~Kova{\v{c}}evi{\'c} and V.~Y.~F. Tan.
\newblock Codes in the space of multisets -- coding for permutation channels
  with impairments.
\newblock {\em IEEE Transactions on Information Theory}, 64(7):5156--5169,
  2018.

\bibitem{land2006information}
Ingmar Land and Johannes Huber.
\newblock {\em Information combining}.
\newblock Now Publishers Inc, 2006.

\bibitem{land2005bounds}
Ingmar Land, Simon Huettinger, Peter~A Hoeher, and Johannes~B Huber.
\newblock Bounds on information combining.
\newblock {\em IEEE Transactions on Information Theory}, 51(2):612--619, 2005.

\bibitem{lenz2019anchor}
A.~Lenz, P.~H. Siegel, A.~Wachter-Zeh, and E.~Yaakobi.
\newblock Anchor-based correction of substitutions in indexed sets.
\newblock In {\em IEEE International Symposium on Information Theory}, pages
  757--761. IEEE, 2019.

\bibitem{lenz2019coding}
A.~Lenz, P.~H. Siegel, A.~Wachter-Zeh, and E.~Yaakobi.
\newblock Coding over sets for {DNA} storage.
\newblock {\em IEEE Transactions on Information Theory}, 66(4):2331--2351,
  2019.

\bibitem{lenz2019upper}
A.~Lenz, P.~H. Siegel, A.~Wachter-Zeh, and E.~Yaakobi.
\newblock An upper bound on the capacity of the {DNA} storage channel.
\newblock In {\em IEEE Information Theory Workshop}, pages 1--5. IEEE, 2019.

\bibitem{lenz2020achieving}
A.~Lenz, P.~H. Siegel, A.~Wachter-Zeh, and E.~Yaakohi.
\newblock Achieving the capacity of the {DNA} storage channel.
\newblock In {\em IEEE International Conference on Acoustics, Speech and Signal
  Processing}, pages 8846--8850. IEEE, 2020.

\bibitem{lenz2020achievable}
A.~Lenz, L.~Welter, and S.~Puchinger.
\newblock Achievable rates of concatenated codes in {DNA} storage under
  substitution errors.
\newblock In {\em International Symposium on Information Theory and Its
  Applications}, pages 269--273. IEEE, 2020.

\bibitem{marshall1979inequalities}
A.~W. Marshall, I.~Olkin, and B.~C. Arnold.
\newblock {\em Inequalities: {T}heory of majorization and its applications},
  volume 143.
\newblock Springer, 1979.

\bibitem{marton1986simple}
K.~Marton.
\newblock A simple proof of the blowing-up lemma.
\newblock {\em IEEE Transactions on Information Theory}, 32(3):445--446, 1986.

\bibitem{marton1996bounding}
K.~Marton.
\newblock Bounding $\bar{d}$-distance by informational divergence: {A} method
  to prove measure concentration.
\newblock {\em Annals of probability}, 24(2):857--866, 1996.

\bibitem{merhav2010statistical}
N.~Merhav.
\newblock Statistical physics and information theory.
\newblock {\em Foundations and Trends in Communications and Information
  Theory}, 6(1-2):1--212, 2009.

\bibitem{mitzenmacher2006theory}
M.~Mitzenmacher.
\newblock On the theory and practice of data recovery with multiple versions.
\newblock In {\em IEEE International Symposium on Information Theory}, pages
  982--986. IEEE, 2006.

\bibitem{mitzenmacher2017probability}
M.~Mitzenmacher and E.~Upfal.
\newblock {\em Probability and computing: Randomization and probabilistic
  techniques in algorithms and data analysis}.
\newblock Cambridge University Press, 2017.

\bibitem{neiman1964some}
M.~S. Neiman.
\newblock Some fundamental issues of microminiaturization.
\newblock {\em Radiotekhnika}, 1(1):3--12, 1964.

\bibitem{organick2018random}
L.~Organick, S.~D. Ang, Y.~Chen, R.~Lopez, S.~Yekhanin, K.~Makarychev, M.~Z.
  Racz, G.~Kamath, P.~Gopalan, and B.~Nguyen.
\newblock Random access in large-scale {DNA} data storage.
\newblock {\em Nature biotechnology}, 36(3):242--248, 2018.

\bibitem{polyanskiy2016wasserstein}
Y.~Polyanskiy and Y.~Wu.
\newblock Wasserstein continuity of entropy and outer bounds for interference
  channels.
\newblock {\em IEEE Transactions on Information Theory}, 62(7):3992--4002,
  2016.

\bibitem{raginsky2018concentration}
M.~Raginsky and I.~Sason.
\newblock {\em Concentration of Measure Inequalities in Information Theory,
  Communications, and Coding}, volume~10.
\newblock Now Foundations and Trends, 2013.

\bibitem{sala2017exact}
F.~Sala, R.~Gabrys, C.~Schoeny, and L.~Dolecek.
\newblock Exact reconstruction from insertions in synchronization codes.
\newblock {\em IEEE Transactions on Information Theory}, 63(4):2428--2445,
  2017.

\bibitem{sayir2016codes}
J.~Sayir.
\newblock Codes for efficient data storage on {DNA} molecules.
\newblock In {\em Talk at Inform., Inference, and Energy symposium, Cambridge,
  U.K.}, 2016.

\bibitem{shomorony2021dna}
I.~Shomorony and R.~Heckel.
\newblock {DNA}-based storage: {M}odels and fundamental limits.
\newblock {\em IEEE Transactions on Information Theory}, 67(6):3675--3689,
  2021.

\bibitem{sima2021coding}
J.~Sima, N.~Raviv, and J.~Bruck.
\newblock On coding over sliced information.
\newblock {\em IEEE Transactions on Information Theory}, 67(5):2793--2807,
  2021.

\bibitem{song2020sequence}
W.~Song, K.~Cai, and K.~A.~S. Immink.
\newblock Sequence-subset distance and coding for error control in {DNA}-based
  data storage.
\newblock {\em IEEE Transactions on Information Theory}, 66(10):6048--6065,
  2020.

\bibitem{sutskover2005extremes}
Ilan Sutskover, Shlomo Shamai, and Jacob Ziv.
\newblock Extremes of information combining.
\newblock {\em IEEE Transactions on Information Theory}, 51(4):1313--1325,
  2005.

\bibitem{tang2021error}
Y.~Tang and F.~Farnoud.
\newblock Error-correcting codes for noisy duplication channels.
\newblock {\em IEEE Transactions on Information Theory}, 67(6):3452--3463,
  2021.

\bibitem{tse2005fundamentals}
D.~Tse and P.~Viswanath.
\newblock {\em Fundamentals of wireless communication}.
\newblock Cambridge University Press, 2005.

\bibitem{yazdi2015rewritable}
S.~M. H.~T. Yazdi, Y.~Yuan, J.~Ma, H.~Zhao, and O.~Milenkovic.
\newblock A rewritable, random-access {DNA}-based storage system.
\newblock {\em Scientific reports}, 5(1):1--10, 2015.

\bibitem{zhang2007estimating}
Z.~Zhang.
\newblock Estimating mutual information via {K}olmogorov distance.
\newblock {\em IEEE Transactions on Information Theory}, 53(9):3280--3282,
  2007.

\end{thebibliography}

\end{document}